\documentclass[aps,rmp,reprint,amsmath,amssymb,graphicx,longbibliography,tightenlines]{revtex4-1}
\def\figwidth {0.9\columnwidth} 
 
\usepackage{xcolor}
\definecolor{darkblue}{rgb}{0.1,0.1,.7}
\usepackage[colorlinks, linkcolor=darkblue, citecolor=darkblue, urlcolor=darkblue, linktocpage]{hyperref} 
\usepackage{graphicx}
\usepackage{subfigure}
\usepackage{ytableau}
\usepackage[none]{hyphenat}
\usepackage{cancel}

\usepackage{slashed}
\usepackage{dsfont}
\def\unit{\mathds{1}} 

\def\red{\color [rgb]{0.9,0.1,0.1}}

\newcommand{\reef}[1]{(\ref{#1})}

\def\eps{\epsilon}
\newcommand{\beq}{\begin{equation}} 
\newcommand{\eeq}{\end{equation}}
\def\del {\partial} 
\def\nn{\nonumber} 
\def\bi{\mathbf{i}}
\def\bj{\mathbf{j}}

\def\bn{\mathbf{n}}
\def\Geg{{\rm Geg}}
\def\bZ {\mathbb{Z}} 
\def\bR {\mathbb{R}}

\def\calO {{\cal O}}
\def\cO{{\cal O}}
 
\def\calD {{\cal D}} 
\def\calF {{\cal F}} 
\def\calN {{\cal N}}

\def\calP {{\cal P}}

\def\calJ {{\cal J}}

\def\calT {{\cal T}} 
\def\half{{\textstyle\frac 12}}

\def\ge{\geqslant}
\def\le{\leqslant}
\def\geq{\geqslant}
\def\leq{\leqslant}
\newcommand{\diffop}[2]{\ifthenelse{\equal{#2}{1}}{\frac{\mrm{d}}{\mrm{d} #1}}{\frac{\mrm{d}^#2}{\mrm{d} #1^#2}}}

\def\1{{\ds 1}}

\def\Z{\hbox{$\bb Z$}}
\def\N{\hbox{$\bb N$}}

\def\bb{
\font\tenmsb=msbm10
\font\sevenmsb=msbm7
\font\fivemsb=msbm5
\textfont1=\tenmsb
\scriptfont1=\sevenmsb
\scriptscriptfont1=\fivemsb
}

\def\<{\langle}
\def\>{\rangle}
\def\cO{{\cal O}}

\def\De{\Delta}

\def\e{\epsilon}
\def\s{\sigma}

\usepackage{marginnote}
\usepackage[normalem]{ulem}

\begin{document}
\count\footins = 1000

\title{The Conformal Bootstrap: Theory, Numerical Techniques, and Applications}

\author{David Poland}
\affiliation{\mbox{Department of Physics, Yale University, 217 Prospect St, New Haven, CT 06511}}
\author{Slava Rychkov} 
\affiliation{\mbox{Institut des Hautes \'Etudes Scientifiques, Bures-sur-Yvette, France}
\\
\& \\
\mbox{Laboratoire de physique th\'eorique,\\ D\'epartement de physique de l'ENS},
\mbox{\'Ecole normale sup\'erieure, PSL University,
Sorbonne Universit\'es, UPMC Univ.~Paris 06}, \mbox{CNRS, 75005 Paris, France}
}
\author{Alessandro Vichi}
\affiliation{\mbox{Institute of Physics, \'Ecole Polytechnique F\'ed\'erale de Lausanne, Switzerland}}

\date{\today{}}

\begin{abstract}
Conformal field theories have been long known to describe the fascinating universal physics of scale invariant critical points. They describe continuous phase transitions in fluids, magnets, and numerous other materials, while at the same time sit at the heart of our modern understanding of quantum field theory. For decades it has been a dream to study these intricate strongly coupled theories nonperturbatively using symmetries and other consistency conditions. This idea, called the conformal bootstrap, saw some successes in two dimensions but it is only in the last ten years that it has been fully realized in three, four, and other dimensions of interest. This renaissance has been possible both due to significant analytical progress in understanding how to set up the bootstrap equations and the development of numerical techniques for finding or constraining their solutions. These developments have led to a number of groundbreaking results, including world record determinations of critical exponents and correlation function coefficients in the Ising and $O(N)$ models in three dimensions. This article will review these exciting developments for newcomers to the bootstrap, giving an introduction to conformal field theories and the theory of conformal blocks, describing numerical techniques for the bootstrap based on convex optimization, and summarizing in detail their applications to fixed points in three and four dimensions with no or minimal supersymmetry.  
\end{abstract}


\maketitle

\tableofcontents{}

\section{Introduction}
\label{sec:intro}

For most physical systems, the first step to qualitative understanding is to identify their characteristic scales (length, energy, etc.), through which everything else can be expressed via approximate dimensional analysis. However, there exist theories for which this familiar approach does not work, and which are therefore harder to understand intuitively. These are scale invariant theories, which by definition look the same at all distances and energies, and hence do not possess any characteristic scale{s}.

Scale invariant theories are important in physics, because they arise naturally in the theory of critical phenomena. One experimental manifestation of scale invariance in critical phenomena is critical opalescence, first observed near the critical point of CO$_{2}$ by \textcite{Andrews:CriticalPoint}, and interpreted as a sign of density fluctuations occurring over many distance scales by \textcite{Smoluchowski:CO}. The exact solution of the two-dimensional (2d) Ising model by \textcite{Onsager:Ising} also made it possible to see the emergence of scale invariance at the ferromagnet-paramagnet critical point. Nowadays it is understood that all critical points are described by scale invariant theories. This has been incorporated into Wilson's renormalization group (RG) theory of phase transitions, introduced in \textcite{Wilson:1973jj} and \textcite{Wilson:1983dy}, according to which continuous phase transitions are described by the fixed points of RG flows, and are therefore scale invariant. 

Formally, scale invariance is expressed as invariance under a rescaling (dilatation) of all coordinates by a uniform factor $x\to \lambda x$. Another interesting class of transformations of space are \emph{conformal transformations}, defined as transformations preserving angles. Thus conformal transformations are required to look locally at each point as a rotation accompanied by a dilatation, although the rescaling factor can be $x$-dependent. Conformal transformations have been studied by mathematicians since the 19th century \cite{Monge}. They first entered into physics when \textcite{Bateman} and \textcite{Cunningham} showed that Maxwell's equations are conformally invariant (they are also trivially scale invariant because of the masslessness of the photon).\footnote{See \textcite{Kastrup:2008jn} for the early history of conformal transformations.}

With conformal invariance being thus a natural extension of scale invariance, one may wonder if scale invariant theories describing critical points in fact possess full conformal invariance. That this should be the case was first conjectured by \textcite{Polyakov:1970xd}. Since then several theoretical arguments have been given for why scale invariance should generically imply conformal invariance.\footnote{See \textcite{Polchinski:1987dy} and references therein, 
as well as the recent review {by} \textcite{Nakayama:2013is}. The question is subtle because in fact rare examples of scale invariant and not conformally invariant theories do exist \cite{Riva:2005gd}, although it is fully understood how they evade the general expectation.} By now it is understood that most physically relevant scale invariant theories are conformally invariant, and hence referred to as `Conformal Field Theories', or CFTs.

In addition to their role in the theory of critical phenomena, CFTs are also extremely important for the study of quantum field theories (QFTs). For this discussion QFTs may be Euclidean $d$-dimensional field theories, relevant for statistical physics, or field theories in {Lorentzian} signature, which are relevant for high-energy physics and quantum condensed matter.\footnote{\label{footnote:dimension}In this review the space dimension $d$ will denote the total number of coordinate dimensions, including time if one works in {Lorentzian} signature.} From the modern perspective, {large classes of QFTs can be seen as RG flows which emerge from one CFT (called the UV fixed point) at short distances and flow to {either} another nontrivial CFT (called the IR fixed point) or a massive phase at long distances.}\footnote{Some QFTs cannot be viewed as coming from a CFT in the UV, for example RG flows involving 3d gauge fields.} In this sense CFTs can be called signposts in the space of general QFTs. The quest to classify and solve CFTs is a major goal of modern theoretical physics. 

The study of CFTs was initiated in the late 1960s, focusing mostly on formal properties of these theories.\footnote{We will not attempt here a full historical account. Early pioneering contributions included \textcite{Mack:1969rr}, \textcite{Polyakov:1970xd}, \textcite{Ferrara:1971vh}, \textcite{Migdal:1972tk}, \textcite{Parisi:1972zm}, \textcite{Ferrara:1973vz}, \textcite{Ferrara:1973yt}, \textcite{Polyakov:1974gs}, \textcite{Ferrara:1974nf}, \textcite{Ferrara:1974ny}, \textcite{Ferrara:1974pt}, \textcite{Mack:1975jr}, and \textcite{Dobrev:1977qv}.} This early work was done in general dimension $d$, where the group of conformal transformations is finite dimensional, while it is infinite-dimensional in $d=2$ where any holomorphic map gives rise to a conformal transformation. The importance of this special case was realized by \textcite{Belavin:1984vu}. Using the infinite-dimensional conformal symmetry, they solved the 2d minimal models---an infinite sequence of CFTs describing the critical points of the 2d Ising model and other lattice models such as the 3-state Potts model. 

Of course many 2d models can be exactly solved directly on the lattice \cite{Baxter:1982zz}, starting with the above-mentioned Onsager solution of the 2d Ising model. The approach of \textcite{Belavin:1984vu} was different in that it allowed for a solution of critical theories using the constraints of conformal symmetry alone, with minimal or no microscopic input. The crucial idea to find these solutions was the conformal bootstrap, first described by \textcite{Ferrara:1973yt} and \textcite{Polyakov:1974gs}. The conformal bootstrap combines {conformal invariance} with the existence of the operator product expansion (OPE), another powerful concept going back to \textcite{Wilson:1969zs} and \textcite{Kadanoff:1969zz}. This leads to mathematical consistency conditions on the CFT parameters, which were enough to solve the 2d minimal models. 

Following these developments, CFT has become an indispensable tool in the theory of 2d critical phenomena.\footnote{\label{note:2dLit}Many excellent 2d CFT reviews include \textcite{Cardy-LH}, \textcite{Ginsparg}, \textcite{DiFrancesco:1997nk}, and \textcite{Henkel}.} On the other hand, applications of the conformal bootstrap in higher dimensions lagged behind. For example, 3d continuous phase transitions are traditionally studied using the {RG} by starting from a microscopic action and looking for a fixed point. Often the 3d fixed point of interest is strongly coupled, requiring one to deform the theory artificially in order to do perturbation theory in a small parameter, with the {large-$N$} expansion (see e.g.~\textcite{Moshe:2003xn} for a review) and the $\epsilon$-expansion \cite{Wilson:1971dc} being two prime examples. While these theoretical approaches have undoubtedly scored some successes in describing the experimental data, one may wonder what a fully nonperturbative approach such as the conformal bootstrap has to say about this problem.

A period of renewed interest in the conformal bootstrap started following \textcite{Rattazzi:2008pe}. This work proposed a numerical method, based on linear programming, which made it possible to extract concrete predictions from the conformal bootstrap equations. The method was applicable in higher dimensions (as well as in 2d). {Also, an advantage of the method was that rigorous} predictions could be extracted without having to fully solve the equations.\footnote{Such full solutions in $d>2$ are still beyond reach except in very special cases, see Sec.~\ref{sec:explicit}.} Since then the method was greatly improved and many interesting results were obtained, mostly for conformal field theories in 3d and 4d, but also in other dimensions. One flagship result of this line of research is the world's most precise determination of the critical exponents of the critical 3d Ising model (see \textcite{Kos:2016ysd} for the current world record results). Our purpose is to review these developments, focusing on applications to the most interesting 3d and 4d CFTs considered so far, as well as to give an overview of the theoretical and numerical techniques which proved useful for these applications.

\subsection{Outline}

Due to the overwhelming number of results in various incarnations of the conformal bootstrap, our review will necessarily be limited in scope. Let us briefly outline the topics that we will cover. We begin in Sec.~\ref{sec:informal} with an informal overview of the conformal bootstrap. Sec.~\ref{sec:cft} provides a concise introduction to the conformal field theory techniques that are needed to set up the bootstrap in $d$ dimensions. We follow in Sec.~\ref{sec:numerical} with an overview of the various numerical methods that have been employed in studies of the bootstrap. Secs.~\ref{sec:appl} and~\ref{sec:appl4d} review results obtained from applying these methods to 3d and 4d CFTs. Sec.~\ref{sec:4Dsusy} reviews results obtained with the stronger assumptions of 4d $\mathcal{N}=1$ or 3d $\mathcal{N}=2$ superconformal symmetry. We comment on applications to nonunitary models in Sec.~\ref{sec:nonunitary}. Notably absent from our main review are CFTs in other dimensions (e.g., $d=2$ or $d>4$), CFTs with extended supersymmetry, analytical progress in the bootstrap, logarithmic and nonrelativstic CFTs, and other related topics. We finish with a brief overview of progress in these related lines of research in Sec.~\ref{sec:other} and give some concluding words in Sec.~\ref{sec:outlook}.

\section{Conformal bootstrap: informal overview}
\label{sec:informal}

In this section we will give a brief outline of the conformal bootstrap approach to critical phenomena in $d$ dimensions.
We will be rather informal in this section, while in the subsequent sections the same material will be treated in more depth and at a higher level of rigor. For another short introduction to these matters, see \textcite{Poland:2016chs}. For longer pedagogical introductions see \textcite{Rychkov:2016iqz,Simmons-Duffin:2016gjk}.

As a simple physical setup where these methods would be applicable, we can consider a statistical physics system in $d$ spatial dimensions which is (a) in thermodynamic equilibrium and (b) at a temperature corresponding to a continuous phase transition (so that the correlation length is infinite). Suppose that we are interested in equal-time correlation functions of some local quantities characterizing this system:
\beq
\langle \calO_1(x_1)\ldots \calO_n(x_n) \rangle\,,
\label{eq:corr}
\eeq
where $x_i$ are positions in $\mathbb{R}^d$. For example, one can think of the 3d Ising model at the critical point, with $\calO_i(x)$ the local magnetization, local energy density, etc. In general, the $\calO_i(x)$ are called local operators. 

We are interested in the behavior of the correlators~\reef{eq:corr} at distances large compared to any microscopic (such as lattice) scale $a$. According to Wilson's RG theory, continuous phase transitions are fixed points of RG flows, which means that the long-distance behavior of~\reef{eq:corr} will have scale invariance (as well as rotation and translation invariance). Using scale invariance, we can formally extend the long-distance behavior of these correlators from distances $|x_i-x_j|\gg a$ to arbitrary short distances. In what follows we work in the so-defined \emph{continuous limit} theory, which is exactly scale invariant at all distances from $0$ to $\infty$.\footnote{However, in this paper we will not consider behavior of correlators at coincident points.} 

As discussed in the introduction, we expect that the critical theory is also conformally invariant (i.e., a CFT). This means that for any \emph{conformal transformation} of $d$-dimensional space $x\to x'$ (see Sec.~\ref{sec:conformaltransformations} for the definition), Eq.~\reef{eq:corr} is related to the same correlation function evaluated at points $x_1',\ldots,x_n'$. This invariance property (or covariance) of correlation functions is expressed as a transformation rule for local operators, in the next section appearing in Eq.~\reef{eq:fieldrepresentation}. For scalar operators, we have
\beq
\calO(x')=\Omega(x)^{-\Delta_\calO} \calO(x)\,,
\label{eq:invsc}
\eeq
 where $\Omega(x)=|\partial x'/\partial x|^{1/d}$ is the $x$-dependent scale factor of the conformal transformation, and $\Delta_\calO$ is a fixed parameter characterizing the operator $\calO$, called its \emph{scaling dimension}.\footnote{To be precise, such transformation rules hold for \emph{primary} local operators, a subtlety which will not play a role in this informal discussion.}

\textcite{Polyakov:1970xd} noticed that invariance under \reef{eq:invsc} strongly restricts two-point (2pt) and three-point (3pt) correlation functions. The 2pt function is nonzero only for identical {operators} 
and can be normalized to one:
\beq
\langle \calO_i(x_1)\calO_j(x_2)\rangle =\delta_{ij}{|x_{1}-{x_2}|^{-2\Delta_i}}\,,
\label{eq:2pt}
\eeq
while the 3pt function is fixed up to a numerical coefficient:
\beq
\langle \calO_1(x_1)\calO_2(x_2)\calO_3(x_3)\rangle =\frac{\lambda_{123}}{|x_{12}|^{h_{123}} |x_{13}|^{h_{132}} |x_{23}|^{h_{231}}}\,,
\label{eq:3pt}
\eeq
where $x_{ij} \equiv x_i-x_j$ and $h_{ijk} \equiv \Delta_i+\Delta_j-\Delta_k$. Similar equations hold for operators with indices, see Sec.~\ref{sec:correlations}. 

The set of numerical parameters $\Delta_i$ and $\lambda_{ijk}$ appearing in \reef{eq:2pt} and \reef{eq:3pt} is called the \emph{CFT data}. It turns out that the CFT data determine not only 2pt and 3pt functions, but are also sufficient to compute all local observables in CFTs in flat space, by which we mean all correlation functions of local operators, including four-point (4pt) and higher-order correlation functions.\footnote{It should be mentioned that CFTs also possess nonlocal observables in addition to the local ones, which are not necessarily determined by the OPE data. For example, one can probe a CFT by extended operators, such as boundaries or defects, or put it in a space of nontrivial geometry or topology. In this review we will focus on the local observables, although the bootstrap philosophy can also be useful in the study of some nonlocal observables; see~Sec.~\ref{sec:bdry} for boundaries and defects and Sec.~\ref{sec:other} for the modular bootstrap.} 
 
To see this, one uses the OPE, which says that we can replace the insertion of two nearby local operators inside a correlation function by a series of single local operators:
\beq
\calO_i(x_1)\calO_j(x_2)=\sum_k f_{ijk} \calO_k(y)\,.
\label{eq:OPE}
\eeq
The coefficients of the series $f_{ijk}$ may and will depend on the relative positions of the operators $\calO_i$, $\calO_j$, $\calO_k$, and on their quantum numbers. Crucially, however, these coefficients are not supposed to depend on which other operators appear in the correlation function, as long as they are sufficiently far away from $x_1,x_2,y$. The precise criterion in the CFT context will be given in Eq.~\reef{eq:converge}.

Notice the freedom in where we put operators appearing on the r.h.s.~of the OPE: we can choose $y=\half(x_1+x_2)$, $y=x_1$, or any other point nearby. Different choices of $y$ can be related by Taylor-expanding $\calO_k$, and thus can be compensated by changing {the} coefficients of derivatives of $\calO_k$ in the OPE. In what follows we will group the operator $\calO_k$ together with all its derivatives, formally thinking of $f_{ijk}$ as an infinite power series in $\del_y$. 

There are two things that make OPE in conformal field theories more powerful than in a generic QFT. Firstly, compatibility of the OPE with conformal invariance determines the functions $f_{ijk}$ up to a numerical prefactor, coinciding with the 3pt function coefficient $\lambda_{ijk}$ (for this reason it is also called an OPE coefficient):
\beq
f_{ijk}(x_1,x_2,y,\del_y) =  \lambda_{ijk} \hat{f}_{ijk}(x_1,x_2,y,\del_y)\,.
\eeq
The reduced functions $\hat f_{ijk}$ depend only on {the} operator dimensions $\Delta_i,\Delta_j,\Delta_k$, the spins of these operators (which are kept implicit in this informal discussion), and on the space dimension $d$.\footnote{Note that here we are assuming the normalization in Eq.~(\ref{eq:2pt}).}

Secondly, the OPE in conformal theories has a finite radius of convergence, which is determined by the distance to the next operator insertions. For example, in the correlator of Eq.~\reef{eq:npt} given below, the OPE will converge if 
\beq
|x_1-y|,\,|x_2-y|<\min_{i=3\ldots n} |x_i-y|\,,
\label{eq:converge}
\eeq
i.e.~if there exists a sphere centered at $y$ and separating $x_1$, $x_2$ from any other operator insertion.

Because of these two reasons we can compute any correlation function recursively using the OPE, provided that we know the CFT data. For example, suppose we want to compute the $n$-point function
\beq
\langle \calO_1(x_1) \calO_2(x_2) \calO_3(x_3)\ldots \calO_n(x_n)\rangle\,.
\label{eq:npt}
\eeq
Applying the OPE to $\calO_1(x_1) \calO_2(x_2)$, we reduce this correlator to a sum of correlators containing $n-1$ operators
\beq
\langle \calO_k(y) \calO_3(x_3)\ldots \calO_n(x_n)\rangle\,.
\eeq
Proceeding in this way, we will eventually get down to 2pt functions, which are determined by the CFT data. The only parameters which will enter this computation are the operator positions and quantum numbers, the CFT data, and the space dimension $d$.\footnote{Notice that although the presented scheme solves the problem of computing $n$-point functions in principle, it is not trivial to do in practice. For 4pt functions, the necessary techniques will be presented in Sec.~\ref{sec:cb}.}
 
\begin{figure}[t!]
    \includegraphics[width=0.49\textwidth]{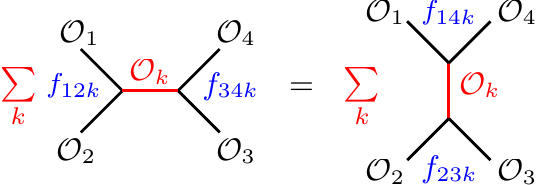}
    \caption{\label{fig:bootstrap}
   Crossing relation for the 4pt function $\<\calO_1 \calO_2 \calO_3 \calO_4\>$.}
  \end{figure}

Consider now the case of a 4pt function (Eq.~\reef{eq:npt} with $n=4$) 
 and compute it in two different ways. The first way is to apply the OPE to the pairs of operators $\calO_1 \calO_2$ and $\calO_3\calO_4$. This reduces the 4pt function to an infinite sum of 2pt functions of operators which appear in these OPEs. A second way is to apply the OPE to the pairs $\calO_1\calO_4$ and $\calO_2\calO_3$. Since we are dealing with the same 4pt function, the two expansions must agree in their overlapping regions of convergence. This \emph{crossing relation} represents a consistency condition on the CFT data and is illustrated in Fig.~\ref{fig:bootstrap}.

The main idea of the conformal bootstrap is that by imposing the crossing relation, we should be able to significantly winnow down the set of all possible CFT data. In the subsequent sections of this review, we will see how the crossing relation can be written in a mathematically manageable form, and how numerical algorithms can be applied to extract from it concrete constraints. 

Ideally, if we impose crossing relations for \emph{all} 4pt functions of the theory, we will be left with the CFT data corresponding to the actually existing critical theories. In practice, it has so far been possible to impose crossing relations on only a handful of 4pt functions at a time. However, we will see that even this limited procedure produces nontrivial constraints, which are in some cases surprisingly strong.

 \subsection{Universality and the role of microscopic input}
\label{sec:universality}
A fundamental concept in the theory of critical phenomena is universality: all continuous phase transitions can be grouped into universality classes which share the same critical exponents. This is neatly explained in Wilson's RG theory: two phase transitions will fall into the same universality class if they are described by the same fixed point. On the other hand, the conformal bootstrap provides a different perspective on the same phenomenon: each universality class corresponds to a different CFT, with a different set of CFT data. 

These two points of view are clearly complementary, and it is important to establish the correspondence between them. Consider for example the critical exponents. In RG theory they can be related to the eigenvalues $\lambda^{y_i}$ of the RG transformation linearized around the fixed point, where $\lambda>1$ is the RG rescaling factor. As is well known, these eigenvalues are simply related to the scaling dimensions of the local operators: $y_i=d-\Delta_i$. Thus, information about the critical exponents can be easily extracted from CFT data, and agreement of their values between an RG fixed point and a CFT may give us confidence that the two describe the same critical universality class.

There are however three more fundamental structural characteristics which can be used to identify universality classes, even before considering the numerical values of critical exponents. These characteristics may not be sufficient to uniquely classify the different CFTs, but they will give us a convenient starting point.

\emph{1. The global (or internal) symmetry group.} It can be discrete, as for the $\bZ_2$ symmetry of the Ising model, or continuous, as for the $O(N)$ models. In RG studies, the global symmetry group is specified by considering an RG flow in the space of microscopic theories described by an action possessing a given symmetry. The global symmetry group for a CFT is the same group $G$ as for the corresponding RG fixed point, although it is specified in a different way: by demanding that each local operator transform in an irreducible representation of $G$ and that OPE coefficients respect this symmetry structure.

We note in passing that unlike the global symmetry, the presence of a \emph{gauge symmetry} in a microscopic description does not manifest itself in the conformal bootstrap, because physically observable local CFT operators are gauge invariant.\footnote{Gauge symmetries can make themselves known more indirectly, through anomaly coefficients which show up in the correlation functions of local operators or the existence of higher-form symmetries.}

\emph{2. The number of relevant singlet scalars.} The number of scalar operators which are relevant (i.e., have dimension $\Delta_i<d$) and transform as singlets under the global symmetry determines whether the universality class has critical as opposed to multicritical behavior. This will be discussed in more detail in Sec.~\ref{sec:multicrit}. Here it suffices to note that this number is easy to identify from both the RG and CFT perspectives.

\emph{3. Unitarity.} Unitarity is of course a required property when quantum mechanics is involved, which is the case for theories of interest to high-energy physics and quantum condensed matter. Many universality classes of interest to statistical physics also happen to be unitary.\footnote{In statistical physics, the role of unitarity is played by its Euclidean counterpart called reflection positivity.} Importantly, the existence of unitarity can be established at the microscopic scale, and is then inherited by the RG fixed point. In the CFT description, unitarity is imposed via lower bounds on the operator dimensions and reality constraints on the OPE coefficients, see Sec.~\ref{sec:unitarity}.
 
Finally, let us comment on the OPE coefficients $\lambda_{ijk}$. From the CFT point of view, they are an integral part of the CFT data, on par with the scaling dimensions. In the conformal bootstrap approach, the crossing relation involves both $\lambda_{ijk}$ and $\Delta_i$. In the examples below, when we are able to determine the $\Delta_i$'s to some accuracy (as for the 3d Ising and the $O(N)$ models), we can typically determine the $\lambda_{ijk}$'s to a comparable accuracy. This can be contrasted with the RG approach, where the OPE coefficients do not appear to play such a fundamental role, and they have received relatively little attention.

\section{Conformal field theory techniques in $d$ dimensions}
\label{sec:cft}
In this section we review the theory techniques that form the backbone of the conformal bootstrap. These include conformal symmetry, operators and their correlation functions, unitarity and reflection positivity, conformal blocks, and the way they enter crossing relations (also in the presence of global symmetries).

\subsection{Conformal transformations}
\label{sec:conformaltransformations}

The content of this section is standard textbook material. We will only mention a few fundamental results and set up {our} conventions. For more details see e.g.~\textcite{Rychkov:2016iqz,Simmons-Duffin:2016gjk}.\footnote{Other expository sources about CFTs in $d>2$ dimensions containing material of interest to this review are \textcite{Ferrara:1973eg}, \textcite{Cardy:1987cs}, \textcite{Fradkin:1996is}, \textcite{DiFrancesco:1997nk}, \textcite{Qualls:2015qjb}, and and \textcite{Osborn-notes}.}

We consider CFTs in flat Euclidean or Lorentzian space with coordinates $x^{\mu}$ and metric $\eta_{\mu\nu}$.\footnote{Set $\eta_{\mu\nu}\to\delta_{\mu\nu}$ if interested uniquely in the Euclidean signature.}
Conformal transformations are diffeomorphisms $x\to x'$ which locally look like a rotation $\Lambda^{\mu}{}_{\nu}(x)$ combined with a rescaling $\Omega(x) \ge 0$ (also called a dilatation), which means that the Jacobian takes the form
\begin{eqnarray}
\label{eq:Jac}
\frac{\del x'^{\mu}}{\del x^{\nu}}=\Omega(x)\Lambda^{\mu}{}_{\nu}(x),\quad 
\eta_{\rho\sigma}\Lambda^{\rho}{}_{\mu}(x) \Lambda^{\sigma}{}_{\nu}(x)=\eta_{\mu\nu}\,.\qquad
\end{eqnarray}
Alternatively, the same condition can be expressed by saying that the transformation preserves angles, or that it leaves the metric invariant up to an overall factor. 

In any dimension $d\geq3$,\footnote{See footnote~\ref{footnote:dimension}.} the case of primary interest for this review, a theorem of Liouville says that any conformal transformation can be obtained by composing 4 types of basic transformations: translations and rotations (which by themselves form the Poincar\'e group and have $\Omega=1$), dilatations $x'^{\mu}=\Omega x^{\mu}$ {with $\Omega$ a constant, and inversions $x'^{\mu}=x^\mu/x^2$ which have} $\Omega(x)=1/x^2$.\footnote{As is well known, the 2d case is special. The group of 2d conformal transformations is {infinite dimensional, since any holomorphic function $f(z)$ with $z=x_1+i x_2$ defines a conformal transformation, $z'=f(z)$.} This case has been {subject to} intense study (see footnote \ref{note:2dLit}), and it will be mostly left out {of} this review except for a few comments in Sec.~\ref{sec:other}.} 

The {resulting} conformal group is a Lie group of dimension $(d+1)(d+2)/2$. Its special role in physics and mathematics is explained by the fact that it is actually the largest nontrivial subgroup of diffeomorphisms of $\bR^d$.

The inversion belongs to the component of the conformal group which is disconnected from the identity,
but by composing an inversion, translation, and a second inversion we can define special conformal transformations (SCTs), also called conformal boosts, given by
\begin{eqnarray}
x'^\mu=\frac{ x^\mu - b^\mu x^2}{1-2 {x\cdot b} +b^2 x^2}\,,\quad  \Omega(x)=1-2 {x\cdot b} +b^2 x^2\,,\qquad
\end{eqnarray}
where $b^\mu\in \bR^d$ is an arbitrary constant vector. 

The conformal algebra generators can be obtained by considering the infinitesimal versions of the above-mentioned transformations.
We denote by $M_{\mu\nu}$ and $P_\mu$ the usual Poincar\'e generators, $D$ the dilatation generator, and $K_\mu$ the generators of SCTs. Their nonzero commutation relations are\footnote{We {follow the conventions} of \textcite{Simmons-Duffin:2016gjk}.}
\begin{eqnarray}
\label{eq:algebra}
&&\left[M_{\mu\nu}, M_{\rho\sigma} \right] = \eta_{\nu\rho} M_{\mu\sigma}-\eta_{\mu\rho} M_{\nu\sigma} +\eta_{\nu\sigma} M_{\rho\mu}-\eta_{\mu\sigma} M_{\rho\nu}\,,\nonumber\\
&&\left[M_{\mu\nu}, P_{\rho} \right] = \eta_{\nu\rho} P_\mu -\eta_{\mu\rho} P_\nu \,,\nonumber\\
&&\left[M_{\mu\nu}, K_{\rho} \right] = \eta_{\nu\rho} K_\mu -\eta_{\mu\rho} K_\nu \,,\\
&&\left[D, P_{\mu} \right] =  P_\mu\,, \nonumber\\
&&\left[D, K_{\mu} \right] =- K_\mu\,, \nonumber\\
&&\left[K_{\mu}, P_{\nu} \right] = 2\eta_{\mu\nu}D- 2M_{\mu\nu} \,.   \nonumber
\end{eqnarray} 

In Euclidean signature, the conformal algebra is isomorphic to the algebra of $SO(d+1,1)$.\footnote{In Lorentzian signature it is $SO(d,2)$.} This is shown by the mapping
\begin{gather}
\label{eq:SODbasis}
\calJ_{d+1\,\mu}= \left(P_\mu-K_\mu\right)/2\,,\quad \calJ_{d+2\,\mu}= \left(P_\mu+K_\mu\right)/2\,,\\
 \calJ_{\mu\nu} = M_{\mu\nu}\,, \quad \calJ_{d+1\, d+2}= D\,, \nonumber
\end{gather} 
which satisfies the $SO(d+1,1)$ commutation relations 
\begin{eqnarray}
	\label{eq:SODalgebra}
	&&\left[\calJ_{AB}, \calJ_{CD} \right] = \\
	&&\hspace{3em}\eta_{BC} \calJ_{AD}-\eta_{AC} \calJ_{BD} +\eta_{BD} \calJ_{CA}-\eta_{AD} \calJ_{CB}\,,\nonumber
\end{eqnarray} 
where $\eta_{AB}$ is the {Lorentzian} metric on $\mathbb{R}^{d+1,1}$.

\subsection{Operators: primaries and descendants}
\label{sec:primaries}

Our main objects of study will be correlation functions of local operators. Conformal symmetry places constraints on these correlators, expressed as {covariance} properties when the operators are transformed in a certain way. Our goal here will be to present these transformations, {the form of which} is determined by representation theory of the conformal group.

Following \textcite{Mack:1969rr}, we can restrict to operators inserted at $x=0$, since the transformation properties at any other point can be obtained by applying a translation,
\beq
\label{eq:Pacts}
\mathcal O(x) = e^{x^\mu P_\mu}\mathcal O(0)e^{-x^\mu P_\mu}\,,
\eeq
and the commutation relations of Eq.~(\ref{eq:algebra}). Then, we only have to specify the action of the stabilizer group of the origin, generated by $M_{\mu\nu}$, $D$, and $K_\mu$.
We will assume that $\calO(0) {\equiv} \mathcal O_{\Delta,r}^{\bi}(0)$ forms a finite-dimensional irreducible representation (irrep) $r$ of the rotation group (with indices $\bi$), and is characterized by the dilatation eigenvalue $\Delta$, called its scaling dimension:
\begin{eqnarray}
\label{eq:fieldrepresentation}
&&\left[D, \mathcal O_{\Delta,r}^{\bi}(0) \right] =  \Delta \mathcal O_{\Delta,r}^{\bi} (0) \,, \nonumber\\
&&\left[M_{\mu\nu}, \mathcal O_{\Delta,r}^{\bi} (0) \right] =  (R_{\mu\nu})^{\bi}{}_{\bj} \mathcal O_{\Delta,r}^{\bj} (0) \,.
\end{eqnarray} 
Here $R_{\mu\nu}$ are generators of the representation $r$ of $SO(d)$ (or its {double cover $Spin(d)$} for spinor representations). 

According to the conformal algebra in Eq.~(\ref{eq:algebra}), the generators $P_\mu$ and $K_\mu$ act as raising and lowering operators for $D$, generating what we call the conformal multiplet of $\calO$. In physically interesting theories the spectrum of the dilatation operator is real and bounded from below,\footnote{As discussed in Sec.~\ref{sec:unitarity}, in unitary theories this property can be shown rigorously.} so the conformal multiplet must contain an operator of lowest dimension. Without loss of generality we assume that $\calO(0)$ is this lowest operator, so that
\begin{eqnarray}
\label{eq:fieldrepresentationK}
\left[ K_{\mu} , \mathcal O_{\Delta,r}^{\bi}(0) \right] = 0 \,.
\end{eqnarray} 
An operator satisfying this condition is called the \emph{primary operator} of the conformal multiplet.\footnote{This is called a quasiprimary in {the context of 2d CFTs}.} All other operators in the multiplet are called descendants and are obtained from the primary by acting $n\ge 1$ times with $P_\mu$, which means that they are simply its derivatives.\footnote{Explicitly $[P_\mu, \mathcal O_{\Delta,r}^{\bi}(x) ] = \partial_\mu \mathcal O_{\Delta,r}^{\bi}(x)$. Often $n$ is called the level of the descendant.}

Eqs.~\reef{eq:fieldrepresentation} define the main quantum numbers characterizing the operator: its scaling dimension $\Delta$ and its irrep $r$ under the rotation group. In practice it is important to know the transformation rules of an operator $\calO(x)$ under general infinitesimal or finite conformal transformations and for any $x$. These rules can be determined uniquely from Eqs.~(\ref{eq:Pacts}, \ref{eq:fieldrepresentation}, \ref{eq:fieldrepresentationK}). Infinitesimal transformations take the form of first-order differential operators, see e.g.~\textcite[{section 3.1.2}]{Rychkov:2016iqz}. Here we will just give the explicit form for the finite transformations in terms of the parameters of Eq.~\reef{eq:Jac}:
\begin{equation}
\mathcal O'{}^{\bi}_{\Delta,r}{}(x') = \calF^{\bi}{}_{\bj} \mathcal O_{\Delta,r}^{\bj}(x)\,,\quad \calF=\frac{1}{\Omega(x)^{\Delta}} \mathcal R[\Lambda^{\mu}_{\,\,\nu}(x)]\,,
\label{eq:transf}
\end{equation}
where $\mathcal R[\Lambda^{\mu}_{\,\,\nu}(x)]$ is the matrix representing the finite rotation $ \Lambda^{\mu}_{\,\,\nu}(x)$ in the representation $r$.\footnote{If $r$ is a spinorial representation then $\Lambda^{\mu}_{\,\,\nu}$ specifies $\mathcal R$ only up to a sign, and this sign has to be chosen consistently for all operators in a correlator.} This equation generalizes Eq.~\reef{eq:invsc} for scalar operators.\footnote{Although we write the left hand side (l.h.s) as $\calO'$ (as is customary), it is important to remember that $\calO$ and $\calO'$ represent the same operator.}

{The scaling dimensions of primary operators comprise the \emph{spectrum} of the theory. In $d\ge 3$, the spectrum is typically discrete.\footnote{The only exceptions known to us are discussed in \textcite{Levy:2018bdc}. They are nonunitary.} Discreteness of the spectrum follows e.g.~from the requirement of a well-defined thermal partition function \cite{Simmons-Duffin:2016gjk}.}

\subsection{Correlation functions}
\label{sec:correlations}

Consider now a correlation function of $n$ primaries:
\beq
\mathcal G^{\bi_1\ldots\bi_n}(x_i) = \langle \mathcal O_{\Delta_1 r_1}^{\bi_1}(x_1) \ldots  \mathcal O_{\Delta_n r_n}^{\bi_n}(x_n) \rangle\,.
\label{eq:npointgen} 
\eeq
{For our purposes we will only need to work at non-coincident points, and will not be concerned with possible delta-function-like ``contact terms", which play no role in the numerical conformal bootstrap.} 

Eq.~\reef{eq:transf} implies that this correlator transforms covariantly under the conformal group. Operationally, for any conformal transformation $x\to x'$, correlators at points $x_j'$ and $x_j$ are related by 
\beq
	\label{eq:correlations_covariance}
	\mathcal G^{\bi_1\ldots\bi_n}(x'_j)= \mathcal F^{(1)}{}^{\bi_1}{}_{\bj_1}\cdots 
	\mathcal F^{(n)}{}^{\bi_n}{}_{\bj_n}\mathcal G^{\bj_1\ldots\bj_n}(x_j)\,.
	\eeq
While covariance under translations, rotations, and dilatations is straightforward to understand, it is less intuitive for SCTs, since they act nonlinearly on $x$.

One can classify the most general form of the correlator satisfying Eq.~\reef{eq:correlations_covariance}. This problem has been addressed using different techniques over the years, starting with~\textcite{Polyakov:1970xd}.\footnote{General 3pt functions in 4d were first worked out in~\textcite{Mack:1976pa}.} Two modern efficient methods to obtain such results are the embedding formalism {of \textcite{Costa:2011mg} reviewed in Appendix~\ref{sec:embedding}, and the conformal frame approach described in Sec.~\ref{sec:confframe}, see \textcite{Osborn:1993cr} and \textcite{Kravchuk:2016qvl}.}

We will now state results for the most frequently occurring cases $n=2,3,4$.  
We will focus on scalars $\calO_\Delta$ as well as operators $\calO_{\Delta,\ell}$ transforming in the rank $\ell$ traceless symmetric representation of $SO(d)$. For the latter we will introduce an auxiliary polarization vector $\zeta_\mu$ and consider the contraction
\begin{equation}\label{eq:indexfree}
\mathcal O_{\Delta,\ell}(x,\zeta) = \zeta_{\mu_1}\cdots\zeta_{\mu_\ell}  \mathcal O_{\Delta,\ell}^{\mu_1\ldots\mu_\ell}(x)\,.
\end{equation}
The components of the operator itself can be recovered {by} differentiating in $\zeta$.\footnote{This is called index free notation, see e.g.~\textcite{Dobrev:1975ru} and~\textcite{Costa:2011mg}. Often one imposes $\zeta^2=0$, which sets to zero the ``traces" in e.g.~Eq.~\reef{eq:2points}, but we will not do this here. Index free notation can be generalized to mixed-symmetry tensors and fermions, see e.g.~\textcite{Giombi:2011rz}, \textcite{SimmonsDuffin:2012uy}, \textcite{Li:2014gpa}, \textcite{Costa:2014rya}, and \textcite{Iliesiu:2015qra}.}

\subsubsection{2pt functions}
\label{sec:2pt}

It follows from Eq.~(\ref{eq:correlations_covariance}) that the 2pt function of two operators $\calO_{\Delta_1,r_1}$ and $\calO_{\Delta_2,r_2}$ vanishes unless $\Delta_1=\Delta_2$ and $r_1=r_2^\dagger$.\footnote{Here $\dagger$ means complex conjugation in Lorentzian signature, or taking the dual reflected representation in Euclidean signature, where reflected means replacing generators $R_{1\nu}$ by $-R_{1\nu}$. In 3d all representations are real, so the requirement $r_1 = r_2^{\dagger}$ reduces to $r_1=r_2$, while in 4d if $r_1=(\ell,\bar \ell)$ then $r_2=(\bar\ell,\ell)$.} As a consequence, for every physical operator $\calO_{\Delta,r}$, one can identify an operator $\calO^\dagger_{\Delta,r^\dagger}$ which transforms in the conjugate representation.\footnote{The precise action of Hermitian conjugation on Hilbert space operators depends on the signature and choice of quantization surface. For a detailed discussion see \textcite{Simmons-Duffin:2016gjk}.}

Further, one can almost always work in a basis of operators such that $\calO$ has a nonzero 2pt function only with $\calO^\dagger$, which is usually stated as ``the 2pt function is diagonal".\footnote{Examples of nonunitary conformal theories in which the 2pt functions cannot be so diagonalized occur in logarithmic CFTs, see e.g.~\textcite{Hogervorst:2016itc}. We will not consider them in this review.} 
For example, this is always possible in unitary theories. For operators in real $SO(d)$ representations $r^{\dagger} = r$, like traceless symmetric tensors, we can choose a real operator basis so that $\calO^\dagger = \calO$.

Specializing to traceless symmetric tensors, the 2pt function takes the form\footnote{For the purposes of this review, it is sufficient to consider correlation functions in Euclidean signature. Most equations can also be used in Lorentzian signature, provided that all points are spacelike separated. For timelike separation one needs modifications, such as an $i\eps$ prescription, which we will not discuss.}
\begin{gather}
\langle  \calO_{\Delta,\ell}(x_1,\zeta_1)  \calO_{\Delta,\ell}(x_2,\zeta_2) \rangle = \frac{\left( I_{\mu\nu}(x_{12})\zeta_1^\mu \zeta_2^\nu\right)^\ell-\text{traces} }{(x_{12}^2)^{\Delta}}\,,\nn\\
I_{\mu\nu}(x) =\eta_{\mu\nu} -2 {x_\mu x_\nu}/{x^2}\,, \label{eq:2points}
\end{gather}
where $x_{ij} \equiv x_i-x_j$, and ``traces" are terms proportional to $\zeta_1^2$, $\zeta_2^2$, which are uniquely fixed by the tracelessness of $\calO_{\Delta,\ell}$. This generalizes Eq.~\reef{eq:2pt} for scalars. It is customary to normalize such 2pt functions to unity, with exceptions being conserved currents and the stress tensor, see Sec.~\ref{sec:ward}. The nontrivial part of the correlator is its numerator, which {specifies the dependence on the operator indices}. We will refer to such numerators as ``tensor structures".

If the CFT contains a global symmetry, operators are grouped into global symmetry multiplets $\pi$. In this case Eq.~\reef{eq:2points} still applies to the individual components of the multiplets, with obvious appropriate modifications.\footnote{If $\pi$ is a complex representation, then it is not convenient to use the real operator basis. The nonzero 2pt function will then be between $\calO$ and $\calO^\dagger$ transforming in $\bar\pi$.} We will discuss the consequences of global {symmetries} further in Sec.~\ref{sec:global}.

\subsubsection{3pt functions}
\label{sec:3pt}

Next we turn to 3pt functions, focusing on the case where the first two operators are scalars. Then it turns out that the third operator can only be a traceless symmetric tensor. {Generalizing Eq.~\reef{eq:3pt} for three scalars, the 3pt function takes the form \cite{Mack:1976pa}}
\begin{multline}
\label{eq:3points}
\langle  \mathcal O_{\Delta_1}(x_1)  \mathcal O_{\Delta_2}(x_2)  \mathcal O_{\Delta_3,\ell}(x_3, \zeta) \rangle= \\
\lambda_{123}\, [\left(Z^\mu_{123} \zeta_\mu \right)^\ell -\text{traces}]\, {\bf K}_3\,,
\end{multline} 
where ${\bf K}_3={\bf K}_3(\Delta_i,x_i)$ is given by
\beq
{\bf K}_3 =\frac{1} {(x^2_{12})^{\frac{h_{123}+\ell}2} (x^2_{13})^{\frac{h_{132}-\ell}2}\,
	(x^2_{23})^{\frac{h_{231}-\ell}2}}\,,
\eeq
$h_{ijk} \equiv \Delta_i+\Delta_j-\Delta_k$, and $Z_{123}^{\mu}= \frac{x^\mu_{13}}{x_{13}^2}-\frac{x^\mu_{23}}{x_{23}^2}$. This 3pt function is unique up to the overall coefficient $\lambda_{123}$. Notice that as defined,
\beq
\label{eq:flipsign}
\lambda_{123} = (-1)^\ell \lambda_{213}\,,
\eeq
while if $\ell=0$ we can exchange any pair of fields and $\lambda_{123}$ is fully symmetric.
The normalization of these coefficients is unambiguous, since the operators are assumed to be unit-normalized according to Eq.~(\ref{eq:2points}). 
Together with the {spectrum}, the $\lambda$'s constitute the \emph{CFT data}, which distinguish one CFT from another, as discussed in Sec.~\ref{sec:informal}. 

In unitary theories, {the CFT data must satisfy a set of} general well-understood constraints, see Sec.~\ref{sec:unitarity}. {Significantly more} nontrivial constraints on the CFT data come from the crossing relations to be discussed in Sec.~\ref{sec:crossing}.

For operators in three general $SO(d)$ representations{, the 3pt} functions take a form more complicated than \reef{eq:3points}. They are also in general not unique, although for any three representations there is at most a finite-dimensional space of allowed tensor structures. The problem of their construction has been completely solved in the most physically important cases of $d=3$~\cite{Costa:2011mg,Iliesiu:2015qra} and $d=4$~\cite{Elkhidir:2014woa}. For general $d$ there are partial results, e.g.~\textcite{Costa:2011mg} for 3pt functions of traceless symmetric tensors,~\textcite{Costa:2016hju} for traceless mixed-symmetry tensors, and~\textcite{Kravchuk:2016qvl} for a general approach to classifying the structures.

\subsubsection{4pt functions}

Finally let us consider 4pt functions, which as mentioned in Sec.~\ref{sec:informal} play {a} fundamental role in the conformal bootstrap. Focusing here on the case of scalars, the 4pt function must take the general form
\beq
\label{eq:4p}
\langle  \mathcal O_{\Delta_1}(x_1)  \mathcal O_{\Delta_2}(x_2)  \mathcal O_{\Delta_3}(x_3)  \mathcal O_{\Delta_4}(x_4)\rangle =  g(u,v)\, \mathbf{K}_4\,.
\eeq
The factor $\mathbf{K}_4=\mathbf{K}_4(\Delta_i,x_i)$ is given by
\beq
\label{eq:K4}
\mathbf{K}_4 = 
\frac{1}{(x_{12}^2)^{\frac{\Delta_1+\Delta_2}2}(x_{34}^2)^{\frac{\Delta_3+\Delta_4}2}} \left(\frac{x^2_{24}}{x^2_{14}}\right)^{\frac{\Delta_{12}}2}
\left(\frac{x^2_{14}}{x^2_{13}}\right)^{\frac{\Delta_{34}}2}
\,, 
\eeq
where {$\Delta_{ij} \equiv \Delta_i - \Delta_j$}. This factor by itself transforms under conformal transformations as {prescribed by Eq.~(\ref{eq:correlations_covariance}). }
The remaining part of the correlator, $g(u,v)$, must be a function of two \emph{cross ratios} $u,v$:
\begin{eqnarray}
\label{eq:uv}
u=\frac{x_{12}^2 x_{34}^2}{x_{13}^2 x_{24}^2}\,, \qquad v=\frac{x_{14}^2 x_{23}^2}{x_{13}^2 x_{24}^2}\,,
\end{eqnarray}
which are invariant under all conformal transformations.

While no further information about $g(u,v)$ can be obtained from conformal invariance alone, it can in fact be computed in terms of the CFT data using additional tools such as the OPE and conformal blocks. This will be discussed in Secs.~\ref{sec:OPE} and~\ref{sec:cb}.

\subsubsection{Conformal frames} 
\label{sec:confframe}
Here we will give a more group theoretical intuition of the number of degrees of freedom contained in a given correlator, and in particular of why conformal invariance fixes 2pt and 3pt functions up to a few constants, but allows arbitrariness in 4pt functions. 
 
Given a set of $n$ points, we can make use of conformal transformations to arrange them in convenient configurations. {For instance, given} 3 arbitrary points we can find a conformal transformation which maps them to $x_{1,2,3}=0,\hat{e},\infty$, where $\hat{e}$ is a fixed unit vector.

\begin{figure}[t]
	\begin{centering}
		\includegraphics[width=0.6\columnwidth]{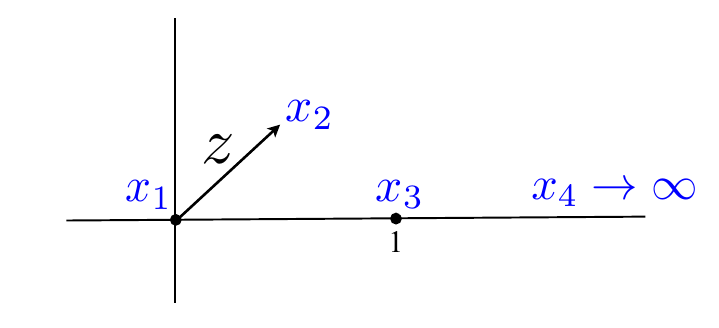}
		\caption{\label{fig:z-frame}
			(Color online) Conformal frame defining the $z$ coordinate. Figure from \cite{Hogervorst:2013sma}.
		}
	\end{centering}
\end{figure}

For 4 points, we can first find a conformal transformation fixing 3 of them as above, and then rotate around the axis to put the fourth point into a fixed plane (we assume that $d\ge 2$). The resulting configuration can be parametrized in Euclidean signature as ($\mathbf{0} \equiv \mathbf{0}_{d-2}$)\footnote{We define $\calO(\infty)$ {by taking} the limit of $|x_4|^{2\Delta_\calO} \calO(x_4)$ {as} $x_4\to\infty$, {which yields a finite value for the correlation function.}}
\begin{align}
x_1 &= (0,0,\mathbf{0})\,,\quad x_2 = (\sigma,\tau,\mathbf{0})\,,\nn\\
x_3 &= (1,0,\mathbf{0})\,,\quad  x_4 = (\infty,0,\mathbf{0})\,.
	\label{eq:conformalframe1}
\end{align}
It is customary to define (see Fig.~\ref{fig:z-frame})
\beq
\label{eq:z}
z=\sigma+i\tau,\quad \bar{z}=\sigma-i\tau\,,
\eeq 
which are complex conjugate variables if we are working in the Euclidean.\footnote{Notice that we can analytically continue to {the Lorentzian} via $\tau\to i t$, and then $z$ and $\bar z$ become independent real variables, but this will not play a role in this review.} The conformal cross ratios can be expressed in terms of $z$, $\bar{z}$ as
\begin{eqnarray}
\label{eq:zzb}
	u= z\bar{z} \,, \qquad v= (1-z)(1-\bar{z})\,.
\end{eqnarray}

A choice of points $x_i$, as in Eq.~(\ref{eq:conformalframe1}), is called a \emph{conformal frame}. It can be thought of as a gauge fixing of most or all of the conformal symmetry. By construction, any coordinate configuration can be reduced to the conformal frame form. Therefore, the knowledge of a correlation function in the conformal frame is sufficient to reconstruct it at any other point through its covariance properties {\cite{Osborn:1993cr}}. The functional forms of 2pt and 3pt functions are fixed because their conformal frames do not contain any free parameters. The 4pt conformal frame
\reef{eq:conformalframe1} has 2 real parameters, explaining the functional freedom of the conformal 4pt function. See Sec.~\ref{sec:radial} for another frequently used conformal frame.

Conformal frames provide a way to construct conformal correlators {which is sometimes more convenient than the embedding formalism described in App.~\ref{sec:embedding}.} This method can also be used to classify the allowed tensor structures. {An important role is then played by the stabilizer group, defined as the set of conformal transformations leaving the conformal frame configuration invariant. It is $SO(d-1)$ for 3pt functions and $SO(d-2)$ for 4pt functions. One classifies tensor structures invariant under the stabilizer group, and each of them lifts to an independent conformally invariant tensor structure \cite{Kravchuk:2016qvl}. This method is particularly useful when dealing with 4pt functions of tensor operators: it does not overcount tensor structures, which may happen in the embedding formalism unless special care is taken.}

\subsection{Operator product expansion}
\label{sec:OPE}

Our point of view on the origin and the role of {the} Operator Product Expansion (OPE) in CFT is the one pedagogically reviewed in \textcite{Rychkov:2016iqz,Simmons-Duffin:2016gjk}. Here we will present the main logic and set some conventions.

The key idea is that of radial quantization, which says that we can represent {Euclidean} CFT correlation functions as scalar products of states $\<\Psi_{\rm out}|\Psi_{\rm in}\>$ which live on a sphere of radius $R$. The state $|\Psi_{\rm in}\>$ is generated by operators in the interior of the sphere, while $\<\Psi_{\rm out}|$ by those in the exterior. Once we replace the interior operators by the state $|\Psi_{\rm in}\>$, in a scale invariant theory we can scale the radius of the sphere to zero. Thus any state $|\Psi_{\rm in}\>$ can be expanded in a basis of local operators inserted at the center of the sphere. This is called the state-operator correspondence.

The OPE, written schematically in \reef{eq:OPE}, is just the special case of the above when there are two operators at points $x_1$ and $x_2$ inside the sphere centered at $y$. We also see the origin of the OPE convergence criterion \reef{eq:converge}, since we need to have a separating sphere to start the 
argument.\footnote{See \textcite{Pappadopulo:2012jk} for a detailed discussion of OPE convergence in CFT.}

As discussed in Sec.~\ref{sec:informal}, the next step is to group primaries and descendants in the OPE and to impose the constraints of conformal invariance. This gives the ``conformal OPE":
\beq
\calO_{\Delta_i}(x_1)\calO_{\Delta_j}(x_2) = \sum_k \lambda_{ijk} \hat{f}_{ijk}(x_1,x_2,y,\del_y) \calO_{\Delta_k}(y)\,.
\label{eq:confOPE}
\eeq
The differential operator $\hat{f}_{ijk}$ is fixed by conformal invariance. It can be determined by demanding that the conformal OPE reproduce the 3pt function $\<\calO_{\Delta_i}(x_1)\calO_{\Delta_j}(x_2)\calO_{\Delta_k}^{\dagger}(x_3)\>$, whose form is by itself fixed by conformal invariance {up to the constants $\lambda_{ijk}$.}

Any $SO(d)$ (or $Spin(d)$) index which the operators $\calO_{\Delta_i}$ may have are left implicit in \reef{eq:confOPE}. Depending on their representations, there may be several allowed 3pt function tensor structures, and then each structure comes with its own OPE coefficient and {a corresponding conformally-invariant} differential operator $\hat{f}_{ijk}$. In the most frequently occurring case where {$\calO_{\Delta_i}$ and $\calO_{\Delta_j}$ are scalars and $\calO_{\Delta_k,\ell}$ a spin-$\ell$} traceless symmetric tensor there is just one OPE coefficient.

While it is important to know that the conformal OPE exists and converges, it turns out that in practice one rarely needs {its} full explicit form.\footnote{For some cases when the explicit conformal OPE has been worked out, see \textcite{Ferrara:1971vh}, \textcite{Ferrara:1973eg}, \textcite{Mack:1976pa}, and \textcite{Dolan:2000ut}.} For example, conformal block computations can be organized in ways which avoid explicit knowledge of the full OPE, see Sec.~\ref{sec:cb}. For this reason, one frequently writes only the ``leading OPE", i.e.~the primary term. 

For example, in the above-mentioned case of two scalars and a traceless symmetric tensor, the leading OPE has the form (specializing to $x_1=x$, $x_2=y=0$):
\begin{multline}
 \calO_{\Delta_i}(x)\calO_{\Delta_j}(0)\supset \lambda_{ijk}
 \frac{x_{\mu_1}\cdots x_{\mu_\ell}}{(x^2)^{(h_{ijk}+\ell)/2}}\calO^{\mu_1\ldots\mu_\ell}_{\Delta_k,\ell}(0) + \ldots\,.
 \label{eq:OPEnorm}
 \end{multline}
This reproduces the leading asymptotics of the 3pt function \reef{eq:3points} in the limit $x\to0$ with $x_3$ fixed, including the normalization, provided that the 2pt function of {$\calO_{\Delta_k,\ell}$} is unit-normalized as in \reef{eq:2points}. Occasionally we will schematically write such a leading OPE as $\calO_{\Delta_i} \times \calO_{\Delta_j} \supset \calO_{\Delta_k,\ell}$, but the form \reef{eq:OPEnorm} should always be understood.

As explained in Sec.~\ref{sec:informal}, any $n$-point function can be computed from the CFT data by repeated application of the OPE. The 4pt function case, of primary importance for the bootstrap, will be discussed in Sec.~\ref{sec:cb}.

\subsection{Constraints from unitarity}
\label{sec:unitarity}

Here we will review the notion of a unitary CFT, focusing on the constraints on CFT data arising for such theories which make the bootstrap especially powerful. 

Unitary CFTs can be considered both in Lorentzian and Euclidean signature. They are characterized in the latter by a property called reflection positivity.\footnote{We will often abuse terminology and refer to ``reflection positivity" as ``unitarity" in the context of Euclidean CFTs or when the signature is ambiguous.} On the other hand, nonunitary CFTs are generally expected to make sense only in Euclidean signature. They will be discussed briefly in Sec.~\ref{sec:nonunitary}.

Unitary theories allow for quantization in a Hilbert space with a positive-definite norm. In the quantization by planes normally used in Euclidean QFT, an ``in" state $|\Psi\>$ is generated by $n$ local operators $\calO_i$ inserted in the half-space $x_1<0$, and an ``out" state $\<\Psi|$ is generated by reflected operators $\calO_i^\dagger$, inserted at $x_1>0$ at mirror-symmetric positions.\footnote{\label{note:Theta}For reflected tensor operators, each tensor component is multiplied by a factor $\Theta = (-1)^{N_\perp}$ where $N_\perp$ is the number of tensor indices perpendicular to the reflection plane.} Unitarity implies that the norm $\<\Psi|\Psi\>$ must be non-negative. This norm is just a $2n$-point function in a particular kinematic configuration, and its positivity is called (Osterwalder-Schrader) reflection positivity.
 
Analogously, in the radial quantization usually used for CFTs, an ``in" state $|\Psi\>$ is generated by local operators $\calO_i$ inserted at positions $x_i$ inside the unit sphere $(i=1,\ldots,n)$, and a conjugate ``out" state $\<\Psi|$ is generated by operators $\calO_i^\dagger$ inserted at positions related by an inversion transformation $x_i'=x_i/x^2$. The norm $\<\Psi|\Psi\>$, which is just an inversion-symmetric $2n$-point function, must be non-negative, a property we {will call} ``inversion positivity".\footnote{If {the} $\calO_i$ are not scalars, their indices at the inverted positions are contracted with the $I_{\mu\nu}$ tensors {defined in} Eq.~\reef{eq:2points}, as in \cite[{Eq.~(110)}]{Simmons-Duffin:2016gjk}.} In CFTs, both forms of positivity are equivalent,\footnote{By a conformal transformation, radial quantization may be mapped onto a ``North-South quantization", {relating} ``inversion positivity" to the usual reflection positivity~\cite{Rychkov:2016iqz}.} and they are both useful depending on circumstances.

\subsubsection{Unitarity bounds}
\label{sec:UB}

We already get a simple and powerful constraint by considering radial quantization states $|\Psi\>$ produced by a local operator $\calO$ acting at the origin. In this case the conjugate operator is inserted at infinity. For {a} primary $\calO$ we recover that its 2pt function must have positive normalization and hence can be normalized to one as in \reef{eq:2points}. Additional constraints arise from considering descendants of $\calO$. The conformal algebra computes the norms of descendants as polynomials in the primary dimension $\Delta$. Imposing that all descendants have a non-negative norm gives a lower bound on $\Delta$. This 
``unitarity bound" depends on the representation $r$ of $SO(d)$ (or its double cover {for spinor representations}) in which the primary transforms.\footnote{Standard CFT references are \textcite{Ferrara:1974pt}, \textcite{Mack:1975je}, and \textcite{Minwalla:1997ka}. An early physics reference is \textcite{doi:10.1063/1.1705183}. In the mathematics literature, these bounds were derived by \textcite{jantzen_kontravariante_1977}, although the relevance of this work for physics was realized only recently \cite{Penedones:2015aga,Yamazaki:2016vqi}. See also \textcite{Rychkov:2016iqz,Simmons-Duffin:2016gjk} for a review. Unitarity bounds can be equivalently derived by studying the positivity of the Fourier transform of the 2pt function analytically continued to Lorentzian signature (the Wightman function), see \textcite{Ferrara:1974pt}, \textcite{Mack:1975je} (in the sufficiency part of the argument), as well as \textcite{Grinstein:2008qk} for a recent exposition emphasizing physics.} \footnote{In Lorentzian signature, operators satisfying the unitarity bounds correspond to the unitary representations of the universal covering group of the Lorentzian conformal group $SO(d,2)$ having positive energy. Notice that in Euclidean signature operators satisfying the unitarity bounds have no relation to the representation of the Euclidean conformal group $SO(d+1,1)$ which are unitary in the usual mathematical sense of the term. This is already clear from looking at the principal series unitary representations of $SO(d+1,1)$ which have complex scaling dimensions $d/2+i\bR$.}

In 3d, the representation $r$ is labeled by a half-integer $j$, with $j=\ell$ for traceless symmetric spin-$\ell$ tensors. The unitarity bounds are
\begin{align}
	d=3:\quad &\Delta\ge 1/2 &&(\text{scalar, } j=0)\,,\nn\\
	&\Delta\ge 1 &&(\text{smallest spinor, } j=1/2) \,,\\
	&\Delta\ge j+1\  &&(j>1/2)\,.\nn
	\end{align}

In 4d, we can label the representation $r$ by two integers $(\ell,\bar \ell)$, with traceless symmetric spin-$\ell$ tensors having $\ell=\bar\ell$.\footnote{It is also common in the literature to label by half-integers $j=\ell/2,\, \bar{j}=\bar{\ell}/2$.} The unitarity bounds then read
\begin{align}
	d=4:\quad &\Delta\ge 1&&(\text{scalar, } \ell=\bar\ell=0)\,,\nn\\
	&\Delta\ge \half \ell+1 &&(\ell>0, \bar\ell=0) \,,\label{eq:4dunitarity}\\
	&\Delta\ge \half(\ell+\bar\ell)+2 &&(\ell \bar\ell\ne0)\,.\nn
\end{align}

For the 5d and 6d unitarity bounds see \textcite{Minwalla:1997ka}. For some representations occurring in all dimensions the unitarity bounds can be written in dimension-independent form as follows:
\begin{align}
	 &\Delta\ge \half(d-2)&&(\text{scalar})\,,\nn\\
	&\Delta\ge \half(d-1) &&(\text{smallest spinor}) \,,\\
	&\Delta\ge \ell+d-2 &&(\text{traceless symmetric, spin }\ell\ge 1)\,.\nn
\end{align}

As a final comment, in physics literature the unitarity bounds are often derived by imposing positivity of the descendant norm{s} on the first (and the second, for scalars) level. It is a nontrivial fact that no further constraints arise from higher levels. See \textcite[{Tables 3 and 5}]{Bourget:2017kik} for a review of rigorous mathematical results for unitary bounds in any $d$.

\subsubsection{OPE coefficients}

\begin{figure}[t]
	\begin{centering}
		\includegraphics[width=0.6\columnwidth]{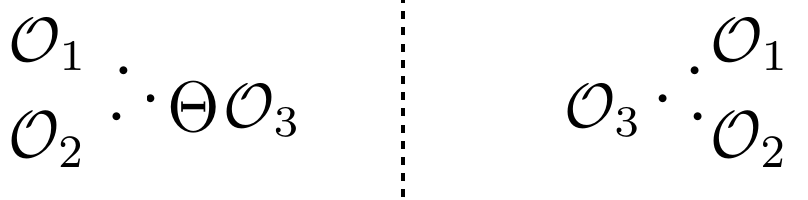}
		\caption{\label{fig:6pt}
       Positivity of this 6pt function implies reality of the 3pt function coefficient $\lambda_{123}$, see the text.
		}
	\end{centering}
\end{figure}

Unitarity also gives reality constraints on OPE coefficients of real operators. Consider the 3pt function \reef{eq:3points} between two scalars and a traceless symmetric tensor, assuming all three operators are real. Then the 3pt function coefficient must be real:
\beq
\label{eq:reality}
\lambda_{123}\in \bR\,.
\eeq
To argue this, we can consider a 6pt function $\<\calO_1 \calO_2 (\Theta \calO_3) \calO_3 \calO_2 \calO_1\>$, with the operators arranged mirror-symmetrically against a plane into two compact groups positioned a large distance from each other (see Fig.~\ref{fig:6pt}). Here $\Theta$ is the reflection factor mentioned in footnote \ref{note:Theta}. Reflection positivity implies that this 6pt function should be real and positive.\footnote{For this argument we are thus using the standard Osterwalder-Schrader reflection positivity and not the ``inversion-positivity".} On the other hand, by cluster decomposition this 6pt function is equal to the product of two distant 3pt functions, which is easily seen to be $\lambda_{123}^2$ times a positive number. So \reef{eq:reality} follows. We stress that this conclusion holds for both even and odd $\ell$.\footnote{In essence we argued that the complex conjugate of a 3pt function is equal to the 3pt function of conjugate fields at reflected positions. This (for general $n$-point functions) is sometimes taken as an additional axiom for unitary theories, encoded by the equation $\calO(\tau,\mathbf{x})^\dagger = \calO^\dagger(-\tau,\mathbf{x})$ valid in Euclidean quantization by planes. Upon analytic continuation to Lorentzian signature, this leads to commutativity of operators at spacelike separation, used to prove reality of OPE coefficients in \textcite{Rattazzi:2008pe}. Our 6pt argument shows that this axiom is not independent but follows from reflection positivity and cluster decomposition.}

It was important for the above argument that the tensor structure entering \reef{eq:3points} was parity invariant (i.e., it did not involve the $\eps$-tensor). This argument can be generalized to OPE coefficients for general 3pt tensor structures. The OPE coefficients of tensor structures must be purely imaginary or real depending on whether they involve the $\epsilon$-tensor or not. One must similarly be careful with OPE coefficients involving spinors.

Consider now the 4pt function $\<\calO_2 \calO_1 \calO_1 \calO_2\>$ where $\calO_1$ and $\calO_2$ are real scalars and the point configuration is reflection-symmetric or inversion-symmetric. This 4pt function should be non-negative as a basic consequence of unitarity, and Eq.~\reef{eq:reality} implies that a more nuanced statement is true: the individual contribution of every primary $\calO$ to this 4pt function is non-negative, see Eq.~\reef{eq:CBdec} below. This can be generalized to external operators in general $SO(d)$ (or $Spin(d)$) representations, including the case when there are multiple 3pt function tensor structures.

To summarize, the unitarity bounds say that the CFT Hilbert space has a positive-definite norm, and the OPE coefficient reality constraints say that {the} OPE preserves this positive-definite structure. If the CFT data satisfies both of these constraints, we are guaranteed that the CFT will be unitary. The bootstrap {obtains} further constraints on CFT data by combining unitarity with crossing relations.

\subsubsection{Averaged null energy condition}
\label{sec:ANEC}

In a QFT in Lorentzian signature, we can consider the integral of the stress tensor component $T_{++}$ along a light ray: the light-like direction $x^+$ {with} all other coordinates fixed to zero. The averaged null energy condition (ANEC) says that this light-ray operator has a non-negative expectation value in any state:\footnote{Such conditions were first introduced in general relativity, with integration along a null geodesic, in connection with singularity theorems and wormholes. Here we focus on the ANEC in flat space, first discussed by \textcite{Klinkhammer:1991ki}.}
\beq
\label{eq:ANEC}
\<\Phi|\int_{-\infty}^\infty dx^+\,T_{++} |\Phi\>\ge 0\,.
\eeq
The ANEC should hold in any unitary QFT. Two general proofs of the ANEC were given recently, one via quantum information \cite{Faulkner:2016mzt}, and one by causality \cite{Hartman:2016lgu}.\footnote{See also \textcite{Kravchuk:2018htv} for a recent discussion of light-ray operators in Lorentzian CFTs and an alternative proof of the ANEC.} 
Specializing to CFTs, the causality argument makes it clear that the ANEC is not an extra assumption but follows from other CFT axioms such as unitarity, the OPE, and crossing relations for correlation functions involving $T_{\mu\nu}$.\footnote{This is also suggested by the fact that {bounds following from the ANEC can be reproduced in the numerical bootstrap, see Sec.~\ref{sec:JandT}.}}  Notice however that any results following from the ANEC will require the existence of a local stress-tensor operator.

Choosing $|\Phi\>$ in \reef{eq:ANEC} to be generated by a local operator $\calO$ acting on the vacuum, the ANEC leads to positivity constraints on 3pt functions $\<\calO T_{\mu\nu} \calO\>$ called  ``conformal collider bounds" \cite{Hofman:2008ar}.\footnote{Conformal collider bounds in general dimensions for states created by the stress tensor or global symmetry currents were obtained in \textcite{Buchel:2009sk} and \textcite{Chowdhury:2012km}. A proof of these bounds independent from the ANEC was given in \textcite{Hofman:2016awc}; see also \textcite{Hartman:2015lfa,Hartman:2016dxc}. Other generalizations of these bounds have been explored in~\textcite{Li:2015itl}, \textcite{Komargodski:2016gci}, \textcite{Chowdhury:2017vel}, \textcite{Cordova:2017zej}, \textcite{Meltzer:2017rtf}, and \textcite{Cordova:2017dhq}. Sum rules involving the same coefficients were also recently presented in~\textcite{Witczak-Krempa:2015pia}, \textcite{Chowdhury:2016hjy}, \textcite{Chowdhury:2017zfu}, and \textcite{Gillioz:2016jnn,Gillioz:2018kwh}.  } 

Recently, \textcite{Cordova:2017dhq} used the ANEC to argue that primaries of high chirality (large $|\ell-\bar\ell|$) in unitary 4d CFTs should satisfy unitarity bounds stronger than \reef{eq:4dunitarity}. From partial checks for $\bar\ell=0,1$, they conjecture the general bound (assuming $\ell\ge \bar\ell$)
\beq
\label{eq:Cordova}
\Delta \ge \ell\,.
\eeq
If $\bar\ell=0$ this becomes stronger than \reef{eq:4dunitarity} for $\ell>2$ and for $\ell>\bar\ell+4$ otherwise. This can be viewed as a CFT strengthening of the theorem of \textcite{Weinberg:1980kq}.

\subsection{Conformal blocks}
\label{sec:cb}

Conformal blocks are of capital importance for the bootstrap. Their theory was initiated in 1970s \cite{Ferrara:1973vz,Ferrara:1974nf,Ferrara:1974ny}{,} with further advances in the early 2000s \cite{Dolan:2000ut,DO2} which were crucial for the bootstrap revival. Recently it experienced further rapid developments, and here we will review its current state.

Consider a 4pt function of four primary \emph{scalar} operators $\phi_i(x_i)$ with $i=1,\ldots,4$ (see Sec.~\ref{sec:spinning} for the general case of external operators with spin). As mentioned in Sec.~\ref{sec:informal}, this 4pt function can be computed by applying the OPE of Eq.~(\ref{eq:OPE}) to {two} pairs of fields. For definiteness we fix here the pairing $\phi_1(x_1) \phi_2(x_2)$ and $\phi_3(x_3) \phi_4(x_4)$. This is referred to as ``the (12)-(34) OPE channel", to distinguish it from other pairings which will play a role when we discuss crossing. This gives an expansion
\beq
\label{eq:4pf_scalars}
\langle \phi_1(x_1) \phi_2(x_2) \phi_3(x_3) \phi_4(x_4) \rangle =
\sum_\calO \lambda_{12\calO}\lambda_{34\calO}\,{\rm W}_\calO \,,
\eeq
where ${\rm W}_\calO \equiv {\rm W}_\calO(x_i)$ are the \emph{conformal partial waves} (CPWs) given by 
\begin{gather}
	\label{eq:WOdef}
{\rm W}_\calO
 =\hat{f}_{12\calO}(x_1,x_2,y,\del_y) \hat{f}_{34\calO}(x_3,x_4,y',\del_{y'}) \langle\calO(y)\calO(y')\rangle\,.
\end{gather}
Since the 2pt function is diagonal,  the summation is over the same operator $\calO$ in both OPEs.
 It follows from conformal invariance of the OPE that each CPW transforms under the conformal transformations in the same way as the 4pt function itself, see e.g.~\textcite{Costa:2011dw}. It is then conventional to split off the factor ${\bf K}_4$ defined in Eq.~(\ref{eq:K4}), so that we finally have
\beq
{\rm W}_{\calO} =  g^{\Delta_{12},\Delta_{34}}_{\Delta_\calO,\ell_\calO}(u,v)\,{\bf K}_4\,,
\label{eq:CPW}
\eeq
where $g^{\Delta_{12},\Delta_{34}}_{\Delta_\calO,\ell_\calO}(u,v)$ is called the conformal block.\footnote{We make {a} distinction between CPWs and conformal blocks following {the} conventions of \textcite{Costa:2011dw}. In part of the literature these two terms are used interchangeably.} {It represents the contribution of a primary $\calO$ and all of its descendants to the 4pt function.}
As shown{,} it depends on the dimension and spin of the exchanged traceless symmetric primary $\calO$, and also on the dimension differences $\Delta_{12}$, $\Delta_{34}$ of the external scalars.\footnote{Sometimes we will omit the latter dependence, if {it is} clear from the context.}
{Comparing with Eq.~(\ref{eq:4p}), we thus have:}
\beq
\label{eq:CBdec}
g(u,v)=  \sum_\calO \lambda_{12\calO}\lambda_{34\calO}\,g^{\Delta_{12},\Delta_{34}}_{\Delta_\calO,\ell_\calO}(u,v)\,.
\eeq
Eqs.~\reef{eq:4pf_scalars} and \reef{eq:CBdec} are referred to as the CPW decomposition and the conformal block decomposition.

{Let us briefly discuss the regions of convergence of the considered expansions. If one works in the $z$ conformal frame of Eq.~\reef{eq:conformalframe1} in Euclidean signature, then Eq.~\reef{eq:WOdef} defining the CPWs converges for $|z|<1$, and the conformal block decomposition \reef{eq:CBdec} is also seen to converge in this region, at least if the theory is unitary \cite{Pappadopulo:2012jk}. While this is sufficient for many applications, a stronger convergence result can be established using the $\rho$ frame, see Sec.~\ref{sec:radial} below.}

The above definition of conformal blocks via the conformal OPE is important in principle. In practice, there exist efficient approaches to compute the blocks which avoid needing explicit knowledge of the conformal OPE.\footnote{However, {see \textcite{Dolan:2000ut} and \textcite{Fortin:2016lmf, Fortin:2016dlj}} for direct constructions using the conformal OPE.} They will be described below.

\subsubsection{The Casimir equation}
\label{sec:casimir}

{Let us consider} the following alternative representation of CPWs. 
In radial quantization, as mentioned in Sec.~\ref{sec:OPE}, the above 4pt function is expressed as a scalar product of two states 
\beq
\<\phi_3(x_3)\phi_4(x_4)|\phi_1(x_1)\phi_2(x_2)\>\,
\eeq
living on a sphere separating $x_1,x_2$ from $x_3,x_4$. The CPW then corresponds to inserting an orthogonal projector $\calP_{\Delta,\ell}$ {onto} the conformal multiplet of $\calO_{\Delta,\ell}$:
\beq
\lambda_{12\calO}\lambda_{34\calO}\,{\rm W}_\calO = \<\phi_3(x_3)\phi_4(x_4)|\calP_{\Delta,\ell}|\phi_1(x_1)\phi_2(x_2)\>\,.
\label{eq:CPWrad}
\eeq
For future reference, the projector can be written as
\beq 
\label{eq:proj}
\calP_{\Delta,\ell}=\sum_{\alpha,\beta=\calO,P\calO,PP\calO,\ldots}|\alpha\>G^{\alpha\beta}\<\beta|\,,
\eeq
where $G_{\alpha\beta}=\<\alpha|\beta\>$ is the Gram matrix of the multiplet and $G^{\alpha\beta}$ is its inverse.

Furthermore, consider the quadratic Casimir\footnote{The quartic Casimir operator 
	${\mathcal C}_4 = \frac12 \calJ_{AB}\calJ^{BC}\calJ_{CD}\calJ^{DA}$ has also proved useful in some conformal block studies \cite{DO3, Hogervorst:2013kva}\,.}
\begin{eqnarray}
\mathcal C_2 = \frac12 \calJ_{AB}\calJ^{BA}\,,
\end{eqnarray}
where $\calJ_{AB}$ are the $SO(d+1,1)$ generators, Eq.~(\ref{eq:SODalgebra}). Insert this operator into Eq.~\reef{eq:CPWrad} right after $\calP_{\Delta,\ell}$. The resulting expression can be computed in two ways.
When we act with $\mathcal C_2$ on the left we have
\beq
\calP_{\Delta,\ell}\,{\mathcal C}_2=C_{\Delta,\ell} \calP_{\Delta,\ell}\,,
\eeq
where $C_{\Delta,\ell}$ is the quadratic Casimir eigenvalue: 
\begin{eqnarray}
C_{\Delta,\ell} = \Delta(\Delta-d) + \ell(\ell+d-2)\,.
\end{eqnarray}
On the other hand, the action of $\mathcal C_2$ on the right can be computed using the representation of the conformal generators on primaries as first-order differential operators, mentioned in Sec.~\ref{sec:primaries}.
We conclude that the CPW, and hence the conformal block, satisfies a second-order partial differential equation.\footnote{We followed the presentation in \cite[{section 9.3}]{Simmons-Duffin:2016gjk}. The same conclusion can be reached using the OPE \cite{Costa:2011dw}.} The actual form of this ``Casimir equation" is most conveniently found using the embedding formalism \cite{DO2}. In the $z,\bar{z}$ coordinates of Eq.~(\ref{eq:zzb}) it takes the form
\begin{eqnarray}\label{eq:cb_diffeq}
&&  \mathcal D\, g^{\Delta_{12},\Delta_{34}}_{\Delta,\ell}(z,\bar{z}) =  C_{\Delta,\ell}\,\, g^{\Delta_{12},\Delta_{34}}_{\Delta,\ell}(z, \bar{z})\,, 
\end{eqnarray}
where
\begin{gather}
 \mathcal D = \mathcal D_z + \mathcal D_{\bar{z}} + 2(d-2) \frac{z \bar{z}}{z-\bar{z}} [(1-z)\partial_z - (1-\bar{z})\partial_{\bar{z}} ]\,, \nonumber\\
 \mathcal D_z = \textstyle 2 z^2(1-z)\partial_z^2 -(2+\Delta_{34}-\Delta_{12}) z^2\partial_z + \frac{\Delta_{12}\Delta_{34}}{2}z\,.
\end{gather}

{Moreover}, the leading $z,\bar{z}\to 0$ behavior of the conformal block can be easily determined using the OPE, and this provides boundary conditions for Eq.~\reef{eq:cb_diffeq}. Considering the $x_{12},x_{34}\rightarrow 0$ limit in Eq.~\reef{eq:CPW} and using Eqs. (\ref{eq:2points}) and \reef{eq:OPEnorm}, one obtains\footnote{The limit is worked out carefully in e.g.~\textcite{DO1} or \textcite{Costa:2011dw}.}
\beq
\label{eq:cb_ope}
g^{\Delta_{12},\Delta_{34}}_{\Delta,\ell}(z, \bar{z})
\underset{z,\bar{z}\rightarrow0}{\sim}
\calN_{d,\ell} \, (z\bar{z})^{\frac\Delta2} {\rm Geg}_\ell\left(\frac{z+\bar{z}}{2\sqrt{z\bar{z}}}\right)\,,
\eeq
where ${\rm Geg}_\ell(x)$ is a Gegenbauer polynomial,
\beq
\label{eq:Geg}
{\rm Geg}_\ell(x)=C^{(d/2-1)}_\ell(x)\,,
\eeq
and the normalization factor $\calN_{d,\ell}$ is given by\footnote{Here $(a)_n$ stands for the Pochhammer symbol.}
\beq
\calN_{d,\ell} = \frac{\ell!}{{(-2)^\ell}(d/2-1)_\ell}\,.
\eeq
We warn the reader that many different {normalization choices can be found} in the literature. Different conformal block normalizations correspond to different normalizations of OPE coefficients as compared with the one in Eq.~\reef{eq:OPEnorm}. In this review we will use the above normalization unless {mentioned} otherwise. For the reader's convenience, we have collected some other frequently used normalizations in Table~\ref{tab:cb_norm}.

\begingroup
\squeezetable
\begin{table}[htp]
\begin{center}
\renewcommand{\arraystretch}{2}
\begin{tabular}{|c|c|}
\hline
$\calN_{d,\ell}$ & Reference \\
\hline
 $\frac{\ell!}{(-2)^\ell(d/2-1)_\ell}$ & 
 \begin{minipage}{0.37\textwidth}
 	\textcite{DO1,DO2}, \\
 	\textcite{Rattazzi:2008pe}, \\
 	\textcite{Penedones:2015aga}, {\bf this review} 
 \end{minipage}\\
    \hline
 $ \frac{\ell!}{(d-2)_\ell}$ &  
 \begin{minipage}{0.37\textwidth} 
 	\textcite{DO3},\\ 
 	\textcite{Hogervorst:2013sma}, \\ 
 	\textcite{ElShowk:2012ht,El-Showk:2014dwa}, \\ 
	\textcite{Costa:2016xah}, \\
 	\texttt{JuliBoots} \cite{Paulos:2014vya}, \texttt{cboot} \cite{CBoot}
 \end{minipage}\\
  \hline
 $ \frac{(-1)^\ell \ell!}{4^{\Delta}(d/2-1)_\ell}$ & \begin{minipage}{0.37\textwidth} \textcite{Kos:2014bka,Kos:2015mba,Kos:2016ysd}, \textcite{Li:2017ddj}\\\texttt{PyCFTBoot} \cite{Behan:2016dtz}\end{minipage} \\
 \hline
  $\frac{\ell!}{(d/2-1)_\ell}$ & \begin{minipage}{0.37\textwidth} \textcite{Poland:2011ey}, \textcite{Poland:2015mta} \end{minipage}\\
 \hline
 $ \frac{\ell!}{4^{\Delta}(d-2)_\ell}$ &\begin{minipage}{0.37\textwidth}  \textcite{Kos:2013tga}\\ {\texttt{Mathematica} notebook \cite{DSDnotebookLink}} \end{minipage} \\
 \hline
 $ \frac{(-1)^\ell\ell!}{(d/2-1)_\ell}$ & \textcite{Simmons-Duffin:2016wlq}\\
 \hline
\end{tabular}
\end{center}
\caption{Summary of various conformal block normalizations $\calN_{d,\ell}$, Eqs.~(\ref{eq:cb_ope}, \ref{eq:cb_ope_e1}), used in the literature. \label{tab:cb_norm}}
\end{table}
\endgroup
 
By solving Eq.~(\ref{eq:cb_diffeq}) one can find conformal blocks for even $d$ \cite{DO2}. They are expressed in terms of the basic function
\begin{eqnarray}
k_\beta(x) = x^{\beta/2} {}_2F_1\left(\frac{\beta-\Delta_{12}}2,\frac{\beta+\Delta_{34}}2,\beta;x\right)\,,
\end{eqnarray}
which satisfies
\begin{eqnarray}
\mathcal D_x k_\beta(x) = \frac12\beta(\beta-2) k_\beta(x)\,, \quad k_\beta(x) \underset{x\rightarrow0}{\sim} x^{\beta/2}\,.
\end{eqnarray}
In the simplest case of $d=2$, we have $\calD=\calD_z+\calD_{\bar z}$, so the conformal blocks factorize. They take the form\footnote{A partial case of this result was first found in \textcite{Ferrara:1974ny} by another method. See also \textcite{Osborn:2012vt} for general conformal blocks in 2d. Notice that the 2d global conformal blocks discussed here should be distinguished from the Virasoro conformal blocks.}
\begin{multline} \label{eq:cb_d2}
d=2:\quad g^{\Delta_{12},\Delta_{34}}_{\Delta,\ell}(z,\bar{z}) = \frac{1}{(-2)^\ell(1+\delta_{\ell 0})}\\ \times\left(k_{\Delta+\ell}(z)k_{\Delta-\ell}(\bar{z})+z\leftrightarrow \bar{z}\right)\,.
\end{multline}
Results for higher even $d$ can then be found using recursion relations relating blocks in $d$ and $d+2$ dimensions \cite{DO2}. The important case of $d=4$ reads\footnote{This result was first found in \textcite{DO1} by resumming the OPE expansion.}
\begin{multline} \label{eq:cb_d4}
d=4:\quad  g^{\Delta_{12},\Delta_{34}}_{\Delta,\ell}(z,\bar{z})  =\frac{1}{(-2)^\ell}\\
\times \frac{z\bar{z}}{z-\bar{z}} \left(k_{\Delta+\ell}(z)k_{\Delta-\ell-2}(\bar{z})- z\leftrightarrow \bar{z}\right)\,.
\end{multline}

In odd $d$, general closed-form solutions of the Casimir equation are so far unavailable. 
Sometimes, one can get closed-form solutions along the ``diagonal" $z=\bar{z}$, as e.g.~in $d=3$ for all equal external dimensions \cite[{Eqs.~(3.7-3.10)}]{Rychkov:2015lca}. Other {expressions} along the diagonal{,} valid for any $d${,} can be found in~\textcite{Hogervorst:2013kva}. Using these results as a starting point, one can compute derivatives of conformal blocks orthogonal to the diagonal using the Casimir equation, by the Cauchy-Kovalevskaya method, see Sec.~\ref{sec:rational}. The knowledge of these derivatives is usually sufficient for numerical conformal bootstrap applications. Other techniques used to access the conformal blocks numerically will be discussed below.

Finally, let us mention that conformal blocks have simple transformation properties under the interchange of external operators $1\leftrightarrow 2$ and $3\leftrightarrow 4$ \cite{DO1,DO3}:
\begin{gather}
 g^{\Delta_{12},\Delta_{34}}_{\Delta,\ell}(u/v,1/v) = (-1)^\ell v^{\frac{\Delta_{34}}2 } g^{-\Delta_{12},\Delta_{34}}_{\Delta,\ell}(u,v)\nonumber\\
\hspace{3cm}=  (-1)^\ell v^{-\frac{\Delta_{12}}2 } g^{\Delta_{12},-\Delta_{34}}_{\Delta,\ell}(u,v)\,.
 \end{gather}
This follows from the symmetry of the OPE under the same interchange. As a check, the explicit expressions in Eqs.~(\ref{eq:cb_d2}-\ref{eq:cb_d4}) satisfy these relations.

\subsubsection{Radial expansion for conformal blocks}
\label{sec:radial}

While closed-form expressions for conformal blocks in general $d$ are unknown, there exist rapidly convergent power series expansions. Following \textcite{Hogervorst:2013sma}, we will describe a particular conformal frame used to generate such expansions.

\begin{figure}[t]
\begin{centering}
\includegraphics[width=0.6\columnwidth]{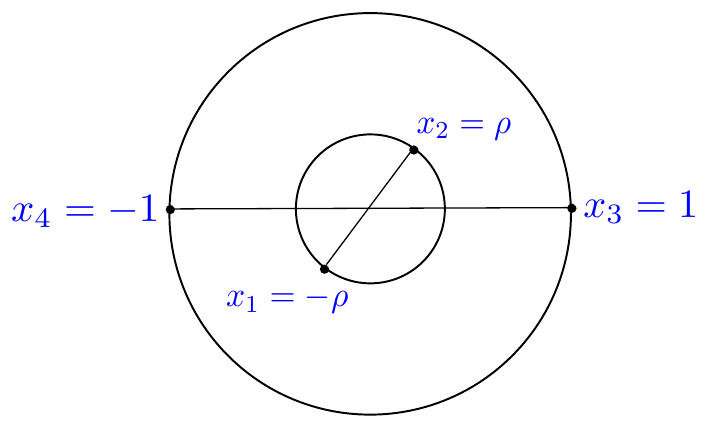}
\caption{\label{fig:radial}
(Color online) Conformal frame defining the radial coordinate. Figure from \cite{Hogervorst:2013sma}.
}
\end{centering}
\end{figure}

Starting from the conformal frame \reef{eq:conformalframe1}, we apply an additional conformal transformation which keeps the four points in the same 2-plane but moves them into a configuration symmetric around the origin as in Fig.~\ref{fig:radial}. So the points $x_1=-x_2$ are now on a circle of radius $r<1$, while $x_3=-x_4$ lie on the unit circle.

Let us call ${\bf n}$ and $\bf{n}'$ the unit vectors pointing to $x_2$ and $x_3$, and introduce the complex \emph{radial coordinate} \cite{Pappadopulo:2012jk}
\beq
\label{eq:reta}
\rho = r e^{i \theta}\,, \qquad {\bf n}\cdot {\bf n}'=  \cos \theta=\eta\,,
\eeq
which is related to the variable $z$ in Eq.~\reef{eq:z} via
\begin{eqnarray}
\rho = \frac{z}{(1-\sqrt{1-z})^2} \,,
\qquad 
z = \frac{4\rho}{(1+\rho)^2}\,.\nonumber
\end{eqnarray}
See \textcite{Hogervorst:2013sma} for why $\rho$ is preferable to $z$ for constructing rapidly convergent power series expansions for conformal blocks. 

In this configuration, the 4pt function is interpreted as a matrix element between two radial quantization states: $\<\phi_3(1,{\bf n}')\phi_4(1,-{\bf n}')|$ and  $|\phi_1(r,-{\bf n})\phi_2(r,{\bf n})\>= r^D|\phi_1(1,-{\bf n})\phi_2(1,{\bf n})\>$. The factor $r^D$, with $D$ the dilatation generator, takes care of the radial dependence.\footnote{$D$ plays the role of the Hamiltonian operator in radial quantization and $\log r$ is time.}

Consider then the conformal partial wave given in Eq.~\reef{eq:CPWrad}.
The conformal multiplet of the operator $\calO_{\Delta,\ell}$ at level $m$ contains descendants $|\Delta+m,j\>$ of spin $j$ varying from $\max(0,\ell-m)$ to $\ell+m$. We need to know the matrix elements between these descendants and the above in and out states. Leaving aside the overall normalization of these matrix elements, their dependence on the unit vector $\bn$
must be proportional to the traceless symmetric tensor $({\bf n}_{\mu_1} \ldots {\bf n}_{\mu_j}-\text{traces})$. Contracting two such tensors for $\bn$ and $\bn'$ {gives}, up to a constant factor, the Gegenbauer polynomial ${\rm Geg}_j({\bf n}\cdot {\bf n'})$ from Eq.~\reef{eq:Geg}.

We conclude that the conformal block has a power series expansion of the form
\beq
\label{eq:radial_exp}
 g^{\Delta_{12},\Delta_{34}}_{\Delta,\ell}(u,v)= r^{\Delta}\sum_{m=0}^\infty   r^{m}
 \sum_{j} w(m,j) \; {\rm Geg}_j(\eta)\,,
\eeq
where $w(m,j)\ne 0$ only for $\max(0,\ell-m)\le j \le \ell+m$. Using unitarity, one can also conclude that $w(m,j)\ge 0$ if $\Delta$ is above the unitarity bound and $\Delta_{12}=-\Delta_{34}$.

Since $z\sim 4\rho$ at small $z$, the OPE limit \reef{eq:cb_ope} becomes
\begin{eqnarray}
\label{eq:cb_ope_e1}
g^{\Delta_{12},\Delta_{34}}_{\Delta,\ell}(r, \eta)\displaystyle \underset{r\rightarrow0}{\sim} 
\calN_{d,\ell}(4r)^{\Delta}\Geg_\ell\left(\eta \right) \,,
\end{eqnarray}
which fixes $w(0,\ell)=\calN_{d,\ell} 4^\Delta$. To find higher $w(m,j)$, one must determine the normalization of the descendant matrix elements and not just their dependence on $\bn,\bn'$. While in principle this can be done using conformal algebra, two more efficient techniques will be discussed below. 

The expansion \reef{eq:radial_exp} converges for $|\rho|<1$, showing that conformal blocks are smooth and real-analytic functions in this region.\footnote{An exception occurs at the origin because of the $r^\Delta$ factor.} The conformal block decomposition \reef{eq:CBdec} can be similarly argued to converge for $|\rho|<1$.\footnote{This can be shown rigorously in unitary CFTs \cite{Pappadopulo:2012jk}. While there are no general results concerning the convergence of conformal block decomposition is nonunitary theories, it appears reasonable to assume that it remains convergent in the same region.} In terms of the $z$ coordinate, this covers the whole complex plane minus the cut $(1,+\infty)$, improving the convergence result argued below Eq.~\reef{eq:CBdec} using the $z$ frame.

\subsubsection{Recursion relation from the Casimir equation}
\label{sec:casimirrecursion}

The first method to find the coefficients $w(m,j)$ is to substitute the expansion \reef{eq:radial_exp} into the Casimir equation.
This gives rise to recurrence relations, obtained in \textcite{Hogervorst:2013sma} and \textcite{Costa:2016xah}, which determine $w(m,j)$ for $m>0$ starting from $w(0,\ell)$.

Namely, defining the functions $ f_{m,j} \equiv r^m \Geg_j(\eta)$, it is straightforward to show that any of the operators $\{r,\eta,\partial_r,\partial_\eta\}$ acting on these functions produces linear combinations of $f_{m\pm1,j\pm1}$. Similarly, the operator $\calD$ in \reef{eq:cb_diffeq}, when written in radial coordinates, maps $f_{m,j}$ into a linear combination of $f_{m+\hat{m},j+\hat{\jmath}}$ functions with suitable shifts. Eq.~\reef{eq:cb_diffeq} then gives rise to a relation which can be economically written in the form \cite{Costa:2016xah}
\begin{eqnarray}
\label{recrelScalar}
\sum_{(\hat{m},\hat{\jmath}) \in \mathcal{S}} c(\hat{m},\hat{\jmath}) \; w(m+\hat{m},j+\hat{\jmath})=0 \ ,
\end{eqnarray}
where the set $\mathcal{S}=\{(0,0),(-1,1),(-1,-1),\ldots\}$ contains $30$ points, all of which but the first have $\hat m<0$. The coefficients $c(\hat{m},\hat{\jmath})$ are known functions of the variables $\Delta_{12}$, $\Delta_{34}$, $\Delta$, $\ell$, $d$, $m$, and $j$ \cite[{attached \texttt{Mathematica} notebook}]{Costa:2016xah}. Using Eq.~\reef{recrelScalar}, the coefficient $w(m,j)$ can then be recursively expressed in terms of $w(m',j)$ with $m'<m$.

\subsubsection{Recursion relation from analytic structure}
\label{sec:analyticrecursion}

The second method exploits the analytic structure in $\Delta$ to obtain a recursion relation for the conformal blocks. A similar approach was first applied by \textcite{Zamolodchikov:1985ie, Zamolodchikov:1987} to the 2d Virasoro conformal blocks, by considering them as meromorphic functions of the central charge $c$ or of the scaling dimension $\Delta$. For conformal blocks of external scalars in arbitrary $d$, this idea was introduced in \textcite{Kos:2013tga,Kos:2014bka}. It was formalized and extended to conformal blocks for external operators with spin in \textcite{Penedones:2015aga}. Here we will explain the external scalar case.

Eqs.~\reef{eq:CPWrad}, \reef{eq:proj} provide a convenient starting point for discussing the analytic structure of a conformal block as a function of the exchanged primary dimension $\Delta$. When $\Delta$ is above the unitarity bound, the Gram matrix $G_{\alpha\beta}(\Delta)$ is positive-definite and invertible. However it turns out that for special values of $\Delta=\Delta_A^*$ at or below the unitarity bounds, the Gram matrix becomes degenerate, in the sense that some states are null (i.e.~have zero norm). The conformal block then develops a pole in $\Delta-\Delta^*_A$. Here we will assume that there are only simple poles, as is true for example in {odd $d$}, while for even $d$ simple poles coalesce into double ones, see below. 

The crucial observation is that the residue of the pole is proportional to another conformal block{:}
\begin{equation}
\label{eq:resid}
g^{\Delta_{12},\Delta_{34}}_{\Delta,\ell}(r,\eta) \sim  \frac{R_A}{\Delta-\Delta^*_A}  g^{\Delta_{12},\Delta_{34}}_{\Delta_A,\ell_A}(r,\eta)\,.
\end{equation}
Namely, we identify the first descendant state $\calO_A^{\rm null}$ of $\calO$ which becomes null when $\Delta\to\Delta_A^*$. Let $\Delta_A=\Delta_A^*+n_A$ be its dimension in this limit, and $\ell_A$ its spin. 
It can be shown that $\calO_A^{\rm null}$ is annihilated by $K_\mu$ when $\Delta\to\Delta_A^*$ and so can be thought of as both a descendant and a primary. Consider then a fictitious primary $\calO_A$ which has quantum numbers $(\Delta_A,\ell_A)$ and which is unit-normalized. It is the conformal block of such a primary, with standard normalization \reef{eq:cb_ope_e1}, that appears in the residue. 

To be more precise, consider the rate with which $\calO_A^{\rm null}$ becomes null as $\Delta\to\Delta_A^*$:
\beq
\label{eq:QA}
\<\calO_A^{\rm null}|\calO_A^{\rm null}\>\sim Q_A (\Delta-\Delta_A^*)\,,
\eeq 
with $Q_A$ some constant. When $\calO_A^{\rm null}$ becomes null, all of its descendants become null too, with the rate proportional to \reef{eq:QA}. Moreover, it can be shown that the Gram matrix in the submultiplet consisting of these descendants is equal to \reef{eq:QA} times the (non-singular) Gram matrix of the multiplet of $\calO_A$, up to corrections of higher order in $\Delta-\Delta_A^*$. This explains why the residue in \reef{eq:resid} involves the whole conformal block of $\calO_A$.\footnote{The Casimir equation gives another argument for why the residue is a conformal block. Near the pole the Casimir equation for the block reduces to the Casimir equation for the residue \cite{Rychkov-OIST}. The Casimir eigenvalue of the null descendant is the same as for the original block (since it's a descendant): $C_{\Delta_*,\ell}=C_{\Delta_A,\ell_A}$. Finally, the boundary condition at $r\to0$ is consistent with the residue being the conformal block.}

The coefficient $R_A$ in $\reef{eq:resid}$ is a product of three factors:
\begin{equation}
 R_A=M_A^{(L)} Q^{-1}_A M_A^{(R)} \ ,
 \label{QMM}
\end{equation}
where $Q_A$ is defined in \reef{eq:QA}, while $M_A^{(L)}$ and $ M_A^{(R)}$ come from the 3pt functions 
 $\langle \phi_3\phi_4| \calO_A^{\rm null} \rangle $ and $\langle  \calO_A^{\rm null} | \phi_1 \phi_2  \rangle$. 

Using information about the poles, we can now write a complete formula for the conformal block. It is convenient to define the regularized conformal block $h_{\Delta,\ell} \equiv h^{\Delta_{12},\Delta_{34}}_{\Delta,\ell}$ by removing a $(4r)^\Delta$ prefactor:
\begin{equation}
g^{\Delta_{12},\Delta_{34}}_{\Delta, \ell}(r,\eta)= (4r)^{\Delta} h_{\Delta, \ell}(r,\eta) \,.
\end{equation}

The {function} $h_{\Delta, \ell}$ has the same poles in $\Delta$ as $g_{\Delta,\ell}$. Moreover it is a meromorphic function of $\Delta$, and is therefore fully characterized by its poles and the value at infinity:
\begin{multline}
\label{eq:recscalar}
h_{\Delta,\ell}(r,\eta)=h_{\infty,\ell}(r,\eta)\\
+\sum_{A} \frac{R_A }{\Delta-\Delta^*_A} (4 r)^{n_A}\,
h_{\Delta^*_A+n_A,\ell_A}(r,\eta)\,.
\end{multline}
Detailed analysis shows that the poles occurring in this equation organize into one finite and two infinite sequences:
\begin{equation}\label{eq:polestable}
\begin{array}{lccc}
A&\Delta^*_A &n_A &\ell_A	 \\ 
\hline
\mbox{I${}_n$ }\ \  (n\in\N)		& 1-\ell-n 	& n 	  &\ell + n \\ 
\mbox{II${}_n$ }\  (1\le n\le \ell) \;\;  	&   \ell+d-1-n  	 & n &\ell - n\\
\mbox{III${}_n$ } (n\in\N) \; 	& \frac{d}2-n  & 2n  &\ell       \\ 
\end{array}
\end{equation}
Using this definition, it is easy to check that the residues of the poles themselves are nonsingular (except in even dimensions, see below).

The $h_{\infty,\ell}$ term and the constants $R_{A}$ are given by \cite{Kos:2014bka, Penedones:2015aga} 
	\begin{align}
			h_{\infty, \ell}(r,\eta)&= \textstyle \frac{\left(1-r^2\right)^{1-\frac{d}2}
			\calN_{d,\ell}\Geg_\ell(\eta)}
		{\left(r^2-2 \eta  r+1\right)^{\frac{1-\Delta_{12}+\Delta_{34}}{2}} \left(r^2+2 \eta  r+1\right)^{\frac{1+\Delta_{12}-\Delta_{34}}{2}}}\,, \nn\\
		R_{\text{I}_n}&=\textstyle \frac{-n (-2)^n }{  (n!)^2} \left(\frac{\Delta_{12}+1-n}{2}\right)_n \left(\frac{\Delta_{34}+1-n}{2}\right)_n \  ,\nn \\
		R_{\text{II}_n}&= \textstyle\frac{-n \,\ell!  }{ (-2)^n (n!)^2 (\ell-n)!   } \frac{(d+\ell-n-2)_n}{\left(\frac{d}2+\ell-n\right)_n
		\left(\frac{d}2+\ell-n-1\right)_n}  	     \label{eq:residues}\\
		&\quad\textstyle\times \left(\frac{\Delta_{12}+1-n}{2}\right)_n  \left(\frac{\Delta_{34}+1-n}{2}\right)_n \ ,\nn \\
		R_{\text{III}_n}&=\textstyle\frac{-n (-1)^{n} \left(\frac{d}2-n-1\right)_{2 n}}{(n!)^2 \left(\frac{d}2+\ell-n-1\right)_{2 n} \left(\frac{d}2+\ell-n\right)_{2 n}}  \nn\\
		&\quad\times\textstyle\left( \frac{\Delta _{12}-\frac{d}2-\ell-n+2}{2} \right)_n \left( \frac{\Delta _{12}+\frac{d}2+\ell-n}{2} \right)_n \nn\\
		&\quad\times\textstyle \left( \frac{\Delta _{34}-\frac{d}2-\ell-n+2}{2} \right) _n \left( \frac{\Delta _{34}+\frac{d}2+\ell-n}{2} \right)_n  \ . \nonumber
	\end{align}

The key property of Eq.~(\ref{eq:recscalar}) is that each pole residue comes with a factor $r^{n_A}$. This means that it can be used as a recursion relation to generate the regularized conformal block as a power series in $r$. Indeed, suppose we want to compute $h_{\Delta,l}(r,\eta)$ up to $O(r^{N})$. We use Eq.~(\ref{eq:recscalar}) keeping all poles with $n_A\le N$, of which there are finitely many. The residues of these poles themselves are needed up to smaller order $O(r^{N-n_A})$, so we get a recursion relation. This is one of the most elegant and efficient currently known methods to compute the conformal blocks {outside of even $d$.}

The described recursion relation is adequate for computing conformal blocks in odd dimensions and also in generic $d$. It cannot be {applied} directly in even $d$, since some simple poles coalesce into double poles. This is not a problem, since even $d$ conformal blocks are known in closed form. Alternatively, one can apply the recursion relation $\eps$ away from an even $d$, and take the limit $\eps\to0$ after the coefficients of the $r$ expansion have been generated. This gives the correct result because the conformal blocks vary analytically with $d$.

\subsubsection{Rational approximation of conformal blocks and their derivatives}
\label{sec:rational}

We will now describe {how to construct} rational approximations to conformal blocks and their derivatives at a given point $(r_*,\eta_*)$, {which permit an efficient numerical evaluation of these quantities as a function of $\Delta$. This will play an important role in the numerical techniques described in Sec.~\ref{sec:numerical}.} Our focus will be on rational approximations to scalar conformal blocks, but later in Sec.~\ref{sec:spinning} we will also describe how they can be extended to blocks for external spinning operators.

A rational approximation for conformal block derivatives at a given point can be obtained by combining the radial expansion (\ref{eq:radial_exp}) and the recursion relation (\ref{eq:recscalar}). It can be expressed in the form
\beq
\label{eq:rationalApprox}
\partial_r^m \partial_\eta^n g_{\Delta,\ell}(r_*,\eta_*) =(4 r_*)^\Delta \left(\frac{P_N^{mn}(\Delta)}{Q_N(\Delta)} + O(r_*^{N-m}) \right)\,.
\eeq
Here $Q_N$ is a polynomial made by the product of poles given in Eq.~(\ref{eq:polestable}) up to order $N$,
\begin{eqnarray}
Q_N(\Delta) 
= 
\prod
_{
	A=(\text{I,II,III})
	{}_n
	,\ n\le N
} 
(\Delta-\Delta_A^*)\,,
\end{eqnarray} 
and $P_N^{mn}(\Delta)$ is a polynomial with ${\rm deg}(P_N^{mn})\leq{\rm deg}(Q_N)+m$. The approximation can be made arbitrarily precise by increasing $N$, at the expense of increasing the order of the polynomials.

In numerical applications it is often desirable to keep the polynomial order relatively small while maintaining a precise approximation. This can be accomplished using a trick introduced in~\textcite{Kos:2013tga}, where one discards some number of poles but compensates by modifying the residues of the kept poles. For example, if one keeps $n$ poles, one can choose their new residues by demanding that the first $n/2$ $\Delta$-derivatives match between the old and new functions at both {the unitarity bound} and $\Delta = \infty$. 

An important property that will be exploited in Sec.~\ref{sec:numerical} is that the denominator $Q_N(N)$ is always positive in unitary theories. This follows from the fact that all the poles are at values of $\Delta$ below the unitarity bound.  

The techniques introduced in the previous sections allow one to compute conformal blocks either in closed form or as {a} power series in the variable $r$. Starting from these expressions one can take a direct approach of first analytically computing the $r$ expansion to order $N$, taking $r,\eta$ derivatives of the resulting expression, and evaluating the result at the point $r_*,\eta_*$. The result can then be recombined to the form in Eq.~(\ref{eq:rationalApprox}). Since the crossing relations will be more simply written in $z,\bar{z}$ coordinates, one then typically converts to $z,\bar{z}$ derivatives at the corresponding point $z_*, \bar{z}_*$ using a suitable transformation matrix. This approach, while somewhat inefficient at large $N$ due to the need to compute the analytical dependence on $\eta$, has been successfully used in the literature, almost always at the crossing symmetric point $z_*=\bar{z}_*=1/2$ which corresponds to $\eta_*=1,r_*=3-2\sqrt{2}$.

{A somewhat more efficient algorithm is the following: \\
(i) Compute the $r$ expansion to order $N$ and take derivatives only along the radial direction $\eta=1$ ($z=\bar{z}$) using either the methods of Sections \ref{sec:casimirrecursion} or \ref{sec:analyticrecursion}.\footnote{In even dimensions one can start from the closed form expression of Sec.~\ref{sec:casimir} evaluated at $\eta=1$, and expand in $r$.} \\
(ii) Convert to $z,\bar{z}$ derivatives along the diagonal $z=\bar{z}$ using a suitable transformation matrix. \\
(iii) Use the Casimir equation to recursively compute derivatives in the transverse direction. }

Let us briefly discuss the last step, also called Cauchy-Kovalevskaya method. Consider the Casimir differential equation, Eq.~(\ref{eq:cb_diffeq}), and express it in the variables $a=z+\bar z,\,\sqrt b=(z-\bar z)$. The radial direction corresponds to $b=0$. Moreover, since conformal blocks are symmetric in $z\leftrightarrow \bar z$, their power series expansion away from the $z=\bar z$ line will contain only integer powers of $b$.
Let us denote the $\partial_a^m\partial_b^n$ derivative of the conformal block evaluated at $z=\bar z=1/2$ by $h_{m,n}$. From step (i) we know $h_{m,0}$ for any $m$.  Then, we can translate the Casimir equation into a recursion relation for $h_{m,n}$ (with $n>0$) in terms of $h_{m,n}$ with lower values of $n$. This recursion relation was obtained in {\cite[{Appendix C}]{ElShowk:2012ht} for $\Delta_{12}=\Delta_{34}=0$, and generalized in \cite[{Eq. (2.17)}]{Behan:2016dtz}.} It has the general structure: 
\begin{align}\label{eq:Cauchy-Kovalevskaya}
	h_{m,n}&=\sum_{m'\le m-1} m (\ldots) h_{m',n} \\
	&+\sum_{m'\le m+2}\left[ (\ldots) h_{m',n-1} + (n-1) (\ldots) h_{m',n-2} \right]\,.\nonumber
\end{align}
Since the Casimir equation is of second order, $m'$ can only take values up to $m+2$. Also the recursion relation for $h_{0,n}$ only involves $h_{m',n'}$ with {$n'<n$}. 
Eq.~(\ref{eq:Cauchy-Kovalevskaya}) is then all that we need to perform step (iii).

We conclude this section by mentioning a few software packages that implement the above efficient algorithm. Their functionality for solving convex optimization problems will be discussed in Sec.~\ref{sec:numerical}, so here we focus on how they compute conformal blocks.

A \texttt{Mathematica} notebook by \textcite{DSDnotebookLink} can be used for general scalar conformal blocks; at step (i) it implements the recursion relation from analytic structure discussed in Sec.~\ref{sec:analyticrecursion}. It also implements the trick of shifting pole residues described above. 

Another \texttt{Mathematica} notebook \textcite{JuliBoots} can also be used for general scalar conformal blocks. At step (i) it implements the recursion relation from the Casimir equation discussed in Sec.~\ref{sec:casimirrecursion}. This notebook accompanies the \texttt{Julia} package \texttt{JuliBoots} for bootstrap computations using linear programming \cite{Paulos:2014vya}.

A \texttt{Python} package \texttt{PyCFTBoot} \cite{Behan:2016dtz} and a \texttt{Sage} package \texttt{cboot} \cite{CBoot} contain integrated functions that compute general scalar conformal blocks derivatives using the above procedure. These two packages are designed as frontends to the semidefinite program solver \texttt{SDPB} \cite{Simmons-Duffin:2015qma}.

\subsubsection{Shadow formalism}
\label{sec:shadow}

Next we will briefly review the shadow formalism, which was historically the very first technique to access the conformal blocks \cite{Ferrara:1973vz}, and it continues to play a role conceptually and also in explicit computations.

Suppose we want to compute the conformal block $g_{\Delta,\ell}$ of a primary operator $\calO_{\Delta,\ell}$ in a scalar 4pt function $\<\phi_1\phi_2\phi_3\phi_4\>$. Consider a primary ``shadow operator" $\widetilde{\calO}_{d-\Delta,\ell}$ which has the same spin $\ell$ as $\calO$ and dimension $d-\Delta$. We stress that this operator is fictitious, it does not belong to the theory as a local operator, and in particular {the fact} that its dimension is below the unitarity bound is of no concern.

The starting point of the shadow formalism is the following integral:
\begin{align}\label{eq:partial_wave}
{\rm U}_{\Delta,\ell}(x_1,x_2,x_3&,x_4) = \int d^d x \langle \phi_1(x_1)\phi_2(x_2)\mathcal O_{\Delta,\ell}^{\mu_1\ldots\mu_\ell}(x)\rangle \nn\\
 &\times\langle \widetilde{\mathcal O}_{d-\Delta,\ell;\mu_1\ldots\mu_\ell}(x)  \phi_3(x_3)\phi_4(x_4)\rangle\,,
\end{align}
where under the integral sign we have a product of the conformal scalar-scalar-(spin $\ell$) 3pt functions in Eq.~\reef{eq:3points}, with the spin-$\ell$ operators having dimensions $\Delta$ and $d-\Delta$.

The function ${\rm U}_{\Delta,\ell}$ has two special properties. First, it conformally transforms in the same way as the 4pt function $\langle \phi_1(x_1)\phi_2(x_2)\phi_3(x_3)\phi_4(x_4)\rangle$.
This is because the product (operator $\times$ shadow) transforms as a dimension $d$ primary scalar, which compensates for the Jacobian in the transformation of $d^d x$. Consequently we can write
\beq
\label{eq:Uf}
{\rm U}_{\Delta,\ell} = f_{\Delta,\ell}(u,v)\,{\bf K}_4 \,,
\eeq
where ${\bf K}_4$ is as in Eq.~(\ref{eq:4pf_scalars}) and $f_{\Delta,\ell}(u,v)$ is some function of $u$ and $v$.

Second, it is straightforward to see that ${\rm U}_{\Delta,\ell}$ is an eigenfunction of the Casimir operator acting at $x_1$, $x_2$, with eigenvalue $C_{\Delta,\ell}$. Since the latter property is also true for the CPW ${\rm W}_\calO$, it is tempting to identify $f_{\Delta,\ell}(u,v)$ in \reef{eq:Uf} with the conformal block (up to a proportionality factor). However, this is not quite true. The point is {that} the conformal blocks of the operator and of its shadow satisfy the same Casimir equation, since their Casimir eigenvalues coincide: $C_{\Delta,\ell}=C_{d-\Delta,\ell}$. For this reason $f_{\Delta,\ell}$ is a linear combination of the block $g_{\Delta,\ell}$ and of the shadow block $g_{d-\Delta,\ell}$; see \cite[{Eq. (3.25)}]{DO3} for the precise relation. 

From the practical viewpoint, the main advantage of the shadow formalism is that the integrand in Eq.~\reef{eq:partial_wave} is quite easy to write. The downside is that the resulting conformal integrals are not always easy to evaluate, and that it is necessary to disentangle the contribution of a proper conformal block from the shadow one. 

Efficient ways to deal with these problems were proposed by \textcite{SimmonsDuffin:2012uy}. First of all, the integrals become much easier to evaluate when written using the embedding formalism. Second, to separate the block from the shadow one uses that they transform differently under a monodromy transformation
\begin{gather}
z\rightarrow e^{2\pi i} z,\quad \bar z=\text{fixed}\,,\\
g_{\Delta,\ell}(z,\bar z) \rightarrow e^{2\pi \Delta i} \,g_{\Delta,\ell}(z,\bar z)\,.
\end{gather}
The wanted conformal block is isolated via a \emph{monodromy projector}, implemented as a proper choice of the integration contour in Eq.~(\ref{eq:partial_wave}).
This prescription allows one to extract integral expressions for generic conformal blocks in arbitrary $d$. In some cases the conformal integrals can be performed exactly, and the results match the known formulas from other techniques.

\subsubsection{Spinning conformal blocks}
\label{sec:spinning}

Although in this review we will mostly deal with scalar 4pt functions, the bootstrap has also been successfully applied to 4pt functions of operators {with spin; e.g., see Secs.~\ref{sec:Fermions} for $j=1/2$ spinors and \ref{sec:JandT} for $\ell=1,2$ tensors in 3d.} Here we will review the theory of the associated conformal blocks, referred to as ``spinning", which present additional difficulties compared to the blocks of external scalars. 

As in the scalar case, spinning conformal blocks correspond to the contribution of an entire conformal multiplet to a 4pt function. They are defined by the equation
\begin{align}
\label{eq:spinning_cb}
&\langle \calO_3 \calO_4 | \mathcal P_{\Delta,r} |\calO_1\calO_2 \rangle 
=\\  
&{\bf K}_4 \sum_{a=1}^{n_3}\sum_{b=1}^{n'_3} \sum_{c=1}^{n_4} \lambda_{12 \mathcal O^\dagger}^{(a)} \lambda_{34 \mathcal O}^{(b)}{\bf T}_{4}^{(c)}(x_i,\zeta_i) {\bf G}_{c,\Delta,r}^{a,b}(\Delta_i,r_i,u,v).\nn
\end{align}

Here, the external operators $\calO_i=\calO_{\Delta_i,r_i}(x_i,\zeta_i)$ are positioned {at} $x_i$ and have their indices contracted with auxiliary polarization vectors (or spinors) $\zeta_i$. They transform in some general $SO(d)$ {(or $Spin(d)$)} representations $r_i$. On the other hand $\mathcal O_{\Delta,r}$ is the exchanged operator (and $\mathcal O_{\Delta,r^\dagger}^\dagger$ its conjugate, see the discussion in Sec.~\ref{sec:2pt}), and $\mathcal P_{\Delta,r}$ is the projector onto its conformal multiplet similar to Eq.~(\ref{eq:proj}).

The prefactor ${\bf K}_4$ is as in Eq.~(\ref{eq:4pf_scalars}); it captures the scaling properties of the 4pt function, leaving everything else dimensionless. Eq.~\reef{eq:spinning_cb} also contains a sum over possible conformally invariant 4pt tensor structures ${\bf T}_4^{(c)}$, and a double sum over possible 3pt function structures
\begin{eqnarray}
\label{eq:3pt-spinning}
&&\langle \mathcal O_{\Delta_1,
r_1}(x_1,\zeta_1) \mathcal O_{\Delta_2,
r_2}(x_2,\zeta_2) \mathcal O^\dagger_{\Delta,
r^\dagger}(x_3,\zeta_3)\rangle =\\
&&\hspace{3em}\displaystyle \sum_{a=1}^{n_3} \lambda_{12 \mathcal O^{\dagger}}^{(a)} {\bf T}_3^{(a)}(x_i,\zeta_i,\{\Delta_1,r_1\},\{\Delta_2,r_2\},\{\Delta,r^{\dagger}\}),\nn
\end{eqnarray}
and similarly for $n'_3$. Finally, the functions ${\bf G}_{c,\Delta,r}^{a,b}(\Delta_i,r_i,u,v)$ are the spinning conformal blocks. 

According to the above definition, when $r$ is not a real representation, both ${\bf G}_{c,\Delta,r}^{a,b}$ and ${\bf G}_{c,\Delta,r^\dagger}^{a,b}$ have to be considered and generally both these blocks are nonzero. We will see a 4d example for $r=(\ell,\ell+p)$ below.

Spinning blocks can be computed by reducing them to ``seed" blocks. Consider the simplest case when the exchanged primary is a traceless symmetric spin $\ell$. To understand the reduction to seeds, the key observation is that the 3pt tensor structures \reef{eq:3pt-spinning} can be produced by differentiating the more elementary scalar-scalar-(spin $\ell$) 3pt functions \reef{eq:3points}. Namely \textcite{Costa:2011dw} showed that there exist ``spinning-up" differential operators $D^{(a)}_{r_1,r_2}$, depending on $x_i$ and $\zeta_i$, such that
\begin{eqnarray}
&&{\bf T}_3^{(a)}(x_i,\zeta_i,\{\Delta_1,r_1\},\{\Delta_2,r_2\},\{\Delta,r\}) =\\
&&\hspace{3em} D^{(a)}_{r_1,r_2} \mathcal {\bf T}_3(x_i,\zeta_3,\{\Delta'_1,0\},\{\Delta'_2,0\},\{\Delta,r\}),\nn
\end{eqnarray}
for a suitable basis of 3pt structures and choice of $\Delta_i'$.\footnote{In a {generic} basis of 3pt structures, e.g.~one that would be naturally constructed using the embedding space or conformal frame formalisms, there would be a linear combination of terms like the r.h.s. with different shifts.} Notice that in the above expression the third point is not affected. Therefore, in the definition (\ref{eq:spinning_cb}), the differential operators do not interfere with the sum over descendants in $\mathcal P_{\Delta,\ell}$. One concludes that the spinning blocks can be obtained by differentiating the scalar blocks:
\begin{multline}
\label{eq:spinning_cb_p0}
{\bf K}_4 (\Delta_i)\sum_{c=1}^{n_4} {\bf T}_{4}^{(c)}(x_i,\zeta_i){\bf G}_{c,\Delta,\ell}^{a,b}(\Delta_i,r_i,u,v) \\
= D^{(a)}_{r_1,r_2}  D^{(b)}_{r_3,r_4} {\bf K}_4 (\Delta_i') g_{\Delta,\ell}^{\Delta_{12}',\Delta_{34}'}(u,v)\,,
\end{multline}
where $g_{\Delta,\ell}^{\Delta_{12}',\Delta_{34}'}(u,v)$ is the scalar conformal block discussed at length in the previous sections, referred to as a seed block in this situation. 

In 3d, traceless symmetric tensors exhaust all bosonic $SO(d)$ representations, and therefore all bosonic spinning blocks can be obtained from scalar seeds via \reef{eq:spinning_cb_p0}.\footnote{Some explicitly worked out cases in 3d are for external operator pairs being (current)-(current) \cite{Costa:2011dw}, scalar-(current or stress tensor) \cite{Li:2015itl}, and (stress tensor)-(stress tensor) \cite{Dymarsky:2017yzx}.} The 3d spinning-up operators were also extended to external spinors and exchanged spin $\ell$ by \textcite{Iliesiu:2015qra}. 

If a representation $r$ does not couple to two scalars, its conformal block cannot be reduced to the scalar seed using this method. One therefore needs more seed blocks for such representations. As an example, consider the half-integer spin representations in 3d. The simplest pair of external operators to which they couple are a scalar $\phi$ and a Majorana fermion $\psi$. The corresponding conformal block $ \langle\phi_3 \psi_4 | \mathcal P_{\Delta,j}|\phi_1 \psi_2\rangle$ for half-integer $j$ can be taken as a seed. It was computed by \textcite{Iliesiu:2015akf}, using recursion relations as in Sec.~\ref{sec:analyticrecursion}, making the list of 3d seeds complete. 

A similar discussion holds in 4d. In this case the complete set of seed blocks corresponds to the representations $r=(\ell,\ell+p)$ and $(\ell+p,\ell)$ appearing in the 4pt function of two scalars, one $(p,0)$ tensor, and one $(0,p)$ tensor:
\beq
\langle \phi_3(x_3)\mathcal O_{\Delta_4,(0,p)}(x_4)| \mathcal P_{\Delta,r}|
 \phi_1(x_1)\mathcal O_{\Delta_2,(p,0)}(x_2) \rangle\,.
\eeq
All of these seeds were computed in closed form by \textcite{Echeverri:2016dun}, making use of the shadow formalism from Sec.~\ref{sec:shadow}.

Once the seeds are known, a relation analogous to (\ref{eq:spinning_cb_p0}) allows one to relate any conformal block to a combination of seed blocks thorough a suitable set of spinning-up operators $D^{(a)}_{r_i,r_j}$. The latter can be nicely written in the embedding formalism discussed in Appendix~\ref{sec:embedding} or one of its generalizations. The precise expressions can be found in \textcite{Costa:2011dw,Iliesiu:2015akf} in 3d or \textcite{Echeverri:2015rwa} in 4d. In 4d there is also available a comprehensive {\tt Mathematica} package {\tt CFTs4D} \cite{Cuomo:2017wme} designed to facilitate general spinning 4d conformal block computations. {Spinors and spinor-tensor correlators in aribtrary dimensions were instead studied in \textcite{Isono:2017grm}}

Let us mention briefly several other ideas which have proved useful when dealing with spinning blocks. \textcite{Karateev:2017jgd} introduced a more general class of ``weight-shifting" operators which act on correlation functions. In addition to reproducing the spinning-up operators as a special case, they have a further interesting consequence: when acting on a conformal block these operators can change the $SO(d)$ {(or $Spin(d)$)} representation of the exchanged state {by utilizing the $6j$ symbols of the conformal group. Through} repetitive use of these operators, it is possible to express any conformal block, including the seeds, in terms of the scalar ones.\footnote{Explicit formulas expressing the seed blocks in 3d and 4d are provided in \textcite{Karateev:2017jgd}.} {These methods also lead to efficient derivations of various recursion relations satisfied by the conformal blocks.}

The Casimir recursion approach from Sec.~\ref{sec:casimirrecursion} was extended to arbitrary external bosonic operators by \textcite{Costa:2016xah}. More recently, \textcite{Kravchuk:2017dzd} considered similar expansions for arbitrary external operators, and related the recursion relation coefficients to the $6j$ symbols of $Spin(d-1)$, which are known in closed form for arbitrary representations in {$d=3,4$}, and for representations entering the seed blocks in arbitrary $d$. He also discusses how to convert from the $z$ to the $\rho$ coordinate, as is needed for practical applications. 

The pole expansion of Sec.~\ref{sec:analyticrecursion} has also been generalized to spinning conformal blocks \cite{Penedones:2015aga,Costa:2016xah}. Although no closed form expressions are known for the analogues of $h_\infty$ and of the residues $R_A$ in Eq.~(\ref{eq:recscalar}), these ingredients can sometimes be found by combining this approach with the spinning-up/weight-shifting operators, as in {\textcite{Iliesiu:2015akf}, \textcite{Dymarsky:2017xzb}, and \textcite{Karateev:2017jgd}.} Commuting these operators with the pole expansion sum, one obtains the expected pole expansion for spinning conformal blocks. By truncating the pole expansion, rational approximations similar to those considered in Sec.~\ref{sec:rational} can then be constructed for each spinning block tensor structure.

Finally, the shadow block technique discussed in Sec.~\ref{sec:shadow} has been used to compute the conformal blocks appearing in 4pt function of two scalars and two identical conserved currents \cite{Rejon-Barrera:2015bpa}.

\subsection{Global symmetry}
\label{sec:global}

A majority of CFTs also possess a global symmetry group $G$, which acts on local operators in a way that commutes with conformal transformations.\footnote{If the CFT arises as an IR fixed point of a gauge theory, we work only with gauge-invariant local operators. So, as mentioned in Sec.~\ref{sec:universality}, the gauge group does not enter into our considerations.}  The conformal multiplet is then characterized by an additional label: an irreducible $G$ representation $\pi$ in which the primary transforms. The cases of interest to physics are when $G$ is a finite discrete group or compact Lie group, or a product thereof.  

The correlator of $n$ primaries will then be as discussed in Sec.~\ref{sec:correlations}, times an extra factor which determines the dependence on indices in $G$-representations. This extra factor is a $(\pi_1\otimes\cdots \otimes\pi_n)^G$ tensor, i.e.~a $G$-invariant tensor belonging to the tensor product representation.

By Schur's lemma, the 2pt function can be nonzero only if $\pi_2=\bar\pi_1$ are conjugate representations (or the same representation if self-conjugate), in which case its form is uniquely determined. The three typical cases of a real, pseudoreal, or complex representation are illustrated by the extra factors being $\delta_{ab}$ for $\pi_1=\pi_2$ a fundamental of $SO(N)$, $i\eps_{ab}$ for $\pi_1=\pi_2$ a fundamental of $SU(2)$, and {$\delta_{a}^{\bar b}$} for $\pi_1$($\pi_2$) (anti)fundamentals of $SU(N)$, $N>2$.\footnote{The indices of global symmetry representations will be denoted either by $a,b,\ldots$ or $i,j\ldots$ depending on the situation.} \footnote{In unitary CFTs, complex representations $\pi$ necessarily occur in conjugate pairs, so it's natural to choose an operator basis so that $\calO$, $\calO^\dagger$ transform in $\pi,\bar\pi$.}

The 3pt function can be nonzero only if $(\pi_1\otimes\pi_2\otimes \pi_3)^G$ is nonempty. This leads to selection rules. For example, if $\pi_1$ and $\pi_2$ are fundamentals of $SO(N)$, then $\pi_3$ can be either a singlet or a rank-2 traceless symmetric or antisymmetric $SO(N)$ tensor. The invariant tensors corresponding to these three possibilities can be readily written down \cite{Rattazzi:2010yc}.

Alternatively one can think in terms of the OPE: $\<\calO_{\pi_1}\calO_{\pi_2}\calO_{\pi_3}\>$ is nonzero if $\calO_{\bar\pi_3}^\dagger$ appears in the OPE $\calO_{\pi_1}\times\calO_{\pi_2}$. The global symmetry structure is given by the Clebsch-Gordan coefficient for $\bar\pi_3$ in $\pi_1\otimes\pi_2$. Notice that the tensor product $\pi_1\otimes\pi_2$ may include several copies of a given representation, in which case there may be several different invariant tensors possible in the 3pt function. This is similar to how conformal tensor structures for 3pt functions of primaries are in general nonunique.

The 4pt functions are proportional to nonzero tensors appearing in $(\pi_1\otimes\pi_2\otimes \pi_3\otimes \pi_4)^G$. In a CPW decomposition like Eq.~\reef{eq:4pf_scalars}, individual CPWs will be proportional to the invariant 4pt tensors obtained by contracting the 3pt tensors from $(\pi_1\otimes\pi_2\otimes \pi)^G$ and $(\pi_3\otimes\pi_4\otimes \bar\pi)^G$, where $\pi$ is the exchanged representation. By basic group theory (decomposing $\pi_1\otimes\pi_2$ and $\pi_3\otimes \pi_4$ into irreducibles and applying Schur's lemma), it's easy to see that any 4pt invariant tensor can be obtained in this way for an appropriate choice of $\pi$.

Another interesting possibility for {additional} symmetry is supersymmetry, in which case the conformal group is enhanced to a superconformal group, primary operators are grouped into supersymmetry multiplets, and conformal blocks are enhanced to superconformal blocks. Later in Sec.~\ref{sec:4Dsusy} we will describe in more detail some of the consequences of superconformal symmetry for the bootstrap.

\subsection{Conserved local currents}
\label{sec:ward}

Next we will turn to the conserved currents associated with conformal or global symmetries. Such currents {are supposed to} exist at the IR fixed points of RG flows starting from a microscopic Lagrangian or from a lattice model with finite-range interactions.\footnote{In a classical or weakly-coupled quantum local field theory, the existence of local conserved currents follows from Noether's theorem. We are not aware of a general Noether's theorem for strongly-coupled theories and lattice models. The existence of local conserved currents in these cases remains a physically-motivated assumption, taken for granted in most of the literature. For an intuitive argument for the existence of a local stress tensor using the RG, see \cite[{section 11.3}]{Cardy:1996xt}.} 

\subsubsection{Stress tensor}

In the axiomatic approach considered here, a local CFT is simply defined as a CFT having a local conserved stress tensor operator $T_{\mu\nu}$. In the operator classification, $T_{\mu\nu}$ is a traceless symmetric spin-2 primary of scaling dimension $d$.\footnote{Conformal invariance allows one to consistently impose conservation of the stress tensor. In technical language, the divergence of the dimension $d$ traceless symmetric spin-2 primary is a null descendant and can be set to zero.}

In local CFTs, the conformal algebra generators \reef{eq:algebra} are obtained by integrating the stress tensor against {a vector field $\eps^\calJ_\nu(x)$, describing the corresponding infinitesimal conformal transformation, over a surface $\Sigma$ surrounding the origin. Thus we have}
\begin{eqnarray}
\label{eq:TwardIdentites}
\calJ = -\int_{\Sigma} dS_\mu  \eps^\calJ_\nu(x) T^{\mu\nu}(x)\,,
\end{eqnarray}
which is independent of the shape of $\Sigma$. See \textcite{Simmons-Duffin:2016gjk} for a detailed review of this way of introducing the conformal algebra.\footnote{However, it should be stressed that there are physically interesting theories which satisfy all CFT axioms except for the existence of the local stress tensor. Examples include defect and boundary CFTs (see Sec.~\ref{sec:bdry} and footnote \ref{note:defects}), and critical points of models with long-range interactions, see~e.g.~\textcite{Paulos:2015jfa} and~\textcite{Behan:2017dwr,Behan:2017emf}.} In particular, the dilatation generator $D$ is given by \reef{eq:TwardIdentites} with $\eps^D_\nu=x_\nu$.

It is conventional to normalize the stress tensor via Eq.~\reef{eq:TwardIdentites}. Namely, inserting the above surface operator in any correlator should have the effect of replacing the operator at the origin by $[\calJ,\calO(0)]$, assuming the other operators are outside of the region enclosed by $\Sigma$. This constraint is called an (integrated) Ward identity.

A frequently occurring case is to consider the 3pt function $\<\calO(0) T_{\mu\nu}(x)\calO(y)\>$ which by the Ward identity should reduce to $\<[\calJ,\calO(0)] \calO(y)\>$ after integration. Since $[\calJ,\calO(0)]$ is known, this provides constraints on the coefficients of various tensor structures in the 3pt function. 

These constraints should be imposed in addition to constraints from conservation of $T_{\mu\nu}$. Vanishing of the divergence is automatic for 2pt functions, while in general it must be imposed on 3pt functions containing $T_{\mu\nu}$, placing constraints on the allowed tensor structures. Such constraints are not independent if the other operators are scalars, but become nontrivial if they have spin, see \textcite{Osborn:1993cr} and \textcite{Costa:2011mg}.\footnote{Some important cases are when $\calO$ is a conserved spin-1 vector or the stress tensor itself. In both these cases there are several tensor structures allowed by conformal invariance and conservation, and only one independent Ward identity, see \textcite{Osborn:1993cr} and \textcite{Dymarsky:2017xzb,Dymarsky:2017yzx}. Ward identity constraints on 3pt functions $\<\psi T \bar\psi\>$ with $\psi$ a fermion were studied in 3d by \textcite{Iliesiu:2015qra} and in 4d by \textcite{Elkhidir:2017iov}. In these cases there are two independent tensor structures allowed by conservation, and their coefficients can both be fixed by considering the Ward identity for $D$ as well as for $P_{\mu}$ or $M_{\mu\nu}$.}

In particular, when $\calO = \phi$ is a scalar, there is just one tensor structure. Using the Ward identity e.g.~for $\calJ=D$ one fixes the OPE coefficient completely. In the notation of \reef{eq:3points} we have \cite{Osborn:1993cr}
\begin{gather}
	\langle\phi(x_1) \phi(x_2) T(x_3,\zeta)\rangle =\lambda_{\phi\phi T} [(Z_{123} \cdot \zeta)^2-\half \zeta^2] \, \mathbf{K}_3\,,\nonumber\\
	\lambda_{\phi \phi T} = - \frac{d \Delta_\phi }{(d-1)S_d},\quad S_d = \frac{2\pi^{d/2}}{\Gamma(d/2)} \,.
		\label{eq:TOO1}
\end{gather}
It can also be shown that the stress tensor does not couple to two scalars of unequal dimension, as the 3pt function structure \reef{eq:3points} is then incompatible with conservation.

Since we normalize via Eq.~\reef{eq:TwardIdentites}, the stress tensor 2pt function will not be unit-normalized but will contain a constant $C_T$ called the central charge:\footnote{This corresponds to one of the central charge definitions in $d=2$. Notice however that in $d>2$, there is no known analogue of the Virasoro algebra interpretation of the central charge.}
\begin{eqnarray}
\label{eq:CT}
\langle T(x_1,\zeta_1)T(x_2,\zeta_2)\rangle = \frac{C_T}{S_d^2}\frac{(\zeta_1 \cdot I \cdot \zeta_2)^2-\frac1d \zeta_1^2\zeta_2^2}{(x_{12}^2)^{d}}\,.
\end{eqnarray}
A similar convention will be set below for conserved spin-1 currents, while the rest of primaries are kept unit-normalized. 

In the normalization Eqs.~\reef{eq:TOO1} and~\reef{eq:CT}, the contribution of the stress tensor to 4pt functions of scalars is given by:\footnote{This is easy to find by rescaling $T_{\mu\nu}$ to match the normalization in Eq.~(\ref{eq:2points}).} 
\begin{eqnarray}
&&\langle\phi(x_1)\phi(x_2) \phi'(x_3)\phi'(x_4)\rangle \supset  p_{d,2}\, g^{0,0}_{d,2}(u,v) \, \mathbf K_4\,,\nonumber\\
&&p_{d,2} = \lambda_{\phi\phi T}  \lambda_{\phi'\phi' T}\frac{S_d^2}{ C_T} = \frac{d^2}{(d-1)^2}\frac{\Delta_\phi \Delta_{\phi'}}{C_T}\,.
\label{eq:lambdaT}
\end{eqnarray}
As usual, the conformal block is normalized according to Eq.~(\ref{eq:cb_ope}). This constraint can play an important role in bootstrap analyses involving multiple 4pt functions, as it implies that the stress tensor contributes to different 4pt functions in a correlated way. 

While outside of 2d there is no analogue of the ``$c$-theorem"~\cite{Zamolodchikov:1986gt} for $C_T$,\footnote{Instead, it is known that in 3d the sphere free energy satisfies an ``$F$-theorem", see \textcite{Jafferis:2011zi}, \textcite{Klebanov:2011gs}, and \textcite{Casini:2012ei}, while in 4d the $a$ anomaly coefficient satisfies an ``$a$-theorem", see \textcite{Cardy:1988cwa}, \textcite{Osborn:1989td}, \textcite{Jack:1990eb}, \textcite{Komargodski:2011vj}, and \textcite{Komargodski:2011xv}.}  the central charge typically scales with the number of degrees of freedom. This is illustrated by the values of the central charge of a free theory containing $n_\phi$ scalars, $n_\psi$ Dirac fermions, and $n_A$ gauge vectors (in 4d only), given by~\cite{Osborn:1993cr}
\beq
C_T= \frac{d}{d-1} n_\phi +2^{\left\lfloor{d/2}\right\rfloor-1} d \, n_\psi +16 \,\delta_{d,4}\, n_A \,.
\eeq

\subsubsection{Global symmetry currents}

The case of a continuous global symmetry in a local CFT is analogous. In this case there are conserved spin-1 currents $J_\mu^A$ which transform in the adjoint representation of $G$ {and have scaling dimension $d-1$.} Global symmetry generators are obtained by integrating $J_\mu^A$ {over} a surface, which defines a normalization for the current and leads to Ward identities. 

For concreteness, consider scalar operator $\phi_i$ with generators $(T^A)_{i}^{j}$  transforming in some representations $r$ of $G$ as well as $\phi^\dagger{}^{j}$ transforming in $\bar r$. We assume that the scalar 2pt function is unit-normalized{, $\langle\phi_i \phi^{\dagger}{}^{j} \rangle \propto \delta_{i}^{j}$,} as discussed in Sec.~\ref{sec:global}. The generators of the global symmetry transformations are {then} $Q^A=-i\int_\Sigma dS^\mu J^A_\mu $ and the associated Ward identity requires $[Q^A,\phi_i] = -(T^A)_{i}^{j} \phi_j$. The 3pt function with $J^A$ is then fixed to be \cite{Osborn:1993cr,Poland:2010wg}
\beq
\label{eq:JOO}
\langle \phi_i(x_1) \phi^{\dagger}{}^{j}(x_2)  J^A(x_3,\zeta)\rangle =-\frac{i}{S_d} (T^A)_{i}^{j} [Z_{123} \cdot \zeta ] \, \mathbf{K}_3 \,.
\eeq
In this normalization one can define a \emph{current central charge} $C_J$ by
\begin{eqnarray}\label{eq:CJ}
&& \langle J^A(x_1,\zeta_1)J^B(x_2,\zeta_2)\rangle =\tau^{AB}\frac{C_J}{S_d^2} \frac{\zeta_1 \cdot I \cdot \zeta_2}{(x_{12}^2)^{d-1}}\,,\hspace{2em} 
\end{eqnarray}
where $\tau^{AB}= {\rm Tr}\left[T^{A}T^{B}\right]$. 

In the end, rescaling $J^A_\mu$ to match the normalization of Eq.~(\ref{eq:2points}), the contribution of a spin-1 conserved current to the scalar 4pt function is
\begin{eqnarray}
&&\langle \phi_i(x_1) \phi^{\dagger}{}^{j}(x_2) \phi_k(x_3) \phi^{\dagger}{}^{l}(x_4) \rangle \supset - \frac{\mathcal T_{ik}^{jl} }{C_J } \, g_{d-1,1}(u,v) \,\mathbf K_4 \,,
\nonumber\\
&&\mathcal T_{ik}^{j l}  =  (\tau^{-1})_{AB}(T^A)_{i}^{j}(T^B)_{k}^{l}\,.
\label{eq:lambdaJ}
\end{eqnarray}
Notice that $\tau^{AB}$ in \reef{eq:CJ}, $(T^A)_{i}^{j}$ in \reef{eq:JOO}, and $\calT_{ik}^{jl}$ are examples of 2pt, 3pt, {and} 4pt $G$-invariant tensors as discussed in Sec.~\ref{sec:global}.

For example, if $\phi$ is a complex scalar charged under a $U(1)$ with charge 1, then $\mathcal T_{ik}^{jl}=\mathcal T=1$. In the case in which $\phi_i$ is in the fundamental representation of $SU(N)$ or $SO(N)$ (where $\bar{r} = r$) we have instead 
\begin{align}
	\mathcal T_{ik}^{jl} &=  \delta_i^{l} \delta_k^{j} - \frac{1}N \delta_i^{j} \delta_k^{l} \qquad  &(G =  SU(N))\,,\\
	\mathcal T_{ijkl} &= \frac12\left(\delta_{il} \delta_{kj} -  \delta_{ik} \delta_{jl}\right) \qquad &(G =  SO(N)) \,.
\end{align}

A note about normalization is in order: once the generators $T^A$ are chosen, the Ward identity fixes the normalization of $J^A$ and determines $C_J$ according to our definition. Clearly, if we use a different generator normalization, then the value of $C_J$ would be modified accordingly. Moreover, once Eq.~(\ref{eq:JOO}) is established, the Ward identity fixes the normalization of any other generator in any other representation. 

Finally, it should be mentioned that while free theories contain higher-spin conserved currents, there exist no-go theorems showing that interacting CFTs in $d\ge3$ dimensions do not have conserved currents of spin $\ell\ge 3$, see \textcite{Maldacena:2011jn} and \textcite{Alba:2013yda}. This can be thought of as a CFT analogue of the Coleman-Mandula theorem for S-matrices.

\subsection{Crossing relations}
\label{sec:crossing}

The main idea of the conformal bootstrap is to constrain CFT data by using the crossing relations for 4pt functions, Fig.~\ref{fig:bootstrap}. Crossing relations are usually analyzed in the conformal frame of Fig.~\ref{fig:z-frame}. Consider the 4pt function of scalar operators in this frame and expand it into conformal blocks in the (12)-(34) and in the (32)-(14) OPE channels, referred to as the s- and t-channels. The two channels are obtained by interchanging points 1 and 3, which transforms $z\to 1-z$. Taking into account the value of the ${\bf K}_4$ factor in both channels, and equating the two CPW decompositions, we get the crossing relation
\begin{multline}
	\label{eq:cross-gen}
		\sum_{\calO}\lambda_{12\calO}\lambda_{34\calO}  \frac{g^{\Delta_{12},\Delta_{34}}_{\Delta_{\calO},\ell_{\calO}}(z,\bar z)}
		{(z\bar{z})^{\frac{\Delta_1+\Delta_2}2}}\\
		=\sum_{\calO'}\lambda_{32\calO'}\lambda_{14\calO'}  \frac{g^{\Delta_{32},\Delta_{14}}_{\Delta_{\calO'},\ell_{\calO'}}(1-z,1-\bar z)}
		{((1-z)(1-\bar{z}))^{\frac{\Delta_3+\Delta_2}2}}\,.
		\end{multline}
Here the sums run over the operators $\calO$ and $\calO'$ which appear in the OPE in the two channels.

One frequently occurring special case is a 4pt function of identical scalars $\<\sigma\sigma\sigma\sigma\>$. Then the crossing relation simplifies because $\calO=\calO'$ and also because we get squares of the OPE coefficients $\lambda_{\sigma\sigma\calO}$. It is customary to write it as
\begin{gather}
\label{eq:crossing_ssss}
\sum_{\cO} \lambda_{\s\s\cO}^2 F^{\Delta_\s}_{\Delta_{\cO},\ell_{\cO}}(z,\bar z)=0\,,
\end{gather}
where
\begin{multline}
	\label{eq:Feq}
F^{\Delta_\s}_{\Delta, \ell}(z,\bar{z}) = ((1-z)(1-\bar{z}))^{\Delta_\s} g^{0,0}_{\Delta,\ell}(z,\bar{z}) \\ 
 - (z\bar{z})^{\Delta_\s} g^{0,0}_{\Delta,\ell}(1-z,1-\bar{z}).
\end{multline}

Among the operators $\calO$ which appear in \reef{eq:crossing_ssss}, a special role is played by the identity operator and (in local CFTs) by the stress tensor, because these are two operators of known dimension whose OPE coefficients are nonzero. In particular the identity operator appears with the coefficient $\lambda_{\sigma\sigma\mathds{1}}=1$. By studying the $z\to0$ limit of the crossing relation, it's easy to show analytically that there should be infinitely many further operators with nonzero $\lambda_{\sigma\sigma\calO}$ \cite{Rattazzi:2008pe}. We will see later on what can be learned about these operators using numerical methods.

Going back to the general case \reef{eq:cross-gen}, it is similarly convenient to rewrite it as follows \cite{Kos:2014bka}. We introduce the functions
\begin{multline}
		\label{eq:Funeq}
	F_{\pm,\De,\ell}^{ij,kl}(z,\bar{z})  = \left((1-z)(1-\bar{z})\right)^{\frac{\De_k+\De_j}{2}} g_{\De,\ell}^{\De_{ij},\De_{kl}}(z,\bar{z}) \\
	\pm  \left(z\bar{z}\right)^{\frac{\De_k+\De_j}{2}} g_{\De,\ell}^{\De_{ij},\De_{kl}}(1-z,1-\bar{z})\,,
\end{multline}
which are symmetric/antisymmetric under $z\to 1-z$, $\bar z\to 1-\bar z$.
We then take the sums and differences of \reef{eq:cross-gen} with the same equation with $z,\bar z$ replaced by $1-z,1-\bar z$. Then \reef{eq:cross-gen} is equivalent to the pair of equations:
\begin{multline}
	\label{eq:crossing_uneq}
	\sum_{\cO} \lambda_{12\calO}\lambda_{34\calO} F^{12,34}_{\mp,\Delta_{\cO},\ell_{\cO}}(z,\bar z) \\
		\pm \sum_{\cO'} \lambda_{32\calO'}\lambda_{14\calO'} F^{32,14}_{\mp,\Delta_{\cO'},\ell_{\cO'}}(z,\bar z) =0\,.
\end{multline}
If all operators are equal, the lower sign case is trivial, and the upper sign reduces to the single correlator crossing relation \reef{eq:crossing_ssss}.

Crossing relations can be imposed at any point $z,\bar z$ where both the s- and t-channels converge. From the discussion in Sec.~\ref{sec:radial}, this is the plane of all complex $z$ minus cuts along $(1,+\infty)$ where the s-channel diverges and $(-\infty,0)$ where the t-channel diverges. As we will see in Sec.~\ref{sec:LP}, the standard choice in numerical studies is to impose crossing in a Taylor expansion around the point $z=\bar z=1/2$, which is well inside this region.

There is also a third u-channel OPE (13)-(24). The u-channel is typically not considered in the numerical bootstrap, because it is not convergent at $z=\bar z=1/2$.\footnote{Although it can be considered when crossing relations are analyzed around another point, e.g.~$u=v=1$ \cite{Li:2017ukc}.} For 4 identical external scalars, the u-channel is automatically satisfied if the s-t channel crossing relation holds \cite{Poland:2010wg}. For nonidentical external operators, the u-channel is important. To impose the u-channel crossing relation, one changes the conformal frame by interchanging the positions of operators 1 and 2 \cite{Rattazzi:2010yc}. The u-channel in the original frame becomes the t-channel in the new frame, and crossing can be imposed around $z=\bar z=1/2$. The s-channel CPW decomposition in the new frame only differs by signs of all odd-spin terms because of \reef{eq:flipsign}.

In the case when the CFT has a global symmetry $G$, the crossing relations were formalized in \textcite{Rattazzi:2010yc}. Consider a 4pt function of scalar operators transforming in $G$ representations $\pi_i$. The exchanged operators $\calO_\pi$ then transform in representations $\pi$ appearing in the tensor product decompositions of $\pi_i \otimes \pi_j$. Each term in the s- and t-channel CPW decompositions comes multiplied with a tensor structure obtained by contracting two 3pt $G$-invariant tensors, as described in Sec.~\ref{sec:global}. We can represent it by a vector $\vec V_\pi$ in the space of 4pt $G$-invariant tensors $(\pi_1\otimes \pi_2 \otimes \pi_3 \otimes \pi_4)^G$.
 (Anti)symmetrizing under $z\to 1-z$, $\bar z\to 1-\bar z$, the crossing relation takes form \reef{eq:crossing_uneq}, with every term multiplied by the corresponding vector $\vec V_\pi$. It is thus a constraint in the space of vector functions. As an explicit example, crossing relations of 4pt functions $\<\phi_a \phi_b \phi_c \phi_f\>$ and $\<\phi_a \phi^{\dagger b} \phi_c \phi^{\dagger f}\>$ for $\phi_a$ a fundamental of $SO(N)$ or $SU(N)$ were found in \textcite{Rattazzi:2010yc}.

A similar vector structure arises when analyzing 4pt functions of operators with Lorentz spin, with the conformally invariant 4pt tensors ${\bf T}_4^{(c)}$ in \reef{eq:spinning_cb} playing the role of the $G$-invariant 4pt tensors in the case of global symmetry. General crossing relations involving both global symmetry and Lorentz indices were formalized in \textcite{Kos:2014bka}.

\subsubsection{Explicit solutions to crossing}
\label{sec:explicit}

Many nontrivial 2d CFTs have exact solutions (e.g.~the minimal models), and the conformal block decompositions of their 4pt functions provide explicit solutions to crossing relations. 
Here we will discuss a few explicit solutions to crossing known in $d>2$. Their existence is important, even though as we will see they come from theories which are not physically the most interesting ones. For example, it is common to check the numerical algorithms against the known explicit solutions to exclude coding errors, before proceeding to study more physically interesting solutions numerically. 

Essentially all explicit solutions in $d>2$ are provided by scale-invariant ``gaussian theories", i.e.~theories coming from a quadratic action written in terms of a fundamental field and not having any massive parameter.\footnote{The only exceptions known to us are the ``fishnet theories" ---  nonunitary bi-scalar field theories integrable in the large-$N$ limit \cite{Gurdogan:2015csr}. Recently some conformally-invariant 4pt functions and their conformal block decompositions were computed in such theories in 4d \cite{Grabner:2017pgm}, and in their nonlocal generalizations to arbitrary $d$ \cite{Kazakov:2018qbr}.}
The correlation functions of such theories are generated by Wick's theorem from the basic 2pt function of the fundamental field. The simplest examples are the massless free scalar and massless free fermion theory, which are conformally invariant in any $d$, and the free abelian gauge theory, conformally invariant in $d=4$. In 4d, explicit conformal block decompositions of 4pt scalar correlation functions $\< \calO \calO \calO \calO \>$ in these theories (for $\calO=\phi$, $\phi^2$, $\bar\psi\psi$, $F_{\mu\nu}^2$) were obtained by \textcite{DO1}. 

Another class of gaussian theories are mean field theories (MFTs), also called generalized free fields~\cite{Heemskerk:2009pn}, \cite[{section 4}]{ElShowk:2011a}. Correlation functions in these theories have the same disconnected structure generated by Wick's theorem as in the above mentioned free theories. The only difference is that the scaling dimension of the fundamental field, fixed to a particular value in free theories, becomes a free parameter in MFT.\footnote{This structure naturally emerges in large-$N$ CFTs as a consequence of large-$N$ factorization. This is particularly transparent in CFTs with holographic duals, since MFT correlation functions are generated by free massive fields in AdS${}_{d+1}$ and the arbitrary scaling dimension is determined by the mass.}  For example, we can consider the MFT of a scalar field $\phi$ of arbitrary dimension $\Delta_\phi$. Such a MFT is unitary as long as $\Delta_\phi$ satisfies the unitarity bound, and reduces to the free massless scalar for $\Delta_\phi=(d-2)/2$. Just like for the usual free theories, the full space of operators in MFTs can be classified by considering normal-ordered products of the fundamental field and its derivatives.\footnote{The OPE $\phi\times\phi$ contains only operators of the schematic form $\phi (\partial^{2})^n\partial^\ell\phi$, which have spin $\ell$ and dimension \mbox{$2\Delta_\phi+2n+\ell$}.} For example there is an operator $\phi^2$ which has dimension $2\Delta_\phi$. 

Although relatively trivial and nonlocal, MFTs satisfy most CFT axioms (except for for the existence of a local stress tensor). As we will see below, they frequently fall inside regions allowed by the bootstrap bounds, so it helps to be familiar with them. Explicit conformal block decompositions of MFT 4pt functions containing scalars were obtained by \textcite{Heemskerk:2009pn} for $d=2,4$ and by \textcite{Fitzpatrick:2011dm} in general $d$.

\section{Numerical methods} 
\label{sec:numerical}

\subsection{Convex optimization and linear programming}
\label{sec:LP}

The CFT crossing relations describe a continuously infinite number of constraints on the CFT data parametrized by the cross ratios. In order to study the crossing relations numerically one must discretize this set of constraints. Starting with~\textcite{Rattazzi:2008pe}, the common approach adopted in numerical studies is to Taylor expand the crossing relations around a point in cross-ratio space, typically taken to be the symmetric configuration $u=v=1/4$ or equivalently $z=\bar{z}=1/2$. If one takes derivatives only up to a certain order $\Lambda$, then one obtains a finite set of constraints. 

Before proceeding, let us highlight that a number of other choices could be made here, e.g.~evaluating the crossing equations at different values of the cross ratios, Taylor expanding around other points, integrating the crossing equations, etc.\footnote{Some of these alternative ideas have been explored by~\textcite[{section 4.2}]{Hogervorst:2013sma}, \textcite{Echeverri:2016ztu}, \textcite{Mazac:2016qev}, \textcite{Li:2017ukc}, and \textcite{Mazac:2018mdx}.} Here we focus on the approach of Taylor expanding around the symmetric configuration since it works well in practice and is the most common approach in the literature. One justification for this choice of the expansion point is that it makes both the direct and the crossed channel in the conformal block expansion to converge maximally fast~\cite{Pappadopulo:2012jk}. However, it is by no means obvious that it is the most efficient way to discretize the crossing relations.

In the case of a single 4pt function of identical scalars $\<\s\s\s\s\>$, the resulting constraints take the form
\begin{eqnarray}\label{eq:crossingvec}
0 &=& \sum_{\cO} \lambda_{\s\s\cO}^2 \vec{F}^{\Delta_\s}_{\Delta_{\cO},\ell_{\cO}},
\end{eqnarray}
where $\vec{F}^{\Delta_\s}_{\Delta_{\cO},\ell_{\cO}}$ can be thought of as a vector with components
\begin{eqnarray}
\label{eq:F}
\left(\vec{F}^{\Delta_\s}_{\Delta_{\cO},\ell_{\cO}}\right)^{mn} &=& \partial_{z}^m \partial_{\bar{z}}^n F^{\Delta_\s}_{\Delta_{\cO}, \ell_{\cO}}(z,\bar{z})\big|_{z=\bar{z}=1/2}\,,
\end{eqnarray}
where we take derivatives of the functions \reef{eq:Feq} and we keep components up to a cutoff $m+n \leq \Lambda$.\footnote{{Since the functions \reef{eq:Feq} are odd under $z\to 1-z$, $\bar{z}\to 1-z$, only components with {$m+n$ odd} lead to nontrivial equations.}} {This computation will thus involve derivatives of conformal blocks up to some finite order.}
 
{Computing the vectors $\vec{F}^{\Delta_\s}_{\Delta,\ell}$ constitutes a nontrivial preliminary step for analyzing Eq.~\reef{eq:crossingvec}. This step is handled starting from one of the many exact or approximate expressions for conformal blocks discussed in Sec.~\ref{sec:cb}. The state-of-the-art approach is to use the rational approximation, see Sec.~\ref{sec:rational}, where available software packages are also described. This approach gives rise to approximate expressions which reproduce $\vec{F}^{\Delta_\s}_{\Delta,\ell}$ with any desired precision. These expressions can be efficiently evaluated ``on the fly", as needed in the {continuous} simplex algorithm from Sec.~\ref{sec:modified-simplex}. They can also be used as an input to the semidefinite programming methods described in Sec.~\ref{sec:SDP}.}

We now proceed to {describe} strategies on how to decide if Eq.~\reef{eq:crossingvec} has solutions, i.e.~if there {exists} some choice of the exchanged CFT spectrum $\{\Delta_{\cO},\ell_{\cO}\}$ and OPE coefficients $\lambda_{\sigma\sigma\calO}$ which makes it satisfied. First, let us remark that Eq.~(\ref{eq:crossingvec}) is a set of linear equations in $\lambda_{\s\s\cO}^2$. This is at the heart of both the linear programming approaches described in this subsection as well as the extremal functional and truncation methods described below. In particular, if one has a candidate CFT spectrum for operators appearing in $\s \times \s$ but does not know the OPE coefficients, one can straightforwardly solve a linear algebra problem to find the coefficients.

\begin{figure}[t!]
    \centering
\includegraphics[width=0.49\textwidth]{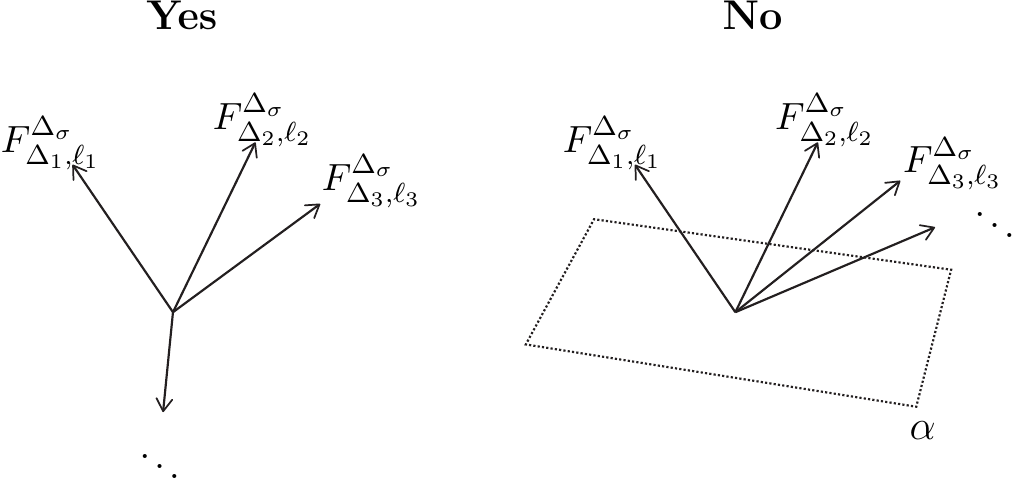}
    \caption{\label{fig:separating-plane}
Left: A case where vectors can sum to zero with positive coefficients. Right: A case where vectors cannot sum to zero with positive coefficients and there exists a separating plane $\alpha$ such that all vectors point on one side of the plane. Figure from \textcite{Poland:2016chs}.}
  \end{figure}

In unitary (or reflection positive) CFTs, Eq.~(\ref{eq:crossingvec}) states that a sum of vectors must add to zero with {\it positive} coefficients, due to $\lambda_{\s\s\cO}$ necessarily being real. For some choices of the CFT spectrum $\{\Delta_{\cO},\ell_{\cO}\}$ this is not possible, as illustrated in Fig.~\ref{fig:separating-plane}. When it is not possible one can identify a separating plane $\alpha$ through the origin such that all vectors point to one side of the plane.\footnote{Some vectors may point in the plane but at least one must point outside of it.}

This observation forms the basis for the first numerical strategy of analyzing the crossing relation \cite{Rattazzi:2008pe}: input some assumption about the CFT spectrum (e.g., a gap in the scalar spectrum with all other operators satisfying unitarity bounds) and numerically search for a separating plane $\alpha$. Equivalently we can say that we are applying a linear functional $\sum_{mn} \alpha_{mn} \partial_{z}^m \partial_{\bar{z}}^n \left[ \cdot \right]\big|_{z=\bar{z}=1/2}$ to the crossing relations and checking if it is possible to derive a contradiction. Concretely, one can look for a vector $\vec{\alpha}$ such that the scalar product is strictly positive on at least one operator whose OPE coefficient is nonzero (this may be the identity, the stress tensor, or any other operator that we assume appears in the OPE):
\beq
\vec{\alpha} \cdot \vec{F}^{\Delta_\s}_{\Delta_{\cO^*},\ell_{\cO^*}} > 0\,,
\eeq
while it is nonnegative for all other $\{\Delta_{\cO},\ell_{\cO}\}$ allowed by our assumptions:
\beq
\vec{\alpha} \cdot \vec{F}^{\Delta_\s}_{\Delta_{\cO},\ell_{\cO}} \geq 0\,.
\eeq
Each inequality $\vec{\alpha} \cdot \vec{F}^{\Delta_\s}_{\Delta_{\cO},\ell_{\cO}} \geq 0$ identifies a half-space and their intersection carves out a convex cone. 

There is still one issue before the vector $\vec{\alpha}$ can be searched for numerically -- a priori there are an infinite number of allowed vectors labeled by all $\{\Delta_{\cO},\ell_{\cO}\}$. 
The first numerical bootstrap studies\footnote{See~\textcite{Rattazzi:2008pe}, \textcite{Rychkov:2009ij}, \textcite{Caracciolo:2009bx}, \textcite{Poland:2010wg}, \textcite{Rattazzi:2010gj,Rattazzi:2010yc}, \textcite{Vichi:2011ux,Vichi:2011zza}, and \textcite{ElShowk:2012ht}.} employed a discretization approach: namely, they discretized the set $\{\Delta_{\cO},\ell_{\cO}\}$ using some small spacing between allowed dimensions so that there are a finite number of linear inequalities satisfied by a finite number of unknown coefficients $\vec{\alpha}$. Then the problem becomes a standard linear programming problem and can be solved using standard algorithms. These include simplex algorithms, where one moves from vertex to vertex on the edge of the feasible region, or interior point algorithms, where one instead traverses the interior of the feasible region. Software packages that have been used {in the past} for this purpose are \texttt{Mathematica}, the \texttt{GNU Linear Programming Kit (GLPK)}, and the \texttt{IBM ILOG CPLEX Optimizer}. {This discretization approach is currently considered to be obsolete, although it retains pedagogical value. More efficient approaches avoiding discretization will be discussed below.}

One can slightly modify the problem in order to place bounds on OPE coefficients \cite{Caracciolo:2009bx}. By isolating one particular contribution $\cO^*$ and again applying a functional $\vec{\alpha}$ one rewrites the equation as
\beq
\label{eq:OPEbound}
\lambda_{\s\s\cO^*}^2 \vec{\alpha} \cdot \vec{F}^{\Delta_\s}_{\Delta_{\cO^*},\ell_{\cO^*}} = -\vec{\alpha} \cdot \vec{F}^{\Delta_\s}_{0,0} - \sum_{\cO} \lambda_{\s\s\cO}^2 \vec{\alpha} \cdot \vec{F}^{\Delta_\s}_{\Delta_{\cO},\ell_{\cO}}.
\eeq
Then by imposing the normalization condition \mbox{$\vec{\alpha} \cdot \vec{F}^{\Delta_\s}_{\Delta_{\cO^*},\ell_{\cO^*}} = 1$} and the positivity constraints \mbox{$\vec{\alpha} \cdot \vec{F}^{\Delta_\s}_{\Delta_{\cO},\ell_{\cO}} \geq 0$} one obtains the upper bound \mbox{$\lambda_{\s\s\cO^*}^2 \leq  -\vec{\alpha} \cdot \vec{F}^{\Delta_\s}_{0,0}$}. The strongest upper bound is obtained by {\it minimizing} $ -\vec{\alpha} \cdot \vec{F}^{\Delta_\s}_{0,0}${, which yields an optimization problem that can be solved with linear programming algorithms, adopting the above-mentioned discretization approach or other methods discussed below}.  Alternatively, one can also seek lower bounds by instead imposing  $\vec{\alpha} \cdot \vec{F}^{\Delta_\s}_{\Delta_{\cO},\ell_{\cO}} \leq 0$ and maximizing $ -\vec{\alpha} \cdot \vec{F}^{\Delta_\s}_{0,0}$ \cite{Poland:2011ey}. However, in general it is not possible to obtain lower bounds on OPE coefficients unless the operator $\cO^*$ is isolated in the allowed spectrum, since one could always imagine that $\cO^*$ has a zero OPE coefficient but operators infinitesimally close to it have nonzero coefficients. 

\subsubsection{Continuous primal simplex algorithm}
\label{sec:modified-simplex}

Instead of looking for a vector $\vec{\alpha}$ with the desired positivity properties, an alternate strategy is to search directly for a set of vectors $\{\vec{F}^{\Delta_\s}_{\Delta_{\cO},\ell_{\cO}}\}$ appearing in Eq.~(\ref{eq:crossingvec}), subject to the positivity conditions $\lambda_{\s\s\cO}^2 \geq 0$. This search can be viewed as a ``primal'' formulation of the linear program, whereas the search for $\vec{\alpha}$ described above can be viewed as the related ``dual'' problem. Note that in this formulation there are a continuously infinite number of possible vectors $\vec{F}^{\Delta_\s}_{\Delta_{\cO},\ell_{\cO}}$ in the search space. \textcite{El-Showk:2014dwa} {developed} a modification of Dantzig's simplex algorithm in order to handle such a continuous search space.\footnote{Such linear programming problems are called `continuous' or `semi-infinite'  \cite{reemtsen_numerical_1998}.} The essential idea is to use Newton's method at each step of the algorithm to identify a vector to add, which is optimal over some continuous interval of scaling dimensions $\left[\Delta_{\text{min}}, \Delta_{\text{max}}\right]$ and discrete set of spins $\left[0, \ell_{\text{max}}\right]$. {For reasons explained in \textcite{El-Showk:2014dwa}, it is necessary to perform computations at a precision higher than machine precision. This continuous simplex algorithm is one of two state-of-the art methods for the conformal bootstrap, the other one being the semidefinite programming method described below.} Three implementations of this algorithm are available: 
a \texttt{C++} code \texttt{SIPSolver} \cite{DSDSIPsolver} and a \texttt{Python/Cython} code \cite{ElShowkRychkov} which were used for the computations in \cite{El-Showk:2014dwa}, as well as a \texttt{Julia} package \texttt{JuliBootS}~\cite{Paulos:2014vya}.

\subsection{Semidefinite programming}
\label{sec:SDP}

While the linear programming techniques described above are adequate for crossing relations of single 4pt functions (possibly charged under some global symmetry), they are more difficult to adapt for systems of crossing relations containing multiple operators. The reason is that the resulting crossing relations for mixed correlators are no longer linear in the positive squares of OPE coefficients.\footnote{However, they can be made linear at the expense of introducing additional continuous parameters~\cite{El-Showk:2016mxr}. {This observation has not yet been implemented and it is not known how it would perform in practice.}} The same issue arises when considering 4pt functions of spinning operators, where multiple 3pt function tensor structures exist. In these situations one can phrase the optimization problem needed to obtain bounds using the language of semidefinite programming rather than linear programming \cite{Kos:2014bka}.\footnote{For a related problem of multiple internal symmetry coupling structures this was observed in \textcite{Rattazzi:2010yc}.}

Another use of semidefinite programming \cite{Poland:2011ey} is to avoid needing to discretize and impose a cutoff on the exchanged operator dimensions appearing in the positivity constraints such as $\vec{\alpha} \cdot \vec{F}_{\Delta_{\cO}, \ell_{\cO}} \geq 0$. We will describe both of these uses of semidefinite programming, as well as how they can be combined, below.

In most applications to the bootstrap, it has proven necessary for numerical stability to solve the semidefinite programs described below at a precision higher than machine precision. The first numerical studies made use of the software \texttt{SDPA-GMP}~\cite{SDPAGMP} (a variant of \texttt{SDPA}~\cite{SDPA}) for this purpose. 
{The state-of-the-art is} an efficient software package \texttt{SDPB}, described in \textcite{Simmons-Duffin:2015qma}, which improves on the \texttt{SDPA}'s primal-dual interior point algorithm primarily by taking advantage of matrix block structure and parallelization.\footnote{{Further development of \texttt{SDPB} is being carried out within the Simons Collaboration on the Nonperturbative Bootstrap (\url{http://bootstrapcollaboration.com/}), and this package will likely remain at the forefront of the numerical bootstrap studies in the coming years.}} In order to set up the problems so that they can be solved by \texttt{SDPB}, recent studies have typically used either \texttt{Mathematica} {notebooks}, or the interfaces \texttt{PyCFTBoot}~\cite{Behan:2016dtz} or \texttt{cboot}~\cite{CBoot}. 

\subsubsection{Mixed correlators}

\label{sec:mixed-correlators}

We will illustrate the use of semidefinite programming for mixed correlators with a  simple example. Consider a system of 4pt functions containing two operators $\s$ and $\e$, where $\s$ is odd under a $\mathbb{Z}_2$ symmetry and $\e$ is even. The resulting system of crossing relations for $\<\s\s\s\s\>$, $\<\s\s\e\e\>$, and $\<\e\e\e\e\>$ takes the form
\cite{Kos:2014bka}
\beq
\label{eq:crossingequationwithv}
 0  =  \sum_{\cO^+} \begin{pmatrix}\lambda_{\s\s\cO} & \lambda_{\e\e\cO}\end{pmatrix} \vec{V}_{+,\De,\ell}\begin{pmatrix} \lambda_{\s\s\cO} \\ \lambda_{\e\e\cO} \end{pmatrix}+ \sum_{\cO^-} \lambda_{\s\e\cO}^2 \vec{V}_{-,\De,\ell}\,,
\eeq
where the components of the vectors $\vec{V}_{\pm,\Delta,\ell}$ run over 5 independent crossing relations,\footnote{In this section we are using vector notation to describe the vector of crossing relations, rather than derivatives.} $\cO_{\pm}$ denote operators even/odd under $\mathbb{Z}_2$ symmetry, and each $\vec{V}_{+,\De,\ell}$ is a 5-vector of $2 \times 2$ matrices:

\beq
\vec{V}_{-,\De,\ell} = \begin{pmatrix} 0  \\ 0 \\ F_{-,\De,\ell}^{\s\e,\s\e}(z,\bar{z}) \\ (-1)^{\ell} F_{-,\De,\ell}^{\e\s,\s\e}(z,\bar{z}) \\ - (-1)^{\ell} F_{+,\De,\ell}^{\e\s,\s\e}(z,\bar{z}) \end{pmatrix},\nonumber
\eeq 
\beq
\vec{V}_{+,\De,\ell} = \begin{pmatrix} \begin{pmatrix}  F^{\s\s,\s\s}_{-,\De,\ell}(z,\bar{z}) & 0 \\ 0 & 0  \end{pmatrix} \\ \begin{pmatrix}  0 & 0 \\ 0 & F^{\e\e,\e\e}_{-,\De,\ell}(z,\bar{z})  \end{pmatrix}\\ \begin{pmatrix}  0 & 0 \\ 0 & 0  \end{pmatrix}  \\ \begin{pmatrix}  0 & \frac12 F^{\s\s,\e\e}_{-,\De,\ell}(z,\bar{z}) \\ \frac12 F^{\s\s,\e\e}_{-,\De,\ell}(z,\bar{z}) & 0 \end{pmatrix} \\ \begin{pmatrix} 0 & \frac12 F^{\s\s,\e\e}_{+,\De,\ell}(z,\bar{z}) \\ \frac12 F^{\s\s,\e\e}_{+,\De,\ell}(z,\bar{z}) & 0  \end{pmatrix} \end{pmatrix}.
\eeq
The appearing functions $F^{ij,kl}_{\pm,\De,\ell}$ are given in \reef{eq:Funeq}. One can then look for bounds by making some assumption about the spectrum and searching for a functional $\vec{\alpha} = \sum_{mn} \vec{\alpha}_{mn} \partial_{z}^m \partial_{\bar{z}}^n \left[\cdot\right] \big|_{z=\bar{z}=1/2}$ satisfying the properties
\begin{align}
&\begin{pmatrix} 1 & 1\end{pmatrix} \vec \alpha \cdot \vec{V}_{+,0,0} \begin{pmatrix} 1 \\ 
1 \end{pmatrix}  > 0 \,, \nn\\
&\vec \alpha \cdot \vec{V}_{+,\De,\ell} \succeq 0 \quad\textrm{for all $\mathbb{Z}_2$-even operators with $\ell$ even,} \nn\\
&\vec \alpha \cdot \vec{V}_{-,\De,\ell}  \ge 0 \quad \textrm{for all $\mathbb{Z}_2$-odd operators.} 
\label{eq:functionalinequalities1}
\end{align}
The novel feature is now that $\vec \alpha \cdot \vec{V}_{+,\De,\ell} \succeq 0$ must be a {\it positive semidefinite} $2 \times 2$ matrix, which makes the search in Eq.~(\ref{eq:functionalinequalities1}) a semidefinite programming problem. Similar structure appears for more general systems of mixed/spinning correlators, where if an exchanged operator has $N$ OPE coefficients appearing in the system then the needed positivity condition will be phrased in terms of positive semidefinite $N \times N$ matrices.

\subsubsection{Polynomial approximations}
\label{sec:poly}
A different use of semidefinite programming, relevant for both single correlators or mixed correlators, is to avoid any discretization of the exchanged operator dimensions~\cite{Poland:2011ey}. We will first explain the idea for single correlators, where one imposes inequalities of the form
\begin{eqnarray}
\sum_{mn} \alpha_{mn} \partial_z^m \partial_{\bar{z}}^n F^{\Delta_\s}_{\Delta,\ell}(z,\bar{z}) \big|_{z=\bar{z}=1/2} &\geq& 0\,.
\end{eqnarray} 
Due to the pole expansion of the conformal blocks described in Sec.~\ref{sec:analyticrecursion}, if one keeps a finite number of poles{,} then by reorganizing $h_{\Delta,\ell}$ into a rational function of $\Delta$, such derivatives can be rewritten in the form (see Sec.~\ref{sec:rational})
\begin{eqnarray}
\partial_z^m \partial_{\bar{z}}^n F^{\Delta_\s}_{\Delta,\ell}(z,\bar{z}) \big|_{z=\bar{z}=1/2} \approx \chi_{\ell}(\Delta) P_{\ell}^{mn}(\Delta)\,,
\end{eqnarray}
where $P_{\ell}^{mn}(\Delta)$ is a {\it polynomial} in $\Delta$, and 
$\chi_{\ell}(\Delta)$ is a positive function for all $\Delta$ and $\ell$ satisfying the unitarity bounds. The degree of the polynomial depends on the number of poles kept in the expansion of the conformal block. Then one simply needs to impose the polynomial inequalities
\begin{eqnarray}
\sum_{mn} \alpha_{mn} P^{mn}_{\ell}(\Delta^{\text{min}}_{\ell}+x) \geq 0
\end{eqnarray}
 for all $x \geq 0$, where the minimum dimension at each spin $\Delta^{\text{min}}_{\ell}$ depends on the assumptions being made. 
 
Such inequalities for polynomials can be rewritten in terms of positive semidefinite matrices following a theorem of~\textcite{Hilbert:1888}. The relevant theorem states that any polynomial $P(x)$ that is nonnegative on the interval $[0,\infty)$ can be written in the form
\beq
P(x) = a(x) + x b(x),
\eeq
where $a(x)$ and $b(x)$ are sums-of-squares of polynomials. Such sums-of-squares can in turn always be expressed in the form
\beq
a(x) = \text{Tr}(A Q_{d_1}(x)), \quad b(x) = \text{Tr}(B Q_{d_2}(x))\,,
\eeq
\\[2pt]
where $Q_{d}(x)\equiv [x]_d [x]_d^T$ is a matrix built out of the monomials $[x]_d = (1,x,\ldots, x^d)^T$, $d_1 = \left\lfloor \frac12 \deg P \right\rfloor$, $d_2 = \left\lfloor \frac12 (\deg P-1) \right\rfloor$, and $A$ and $B$ are some positive semidefinite matrices.

With this rewriting, one needs to search for coefficients $\alpha_{mn}$ and positive semidefinite matrices $A_{\ell}, B_{\ell} \succeq 0$ such that
\begin{multline}
\sum_{mn} \alpha_{mn} P^{mn}_{\ell}(\Delta^{\text{min}}_{\ell}+x) =\\
 \text{Tr}(A_{\ell} Q_{d_1}(x)) + x \text{Tr}(B_{\ell} Q_{d_2}(x)).
\end{multline}
In practice one must also impose a cutoff on the set of included spins $0 \leq \ell \leq \ell_{\text{max}}$. This search, combined with a normalization condition such as $\vec{\alpha} \cdot \vec{F}_{0,0}^{\Delta_\s} = 1$, is now in the form of a semidefinite programming problem.  

As explained in detail in \textcite{Kos:2014bka}, this idea can also be applied to systems of mixed or spinning correlators where exchanged operators have multiple OPE coefficients appearing in the system. In those cases, after truncating the conformal block pole expansions one {imposes} constraints of the form
\begin{gather}
\sum_{mn}\vec{\alpha}_{mn} \cdot
\begin{pmatrix}
\vec{P}^{(11;mn)}_\ell(\Delta) &\dots& \vec{P}_\ell^{(1N;mn)}(\Delta)\\
\vdots & \ddots &\vdots\\
\vec{P}^{(N1;mn)}_\ell(\Delta) & \dots & \vec{P}_\ell^{(NN;mn)}(\Delta)
\end{pmatrix}\succeq 0
\label{eq:matrixpolynomialconstraints1}\\
\textrm{for}\quad \Delta\geq \Delta^{\text{min}}_{\ell}.\nn
\end{gather}
Again there is a theorem that such positive semidefinite matrix polynomials can always be written as sums-of-squares of matrix polynomials. A consequence, worked out in \textcite{Kos:2014bka}, is that each entry can be written as 
\begin{multline}
\sum_{mn} \vec{\alpha}_{mn} \cdot \vec{P}^{(ij;mn)}_\ell(\Delta^{\text{min}}_{\ell}+x) =\\ \text{Tr}(A^{ij}_{\ell} Q_{d_1}(x)) + x \text{Tr}(B^{ij}_{\ell} Q_{d_2}(x))
\end{multline}
in terms of positive semidefinite matrices \mbox{$A_{\ell}^{ij}, B_{\ell}^{ij} \succeq 0$}, 
and the problem is again phrased as a semidefinite programming problem.

\subsection{Bounds and allowed regions}

\label{sec:allowed}
The algorithms described in the previous sections can be used to establish if a given point in the space of CFT data, parametrized by the dimensions of external operators and by assumptions on the exchanged spectrum, belongs to the region allowed by crossing and unitarity. Since the exchanged spectrum contains infinitely many operators, there are infinitely many assumptions one can test. The art of the numerical bootstrap is to choose an interesting assumption, and then to delineate as precisely as possible the allowed region corresponding to this assumption. 

As we will see in the next sections, one of the most frequently asked questions is the following: given an OPE $\mathcal O\times \mathcal O$, derive an upper bound $\Delta_\text{max}$ on the dimension of the first operator appearing in this OPE having specified transformation properties under $SO(d)$ (and eventually under a global symmetry $G$),\footnote{In particular, the existence of a bound with $\Delta_{\max}<\infty$ provides a proof that such an operator exists. See \textcite{Rattazzi:2008pe} and \cite[{section 10.5}]{Simmons-Duffin:2016gjk} for intuitive explanations involving some numerics of why such bounds should exist at all, and \textcite{Hogervorst:2013sma} and \cite[{section 4.3.3}]{Rychkov:2016iqz} for an approximate analytic argument. At present, while the existence of bounds can sometimes be understood via such simple means, their actual values can only be precisely computed using the powerful numerical techniques described in the previous sections. Only in a handful of cases, e.g.~\textcite{Mazac:2016qev} and \textcite{Mazac:2018mdx}, have the best possible bounds been proven analytically.} assuming e.g.~that other operator {dimensions are allowed to take any values} allowed by unitarity. One can answer this question, for instance, as a function of $\Delta_{\mathcal O}$. This defines an allowed region with a boundary $\Delta_\text{max}(\Delta_\mathcal O)$. Similarly, when one obtains an upper (or lower) bound on an OPE {coefficient} as discussed in Sec.~\ref{sec:LP}, this represents the boundary of an allowed region. These boundaries give us a view into the intricate underlying geometry of the space of CFT data allowed by crossing and unitarity.

\subsection{Spectrum extraction}
\label{sec:spectrum-extraction}

A point in the allowed region (see Sec.~\ref{sec:allowed}) is specified by external operator dimensions and by a handful of other numbers characterizing the assumptions, such as gaps on the exchanged operator spectrum. Once we ascertained that a point belongs to the allowed region, in some cases it is important to be able to go one step further and to extract an explicit solution to crossing, i.e.~the whole spectrum of exchanged operator dimensions and their OPE coefficients. The precise way of doing this depends on which algorithm one uses. An important point is that we expect this solution to be {non-unique} inside the allowed region, but it should generically become unique on its boundary (see below).

The spectrum extraction is simplest in the primal simplex method,  Sec.~\ref{sec:modified-simplex}. In this case the spectrum is encoded directly in the set of basic vectors and is available in each step of the algorithm.

In the dual formulation of the linear programming method, one does not have access to the spectrum strictly inside the allowed region. However, one can extract a solution to crossing symmetry from the limiting functional when one approaches a {\it boundary} of this region, by extremizing either an operator dimension or OPE coefficient. This is called the {\it extremal functional method}, introduced in \textcite{Poland:2010wg} and \textcite{ElShowk:2012hu}.

Namely, when approaching the boundary from the disallowed region, the system is on the verge of no longer allowing a separating plane, and the vector on which we require strict positivity is degenerating into the plane. In the case of a single crossing relation where we have imposed strict positivity on the identity operator, as we approach a dimension boundary we can find a vector $\vec{\alpha}$ such that $\vec{\alpha} \cdot \vec{F}^{\Delta_{\s}}_{0,0} \rightarrow 0$, together with the sum rule
\begin{eqnarray}\label{eq:extremal}
	0 &=& \sum_{\cO} \lambda_{\s\s\cO}^2 \vec{\alpha} \cdot \vec{F}^{\Delta_{\s}}_{\Delta_{\cO},\ell_{\cO}},
\end{eqnarray}
where $\vec{\alpha} \cdot \vec{F}^{\Delta_{\s}}_{\Delta_{\cO},\ell_{\cO}} \geq 0$ for all other possible (non-identity) operators in the spectrum. One obtains a similar condition from the OPE coefficient bound in Eq.~(\ref{eq:OPEbound}) if one sets the OPE coefficient to its extremal value $\lambda_{\s\s\cO^*}^2 = -\vec{\alpha} \cdot \vec{F}_{0,0}^{\Delta_{\s}}$. 

In fact, it is easy to see that in order for these sums to hold along the boundary of the allowed region, it is necessary for either $\lambda_{\s\s\cO}^2$ to be zero or for $\vec{\alpha} \cdot \vec{F}^{\Delta_{\s}}_{\Delta_{\cO},\ell_{\cO}}$ to be zero. Thus, the zeroes of $\vec{\alpha} \cdot \vec{F}^{\Delta_{\s}}_{\Delta_{\cO},\ell_{\cO}}$ tell us the scaling dimensions and spins at which the OPE coefficients are allowed to be nonzero.  The resulting {\it extremal spectrum} is generically unique~\cite{ElShowk:2012hu}.

In the above-mentioned primal simplex method, the extremal spectrum is reached from within the allowed region, and is encoded in the set of basic vectors that remain after the algorithm terminates. That this should agree with the dual approach via extremal functionals is guaranteed by the strong duality of linear programs.

The extremal functional method for extracting the spectrum is also applicable when using semidefinite programming. In this case the extremal functional is constructed in the dual formulation.\footnote{It is not understood at present how to formulate an algorithm to extract the extremal spectrum along a dimension bound directly from the allowed region in the context of semidefinite programming. The currently used procedure is to sit in the interior of the space allowed by scaling dimension bounds and extremize an OPE coefficient to find an extremal functional.} Once the extremal spectrum is known, it is straightforward to reconstruct the OPE coefficients of the exchanged operators by either directly solving the bootstrap equations after inputting the extremal spectrum or extracting them from the primal solution of the primal-dual algorithm. {\textcite{Simmons-Duffin:2016wlq} gives a precise algorithm for doing this using functionals output by \texttt{SDPB}, realized in a \texttt{Python} code
	\cite{DSD-spectrum-extraction}.}

An important {open question is to understand} which CFTs are described by spectra which are extremal with respect to some extremization condition. As we will see in subsequent sections, empirically this seems to be the case for a variety of interesting CFTs including the 3d Ising and $O(N)$ models. Although it is not currently understood why it should be so, some speculations are given in Sec.~\ref{sec:Z2-spectrum}.

 \subsubsection{Flow method}
 
 \label{sec:flow}
An interesting idea was proposed in \textcite{El-Showk:2016mxr}, where given one extremal solution one can efficiently ``flow'' along the boundary to reconstruct nearby extremal solutions. The idea is to perturb the extremal spectrum and then impose that the perturbed spectrum is also extremal using (\ref{eq:extremal}) as well as the tangency conditions $\vec{\alpha} \cdot \left(\partial_{\Delta_{\cO}} \vec{F}^{\Delta_{\s}}_{\Delta_{\cO},\ell_{\cO}}\right)$.  By linearizing perturbations of these conditions, the search for a nearby extremal spectrum (or a more precise extremal spectrum) can be efficiently solved using Newton's method. This approach then avoids the use of convex optimization after the initial step of finding an initial extremal solution, and can also be used to flow to nonunitary extremal solutions. This idea was shown to work well in $d=1$ in \textcite{El-Showk:2016mxr,Paulos:2016fap}.\footnote{The code is implemented as a separate module of \texttt{JuliBoots}~\cite{Paulos:2014vya}, available on request from its author.} It appears very promising and it needs to be further explored and extended, especially into higher dimensions.

\subsection{Truncation method}
\label{sec:truncation}

Finally we wish to turn to an idea introduced by \textcite{Gliozzi:2013ysa} and explored in a variety of works,\footnote{See \textcite{Gliozzi:2014jsa}, \textcite{Gliozzi:2015qsa}, \textcite{Gliozzi:2016cmg}, \textcite{Esterlis:2016psv}, \textcite{Hikami:2017hwv,Hikami:2017sbg,Hikami:2018mrf}, \textcite{Li:2017agi,Li:2017ukc}, and \textcite{LeClair:2018edq}.} which we will call the truncation method. The basic idea is to truncate the bootstrap equations to a finite number of operators $\{\Delta_{\s}, \cO_I \}$ with $N$ unknown scaling dimensions.  After normalizing by the identity contribution $f^{\Delta_{\s}}_{\Delta_{\cO_I},\ell_{\cO_I}}(z,\bar{z}) \equiv F^{\Delta_{\s}}_{\Delta_{\cO_I},\ell_{\cO_I}}(z,\bar{z})/\left(-F^{\Delta_{\s}}_{0,0}(z,\bar{z})\right)$, let us write the crossing equations as
\begin{eqnarray}\label{eq:linearsystem}
\sum_{\cO_I} \lambda_{\s\s\cO_I}^2 f^{\Delta_{\s}}_{\Delta_{\cO_I},\ell_{\cO_I}} &=& 1,\nonumber\\
\sum_{\cO_I} \lambda_{\s\s\cO_I}^2 \vec{f}^{\Delta_{\s}}_{\Delta_{\cO_I},\ell_{\cO_I}} &=& 0.
\end{eqnarray}
Here the first equation containing $f^{\Delta_{\s}}_{\Delta_{\cO_I},\ell_{\cO_I}} \equiv f^{\Delta_{\s}}_{\Delta_{\cO_I},\ell_{\cO_I}}(1/2,1/2)$ is viewed as an ``inhomogeneous" equation containing the identity contribution on the right-hand side, and the second ``homogeneous" equation contains the vector of derivatives $\left(\vec{f}^{\Delta_{\s}}_{\Delta_{\cO_I},\ell_{\cO_I}}\right)^{mn} = \partial_{z}^m \partial_{\bar{z}}^n f^{\Delta_\s}_{\Delta_{\cO_I}, \ell_{\cO_I}}(z,\bar{z})\big|_{z=\bar{z}=1/2}$. If one keeps $M$ derivatives with $M > N$ then the system becomes over-constrained, and only has solutions if all of the minors of order $N$ of the linear system vanish,
\begin{eqnarray}\label{eq:det}
\text{det}A_i=0,\quad A_i\subset A=\left[ \left(\vec{f}^{\Delta_{\s}}_{\Delta_{\cO_I},\ell_{\cO_I}}\right)^{mn} \right]_{N\times M}.
\end{eqnarray}
Here the ``rows" of $A$ would run over different choices of $N$ derivatives $mn$ and the ``columns" run over the $N$ unknown scaling dimensions. Note that the set of unknown scaling dimensions will include the external dimension $\Delta_{\s}$ in addition to the $\cO_I$, but may exclude exchanged operators of known dimension, such as the stress tensor of known dimension $\Delta_T = d$. The general strategy is to solve the determinant conditions Eq.~(\ref{eq:det}) to obtain an approximate spectrum of scaling dimensions, and then use the system in Eq.~(\ref{eq:linearsystem}), including the inhomogeneous equation, in order to fix the OPE coefficients.

A big advantage of the truncation approach over the linear and semidefinite programming approaches of the previous sections is that it does not require unitarity, i.e.~it works equally well for any sign of the OPE coefficients. For example, the idea has been successfully applied to the nonunitary Lee-Yang model, as well as to bulk-boundary bootstrap problems where there is no positivity in the coefficients. Another advantage is that it is relatively simple to implement, and the idea can be explored e.g.~using fairly simple \texttt{Mathematica} notebooks. 

On the other hand, we also see several disadvantages with this approach in its current incarnation. One is that the resulting spectrum can have a strong sensitivity to the set of included operators (e.g., the choices of spins) and to the set of derivatives included. It is also very difficult to assign reliable errors to the spectrum output from the method.\footnote{Comparison with the rigorous results obtained using the linear and semidefinite programming methods, when possible, shows that the published truncation method errors are often underestimated.} Thus, it would be desirable to find ways to make the approach more systematic with errors under control. Some steps in this direction were recently taken in \textcite{Li:2017ukc}. Applications to the boundary bootstrap also seem to be less sensitive to these issues~\cite{Gliozzi:2015qsa,Gliozzi:2016cmg}. 

Another issue is that simple implementations of numerical studies of the nonlinear determinant conditions~(\ref{eq:det}), such as using the iterative Newton method implemented in \texttt{Mathematica}'s \texttt{FindRoot} function, do not scale very well with increasing the number of operators and the method likely needs a more efficient numerical implementation in order to push beyond $\sim 10$ operators.\footnote{One can view the flow method described in Sec.~\ref{sec:flow} as a kind of more efficient implementation where additional extremality conditions have been added.}  

Notice that since we are truncating the spectrum, we cannot generally expect to find \emph{exact} solutions of Eq.~\reef{eq:det}. On the other hand, the set of determinant conditions is in fact redundant because of the Pl\"ucker relations satisfied by the minors of a matrix, see \textcite{Hikami:2017hwv}. A cleaner numerical formulation can be obtained by replacing Eq.~\reef{eq:det} with the problem of minimizing the smallest singular value of the matrix $A$~\cite{Esterlis:2016psv,LeClair:2018edq}.

Finally, similarly to the extremal spectra methods above, it is not clear which CFT spectra are ``truncable" in the sense that they can be found with this approach.

\section{Applications in $d=3$}
\label{sec:appl}

In this section we turn to applications of the numerical bootstrap techniques to CFTs in $d=3$ dimensions. Our discussion is organized as follows. We start in Sec.~\ref{sec:multicrit} by presenting a general bound on critical vs.~multicritical behavior in unitary 3d CFTs. In the following Secs.~\ref{sec:Z2} and \ref{sec:ON} we discuss bootstrap bounds which can be derived under the assumption of a $\bZ_2$ or $O(N)$ global symmetry. Applications to the most famous 3d CFTs realizing these symmetries---the critical 3d Ising and $O(N)$ models---will be emphasized.

In Sec.~\ref{sec:Fermions} we describe bounds on CFTs with fermionic operators, such as the IR fixed point of the Gross-Neveu-Yukawa models. In Sec.~\ref{sec:QED3} we discuss what the bootstrap currently has to say about CFTs realizable as IR fixed points of 3d QED coupled to matter. In Sec.~\ref{sec:JandT} we discuss recent bootstrap studies which implement crossing symmetry constraints on 4pt functions of stress tensors and conserved currents. These results are very general as they apply to any local 3d CFT.  Finally, in Sec.~\ref{sec:targets} we highlight some targets that may be interesting to look at in future numerical bootstrap studies. 

While most of the results will be phrased in a way which is highly model independent and in the language of conformal field theory, we hope that we can emphasize the physical interpretation of the various assumptions that are being made. {All} of the results summarized in this section have been obtained under the assumption of unitarity. Nonunitary CFTs, which can be studied e.g.~using the truncation method, are discussed separately in Sec.~\ref{sec:nonunitary}.

Finally, let us remind the reader that all of the numerical results that we summarize have been obtained using a variety of methods for numerically computing conformal blocks, with different choices of tolerance parameters in linear/semidefinite programs, with different choices of the cutoff $\Lambda$ on the number of derivatives applied to the crossing relation, etc. To keep our discussion readable, we will in most cases suppress these details, which can be found described in the original studies.

\subsection{Bounds on critical vs multicritical behavior}

\label{sec:multicrit}

As discussed in Sec.~\ref{sec:universality}, two basic structural characteristics of any CFT are the global symmetry group $G$ and the number of relevant scalar operators $S$ which are singlets under $G$. For this discussion we will view discrete spacetime symmetries such as {the} spatial parity $P$, if preserved, as a part of $G$.

The importance of the number $S$ becomes clear when we try to reach the CFT as an IR fixed point of an RG flow starting from a microscopic description, which for this discussion we will assume has the full symmetry $G$. Later in Sec.~\ref{sec:enhancement} we will comment on the situation of emergent symmetries which are not present in the microscopic description.

It follows from basic RG theory that the RG flow can reach the IR fixed point without any fine tuning if and only if $S=0$. We will call such CFTs ``self-organized" by loose analogy with what happens in self-organized criticality~\cite{Bak}. Examples include QED${}_3$ and QCD$_4$ in the conformal window, to be discussed in Secs.~\ref{sec:QED3} and~\ref{sec:4Dwindow}.

On the other hand, if relevant singlet scalars are present ($S>0$), then reaching the fixed point requires fine-tuning $S$ parameters in the microscopic Lagrangian. A common case is when $S=1$, as is realized for the critical Ising model and $O(N)$ models discussed below. Then we say that we have a critical point. Finally, the case $S>1$ is classified as a multicritical point.\footnote{In microscopic realizations which do not have the full symmetry $G$, one must tune a number of parameters equal to the number of relevant singlets under the microscopic symmetry.}

The simplest example of a 3d multicritical point is the free scalar field $\phi$. It has a $\bZ_2$ global symmetry acting as $\phi\to-\phi$, with two relevant singlet scalars, $\phi^2$ and $\phi^4$, of dimension 1 and 2 respectively. A third singlet scalar $\phi^6$ has dimension exactly 3 and is marginal (it is actually marginally irrelevant). This CFT describes a tricritical 3d Ising model.
Many nontrivial multicritical fixed points can be realized in systems of multiple interacting scalar fields.

\begin{figure}[t!]
    \includegraphics[width=\figwidth]{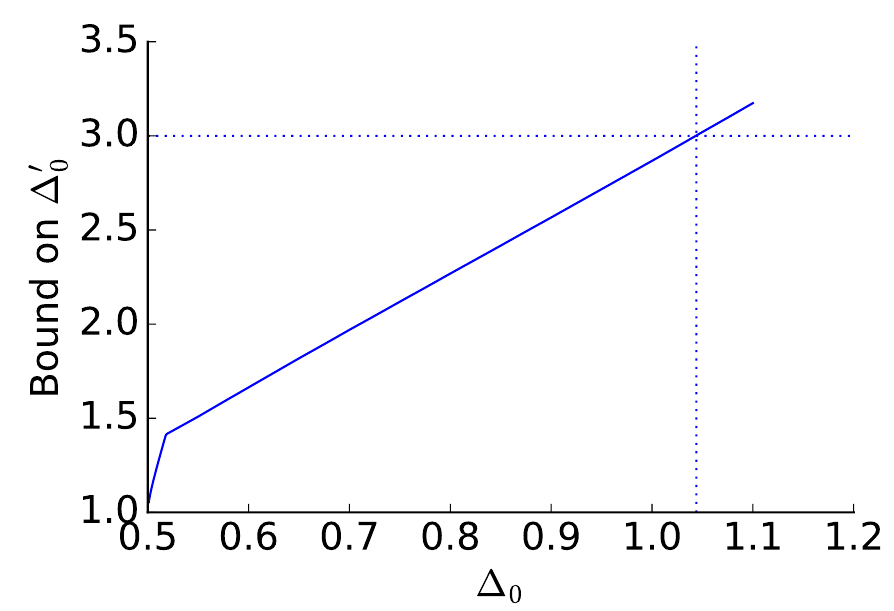}
    \caption{\label{fig:multicrit}
   (Color online) Upper bound on the dimension of the second singlet scalar $\calO_0'$ as a function of the dimension of the first $\calO_0$ \cite[{Supplementary Material}]{Nakayama:2016jhq}.}
  \end{figure}

Suppose that we know that we have a critical point, but not a multicritical point, i.e.~that there is one and only one singlet scalar, call it $\calO_0$. The OPE of $\calO_0$ with itself has the schematic form
\beq
\label{eq:O0}
\calO_0\times \calO_0\sim \mathds{1}+ \lambda \calO_0 +\lambda' \calO_0'+\ldots
\eeq
where since $\calO_0$ is a singlet it can appear on the r.h.s., {we denote the next singlet scalar as $\calO_0'$}, and $\ldots$ stands for all other operators. In this setup, \textcite{Nakayama:2016jhq} used the numerical bootstrap to derive an upper bound on the dimension of $\calO_0'$ as a function of dimension of $\calO_0$, shown in Fig.~\ref{fig:multicrit}. From this plot, the requirement $\Delta_{\calO_0'}>3$ translates into the lower bound 
\beq
\label{eq:multicrit}
\Delta_{\calO_0}>1.044\quad \text{for any critical 3d CFT.}
\eeq 
In terms of the critical exponent $\nu=1/(3-\Delta_{\calO_0})$, this means that $\nu>0.511$ for any critical (but not multicritical) 3d fixed point described by a unitary CFT.

Among all critical 3d fixed points that we know of, the lowest $\Delta_{\calO_0}\approx 1.41$ is realized in the critical Ising model, see below. This satisfies the above general bound \reef{eq:multicrit} by a large margin.

\subsection{$\bZ_2$ global symmetry} 

\label{sec:Z2}
\subsubsection{General results}
\label{sec:Z2-general}

We are not aware of any unitary 3d CFTs which do not possess any global symmetry.\footnote{The Lee-Yang model has no global symmetry but is nonunitary, see Sec.~\ref{sec:nonunitary}. {In any CFT with a global symmetry $G$, the singlet sector is closed under OPE. From the bootstrap point of view the singlet sector can be studied in isolation, results of Sec.~\ref{sec:multicrit} being an example, and would appear as a perfectly consistent CFT with no global symmetry. Dealing only with local operators, we do not consider this construction as defining a complete CFT, as the singlet sector can in principle be extended back by including the other sectors (although it is not known how to decide in practice whether such an extension is possible by looking at the correlators of the singlet sector).}} Actually, most 3d CFTs have \emph{continuous} global symmetries. Here we will start by considering the effect of having a discrete $\bZ_2$,\footnote{The bounds described in this section will also hold if the $\bZ_2$ is taken to be a parity or time-reversal symmetry.} which may be a full symmetry as for the 3d Ising model, or a subgroup of a larger group.\footnote{Another physically important discrete symmetry is cubic symmetry, see Sec.~\ref{sec:O3} and footnote~\ref{note:B3}.} 

In the CFT context, a $\bZ_2$ symmetry imposes selection rules on the possible operators appearing in different OPE channels. Let us take a {$\bZ_2$-odd scalar} operator $\sigma$ and consider the $\sigma \times \sigma$ OPE. It can only contain $\bZ_2$-even operators: 
\beq
\sigma \times \sigma \sim \unit + \lambda_{\s\s\e} \e + \lambda_{\s\s T} T^{\mu\nu} + \ldots .
\eeq
Here, $\unit$ is the identity operator, $\e$ is the leading $\bZ_2$-even scalar, $T^{\mu\nu}$ is the stress-energy tensor, and so on. In particular, unlike in \reef{eq:O0}, $\sigma$ does not appear in the OPE.

 \begin{figure}[t!]
    \centering
\includegraphics[width=\figwidth]{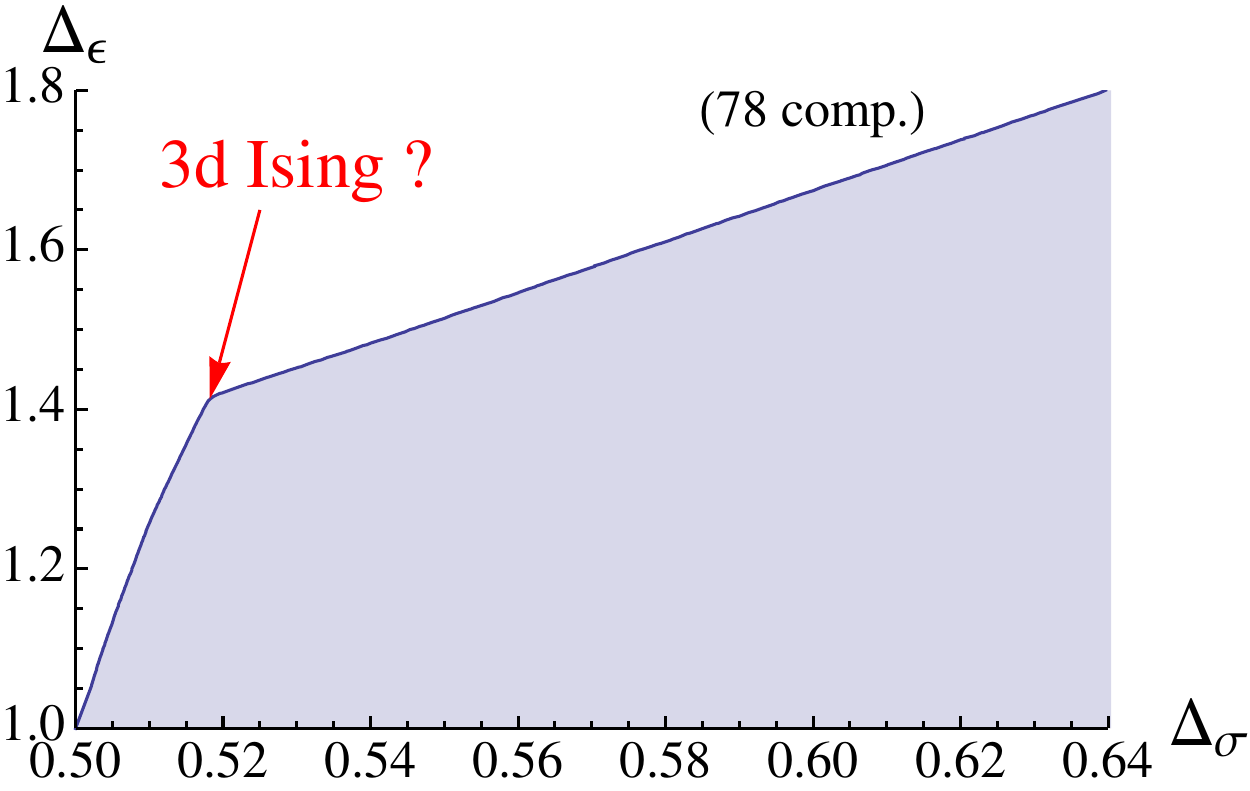}
    \caption{\label{fig:Z2-epsbound}
(Color online) Upper bound on $\Delta_\epsilon$ as a function of $\Delta_\sigma$ in 3d CFTs~\cite{ElShowk:2012ht}.}
  \end{figure}

In this setup, we would like to ask what is the maximal allowed value of $\Delta_\eps$. A numerical bootstrap analysis of the 4pt function function $\<\s\s\s\s\>$ (via linear or semidefinite programming) produces an upper bound on $\Delta_{\e}$ as a function of $\Delta_{\s}$, shown in Fig.~\ref{fig:Z2-epsbound}.\footnote{\textcite{Nakayama:2016jhq} observed empirically that the bounds in Figs.~\ref{fig:multicrit} and~\ref{fig:Z2-epsbound} coincide. A priori one may have expected a stronger bound in Fig.~\ref{fig:Z2-epsbound} due to the extra constraint of not allowing $\sigma$ in the r.h.s.~of the OPE.} The point $\{1/2,1\}$ corresponds to the theory of a free massless scalar while the point $\sim\{0.518, 1.413\}$, sitting near a discontinuity in the boundary, corresponds to the critical 3d Ising model which we discuss further below. Other theories that live in the interior of this region are the critical $O(N)$ models (see Sec.~\ref{sec:ON}), where we can identify $\s$ with a component of the $O(N)$ fundamental $\phi_i$ and $\e$ with a component of the $O(N)$ symmetric tensor $t_{ij}$, as well as the line of mean field theory CFTs with $\Delta_{\e} = 2 \Delta_{\s}$ (see Sec.~\ref{sec:explicit}).

 \begin{figure}[t!]
    \centering
\includegraphics[width=\figwidth]{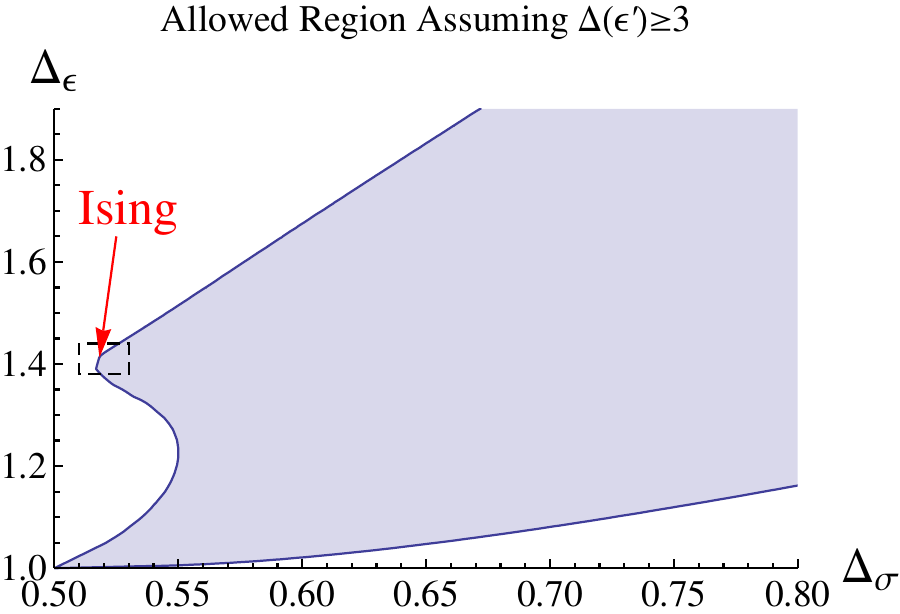}
    \caption{\label{fig:Z2-epsgap3bound}
(Color online) Allowed region in the $\{\Delta_\s,\Delta_\e\}$ plane under the assumption that $\e$ is the only relevant scalar \cite{ElShowk:2012ht}.}
  \end{figure}

A particularly physically interesting class of $\bZ_2$-symmetric CFTs are those with only one relevant $\bZ_2$-even operator (i.e., they have $S=1$). If the microscopic realization of the theory preserves the $\bZ_2$ symmetry, then this condition ensures that only one parameter must be tuned in order to reach the critical point. This allowed region in $\{\Delta_\s,\Delta_\e\}$ space was also computed in \textcite{ElShowk:2012ht} from the $\<\s\s\s\s\>$ correlator, assuming that all scalars aside from the contribution at $\Delta_{\e}$ are irrelevant. This region is shown in Fig.~\ref{fig:Z2-epsgap3bound}, with the assumption having the effect of carving into the allowed region from both the left and from the bottom.

  \begin{figure}[t!]
    \centering
\includegraphics[width=\figwidth]{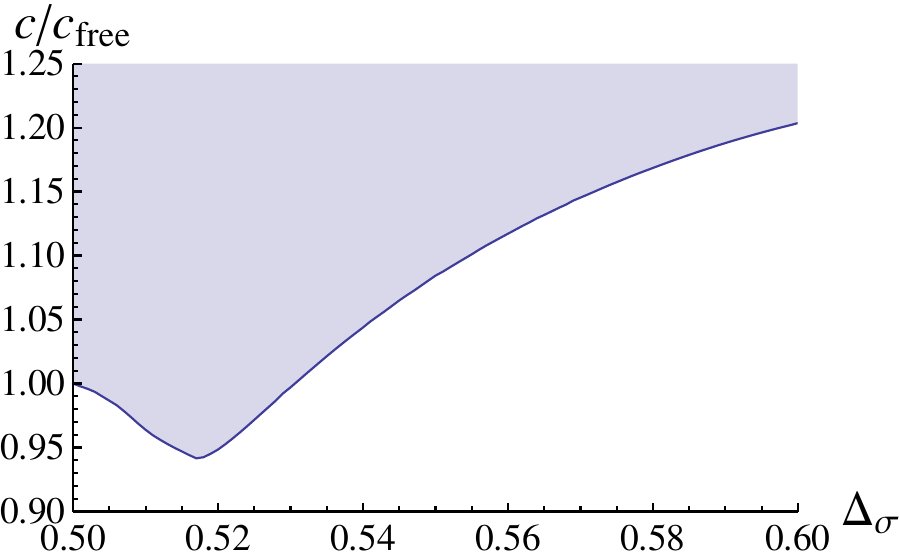}
    \caption{\label{fig:Z2-cbound}
    (Color online) Lower bound on the central charge as a function of $\Delta_\sigma$ \cite{ElShowk:2012ht}.}
  \end{figure}

Another general result from this 4pt function is a lower bound on the central charge $C_T$ shown in Fig.~\ref{fig:Z2-cbound}, obtained by computing an upper bound on the coefficient $\lambda_{\s\s T} \propto \frac{\Delta_{\s}}{\sqrt{C_T}}$ (see Sec.~\ref{sec:ward} and Eq.~\reef{eq:lambdaT}). As $\Delta_{\s} \rightarrow 1/2$, the lower bound on $C_T$ approaches the free scalar value, while near the critical 3d Ising dimension $\Delta_{\s} \sim 0.518$, the lower bound on $C_T$ is seen to have a minimum. This particular bound was computed with the mild assumption $\Delta_{\e} \geq 1$, so it is applicable to any theory living in the allowed region seen in Fig.~\ref{fig:Z2-epsbound}. 

   \begin{figure}[t!]
    \centering
\includegraphics[width=\figwidth]{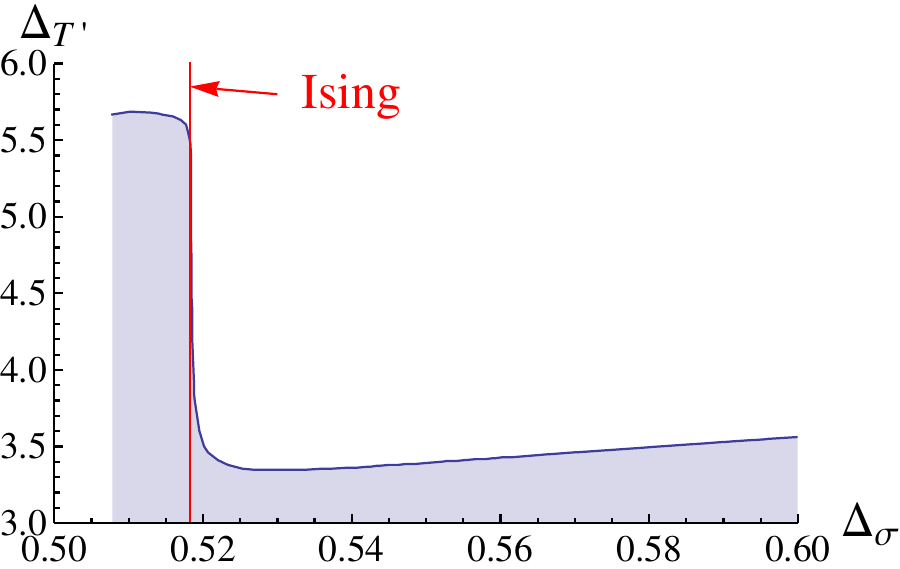}
    \caption{\label{fig:Z2-spin2bound}
    (Color online) Upper bound on the dimension $\Delta_{T'}$ of the first $\bZ_2$-even spin-2 operator after the stress tensor, as a function of $\Delta_\sigma$ \cite{ElShowk:2012ht}.}
  \end{figure}

   \begin{figure}[t!]
    \centering
\includegraphics[width=\figwidth]{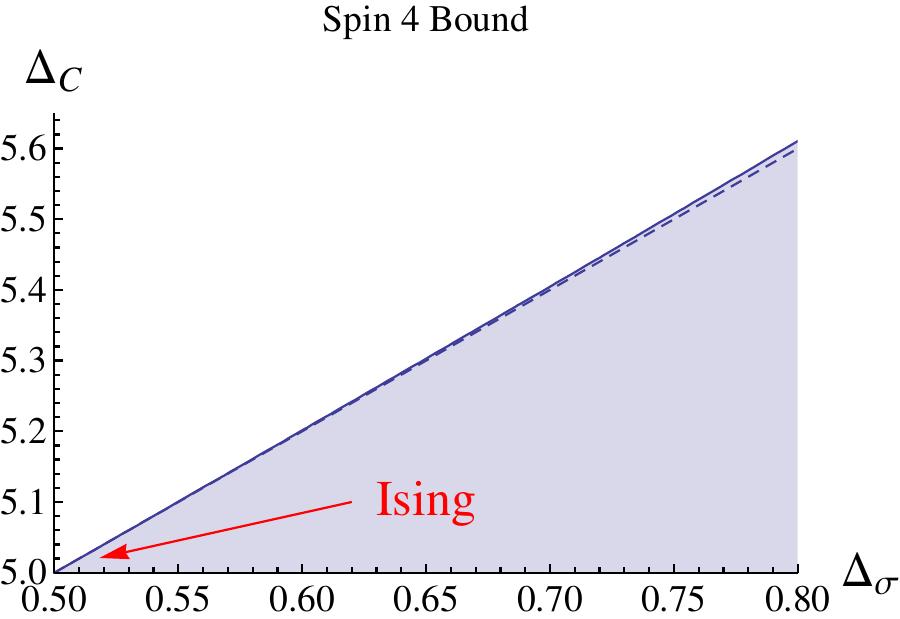}
    \caption{\label{fig:Z2-spin4-bound}
   (Color online) Upper bound on the dimension of the leading $\bZ_2$-even spin-4 operator \cite{ElShowk:2012ht}.}
  \end{figure}

Before we move on to discussing what can be learned from systems of several 4pt functions, we would like to highlight that upper bounds on the leading unknown scaling dimension in other channels can also be computed and are often quite strong. For example, an upper bound on the leading unknown spin-2 dimension $\Delta_{T'}$ (the first $\bZ_2$-even spin-2 operator after the stress tensor) is shown in Fig.~\ref{fig:Z2-spin2bound}, and an upper bound on the leading spin-4 dimension $\Delta_C$ is shown in Fig.~\ref{fig:Z2-spin4-bound}. The bound on $\Delta_{T'}$ shows a sharp jump near the critical 3d Ising value, while no such transition is seen in the bound on $\Delta_C$ (which is close to being saturated by MFT: $\Delta_C = 2\Delta_{\sigma} + 4$). The jump in $\Delta_{T'}$ shows that it is possible for the low-dimension spin-2 operator present in the spectrum for $\Delta_{\s} \gtrsim 0.52$ to decouple at smaller values {of $\Delta_{\s}$}. We discuss operator decoupling phenomena further in Sec.~\ref{sec:Z2-spectrum} .

   \begin{figure}[t!]
    \centering
\includegraphics[width=\figwidth]{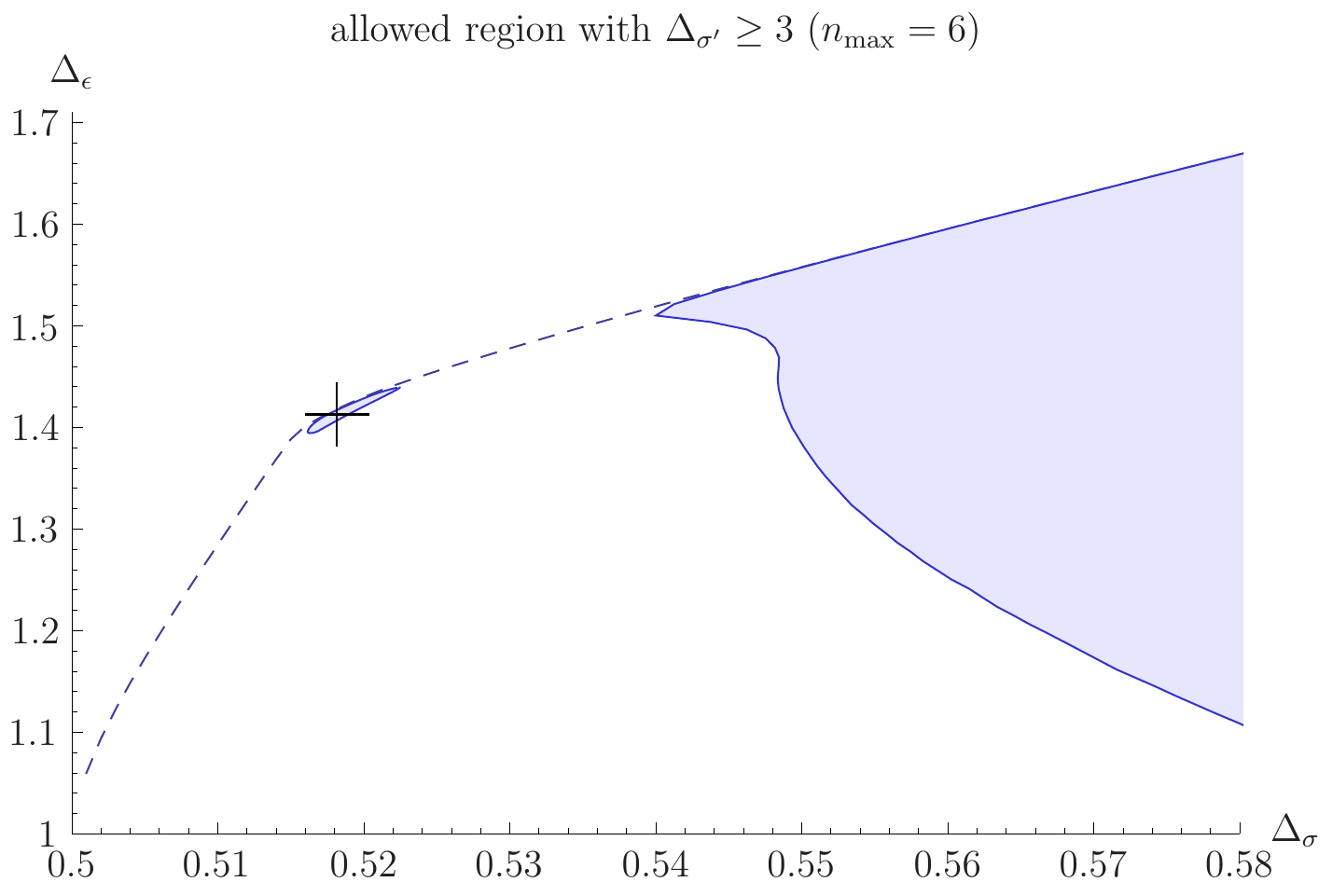}
    \caption{\label{fig:Z2-mixed-sigpgap}
    (Color online) Allowed region following from the analysis of three 4pt functions assuming $\Delta_{\sigma'}\ge 3$ with no assumption on $\Delta_{\eps'}$ \cite{Kos:2014bka}.}
  \end{figure}

Next one can ask what is the effect of adding constraints from other 4pt functions. So far the main system that has been studied in the literature is $\{\<\s\s\s\s\>, \<\s\s\e\e\>, \<\e\e\e\e\>\}$, though other systems may also prove interesting. An advantage of including the correlator $\<\s\s\e\e\>$ is that it allows one to probe the $\bZ_2$-odd operators appearing in the OPE: 
\beq
\s \times \e \sim \lambda_{\s\e\s} \s + \lambda_{\s\e\s'} \s' + \ldots .
\eeq
In \textcite{Kos:2014bka} it was found that with no assumptions this system leads to an allowed region identical to Fig.~\ref{fig:Z2-epsbound}, while by inputting the assumption of a single relevant $\bZ_2$-odd operator (i.e., $\Delta_{\s'} \geq 3$) it leads to the allowed region shown in Fig.~\ref{fig:Z2-mixed-sigpgap}. In this plot one can see a detached ``island" containing the critical Ising model as well as a ``bulk" region further to the right. This ``bulk" region has so far not been systematically explored in the literature: it would be very interesting to understand what other CFTs lie inside of it.

     \begin{figure}[t!]
    \centering
\includegraphics[width=\figwidth]{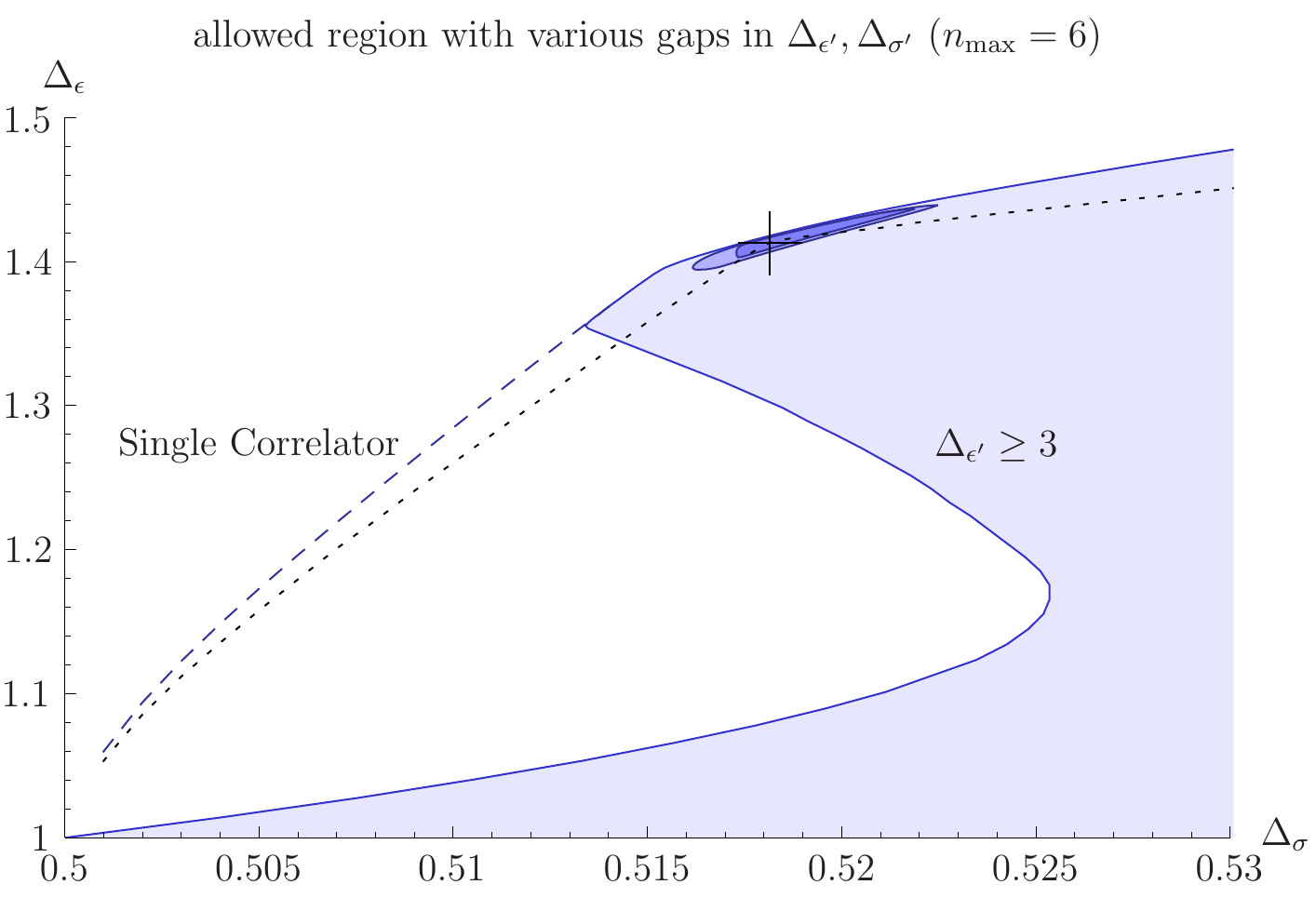}
    \caption{\label{fig:Z2-mixed-differentgaps}
    (Color online) This plot assumes $\Delta_{\epsilon'} \geq 3$ (light blue), $\Delta_{\sigma'} \geq 3$ (medium blue), or both gaps simultaneously (dark blue). Figure from \cite{Kos:2014bka}.}
  \end{figure}

In Fig.~\ref{fig:Z2-mixed-differentgaps} we also show the difference between assuming $\Delta_{\e'} \geq 3$, $\Delta_{\s'} \geq 3$, and both assumptions simultaneously. One can see that the assumption of a gap in the $\bZ_2$-odd spectrum is primarily responsible for creating the detached region. In the next section we describe the connection to the critical Ising model in more detail, as well as the techniques and additional inputs that can be used to make this detached island as small as possible.

\subsubsection{Critical Ising model}
\label{sec:Z2-Ising}

Perhaps the most well-known 3d CFT is the critical 3d Ising model. The study of this model has a long history~\cite{DombGreenVol3}, in part because it describes critical behavior in uniaxial magnets, liquid-vapor transitions, binary fluid mixtures, the quark-gluon plasma, and more \cite{Pelissetto:2000ek}. While these applications are predominantly for systems in three spatial dimensions at finite temperature, described by a 3d Euclidean CFT, the critical Ising model can also be realized as a Lorentzian (2+1)d quantum critical point~\cite{Fradkin-Susskind,Henkel-Ising3d}. Here we work in the Euclidean signature; the Lorentzian version is obtainable by Wick rotation and has the same set of CFT data.

In its original formulation as a model of ferromagnetism, the 3d Ising model is described using a set of spins $s_i=\pm 1$ on a cubic lattice in $\bR^3$ with nearest neighbor interactions, with partition function
\beq
\label{eq:Ising}
Z = \sum_{\{s_i\}} \exp\Bigl(- J \sum_{\<ij\>} s_i s_j \Bigr).
\eeq
At a critical value of the coupling $J$, the model becomes a nontrivial CFT at long distances. Notice that the lattice model has a manifest $\bZ_2$ symmetry under which $s_i \rightarrow -s_i$. This symmetry is inherited by the CFT, which contains local operators that are either even or odd under its $\bZ_2$ global symmetry.

Another microscopic realization is in terms of a continuous scalar field theory in 3 dimensions, with action
\beq
S = \int d^3 x \left( \frac12 (\partial \sigma)^2 + \frac12 m^2 \sigma^2 + \frac{1}{4!} \lambda \sigma^4 \right),
\eeq
which also has a $\bZ_2$ symmetry under which $\sigma \rightarrow -\sigma$. Because both $m^2$ and $\lambda$ describe relevant couplings, this theory is described by a free scalar at short distances but has nontrivial behavior at long distances. At a critical value of the dimensionless ratio $m^2/\lambda^2$ the long-distance behavior is described by a CFT, which is the same as for the above lattice model.

From the conformal bootstrap perspective, the Ising CFT has a $\bZ_2$ global symmetry, one relevant $\bZ_2$-odd scalar operator $\sigma$, and one relevant $\bZ_2$-even scalar operator $\epsilon$. This is evident from experimental realizations, where $\bZ_2$-preserving microscopic realizations require one tuning (e.g., tuning the temperature in uniaxial magnets) and $\bZ_2$-breaking microscopic realizations require two tunings (e.g., tuning both temperature and pressure in liquid-vapor transitions).\footnote{The existence of $\bZ_2$-breaking liquid-vapor experimental realizations, allowing one to get $\bZ_2$ as an emergent symmetry and predict the total number of relevant scalars, is a {nice feature of} the Ising model which does not have analogues for the $O(N)$ models.}   Note that the assumption that the only relevant scaling dimensions are $\Delta_{\s}$ and $\Delta_{\e}$ is the same assumption that went into producing the dark blue detached region of Fig.~\ref{fig:Z2-mixed-differentgaps}.

\begin{figure}[t!]
    \centering
\includegraphics[width=\figwidth]{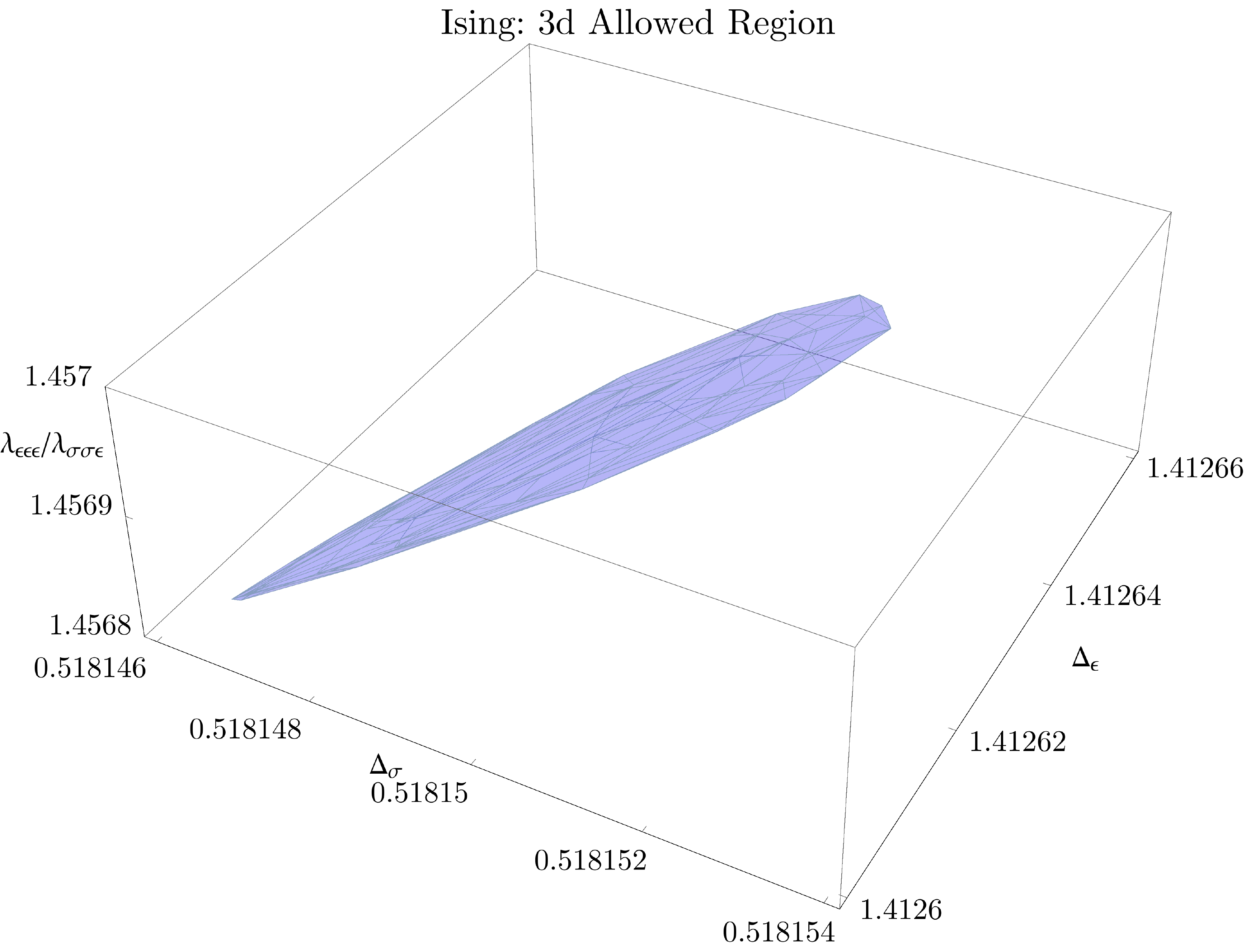}
    \caption{\label{fig:Z2-3dIsingIsland}
    (Color online) Allowed region in the $\{\Delta_\sigma,\Delta_\e,\lambda_{\e\e\e}/\lambda_{\s\s\e}\}$ space obtained in \textcite{Kos:2016ysd}. 
  }
  \end{figure}

\begin{figure}[t!]
    \centering
\includegraphics[width=\figwidth]{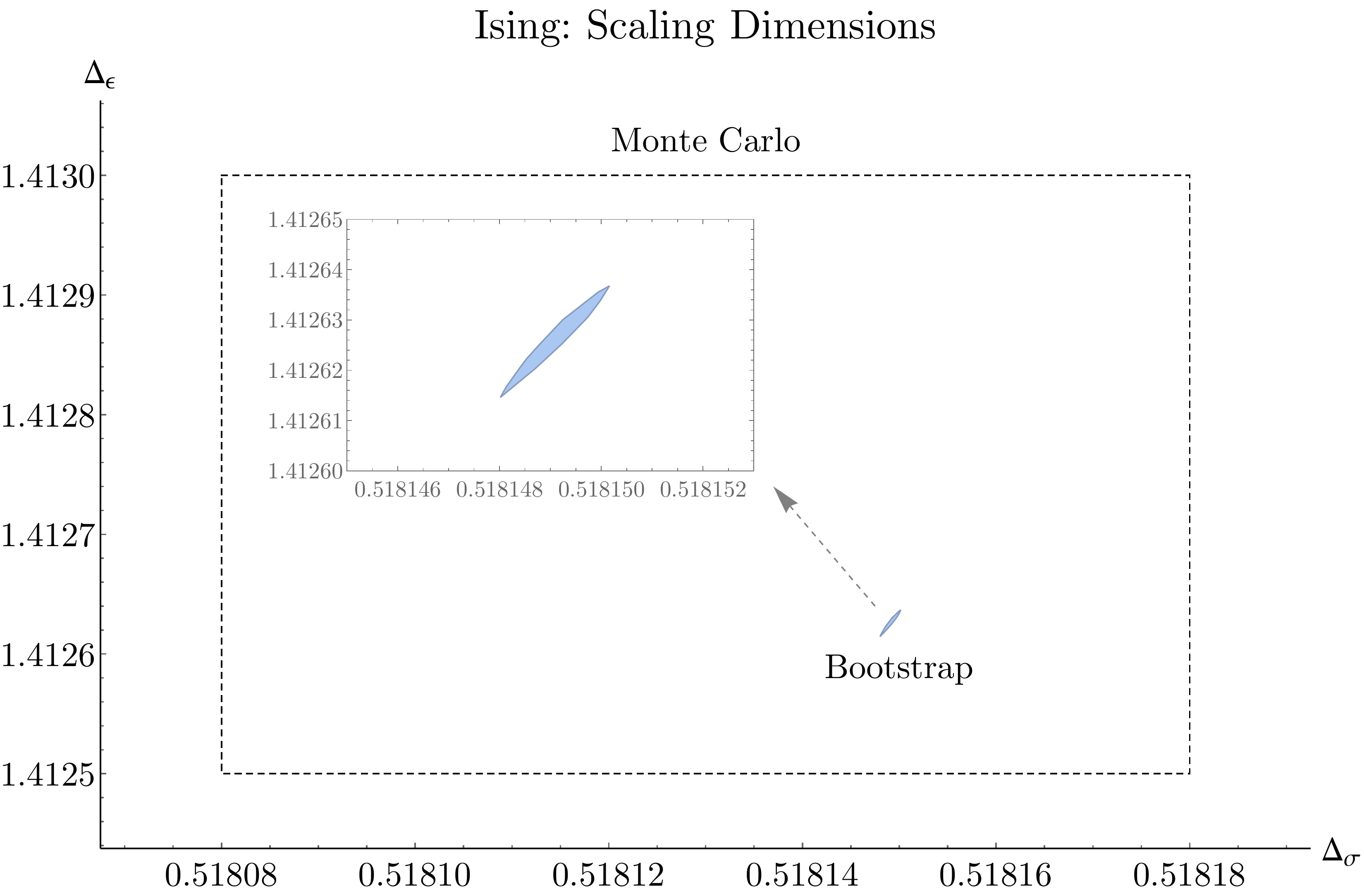}
    \caption{\label{fig:Z2-IsingIsland}
    (Color online) Projection of the 3d region in Fig.~\ref{fig:Z2-3dIsingIsland} on the $\{\Delta_\sigma,\Delta_\e\}$ plane and its comparison with a Monte Carlo prediction for the same quantities \cite{Kos:2016ysd}.
  }
  \end{figure}

\begin{figure}[t!]
    \centering
\includegraphics[width=\figwidth]{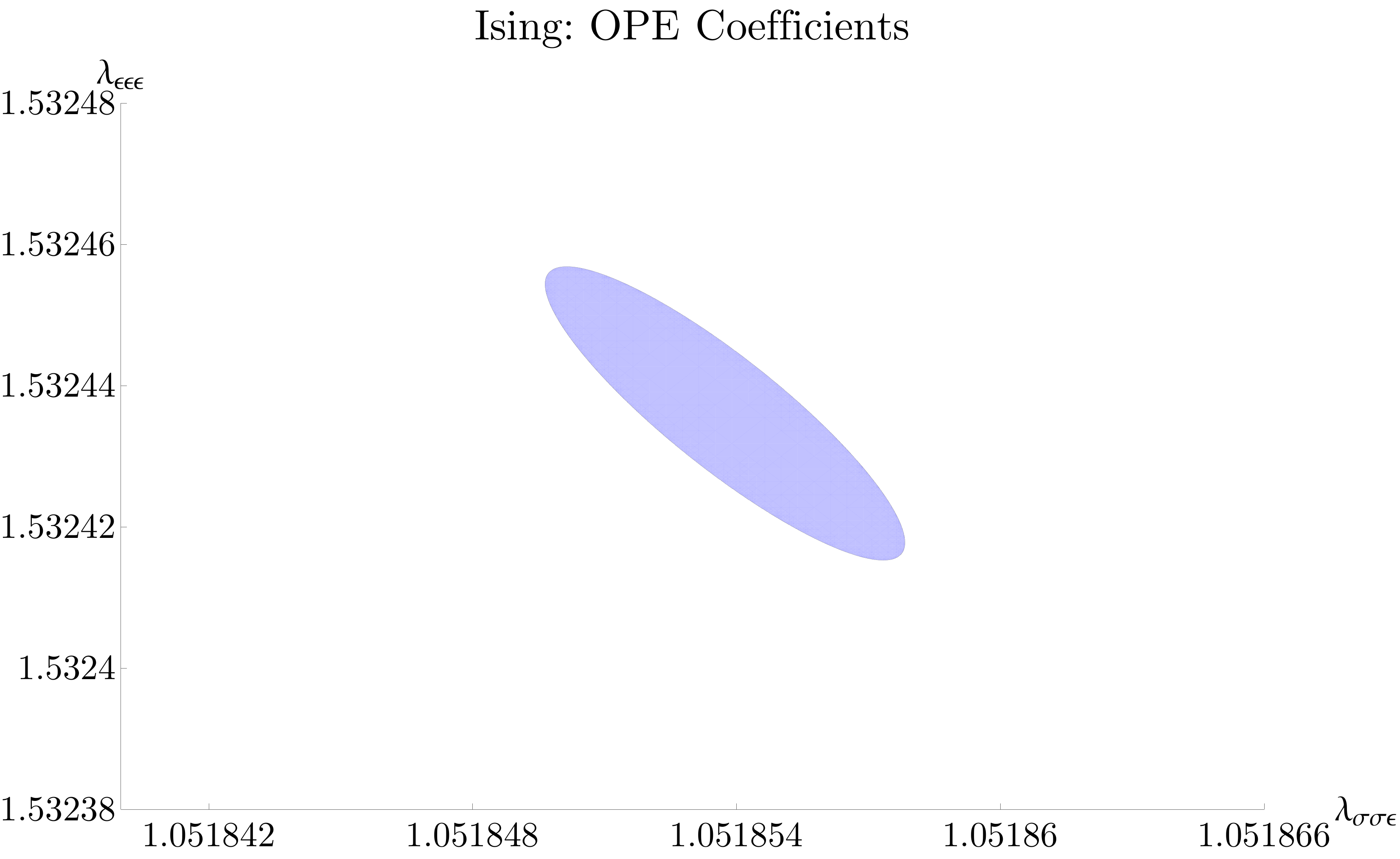}
    \caption{\label{fig:Z2-OPEBound}
    (Color online) Variation of $\lambda_{\e\e\e}$ and $\lambda_{\s\s\e}$ within the allowed region in Fig.~\ref{fig:Z2-3dIsingIsland} \cite{Kos:2016ysd}.
  }
  \end{figure}

\textcite{Kos:2016ysd} pursued a numerical analysis of the mixed-correlator bootstrap system containing $\sigma$ and $\epsilon$ to high derivative order. In addition, they studied the impact of scanning over different possible values of the ratio $\lambda_{\e\e\e}/\lambda_{\s\s\e}$. This scan effectively inputs the information that there is a single operator in the OPE occurring at the scaling dimension $\Delta_{\e}$,  whereas the plot of Fig.~\ref{fig:Z2-mixed-differentgaps} allowed for the possibility of multiple degenerate operator contributions at the dimension $\Delta_{\e}$.\footnote{More precisely, the scan inputs that the outer product of OPE coefficients $\left(\lambda_{\s\s\e} \quad \lambda_{\e\e\e}\right) \otimes \left(\lambda_{\s\s\e} \quad \lambda_{\e\e\e}\right)$ appearing in Eq.~(\ref{eq:crossingequationwithv}) at dimension $\Delta_{\e}$ is a rank 1 matrix, rather than the more generic rank 2 possibility which occurs if there are degenerate contributions.} This led to the three-dimensional allowed region shown in Fig.~\ref{fig:Z2-3dIsingIsland} and its projection to the $\{\Delta_{\s},\Delta_{\e}\}$ plane shown in Fig.~\ref{fig:Z2-IsingIsland}. In addition, for each point in this region the magnitude of the leading OPE coefficients were also bounded, with the result shown in Fig.~\ref{fig:Z2-OPEBound}. These world-record numerical determinations are summarized below in Table.~\ref{tab:lowestdim}.

Finally let us mention that recent studies of the conformal bootstrap for stress-tensor 4pt functions have also made contact with the 3d Ising model. In particular, after inputting  known values of the leading parity-even spectrum, \textcite{Dymarsky:2017yzx} gave a new bound on the leading parity-odd $\mathbb{Z}_2$-even scalar, $\Delta_{\text{odd}} < 11.2$, and constrained the independent coefficient in the stress-tensor 3pt function (parametrized by the variable $\theta$) to be in the range $0.01 < \theta < 0.05$ if $\Delta_{\text{odd}} > 3$ and in a tighter range $0.01 < \theta < 0.018-0.019$ if $\Delta_{\text{odd}}$ is close to saturating its bound. We will discuss these constraints in more detail in Sec.~\ref{sec:JandT}.

\subsubsection{Spectrum extraction and rearrangement}
\label{sec:Z2-spectrum}

We have seen in the previous section the remarkable precision with which the leading scaling dimensions of the critical 3d Ising model can be determined. This raises the immediate question of how well we can extract other operator dimensions and OPE coefficients in the spectrum using bootstrap methods (specifically the strategies described in Sec.~\ref{sec:spectrum-extraction})

\begin{figure}[t!]
    \centering
\includegraphics[width=\figwidth]{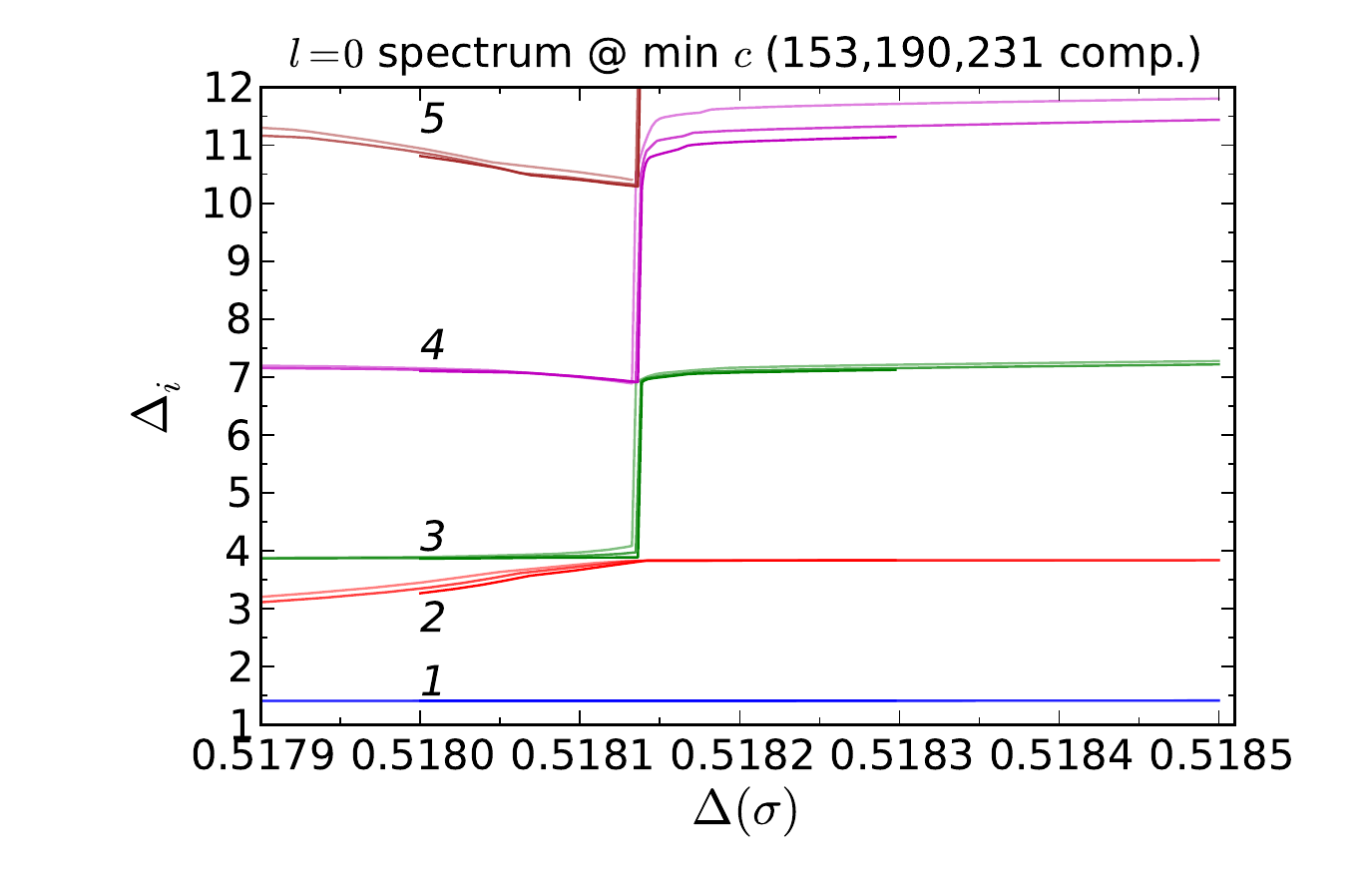}
    \caption{\label{fig:Z2-ct-spin0}
    (Color online) The spectrum of $\bZ_2$-even scalar operators appearing in the solution to crossing minimizing $C_T$ near $\Delta_\sigma$ corresponding to the 3d Ising model \cite{El-Showk:2014dwa}. Line 1 corresponds to the $\eps$ operator and shows little variation on the scale of this plot. All other lines exhibit the spectrum rearrangement phenomenon.
  }
  \end{figure}

Even prior to the mixed-correlator studies mentioned above, \textcite{El-Showk:2014dwa} extracted the spectrum using the primal simplex method strategy, from a solution to crossing for the $\<\s\s\s\s\>$ correlator which minimizes the central charge $C_T$. For example, Fig.~\ref{fig:Z2-ct-spin0} shows the scalar operators in the extracted spectrum as a function of $\Delta_\sigma$ near the 3d Ising model. A fascinating feature of these plots is the bifurcation of operators that occurs at the Ising value of $\De_\s$, which can be interpreted as a decoupling of one of the operators in the spectrum. This ``spectrum rearrangement" phenomenon has yet to be fully understood, but speculatively it could be connected to the nonperturbative equations of motion (i.e., the 3d analogue of the relation $\s \partial^2 \s \sim \s^4$ at the Wilson-Fisher fixed point) or a higher-dimensional extension of the null state conditions in the 2d Ising CFT (see also Sec.~\ref{sec:why}).\footnote{The 2d analogue of Fig.~\ref{fig:Z2-epsbound} also displays a sharp kink exactly at the location of the 2d Ising model \cite{Rychkov:2009ij}, at which the corresponding extremal solution displays a decoupling of states expected from the null state conditions \cite{El-Showk:2014dwa}. The upper bound to the right of the kink can be interpreted as a one-parameter family of unitary 4pt functions which for a discrete sequence of $\Delta_\sigma$'s reduce to the 4pt function of the $\phi_{1,2}$ operator in the higher unitary minimal models, see \textcite{Liendo:2012hy} and \textcite{Behan:2017rca}. While these higher minimal models exhibit further null state conditions, they are not visible in this 4pt function, and hence do not lead to kinks in this bound. 
	}

On the other hand, spectrum extraction using the extremal functional method was applied to the critical 3d Ising model by \textcite{Komargodski:2016auf,Simmons-Duffin:2016wlq}. In the latter work, for a set of 20 trial points distributed within the island of Fig.~\ref{fig:Z2-3dIsingIsland}, $C_T$-minimization was performed and the zeros of the extremal functional were found. While some zeros jump significantly when moving from point to point, many of them are found to be present in all families with tiny variations. About a hundred such ``stable zeros" were identified, and are believed to represent operators which truly exist in the 3d Ising CFT, providing a remarkable view of the spectrum of this theory. The subset of stable operators with dimensions $\De \leq 8$, and their OPE coefficients, are shown in Table~\ref{tab:lowestdim}. 

This approach, while not fully rigorous, is intuitively justified as a means to extend the reach of rigorous analysis which produced the island in Fig.~\ref{fig:Z2-3dIsingIsland}. The errors on stable operator dimensions and OPE coefficients are assigned as standard deviations in the set of trial points. Although these errors are not rigorous, as opposed to rigorous errors implied by Figs.~\ref{fig:Z2-3dIsingIsland}, \ref{fig:Z2-IsingIsland}, and \ref{fig:Z2-OPEBound}, we believe that they represent realistic estimates. In future studies the error estimates can be further checked by enlarging the set of trial points and by extremizing multiple quantities as opposed to just $C_T$.

\begin{figure}[t!]
    \centering
    \scalebox{0.8}{
\includegraphics[width=\figwidth]{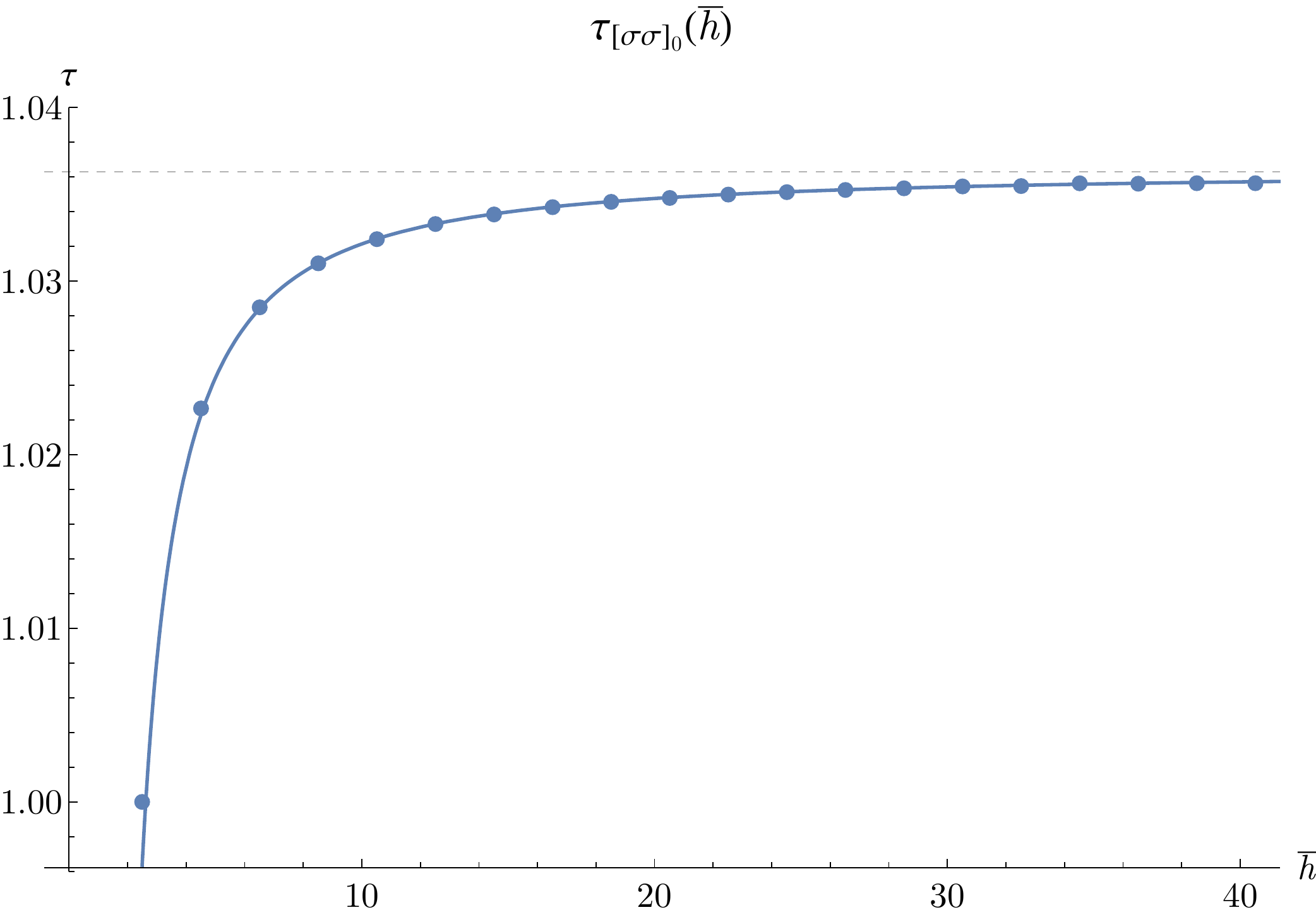}
}
    \caption{\label{fig:tauSigSig0}
    (Color online) Comparison of the extremal functional spectrum with the analytic bootstrap \cite{Simmons-Duffin:2016wlq}; see the text.
  }
  \end{figure}

Results of this approach for the leading towers of low-twist operators (of increasing spin) have also been tested against the analytical bootstrap computations in the lightcone limit\footnote{{See~\textcite{Fitzpatrick:2012yx}, \textcite{Komargodski:2012ek}, \textcite{Alday:2015ewa}, and \textcite{Simmons-Duffin:2016wlq}.}} which yield analytical expressions for the large-spin asymptotics. In Fig.~\ref{fig:tauSigSig0}, the data points extracted from the extremal functional approach show the leading twist ($\tau=\De-\ell$) trajectory in the $\Z_2$-even sector as a function of $\bar{h} = \ell+\tau/2$, while the curve shows the analytical computation, displaying excellent agreement with the data even down to small spins. Similar good agreement was also found with the extracted OPE coefficients and subleading trajectories, as well as in the $\Z_2$-odd sector. 

We also report here the prediction for the central charge from the above $C_T$-minimization over the 20 points in the island \cite{Simmons-Duffin-private1}:
\beq
\label{eq:central-charge}
C^{\rm Ising}_T/C_T^{\text{free boson}} =0.9465389(12)\,,
\eeq
improving the previous $C_T$-minimization determination by \textcite{El-Showk:2014dwa}.\footnote{One can also extract $C_T$ from Table \ref{tab:lowestdim} using $\lambda_{\sigma\sigma T}\propto {\Delta_\sigma}/{\sqrt{C_T}}$. While consistent with \reef{eq:central-charge}, this would have a larger error, because the errors on $\lambda_{\sigma\sigma T}$ and $\Delta_\sigma$ are correlated.}

\begin{table}
\begin{center}
{\small
\begin{tabular}{|c|c|l|l|l|l|l|}
\hline
$\cO$ & $\Z_2$ & $\ell$ & $\De$ &  $f_{\s\s\cO}$ & $f_{\e\e\cO}$ \\
\hline
$\e$ & $+$ & 0 & $1.412625{\bf\boldsymbol(10\boldsymbol)}$ & $1.0518537{\bf\boldsymbol(41\boldsymbol)}$ & $1.532435{\bf\boldsymbol(19\boldsymbol)}$ \\
$\e'$ & $+$ & 0 & $3.82968(23)$ &  $0.053012(55)$ & $1.5360(16)$ \\
& $+$ & 0 & $6.8956(43)$ &  $0.0007338(31)$ & $0.1279(17)$ \\
& $+$ & 0 & $7.2535(51)$ &  $0.000162(12)$ & $0.1874(31)$ \\
$T_{\mu\nu}$ & $+$ & 2 & $3$ &  $0.32613776(45)$ & $0.8891471(40)$ \\
$T'_{\mu\nu}$ & $+$ & 2 & $5.50915(44)$  & $0.0105745(42)$ & $0.69023(49)$ \\
& $+$ & 2 & $7.0758(58)$ & $0.0004773(62)$ & $0.21882(73)$ \\
$C_{\mu\nu\rho\s}$ & $+$ & 4 & $5.022665(28)$ & $0.069076(43)$ & $0.24792(20)$ \\
& $+$ & 4 & $6.42065(64)$  & $0.0019552(12)$ & $-0.110247(54)$ \\
& $+$ & 4 & $7.38568(28)$ & $0.00237745(44)$ & $0.22975(10)$ \\
& $+$ & 6 & $7.028488(16)$  & $0.0157416(41)$ & $0.066136(36)$ \\
\hline
\hline
$\cO$ & $\Z_2$ & $\ell$ & $\Delta$  & $f_{\s\e\cO}$ &-\\
\hline
$\s$ & $-$ & 0 & $0.5181489{\bf\boldsymbol(10\boldsymbol)}$  & $1.0518537{\bf\boldsymbol(41\boldsymbol)}$ &\\
$\s'$ & $-$ & 0 & $5.2906(11)$  & $0.057235(20)$  &\\
& $-$ & 2 & $4.180305(18)$  & $0.38915941(81)$  &\\
& $-$ & 2 & $6.9873(53)$  & $0.017413(73)$  &\\
& $-$ & 3 & $4.63804(88)$ & $0.1385(34)$  &\\
& $-$ & 4 & $6.112674(19)$ & $0.1077052(16)$  &\\
& $-$ & 5 & $6.709778(27)$ & $0.04191549(88)$ &\\
\hline
\end{tabular}
}
\end{center}
\caption{
Stable operators in the critical 3d Ising model with dimensions $\De\leq 8$ \cite{Simmons-Duffin:2016wlq}. Conventional names are shown in the leftmost column when available. Errors in bold are rigorous. All other errors are non-rigorous but, in our opinion, realistic. See Eq.~\reef{eq:central-charge} for the central charge prediction from the same study.
Because we have chosen a different conformal block normalization convention, the OPE coefficients are related to our convention by $\lambda_{ij\cO} = 2^{\ell/2} f_{ij\cO}$ (see Table \ref{tab:cb_norm}). 
}
\label{tab:lowestdim}
\end{table}

\subsubsection{Why kink? Why island?}
\label{sec:why}

One may be wondering why the 3d Ising model happens to live at a kink in Fig.~\ref{fig:Z2-epsbound}. Plausibly, this has to do with the minimality of the spectrum of exchanged operators required to satisfy the crossing relation. In the interior of the allowed region in Fig.~\ref{fig:Z2-epsbound}, the solution to crossing is not unique. When working numerically, a typical solution contains as many operators as the number of derivatives at $z=\bar z=1/2$ one is keeping in \reef{eq:crossingvec}, \reef{eq:F}. On the other hand, when one approaches the boundary of the allowed region in Fig.~\ref{fig:Z2-epsbound}, the nature of the solution changes in that the operators first organize into pairs with nearby dimensions, and the pairs then merge into single operators at the boundary \cite{El-Showk:2014dwa}. Thus the extremal solutions to crossing are quite economical, containing {many} fewer operators than the interior solutions, roughly {by} a half.\footnote{It is a bit more than half because doubling never occurs for operators which remain at the unitarity bound, such as the stress tensor {(if present in the extremal solution)}, and for operators which saturate the gaps that one is imposing. In general, whether doubling occurs in the bulk of the spectrum depends on how many second-order zeros the extremal functional has. If there are too many zeros, then for some of them, called ``singles" in \textcite{El-Showk:2016mxr}, doubling will not occur.  See also Sec.~\ref{sec:flow} for the flow method which uses such considerations to move along the boundary of the allowed region.}

Further reduction of the spectrum occurs at the kink. When one approaches the kink moving along the boundary, squared OPE coefficients of certain operators tend to zero. Further analytic continuation of the solution beyond the kink would be inconsistent with unitarity. Thus two different solution branches meet at the kink \cite{El-Showk:2014dwa}, and the spectrum exhibits rearrangement phenomena mentioned in Sec.~\ref{sec:Z2-spectrum}.

To summarize, that the 3d Ising model lives at a kink suggests that it is a CFT with a particularly minimal spectrum of operators. If this idea can be made precise, perhaps it can pave the way to an exact solution.

Leaving the kink aside, let us discuss the island. It is perhaps not surprising that considering crossing for several 4pt functions the allowed region shrinks compared to what was allowed when considering just one 4pt function. It is however altogether unexpected and remarkable that considering only three 4pt functions, plus a {physically motivated} and robust\footnote{Islands can be also produced for the Ising and other CFTs using a single 4pt function and reasonable assumptions about gaps in the spin-1 and spin-2 operator spectrum \cite{Li:2017kck}. The robustness of these results (i.e.~their independence of the numerical values of the assumed gaps in a certain range) needs further investigation.} assumption of only two relevant operators, allows one to produce the tiny island shown in Figs.~\ref{fig:Z2-mixed-sigpgap} and \ref{fig:Z2-IsingIsland}.

It is not currently understood why this happens. Would the island continue to shrink indefinitely with increasing the number of included derivatives? Or would it stabilize, requiring one to add further correlators to fully fix the CFT? More generally, is it sufficient to include only 4pt functions of relevant operators or are external irrelevant operators also needed to have a unique solution? These are fascinating questions for the future.

We will see many kinks and islands in the subsequent sections of this review, about which similar considerations can be made.

\subsubsection{Nongaussianity}

\label{sec:nongauss}
Since the leading spectrum and OPE coefficients of the critical 3d Ising model are now known to a high degree of precision, it is possible to reconstruct the full 4pt function $\<\s\s\s\s\>$ over a wide range of cross ratios. One can then probe the question of how much this 4pt function deviates from the ``gaussian", i.e.~fully disconnected, form $\<\s\s\s\s\> = \<\s\s\>\<\s\s\> + \text{perms}$. This question is also motivated by the fact that the Ising model contains higher-spin operators with dimensions that deviate by a small amount from those of higher-spin currents, see Fig.~\ref{fig:tauSigSig0}. The first two of these operators are the $\bZ_2$-even spin-4 and spin-6 operators in Table \ref{tab:lowestdim}, of dimension close to 5 and 7 respectively.

\begin{figure}[t!]
    \centering
\includegraphics[width=\figwidth]{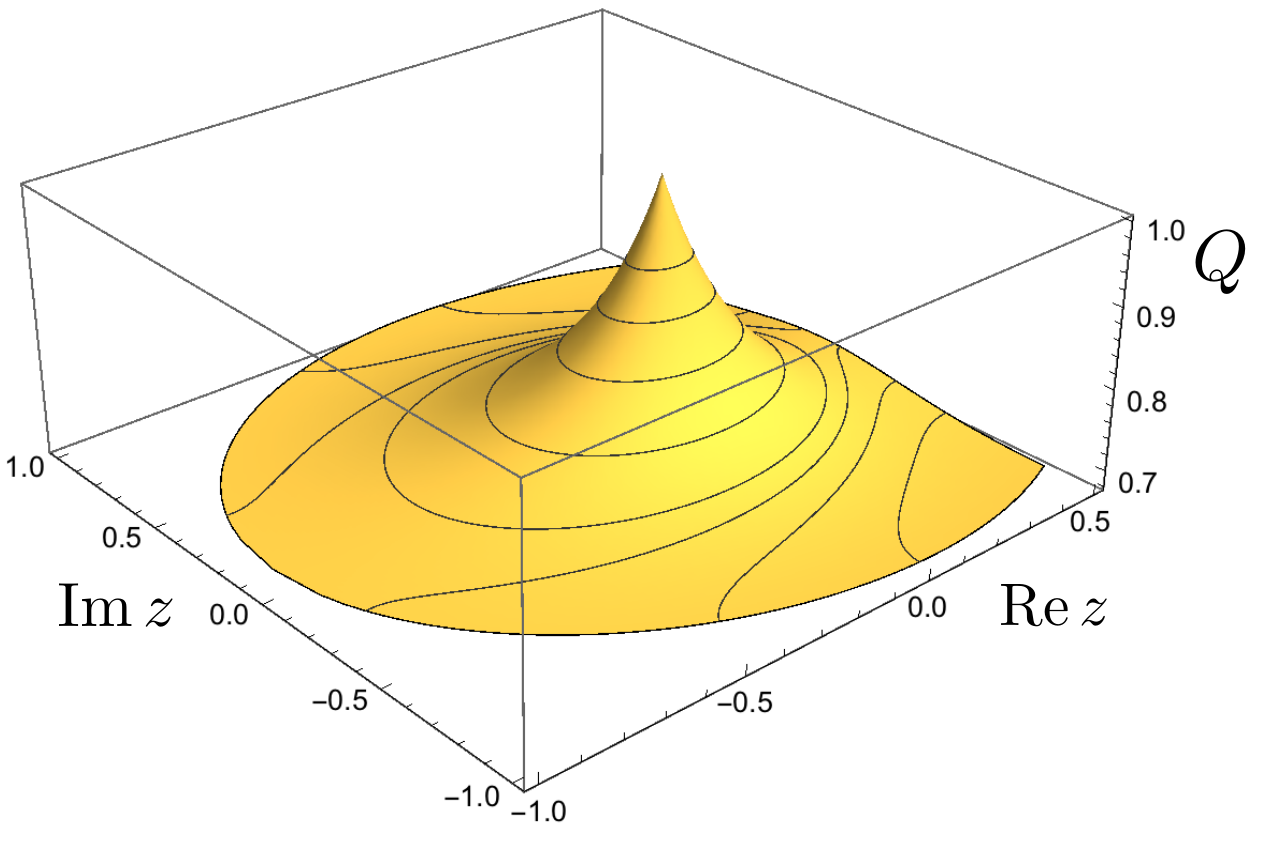}
    \caption{\label{fig:3d}
    (Color online) The nongaussianity ratio $Q$ in the critical 3d Ising model \cite{Rychkov:2016mrc}.
  }
  \end{figure}

\textcite{Rychkov:2016mrc} probed this question quantitatively using bootstrap data to reconstruct the ratio $Q(z,\bar{z}) = \frac{g(z,\bar{z})}{1+(z\bar{z})^{\Delta_\s} + (z\bar{z}/(1-z)(1-\bar{z}))^{\Delta_\s}}$ in the critical 3d Ising model, where the denominator corresponds to the ``gaussian" expectation. A plot of this deviation over a fundamental domain in the complex $z$ plane is shown in Fig.~\ref{fig:3d}. They found e.g. that $Q < 0.75$ over a wide range of cross-ratio space and that it attains a minimum value of $Q_{\min} \approx 0.683$. Thus, any attempt to explain the small anomalous dimensions of higher-spin operators must account for this significant nongaussianity.

\subsubsection{Boundary and defect bootstrap, nontrivial geometries, off-criticality}

\label{sec:bdry}

It is also interesting to study the physics of defects in the 3d Ising model. These include both co-dimension one defects (e.g., flat 2d boundaries or interfaces) and co-dimension two defects (i.e., 1d line defects).\footnote{\label{note:defects}See~\textcite{Gadde:2016fbj}, \textcite{Billo:2016cpy}, \textcite{Lauria:2017wav}, \textcite{Fukuda:2017cup}, \textcite{Rastelli:2017ecj}, \textcite{Herzog:2017xha}, \textcite{Herzog:2017kkj}, and \textcite{Lemos:2017vnx} for some recent general discussions of defects in CFT.} Here we would like to highlight for the reader some recent numerical bootstrap studies of such defects.

Bootstrap constraints in the 3d Ising model in the presence of a flat 2d boundary were first studied using linear programming techniques in \textcite{Liendo:2012hy}, where a number of rigorous bounds were placed on the scaling dimensions and OPE coefficients of boundary operators for different choices of boundary conditions, corresponding to the ``special" and ``extraordinary" transitions, assuming positivity of the bulk channel expansion coefficients. Estimates of the leading boundary data using the truncation method were also computed by \textcite{Gliozzi:2015qsa} and \textcite{Gliozzi:2016cmg}, where precise estimates applicable to the boundary condition of the ``ordinary" transition could also be made. 

Studies have also been performed of the $\mathbb{Z}_2$ twist line defect in the 3d Ising model, constructed on the lattice by reversing the Ising coupling on a semi-infinite half-plane. The 1d boundary of this half-plane then yields the twist line defect, which can also be defined in terms of its simple monodromy properties. Local operators living on this defect were studied using both Monte Carlo techniques~\cite{Billo:2013jda} and numerical bootstrap (linear programming) techniques~\cite{Gaiotto:2013nva}, with excellent agreement.  
   
A related line of inquiry is to study CFTs such as the critical 3d Ising model on nontrivial geometries, {the nontrivial case being manifolds not globally conformally equivalent to infinite flat space.}\footnote{CFT correlation functions on manifolds conformally equivalent to flat space, such as the sphere $S^d$ or the ``cylinder" $S^{d-1}\times \bR$, can be obtained from the flat space correlators via a Weyl transformation.} This is motivated in part by the search for a higher-dimensional analogue of modular invariance. One concrete realization has been to study the 3d Ising model on real projective space~\cite{Nakayama:2016cim}. In this case the unknown coefficients in one-point functions of scalar primary operators $\<\mathcal{O}\> \propto A_{\mathcal{O}}$ enter into a variant of the bootstrap equations called the cross-cap bootstrap equations. Numerical truncation studies of the cross-cap bootstrap equations in this work have yielded new nontrivial predictions, e.g. $A_{\epsilon} = 0.667(2)$ and $A_{\epsilon'} = 0.896(5)$ in the critical 3d Ising model on real projective space. Another interesting geometry is $S^1\times \bR^{d-1}$, which corresponds to putting the CFT at finite temperature. This was studied for the 3d Ising model and other higher-dimensional CFTs in \textcite{Iliesiu:2018fao}. \textcite{Gobeil:2018fzy} also discussed a generalization of the conformal block concept relevant for this geometry.

Let us finally mention an interesting study \cite{Caselle:2016mww} which combined the knowledge of the 3d Ising model CFT data acquired by the bootstrap with conformal perturbation theory. They achieved a remarkable agreement with the experimental data describing the 2pt function $\<\sigma\sigma\>$ off criticality, i.e.~at temperatures slightly different from the critical temperature, which corresponds to perturbing the CFT by a $\int d^3x\, \eps(x)$ perturbation.

\subsection{$O(N)$ global symmetry}
\label{sec:ON}

Most known unitary 3d CFTs have a continuous global symmetry, and we now turn to such CFTs. We will focus on bootstrap results obtained by assuming $O(N)$ as a full symmetry or as a subgroup.\footnote{\label{footnote:SON}All bounds in this section are applicable also under a weaker assumption of $SO(N)$ global symmetry, see \cite[{section 2.1.1}]{Kos:2015mba}.} 

\subsubsection{General results}
\label{sec:ON-general}

As discussed in Sec.~\ref{sec:global}, correlation functions of CFT operators that are in irreducible representations of the global symmetry $G$ can be organized using group theory and decomposed into different $G$-invariant tensor structures. Sec.~\ref{sec:crossing} explained how these structures enter the crossing relations. The first numerical analyses of the resulting equations occurred in the context of 4d CFTs,\footnote{See \textcite{Poland:2010wg}, \textcite{Rattazzi:2010yc}, \textcite{Vichi:2011ux}, and \textcite{Poland:2011ey}.} but the group theoretic structure is $d$-independent. The bootstrap for $O(N)$ symmetry in 3d was investigated by \textcite{Kos:2013tga, Kos:2015mba,Kos:2016ysd} and \textcite{Nakayama:2014yia}.

We will start our analysis assuming that the CFT contains an operator $\phi\equiv(\phi_a)_{a=1}^N$ 
in the fundamental representation of $O(N)$, of dimension $\Delta_\phi$. Mimicking the discussion in Sec.~\ref{sec:Z2-general}, we would like to learn about the operators in the OPE $\phi_{a} \times \phi_{b}$. By group theory, operators of even spin $\ell$ in this OPE will transform as $O(N)$ singlets or symmetric traceless tensors of rank 2, while odd-spin operators will transform in the rank-2 antisymmetric representation. 

\begin{figure}
\includegraphics[width=\figwidth]{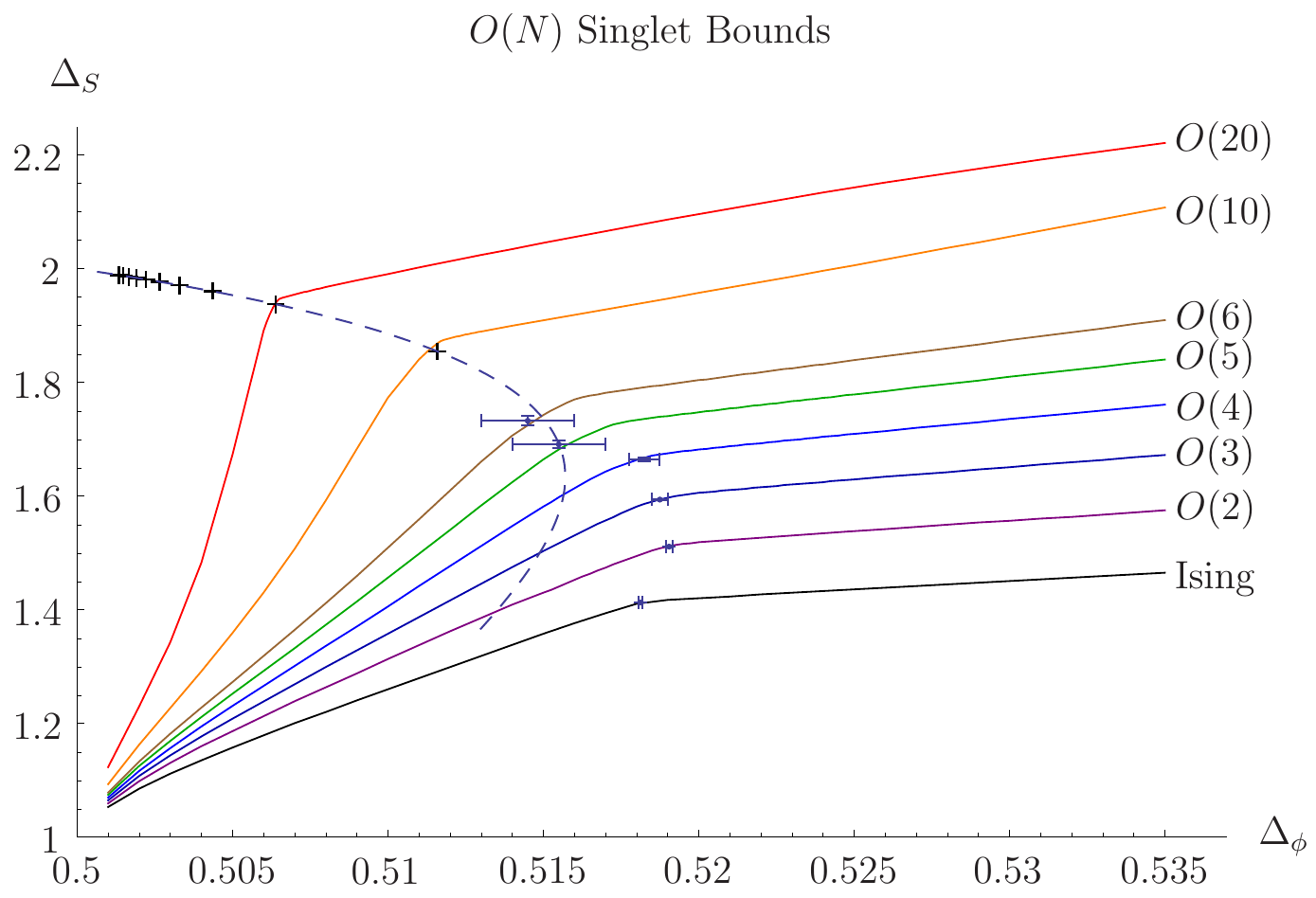}
\caption{\label{fig:ONsinglet}(Color online) Upper bound on the dimension of $s$ \cite{Kos:2013tga}.}
\end{figure}

\begin{figure}
\includegraphics[width=\figwidth]{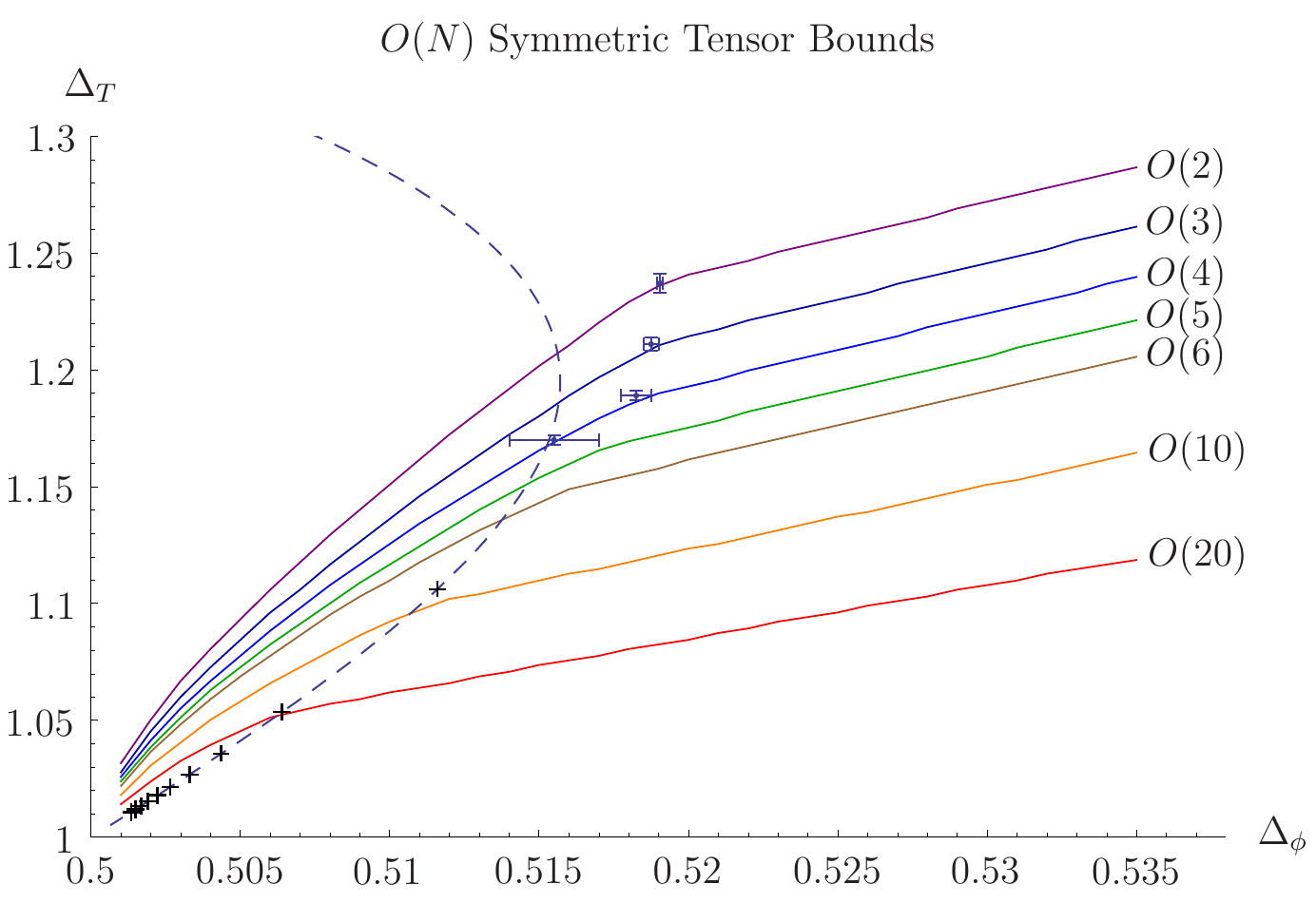}
\caption{\label{fig:ONsymtrace}(Color online) Upper bound on the dimension of $t$ \cite{Kos:2013tga}.}
\end{figure}

From the crossing relations for the 4pt function of $\phi$ one can put upper bounds on the dimensions of various operators. For the lowest dimension scalars ($\ell=0$) in the singlet ($s$) and symmetric traceless tensor ($t$) sector, these bounds are shown in Figs.~\ref{fig:ONsinglet}, \ref{fig:ONsymtrace} as a function of $\Delta_\phi$ for various values of $N$. The ``kinks" in these bounds will be interpreted in the next section. 

\begin{figure}
\includegraphics[width=\figwidth]{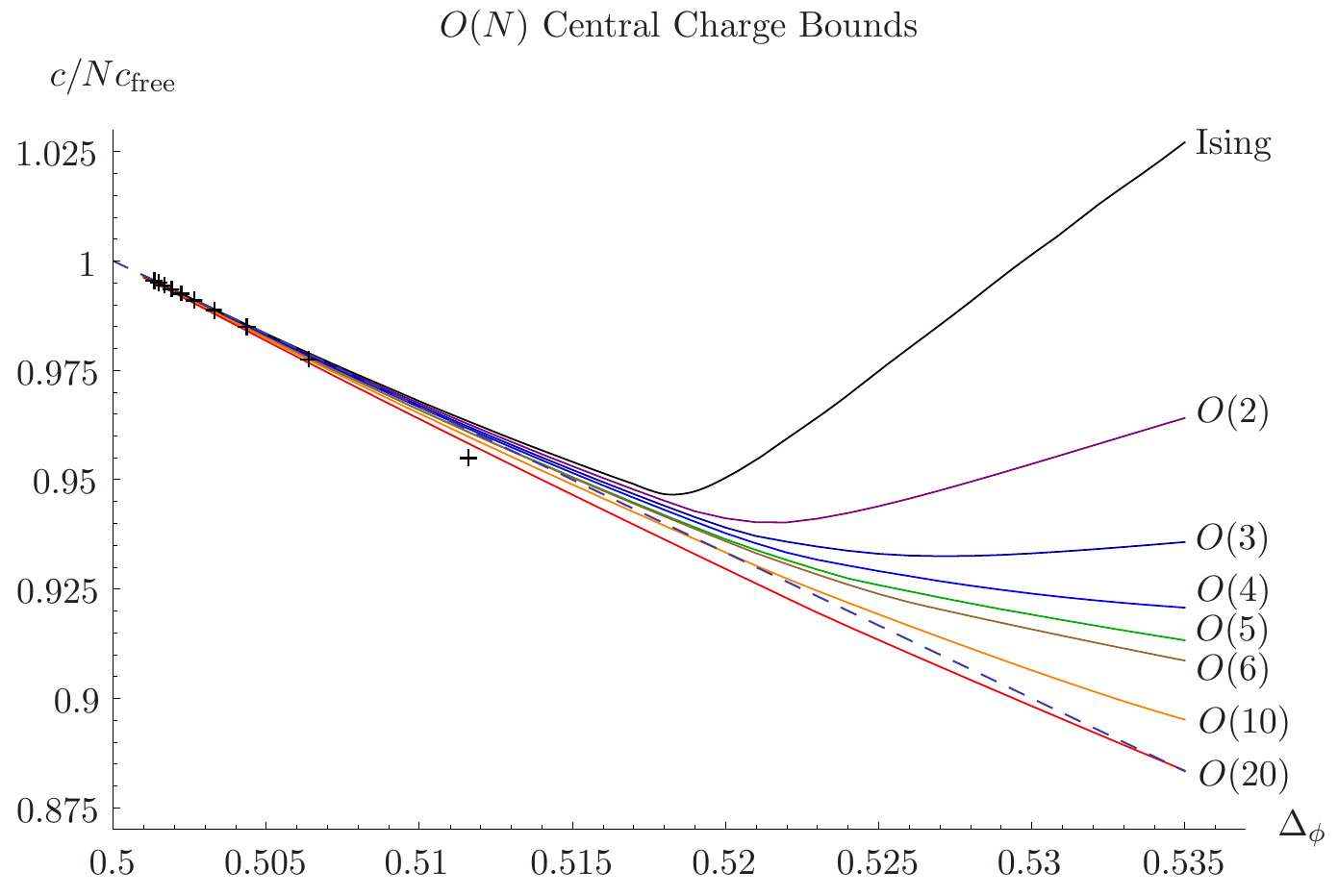}
\caption{\label{fig:ONCT}(Color online) Lower bound on $C_T$ computed under the assumption $\Delta_s, \Delta_t\ge 1$ \cite{Kos:2013tga}.}
\end{figure}

\begin{figure}
\includegraphics[width=\figwidth]{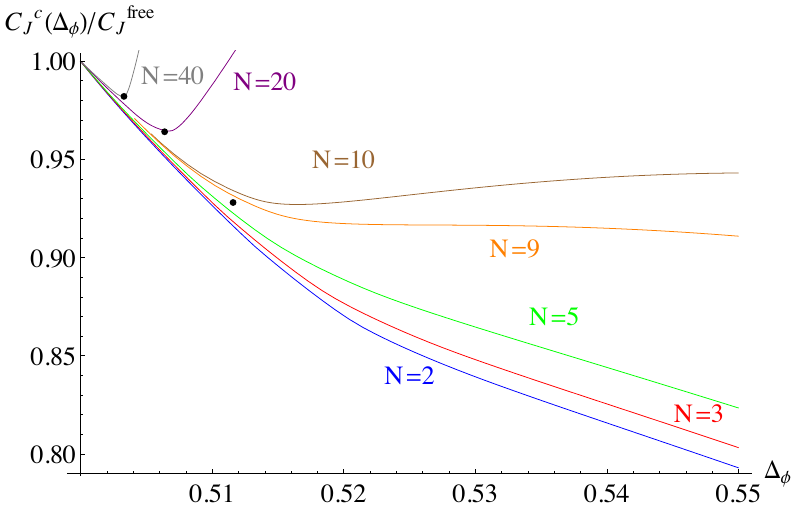}
\caption{\label{fig:ONCJ}(Color online) Lower bound on $C_J$ \cite{Nakayama:2014yia}.}
\end{figure}

The $\phi\times\phi$ OPE also contains two interesting operators of spin $\ell\ge 1$: the stress tensor $T$ and the conserved current $J$. Using the bootstrap one can put lower bounds on their two-point function coefficients $C_T$ and $C_J$ (defined in Sec.~\ref{sec:ward}) given in Figs.~\ref{fig:ONCT}, \ref{fig:ONCJ}. This is done by bounding from above the OPE coefficients $\lambda_{\phi\phi T}\propto \Delta_{\phi}/\sqrt{C_T}$ and $\lambda_{\phi\phi J}\propto 1/\sqrt{C_J}$ {(see Eqs.~(\ref{eq:lambdaT}, \ref{eq:lambdaJ}))}.

Let's discuss the monotonicity of these bounds with $N$. Since $O(N+1)\supset O(N)$, the bounds on $C_T$, $C_J$, and on $\Delta_t$ should get stronger with increasing $N$, and indeed they do (notice that $C_T$ is plotted divided by $N$). Although it may seem counterintuitive that the $\Delta_s$ bound gets weaker with $N$, there is no contradiction. The point is that the symmetric traceless tensor of $O(N+1)$ contains a singlet $\tilde s$ when decomposed with respect to $O(N)$. Therefore the only constraint is that the $O(N)$ singlet bound should be weaker than the $O(N+1)$ symmetric traceless bound, which is satisfied by inspection.

Notice also that the scaling of the $C_T$, $C_J$ bounds with $N$ close to $\Delta_\phi=1/2$ is consistent with the fact that in the theory of $N$ free scalars, $C_T$ grows linearly with $N$ while $C_J$ is constant.

\subsubsection{Critical $O(N)$ model}

The most famous 3d CFT with $O(N)$ symmetry is the critical point of the $O(N)$ lattice model, which is the generalization of Eq.~(\ref{eq:Ising}) to $N$-component spins satisfying the constraint $|\vec{s}|=1$.
This CFT is also known as the Wilson-Fisher fixed point, being an IR fixed point of the $O(N)$ symmetric scalar field theory with quartic interaction~\cite{Wilson:1971dc}. For any integer $N$ this 3d CFT is unitary, given that the microscopic realizations are unitary.\footnote{\label{note:ONnonunitary}Sometimes one discusses analytic continuation of $O(N)$ models to noninteger $N$. These analytic continuations are nonunitary~\cite{Maldacena:2011jn}, and fall outside the range of validity of the linear/semidefinite methods. Although such attempts were made~\cite{Shimada:2015gda}, we would advise caution. Here we will only consider integer $N\ge 2$. We will discuss nonunitary CFTs in Sec.~\ref{sec:nonunitary}.}

It's natural to ask where the critical $O(N)$ models lie in the parameter space of $O(N)$ symmetric CFTs allowed by the general bounds from the previous section. In the Wilson-Fisher description, $\phi_{a}$ is the fundamental scalar field appearing in the Lagrangian, $s=\phi^2$, and $t$ is the traceless part of $\phi_a \phi_b$. Dimensions of these fields have been previously estimated using RG methods (in particular the $\eps$-expansion and the large $N$ expansion), Monte Carlo studies, and experiments. Comparing the $s$ and $t$ bounds and these prior determinations, marked with crosses in Figs.~\ref{fig:ONsinglet}, \ref{fig:ONsymtrace}, one is led to conjecture that the critical $O(N)$ models correspond to the ``kinks". Similar kink-like features are visible in the lower bounds on $C_J$ and $C_T$. In the latter case the kinks can be made sharper by imposing that the $S$ operator saturate the gap, see Fig.~5 in \textcite{Kos:2013tga}. This conjecture can be used to extract values of the $\phi$, $s$, $t$ dimensions and of $C_T$, given in Table 3 of \textcite{Kos:2013tga}. 

We will now discuss how to isolate the critical $O(N)$ models without relying on the kink conjecture. The idea is to exploit the crucial physical feature of these CFTs --- that they possess robust gaps in the operator spectrum. The singlet scalar $s$ corresponds to the temperature deformation of the critical point and is relevant. The next singlet scalar, $s'$, must necessarily be irrelevant (otherwise the critical point would be multicritical), implying the gap $\Delta_{s'}\ge 3$ in the singlet scalar sector. We also expect a gap in the fundamental representation scalar sector. The order parameter $\phi_{a}$ belongs to this sector and is relevant, while most likely the next fundamental scalar is irrelevant: $\Delta_{\phi'}\ge 3$. This can be also deduced using the Wilson-Fisher description, using a nonrigorous but suggestive equation of motion argument~\cite{Kos:2015mba}.

\begin{figure}
\includegraphics[width=\figwidth]{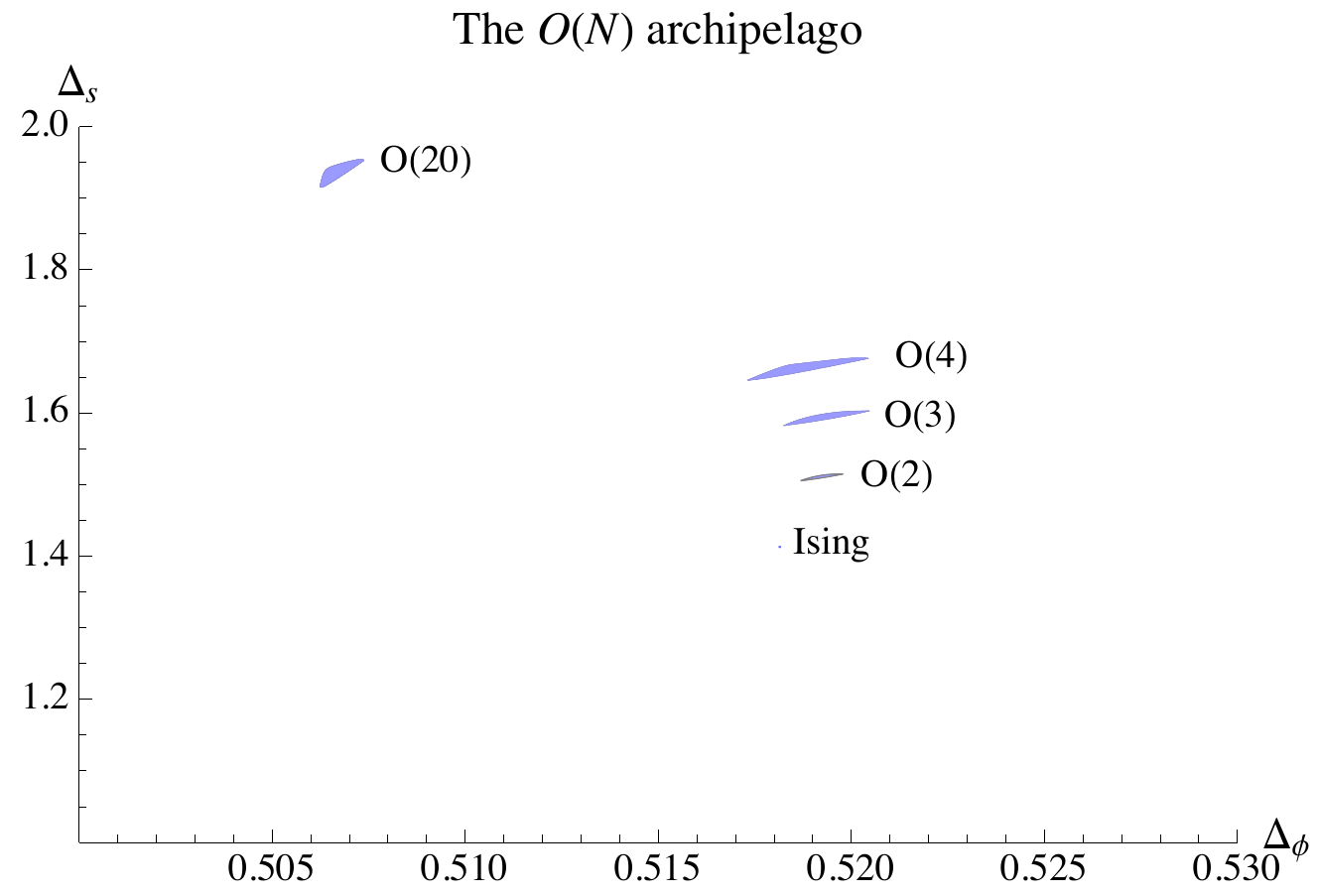}
\caption{\label{fig:ONarchipelago}(Color online) The $O(N)$ archipelago \cite{Kos:2015mba}.}
\end{figure}

\textcite{Kos:2015mba}~studied bootstrap constraints for the system of three correlators 
{$\{\langle \phi_a\phi_b\phi_c\phi_d\rangle$, $\langle \phi_a\phi_b ss\rangle$, $\langle ssss\rangle\}$. } 
Imposing the assumptions $\Delta_{s'}\ge 3$, $\Delta_{\phi'}\ge 3$, they found small allowed regions (``islands") shown in Fig.~\ref{fig:ONarchipelago}. Improved versions of these islands for $O(2)$ and $O(3)$, discussed in the next sections, were subsequently obtained in \textcite{Kos:2016ysd}. It's important to stress that, like in Fig.~\ref{fig:Z2-mixed-sigpgap} for the Ising model, there are disconnected allowed regions outside the shown part of the parameter space; see e.g.~Fig.~\ref{fig:O2singlet} below for the $O(2)$ case. These regions are practically unexplored and they might contain other interesting CFTs.

\subsubsection{$O(2)$ global symmetry}
\def\mywidth{0.6\columnwidth}

\label{sec:O2}

We finally wish to discuss separately specific values of $N$, {starting with $O(2) \supset U(1)$. There are many physically interesting 3d CFTs possessing $O(2)$ or $U(1)$ symmetry.} The most famous of these is the critical $O(N)$ model for $N=2$, also known as the critical $XY$ model. It describes, in particular, the Curie point of easy-plane ferromagnets, and of easy-plane antiferromagnets on bipartite lattices. Here the $O(2)$ symmetry arises as the symmetry of local magnetic moment interactions.

Another frequent appearance of $U(1)$ symmetry in condensed matter physics is as particle number conservation. 
The most famous such $U(1)$ transition is the superfluid transition in ${}^4$He. Another example is the superfluid-insulator quantum phase transition in the (2+1)d Bose-Hubbard model at integer particle density~\cite{Fisher1989}. Both these transitions are also described by the critical $O(2)$ model.

A wide class of CFTs with $U(1)$ global symmetry are IR fixed points of theories which at the microscopic level contain $U(1)$ gauge fields coupled to fermion or scalar matter. The global $U(1)$ symmetry in these theories is topological in origin, and the local operators charged under it are monopoles of the gauge field. These CFTs often appear in condensed matter applications; they will be discussed in more detail in Sec.~\ref{sec:QED3}.

Here we discuss general constraints from the multiple correlator bootstrap for $U(1)$ symmetric CFTs, which have been pursued further than in the general $O(N)$ case discussed above. 

\begin{figure}
\includegraphics[width=\figwidth]{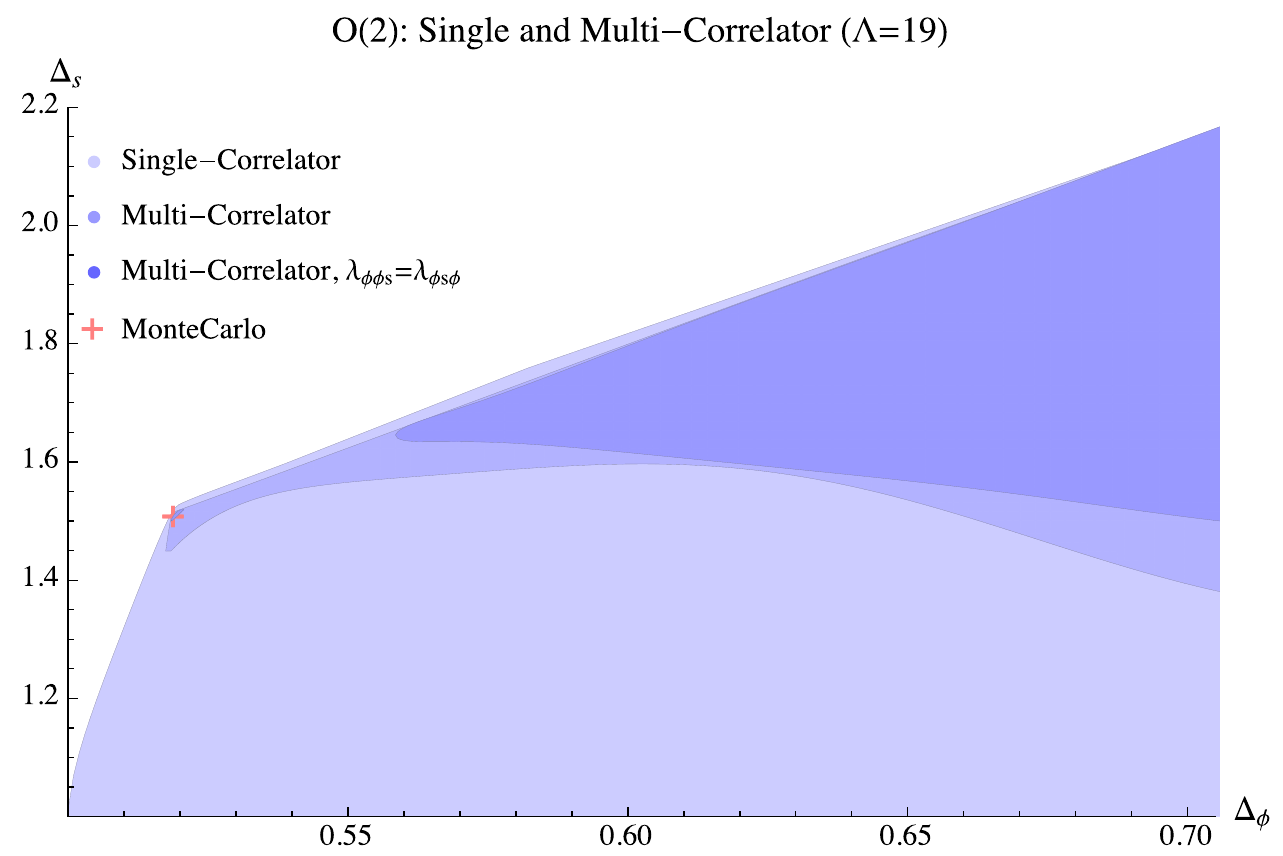}
\caption{\label{fig:O2singlet}(Color online) Allowed regions in the parameter space of {\red $O(2)$ or $U(1)$} symmetric CFTs~\cite{Kos:2015mba}. The strongest constraint (dark blue) has been obtained from the analysis of three correlators $\{\langle \phi \phi\phi\phi\rangle$, $\langle \phi \phi ss\rangle$, $\langle ss ss\rangle\}$, assuming that $s$, $\phi$ are the only two relevant scalars of charge 0, 1, and imposing the OPE coefficient relation $\lambda_{\phi\phi s}=\lambda_{\phi s\phi}$.}
\end{figure}

Operators in $U(1)$ theories are classified by their $U(1)$ charge, with $s, \phi, t$ of the previous section\footnote{Instead of vectors and tensors of $O(2)$ here we use, as is customary, complex fields charged under {$U(1) \subset O(2)$. In the bootstrap studies we describe the distinction between $U(1) = SO(2)$ and $O(2)$ is unimportant, so the constraints will apply to either symmetry (see footnote~\ref{footnote:SON}).}} having charge $0, 1, 2$. Imposing the constraints that there is a unique relevant charge 1 and a unique relevant charge 0 scalar, one gets the allowed region shown in dark blue in Fig.~\ref{fig:O2singlet}. Notice that this region consists of a detached island to which the critical $O(2)$ model belongs, and a further region on the right, similar to Fig.~\ref{fig:Z2-mixed-sigpgap} for the Ising model.

\begin{figure}
\includegraphics[width=\figwidth]{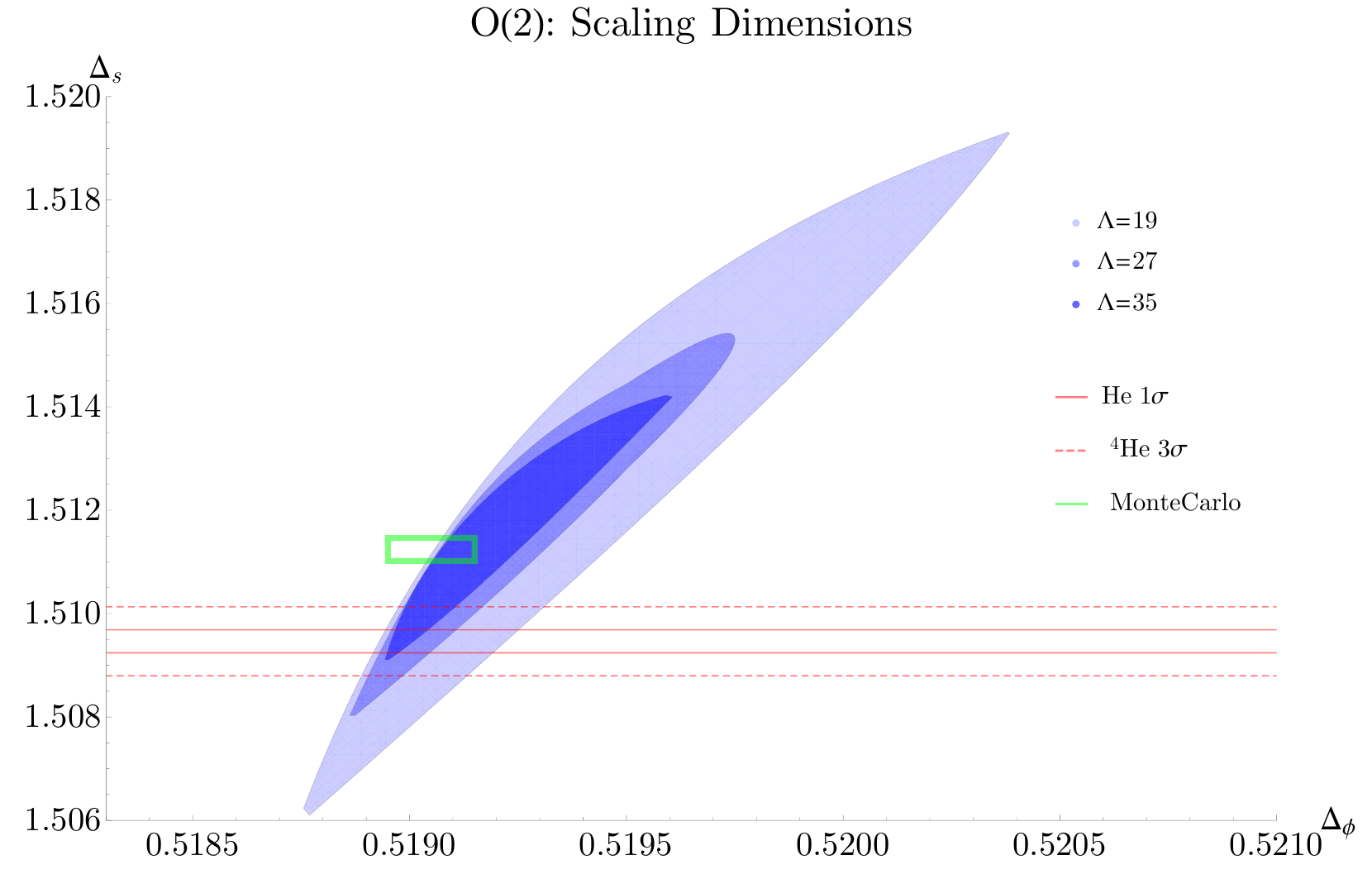}
\caption{\label{fig:O2island}(Color online) Allowed islands in the mixed correlator analysis of {$O(2)$ or $U(1)$} symmetric CFTs after performing a scan over the OPE coefficient ratio $\lambda_{sss}/\lambda_{\phi\phi s}$ \cite{Kos:2016ysd}.}
\end{figure}

The island containing the critical $O(2)$ model has been studied more accurately in \textcite{Kos:2016ysd} by increasing the derivative order and performing a scan over the OPE coefficient ratio $\lambda_{sss}/\lambda_{\phi\phi s}$. This led to the improved constraints shown in Fig.~\ref{fig:O2island}. The resulting dimensions and OPE coefficients are given in Table~\ref{tab:O2O3}. 

In the same table we give the determinations of $\Delta_t$ and $C_J$ obtained in \textcite{Kos:2015mba} by scanning over the allowed island in the $\{\Delta_\phi,\Delta_s\}$ plane, under the {respective assumptions} that $\Delta_{t'}\ge 3$ and $\Delta_{J'}\ge 3$.\footnote{As before, we denote by prime the subleading operator with the same quantum numbers. So $t'$ is the next traceless symmetric scalar after $t$, and $J'$ is the next vector after the conserved current $J_\mu$, transforming in the antisymmetric $SO(N)$ representation.}

These results are compatible but somewhat less precise in the case of $\Delta_\phi$, $\Delta_s$, $\Delta_t$ than other available determinations by lattice and RG methods, see \textcite{Kos:2015mba,Kos:2016ysd} for references. In particular, a further increase in precision is required to resolve the discrepancy between the experimental and theoretical determinations of $\Delta_s$ shown in Fig.~\ref{fig:O2island}. This is an important problem for the future. On the other hand, the bootstrap is currently the only source of information about the OPE coefficients $\lambda_{\phi\phi s}$ and $\lambda_{ss s}$. The central charge $C_J$ is related to the zero-temperature (or high-frequency) conductivity of the quantum critical points described by the critical $O(2)$ model. Although not yet experimentally measured, this parameter has been extensively studied theoretically and numerically in the condensed matter literature. As discussed in \textcite{Kos:2015mba}, the bootstrap currently provides the best determination of $C_J$.

\begin{table}
\begin{tabular}{r|l|l}
& $O(2)$ & $O(3)$ \\
\hline
$\Delta_\phi$ & 0.51926(32) & 0.51928(62) \\
$\Delta_s$ &1.5117(25) & 1.5957(55) \\
$\lambda_{sss}/\lambda_{\phi\phi s} $&1.205(9) & 0.953(25)\\
$\lambda_{\phi\phi s} $&0.68726(65) & 0.5244(11)\\
$\lambda_{sss}$ & 0.8286(60) & 0.499(12) \\
$\Delta_t $& 1.2357(33) & 1.210(6)\\
$C_J/C_J^{\rm free}$ & 0.9050(16) & 0.9065(27)
\end{tabular}
\caption{\label{tab:O2O3} Bootstrap results for the operator dimensions and OPE coefficients in the critical $O(2)$ and $O(3)$ models (see Secs.~\ref{sec:O2}, \ref{sec:O3}).}
\end{table}

\subsubsection{$O(3)$ global symmetry}

\label{sec:O3}

We will now specialize to the case of $O(3)$ global symmetry, focusing on the most famous such CFT which is the critical $O(3)$ model. Apart from describing the critical point of isotropic ferromagnets, the same CFT also describes the $(2+1)$d quantum critical point in coupled dimer antiferromagnets, see \textcite{Sachdev-proc} and references therein. 

\begin{figure}
\includegraphics[width=\figwidth]{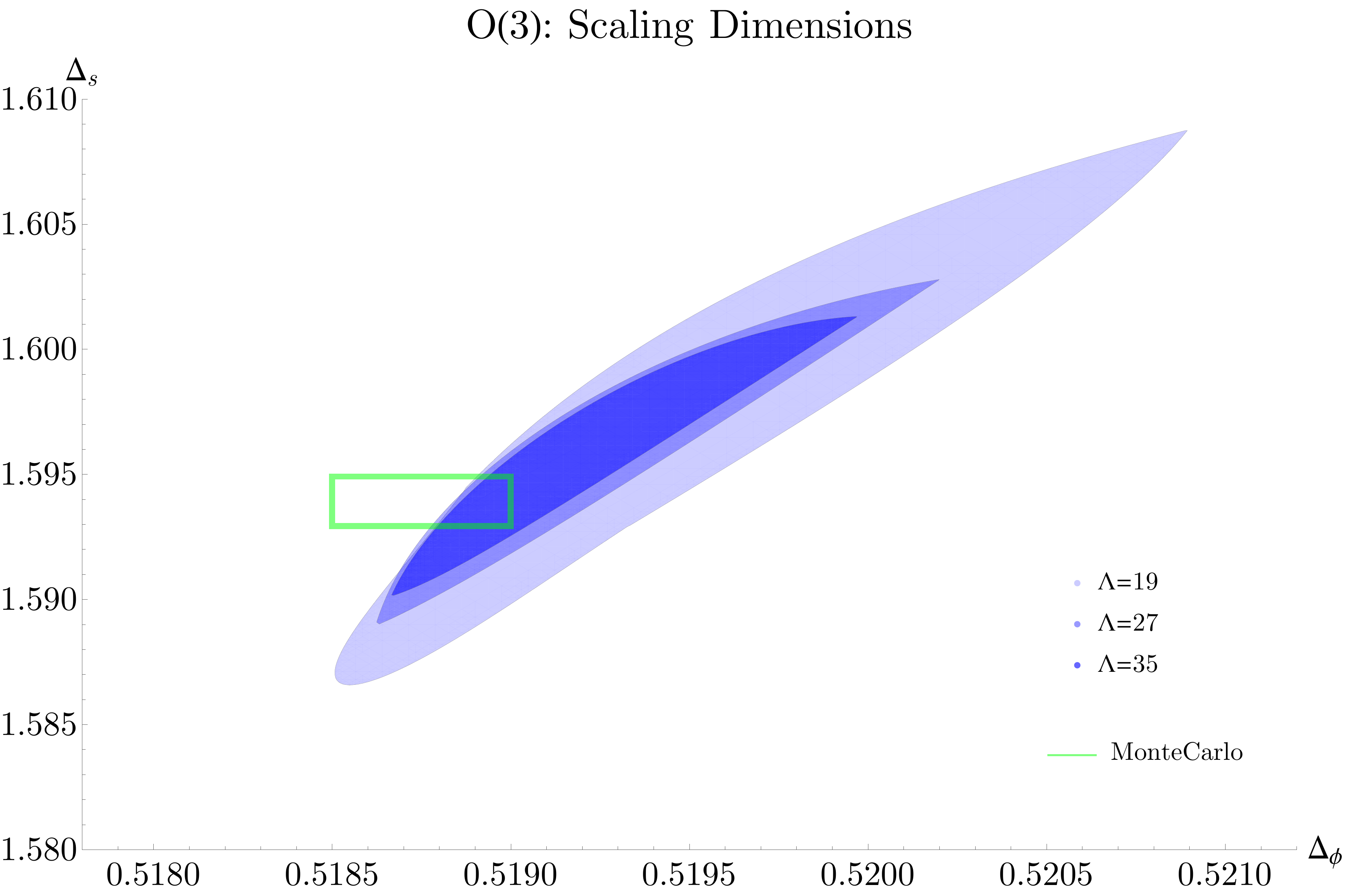}
\caption{\label{fig:O3island}(Color online) The $O(3)$ analogue of Fig.~\ref{fig:O2island} \cite{Kos:2016ysd}.}
\end{figure}

The bootstrap analysis of this theory mimics the $U(1)$ case from the previous section. Under the assumption that $\phi_a$ and $s$ are the only two relevant scalars transforming in the fundamental and trivial representation of $O(3)$,~\textcite{Kos:2016ysd} found an island allowed by the bootstrap constraints, shown in Fig.~\ref{fig:O3island}. The bootstrap determinations of the scaling dimensions and OPE coefficients following from this analysis are given in Table~\ref{tab:O2O3}. As for the $U(1)$ case, the scaling dimension determinations are compatible but somewhat less precise than the best available Monte Carlo and RG results.

One long-standing question about the critical $O(3)$ model concerns its stability with respect to perturbations which may potentially lead to two different CFTs, the so-called cubic and biconical fixed points, which have symmetries $B_3=S_3\rtimes (\mathbb{Z}_2)^3$ and $O(2)\times \mathbb{Z}_2$ respectively. These perturbations take the form $K^{ijkl} \Phi_{ijkl}$, where $\Phi$ is a scalar operator transforming in the rank-4 symmetric traceless representation of $O(3)$, and $K$ is a constant tensor breaking $O(3)$ to one or the other subgroup. RG calculations indicate that the $O(3)$ fixed point is unstable while the cubic and biconical fixed points are stable, with the correction to scaling critical exponents $\omega_{O(3)}= -0.013(6)$ and $\omega_{B_3}=0.010(4)$ or 0.015(2) according to two calculations; see \cite[{sections 11.3, 11.7}]{Pelissetto:2000ek} and references therein. This would imply that $\Delta_\Phi=3+\omega_{O(3)}$ is very weakly relevant. From the bootstrap point of view, $\Phi$ appears in the OPE $t\times t$, and its dimension could be determined by an analysis involving correlators of $t$. This is an interesting problem for the future.\footnote{\label{note:B3}Preliminary investigations of the $\<tttt\>$ bootstrap have produced bounds that still allow $\Phi$ to be irrelevant~\cite{Nakayama:2016jhq, KPSVUnpublished}. See also~\textcite{Rong:2017cow,Stergiou:2018gjj} for recent bootstrap studies of 3d CFTs assuming cubic and related discrete symmetries. The latter finds evidence that there may be two different critical 3d theories with cubic symmetry, one of which is related to the physics of magnets, while the other may describe structural phase transitions in perovskites.}

Let us mention another 3d CFT with an $O(3)$ global symmetry --- the critical Gross-Neveu-Heisenberg (GNH) model~\cite{herbut2009relativistic}, realized as the IR fixed point of a microscopic Lagrangian with Yukawa and quartic couplings
\begin{eqnarray}
g\bar{\Psi} \sigma^i \phi_i \Psi + \lambda (\phi_i^2)^2\,,
     \label{GNH}
 \end{eqnarray}
 where $\phi_i$ is a three-component scalar order parameter, and $\Psi$ is a two-component multiplet of massless Dirac fermions. This CFT is believed to describe the continuum limit of the Hubbard model on the honeycomb and $\pi$-flux lattices~\cite{Sachdev:2010uz}. While clearly distinct from the critical $O(3)$ model, the scalar sector of this theory would be subject to the general $O(3)$ bounds shown in Figs.~\ref{fig:ONsinglet}-\ref{fig:ONCJ}. However the expected value of $\Delta_\phi\approx 0.85$ \cite[{Eq.~(25)}]{toldin2015fermionic} puts it outside of the region explored so far. The fermionic sector of this theory could be constrained by methods from the next section.

\subsection{CFTs with fermion operators}
\label{sec:Fermions}

\subsubsection{Models}

\label{sec:FermionModels}
The preceding sections discussed constraints from crossing relations for 4pt functions of scalar operators. Many 3d or (2+1)d CFTs of theoretical and experimental interest also contain fermionic operators, and here we will discuss what the bootstrap has so far been able to say about them.

Perhaps the simplest example is the family of CFTs described by the Gross-Neveu model at criticality~\cite{Gross:1974jv}.\footnote{This model and its variations are frequently invoked to describe quantum phase transitions in condensed matter systems with emergent Lorentz symmetry in (2+1)d. Some examples of its applications include models for phase transitions in graphene~\cite{herbut2006interactions, herbut2009relativistic}, the Hubbard model on the honeycomb and $\pi$-flux lattice~\cite{toldin2015fermionic}, models of time-reversal symmetry breaking in d-wave superconductors~\cite{vojta2000quantum, vojta2003quantum}, and models of 3-dimensional gapless semiconductors~\cite{Moon:2012rx,Herbut:2014lfa}.} While the critical theory is often described as the UV fixed point in a theory of fermions with 4-fermi interactions $\mathcal{L} \sim (\bar{\psi} \psi)^2$, a better nonperturbative definition is as an IR fixed point in a theory with a scalar field coupled to fermions via Yukawa interactions. The latter Gross-Neveu-Yukawa (GNY) model contains a scalar $\phi$ and $N$ Majorana fermions $\psi_i$:
\beq
    {\cal L}_{\rm GNY} =\frac 12 \sum_{i = 1}^N \bar \psi_i (\slashed{\partial} + g \phi ) \psi_i + \frac 12 \partial^\mu \phi\partial_\mu \phi  + \frac 12 m^2 \phi^2 + \lambda \phi^4 \,.
     \label{GNYLag}
 \eeq
 This model has an $O(N)$ symmetry rotating the fermions. A fixed point can be established perturbatively at large $N$ in a $1/N$ expansion, see e.g.~\textcite{Gracey:1992cp,Gracey:1993kc} and \textcite{Derkachov:1993uw}. This model has also been studied extensively from the perspective of the $\epsilon$-expansion, with recent results by~\textcite{fei2016yukawa,Mihaila:2017ble,Zerf:2017zqi}.
 
An interesting special case is $N=1$ (a single Majorana fermion coupled to a real scalar). It is expected \cite{fei2016yukawa} that this model may contain a fixed point with $\mathcal{N}=1$ supersymmetry. This supersymmetric fixed point has been proposed to described a critical point on the boundary of topological superconductors~\cite{Grover:2013rc}. 

There are variations of this model containing multiple scalar order parameters. One notable example is the $\mathcal{N}=2$ supersymmetric critical Wess-Zumino model, containing a complex scalar related to a 3d Dirac fermion by supersymmetry.\footnote{We will describe some of the implications of supersymmetry and a bootstrap analysis connecting to this model later in Sec.~\ref{sec:4Dsusy}.} This theory has been proposed to describe a critical point on the surface of topological insulators~\cite{Ponte:2012ru,Grover:2013rc}, and a superconducting critical point in (2+1)d Dirac semimetals with an attractive Hubbard interaction \cite{Li:2017dkj}. Another important example is the Gross-Neveu-Heisenberg model, described in Sec.~\ref{sec:O3}.

\subsubsection{General results}
\label{sec:Fermions-general}

We first discuss general results following from the existence of fermionic operators. Specialized bounds where the critical GNY and other models are featured more prominently will be discussed below.

A bootstrap analysis of 4pt functions of identical Majorana fermions $\<\psi\psi\psi\psi\>$ was performed in \textcite{Iliesiu:2015qra} and extended to 4pt functions $\< \psi_i \psi_j \psi_k \psi_l \>$ containing fermions that are vectors under an $O(N)$ symmetry in \textcite{Iliesiu:2017nrv}. These studies both assumed a general (2+1)d CFT with parity symmetry. Tensor structures and conformal blocks for 4pt functions were derived using a spinorial embedding-space formalism also developed in \textcite{Iliesiu:2015qra}, similar in logic to the vectorial embedding space reviewed in Appendix~\ref{sec:embedding}. 

In Fig.~\ref{fig:Fermions-oddScalar} we show general upper bounds on the leading parity-odd and parity-even scalars in the $\psi \times \psi$ OPE, called $\sigma$ and $\eps$ respectively. The bound on $\sigma$ is nearly saturated by the MFT line $\Delta_{\sigma} = 2\Delta_{\psi}$, at least at small values of $\Delta_{\sigma}$. As $\Delta_{\psi} \rightarrow 1$ the bound approaches the free theory value $\Delta_{\sigma} = 2$, where we can identify $\sigma = \bar{\psi} \psi$. On the other hand, there is an abrupt discontinuity in the bound around $\Delta_{\psi} \sim 1.27$ occurring when $\Delta_{\sigma}$ approaches 3. This jump also coincides with a kink in the bound on $\Delta_{\epsilon}$. The interpretation of these features is currently an open question -- it is tempting to speculate that a CFT may live at the top of the jump in the bound on $\Delta_{\sigma}$ and in the kink in the bound on $\Delta_{\epsilon}$ but no concrete candidate CFTs have yet been identified. If it exists, this CFT would appear to have an unusual property of not possessing any relevant scalar deformations.\footnote{Hypothetical theories with this property were recently named ``dead-end" CFTs by~\textcite{Nakayama:2015bwa}. They should be distinguished from ``self-organized" CFTs which do not have any relevant \emph{singlet} scalars as defined in Sec.~\ref{sec:multicrit}.}

 \begin{figure}[t!]
    \centering
\includegraphics[width=\figwidth]{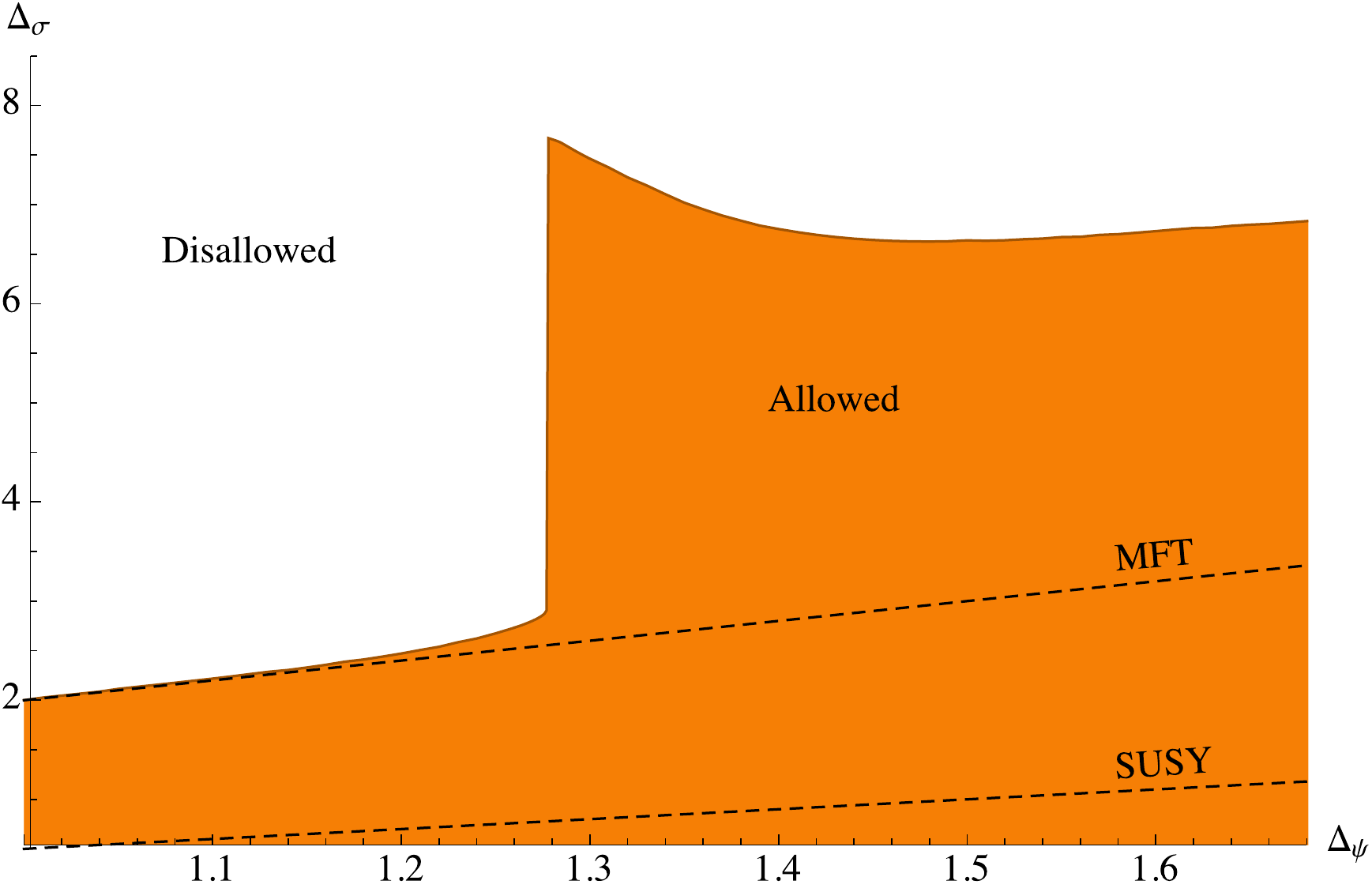}(a)
\includegraphics[width=\figwidth]{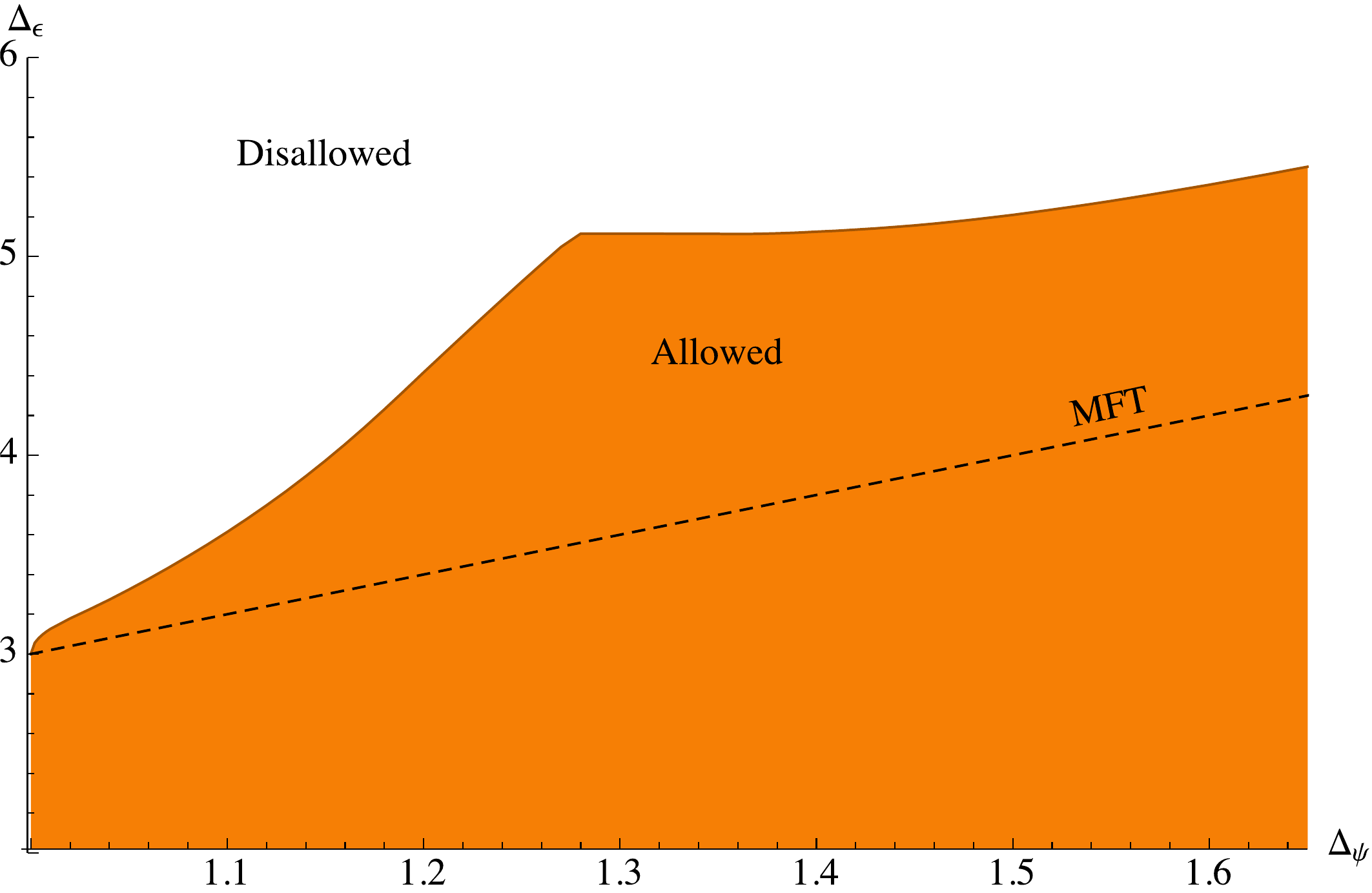}(b)
    \caption{\label{fig:Fermions-oddScalar}
(Color online) Upper bounds on the dimension of (a) the first parity-odd scalar $\sigma$ and (b) the first parity-even scalar $\epsilon$ in the OPE $\psi\times \psi$, as a function of $\Delta_\psi$ \cite{Iliesiu:2015qra}. Here $\psi$ is a Majorana fermion primary operator in a 3d parity-invariant unitary CFT.} 
  \end{figure}

\begin{figure}[t!]
    \centering
\includegraphics[width=\figwidth]{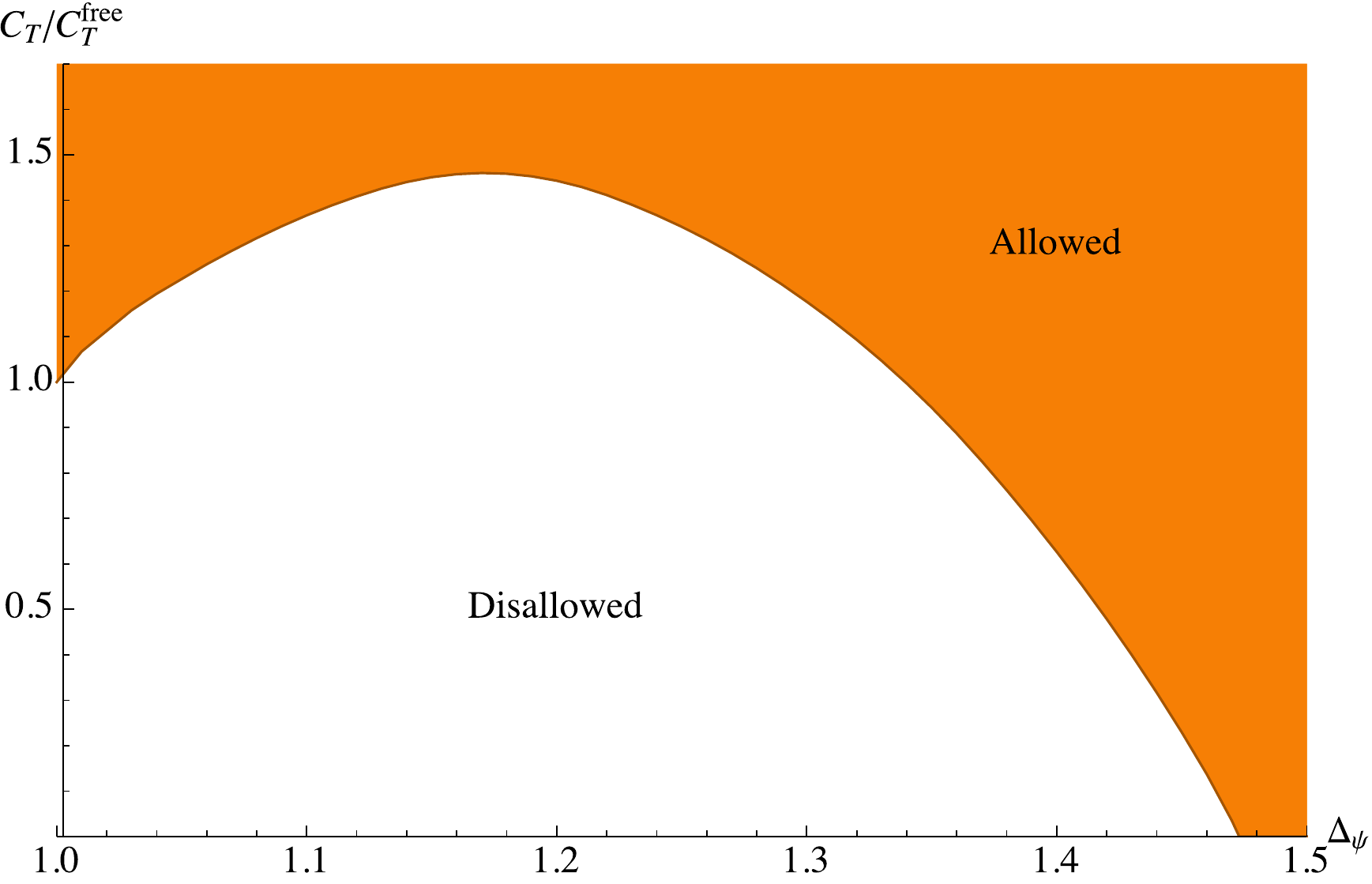}(a)
\includegraphics[width=\figwidth]{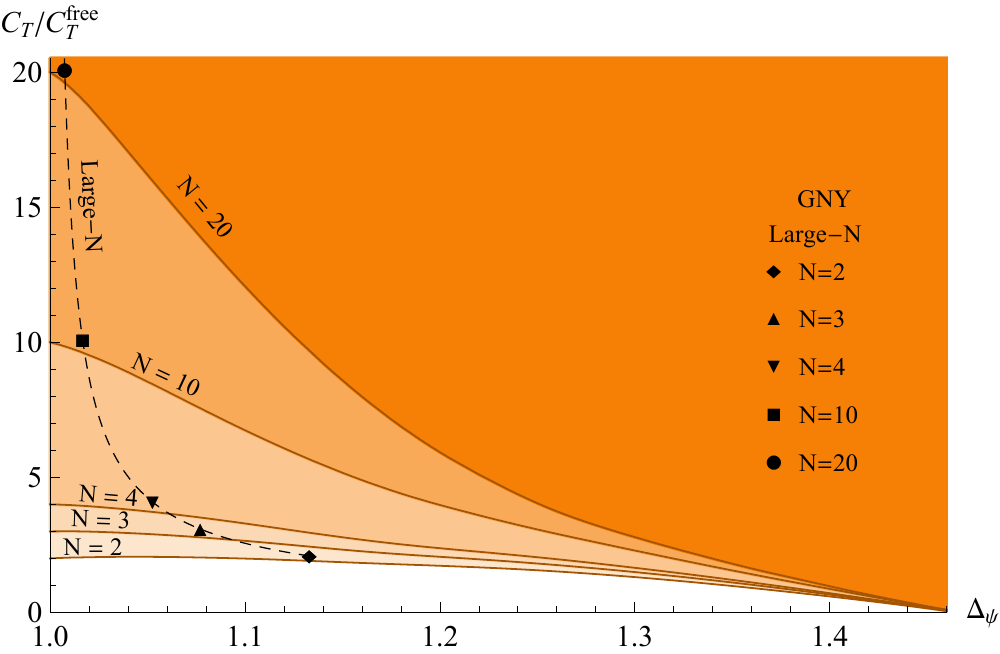}(b)
    \caption{\label{fig:Fermions-centralCharge}
  (Color online) Lower bounds on $C_T$ as a function of $\Delta_\psi$, where $\psi$ is (a) Majorana fermion or (b) a multiplet of Majorana fermions in the fundamental representation of an $O(N)$ global symmetry group \cite{Iliesiu:2015qra,Iliesiu:2017nrv}.}
  \end{figure}

In Fig.~\ref{fig:Fermions-centralCharge} we also show the general lower bounds on the central charge $C_T$ (normalized to its value in the theory of a free Majorana fermion), obtained by bounding the coefficient of the stress-tensor conformal block. These lower bounds approach the free values as $\Delta_{\psi} \rightarrow 1$ and disappear completely for $\Delta_{\psi} \gtrsim 1.47$. In the case of $O(N)$ symmetry they can be seen to grow linearly with $N$ and are compatible with values computed in the $1/N$ expansion of the GNY model. Generalizations to the current central charge $C_J$ for fermions charged under $O(N)$ symmetry were also computed in \textcite{Iliesiu:2017nrv}.

\subsubsection{Gross-Neveu-Yukawa models}
\label{sec:Fermions-GN}

In the critical GNY model at large $N$, $\psi_i$ has dimension $1 + 4/(3\pi^2 N) + \ldots$, while the leading parity-odd scalars in the $\psi_i \times \psi_j$ OPE are the $O(N)$ singlet $\phi$ with dimension $1-32/(3\pi^2 N)+\ldots$ and the $O(N)$ symmetric tensor $\bar{\psi}_i \psi_j$ with dimension $2 + 32/(3\pi^2 N) + \ldots$~\cite[{Table 1}]{Iliesiu:2017nrv}. The accumulation point $(\Delta_{\psi}, \Delta_{\sigma}) \rightarrow (1,1)$ sits well in the interior of Fig.~\ref{fig:Fermions-oddScalar}(a), but by imposing a gap until the second parity-odd scalar, $\Delta_{\sigma'} \geq 2 + \delta$ for different positive values of $\delta$, we have the possibility of obtaining an allowed region that rules out critical GNY models with $N$ sufficiently large.

\begin{figure}[t!]
    \centering
\includegraphics[width=\figwidth]{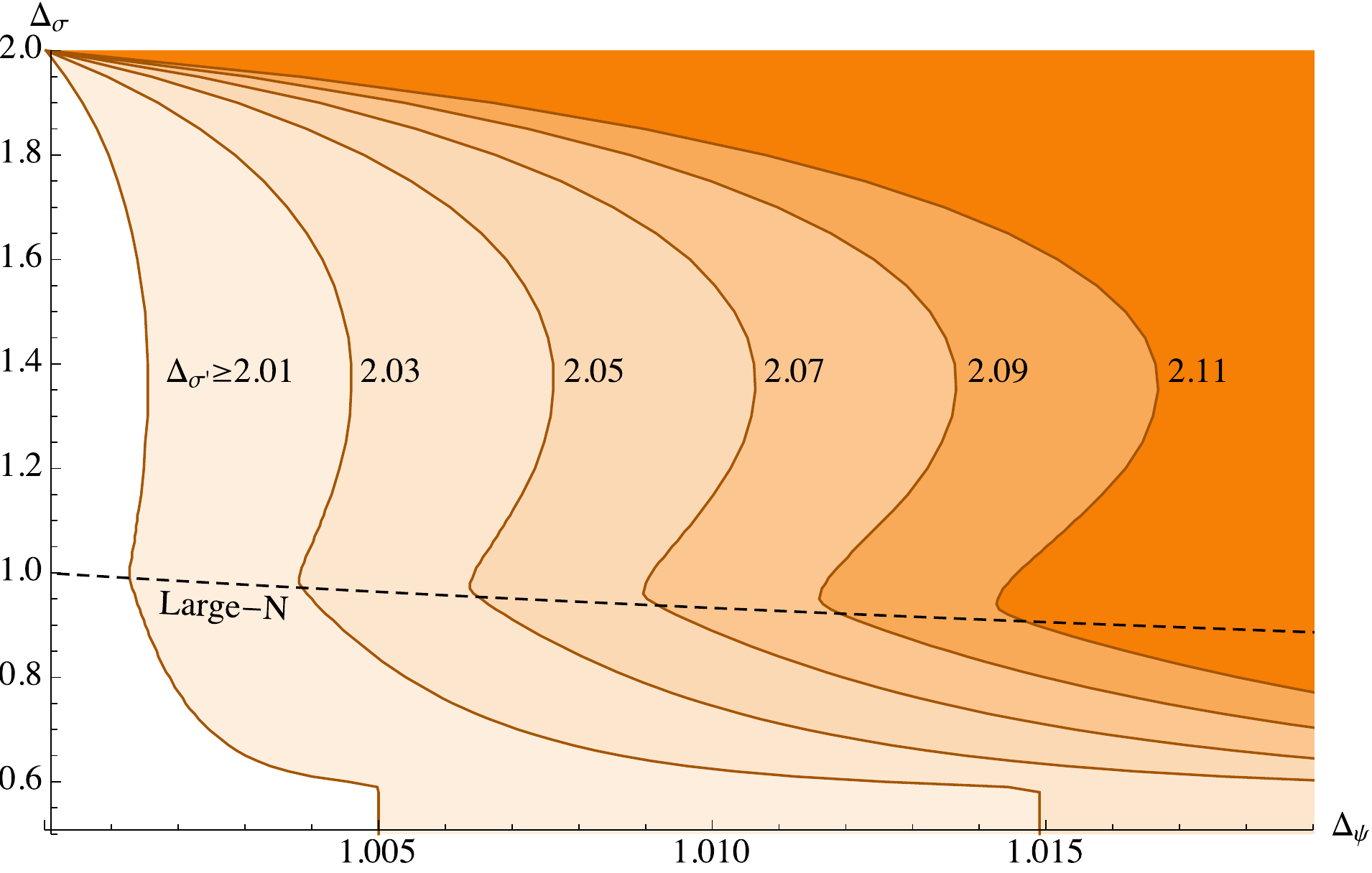}
\includegraphics[width=\figwidth]{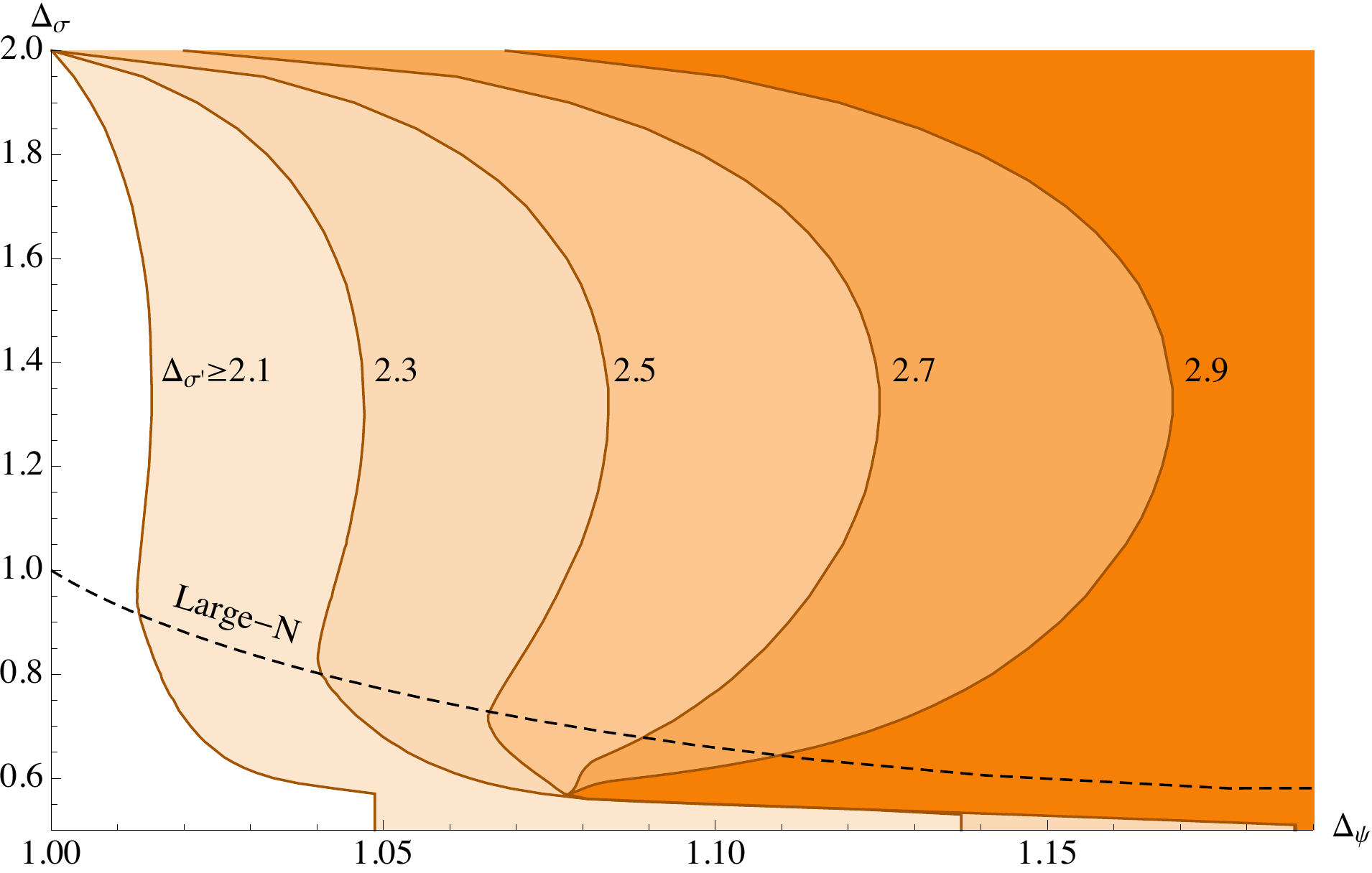}
    \caption{\label{fig:Fermions-grossNeveuSmallSigPrime}
   (Color online) Effect of imposing a gap until the second pseudoscalar $\sigma'$ on the parameter space of 3d parity-invariant CFTs~\cite{Iliesiu:2015qra}.}
  \end{figure}

This is realized in Fig.~\ref{fig:Fermions-grossNeveuSmallSigPrime}, where the effect of gaps ranging from $\Delta_{\sigma'} \geq 2.01$ to $\Delta_{\sigma'} \geq 2.9$ are shown. At very small values of $\delta$ the lower bounds of the allowed regions possess a kink whose location matches very well to the large-$N$ GNY model prediction. At larger values of $\delta$, the precise map between $\delta$ and $N$ is not known but it is plausible that the kinks continue to match to the GNY model even at small values of $N$. However, starting around $\Delta_{\sigma'} \geq 2.3$, a second lower feature also appears in these curves, where they all intersect and have an additional kink at a point near $(1.08, .565)$. 

\begin{figure}[t!]
    \centering
\includegraphics[width=\figwidth]{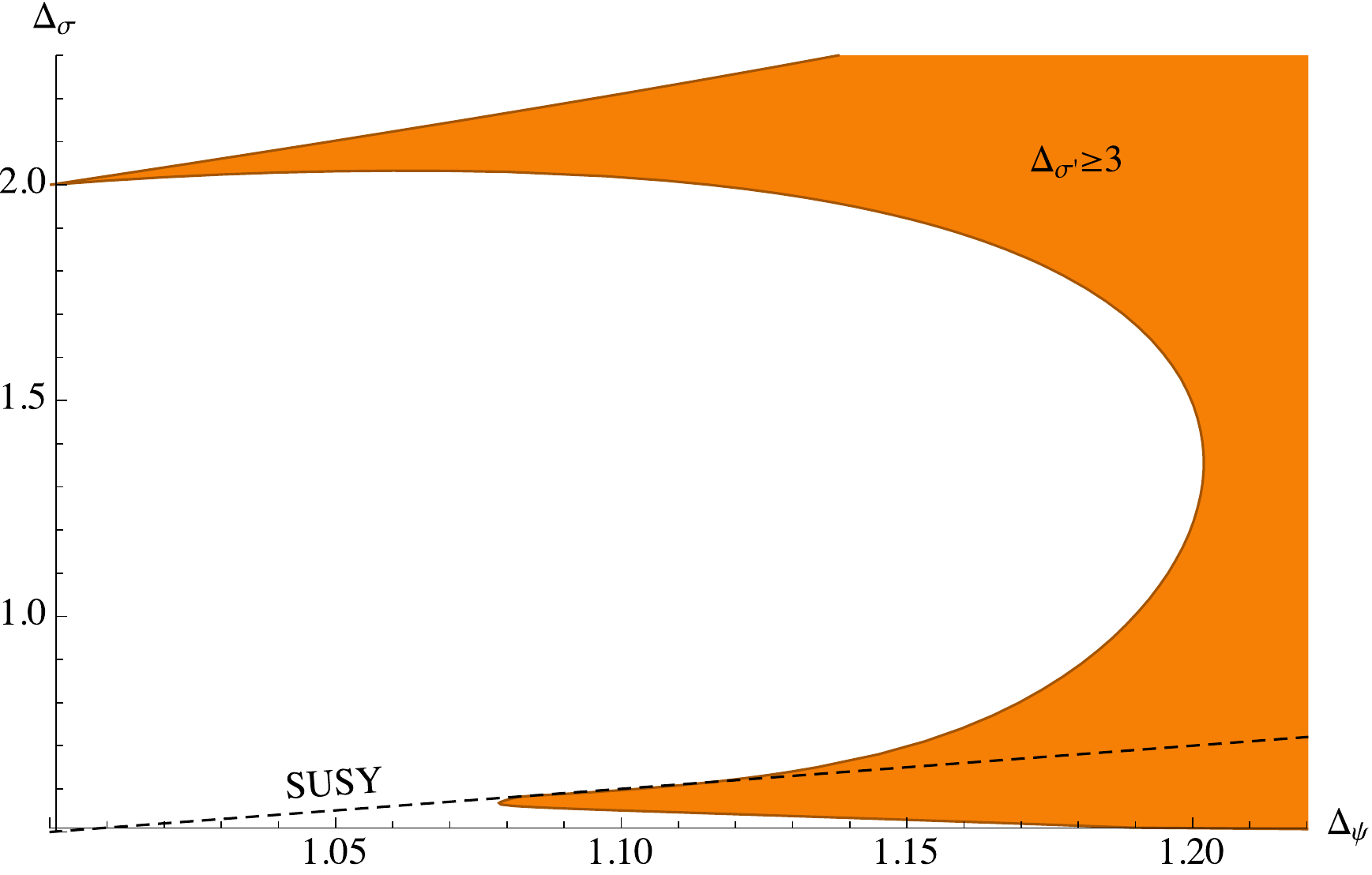}
\includegraphics[width=\figwidth]{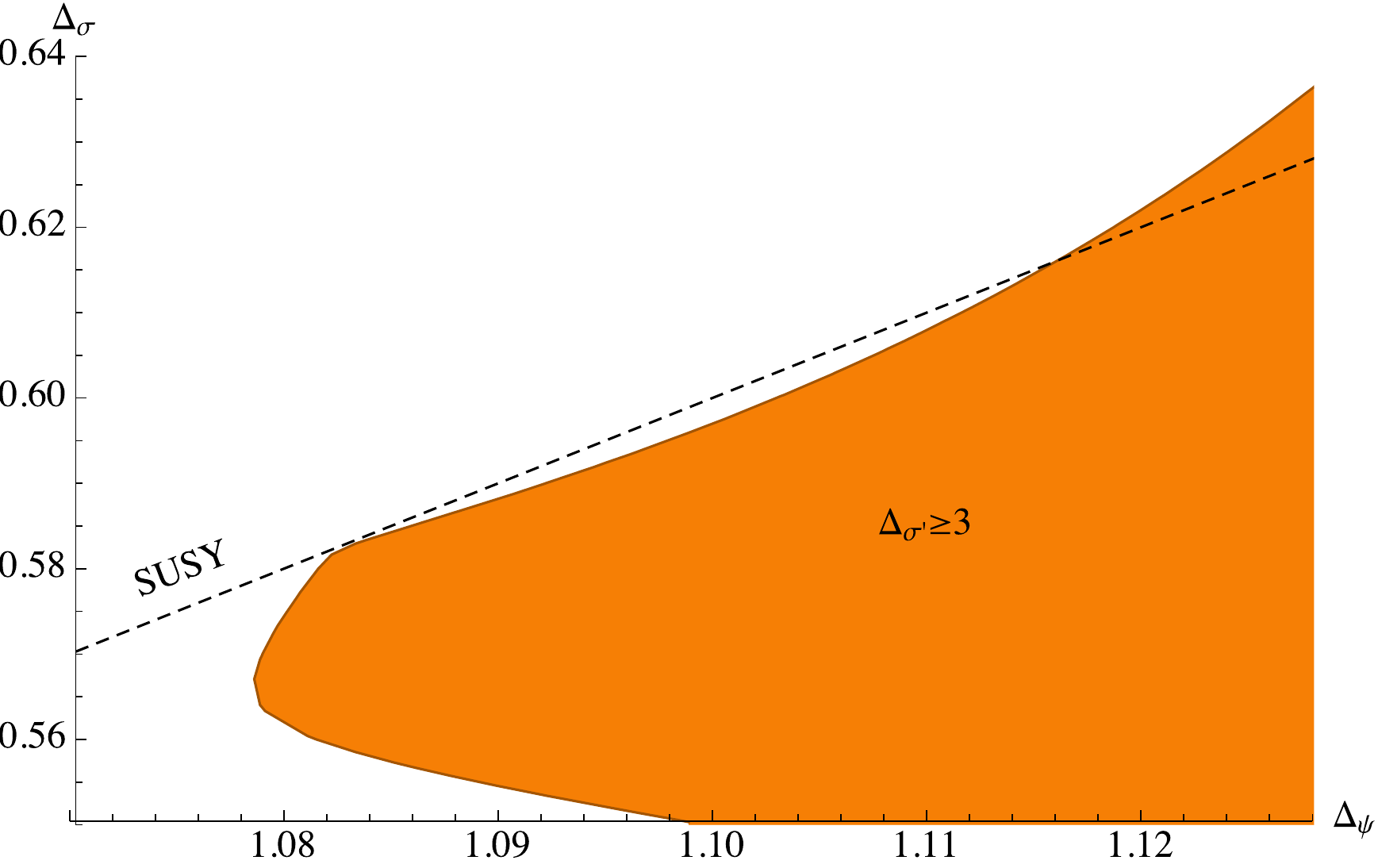}
    \caption{\label{fig:Fermions-oneRelevantOddScalar}
   (Color online) Effect of imposing that there is only one relevant pseudoscalar, $\Delta_{\s'}\ge 3$, in 3d parity-invariant CFTs~\cite{Iliesiu:2015qra}.}
  \end{figure}

This structure of an ``upper" and ``lower" kink can be seen clearly in Fig.~\ref{fig:Fermions-oneRelevantOddScalar}, specialized to the case \mbox{$\Delta_{\sigma'} \geq 3$}. In fact, in this case the line \mbox{$\Delta_{\psi} = \Delta_{\sigma} + 1/2$} expected for theories with supersymmetry comes very close to (but just misses) the upper kink. Thus, it is tempting to conjecture that the $\mathcal{N}=1$ supersymmetric Gross-Neveu-Yukawa model, see Sec.~\ref{sec:FermionModels}, may sit in this feature and has $\Delta_{\sigma'}$ slightly smaller than 3. This picture seems consistent with estimate \mbox{$\Delta_{\sigma} \approx 0.59$} from a Pad\'e-extrapolation of the $\epsilon$-expansion \cite{fei2016yukawa}, as well as with the rigorous lower bound \mbox{$\Delta_\sigma\ge 0.565$} \cite{Bashkirov:2013vya}, which follows using another supersymmetric relation $\Delta_\eps=\Delta_\sigma+1$ together with the bootstrap bound in Fig.~\ref{fig:Z2-epsbound}, applicable with parity playing the role of a $\mathbb{Z}_2$ symmetry.\footnote{\label{footnote:3dN1}Further progress on this CFT was made very recently in~\textcite{Rong:2018okz} and~\textcite{Atanasov:2018kqw}, where it was understood how to obtain an island in the scalar mixed-correlator bootstrap around $\Delta_{\sigma} = 0.584444(30)$. In these studies in addition to relations between scaling dimensions it is important to incorporate nontrivial 3d $\mathcal{N}=1$ superconformal blocks.}  An additional speculation is that the lower feature may coincide with a non-supersymmetric fixed point, called GNY${}^*$ in \textcite{Iliesiu:2017nrv}, which is seen in the $\epsilon$-expansion as a nonunitary fixed point at large $N$, but whose fate at small $N$ and $\epsilon \rightarrow 1$ is not known. 

\begin{figure}[t!]
    \centering
\includegraphics[width=\figwidth]{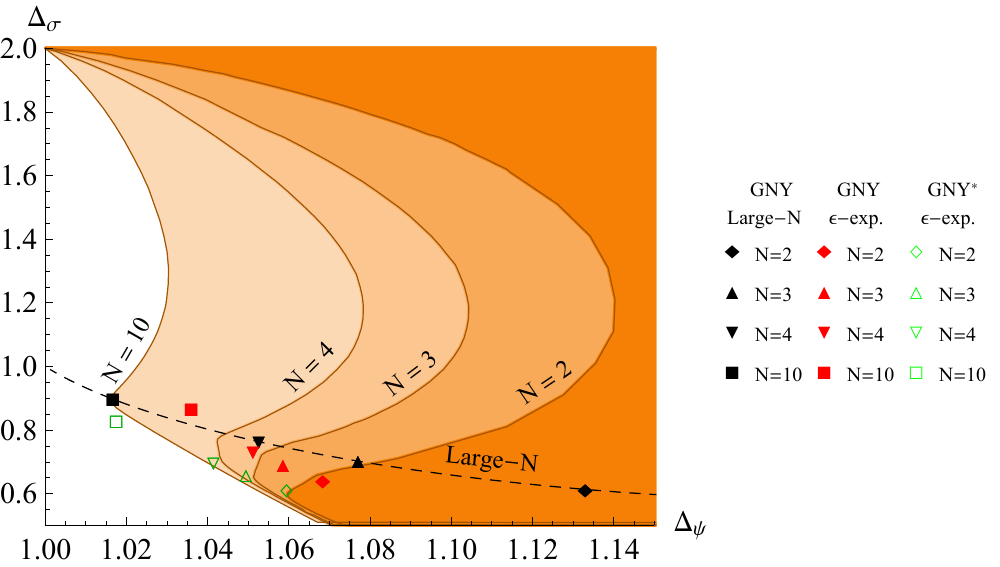}
    \caption{\label{fig:Fermions-ONtwokinks}
   (Color online) Effect of imposing a gap $\Delta_{\s'}\ge 3$ in the singlet pseudoscalar sector of $O(N)$-symmetric fermionic CFTs \cite{Iliesiu:2017nrv}. The kinks at low $N$ may perhaps be identified with the GNY and GNY${}^*$ CFTs.}
  \end{figure}

\begin{figure}[t!]
    \centering
\includegraphics[width=\figwidth]{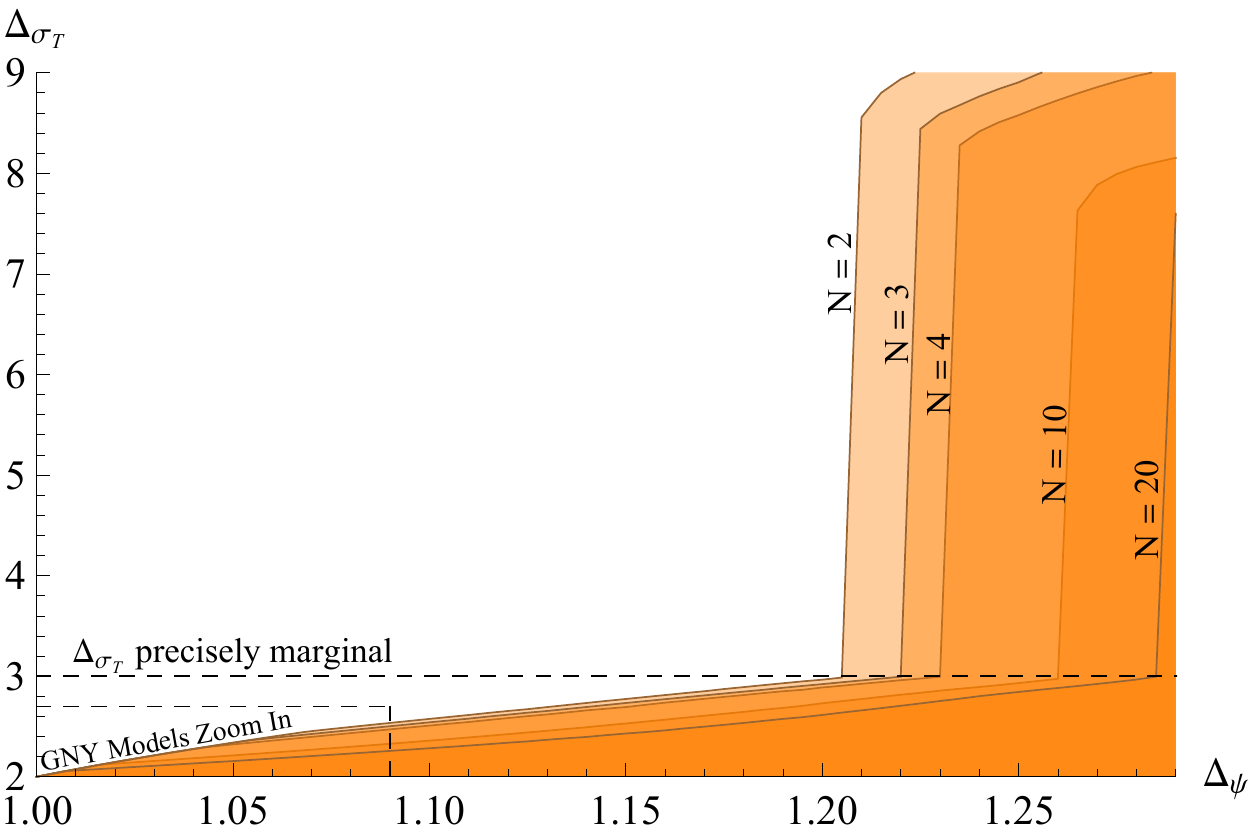}(a)
\includegraphics[width=\figwidth]{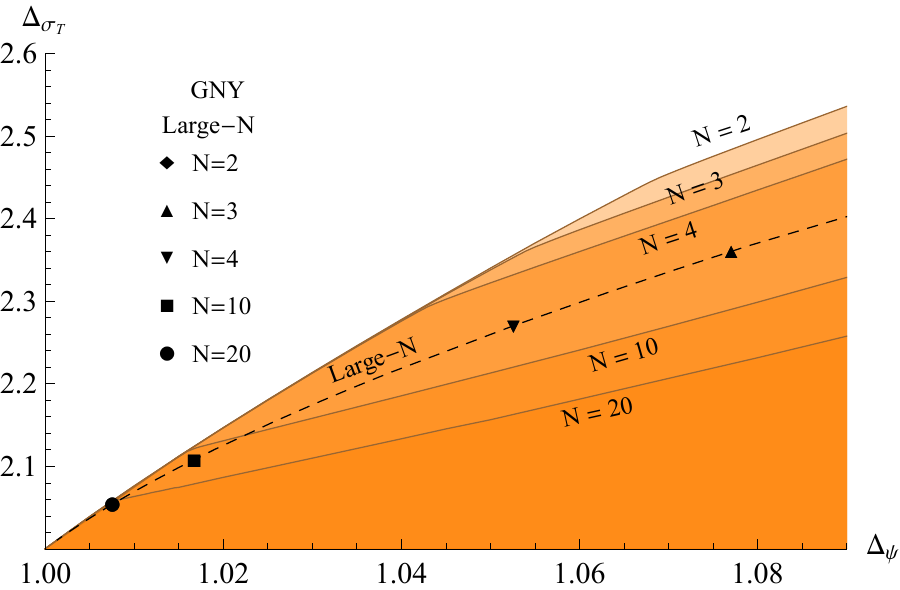}(b)
    \caption{\label{fig:Fermions-SigmaT} 
   (Color online) Upper bounds on the dimension of the symmetric traceless pseudoscalar $\sigma_{T}$ in the OPE $\psi_i\times \psi_j$ in $O(N)$-symmetric fermionic CFTs~\cite{Iliesiu:2017nrv}. Notice the mysterious jumps in the wide view of the bounds (a) when they cross marginality. (b) gives a zoom on the small $\Delta_\psi$ region, where the bounds exhibit kinks, in agreement with the GNY dimensions at large $N$.}
  \end{figure}

Additional evidence for this picture comes from the generalization of the bounds to $O(N)$ symmetry~\cite{Iliesiu:2017nrv}, where one can place independent bounds on different $O(N)$ representations. In Fig.~\ref{fig:Fermions-ONtwokinks} we show computed bounds on the leading singlet dimension $\Delta_{\sigma}$, assuming that the next singlet is irrelevant, $\Delta_{\sigma'} \geq 3$. These bounds also show both an upper and lower kink, which appear not too far from the $\epsilon$-expansion estimates for the GNY and GNY${}^*$ models. In Fig.~\ref{fig:Fermions-SigmaT} we also highlight the bounds on the leading $O(N)$ symmetric tensor $\sigma_T$, which display mysterious and unexplained jumps when $\Delta_{\sigma_T}$ reaches marginality and at smaller values of $\Delta_{\psi_i}$ show a series of kinks which match to the large-$N$ GNY models. Understanding the mechanism behind these jumps is an important open problem, which may be related to the spectrum rearrangement phenomena from Sec.~\ref{sec:Z2-spectrum}.

\subsection{QED${}_3$}

\label{sec:QED3}

Continuing our survey of physically important 3d CFTs, another class of theories are those defined by coupling 3d gauge or Chern-Simons fields to matter. One of the simplest examples is 3d quantum electrodynamics (QED${}_3$), containing a $U(1)$ gauge field coupled to $N_f$ 2-component massless Dirac fermions $\psi_i$. This theory is known {to flow to a nontrivial CFT} at large $N_f$, which can be studied in the $1/N_f$ expansion~\cite{Appelquist:1988sr,Nash:1989xx}. There has been a longstanding question of whether there is a critical value of $N_f$ below which the theory undergoes spontaneous breaking of the $SU(N_f)$ global symmetry. Numerous estimates of the critical value of $N_f$ have been made over the years, see e.g.~\cite[{Table 5}]{Gukov:2016tnp}. 

It would be very interesting to shed light on these theories and the question of the conformal window using bootstrap techniques. One starting point would be to study the bootstrap for 4pt functions of the gauge-invariant fermion bilinears $\bar{\psi}_i \psi_j$.\footnote{Because of gauge symmetry, a single fermion field is unphysical in QED${}_3$ and it would not be legitimate to consider its 4pt functions, unlike in GNY models in Sec.~\ref{sec:Fermions} where $\psi$ was physical.} Basic bootstrap bounds on scalar 4pt functions, of the type discussed in Secs.~\ref{sec:Z2}, \ref{sec:ON}, would apply to this operator, but it is not a priori clear how to isolate QED${}_3$ as compared to other theories with similar scalar operators such as QCD${}_3$. However, this is an under-explored direction and in future studies it may be useful to combine the bootstrap for $SU(N_f)$ adjoints and singlets with additional gap assumptions and bootstrap constraints for other operators.

\subsubsection{Monopole bootstrap for QED${}_3$}
\label{sec:QED3-monopole}

An alternate approach, pursued in \textcite{Chester:2016wrc} and \textcite{Chester:2017vdh}, is to focus on monopole operators. When dealing with a compact $U(1)$ gauge field, these operators create topologically nontrivial configurations of the gauge field having magnetic flux emerging from a spacetime point.\footnote{Thus they could also be called instantons, but the common terminology refers to them as monopoles.} Such operators are charged under a topological $U(1)_T$ global symmetry with symmetry current \mbox{$J_T^{\mu} = \frac{1}{8\pi} \epsilon^{\mu \nu \rho}F_{\nu\rho}$}. Taking the monopole operators to have charge $q \in \mathbb{Z}/2$, the scalar monopoles transform in representations of $SU(N_f)$ corresponding to fully rectangular Young diagrams with $N_f/2$ rows and $2|q|$ columns~\cite{Dyer:2013fja}. Thus, the lightest scalar monopoles $M^I_{\pm1/2}$ are expected to be in $SU(N_f)$ representations with $N_f/2$ fully antisymmetric indices.\footnote{Monopoles with spin transform in other nontrivial flavor representations, see \textcite{Chester:2017vdh}.}

\begin{figure}[t!]
    \centering
\includegraphics[width=\figwidth]{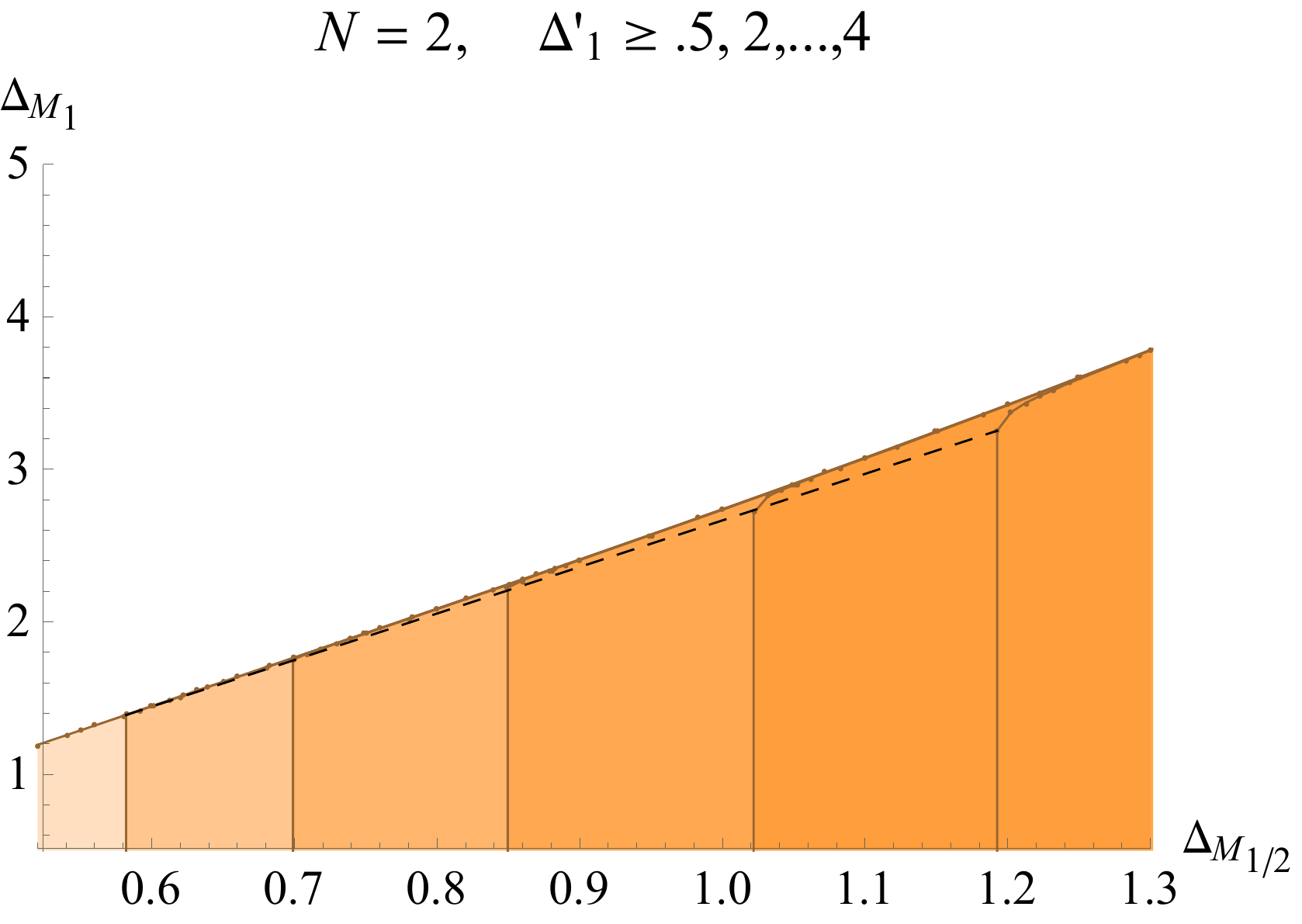}(a)
\includegraphics[width=\figwidth]{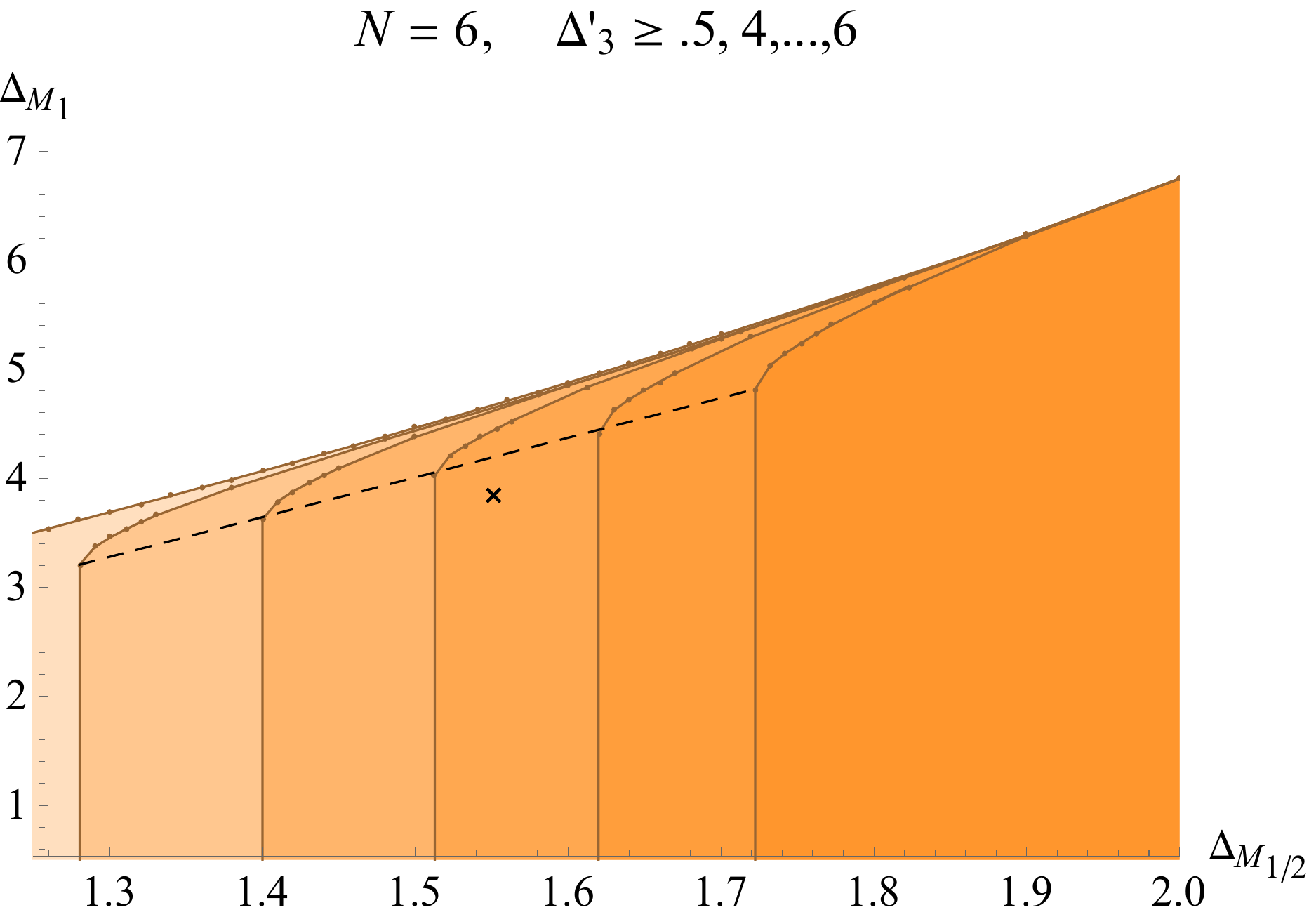}(b)
    \caption{\label{fig:QED3-monopole26} (Color online) Bounds on $\Delta_{M_1}$ in terms of $\Delta_{M_{1/2}}$ in $d=3$ for $N_f=2,6$ (a,b) with various assumptions on the gaps in the uncharged sector in the same $SU(N_f)$ representation as $M_1$ \cite{Chester:2016wrc}.}
  \end{figure}

\begin{figure}[t!]
    \centering
\includegraphics[width=\figwidth]{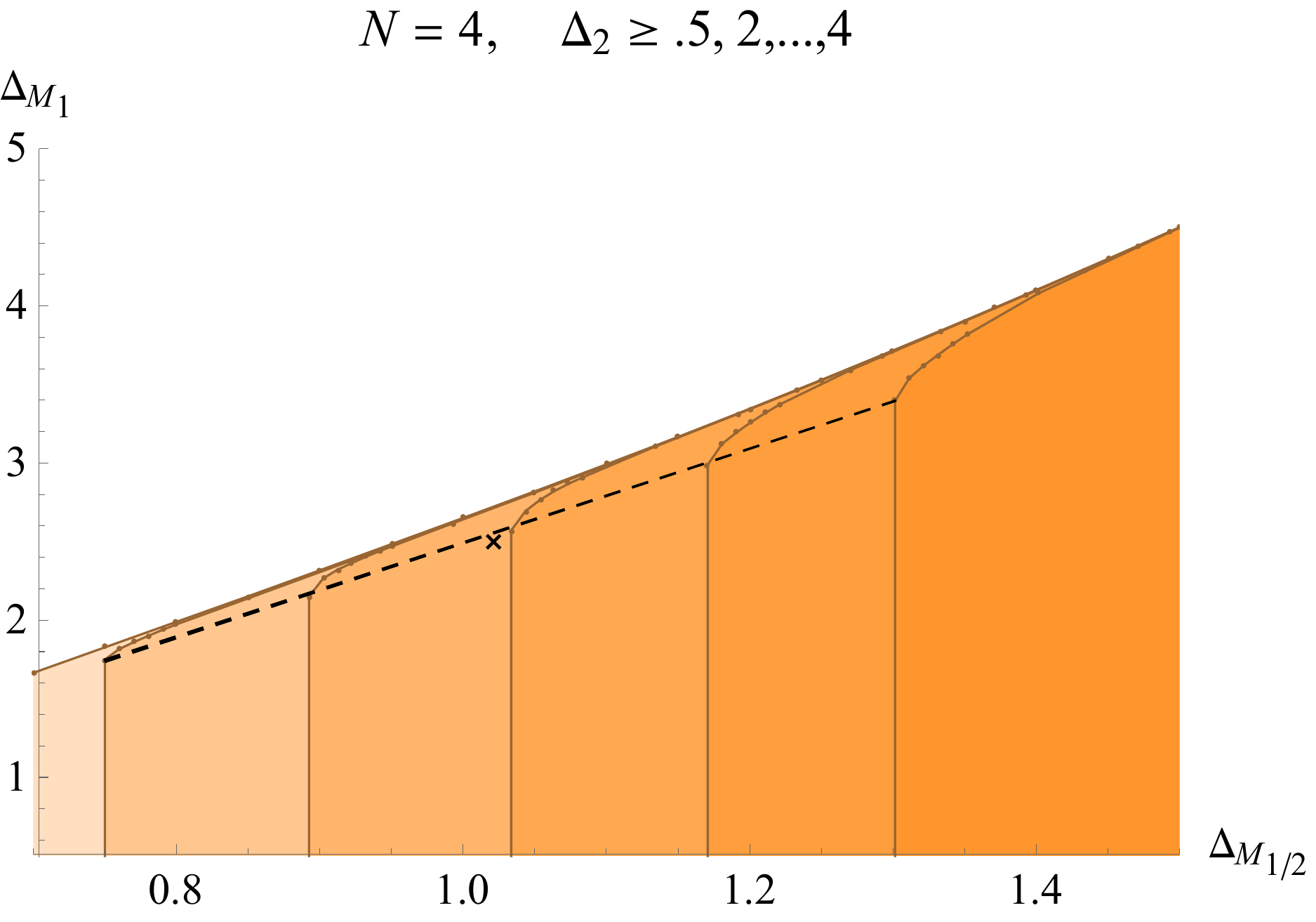}(a)
\includegraphics[width=\figwidth]{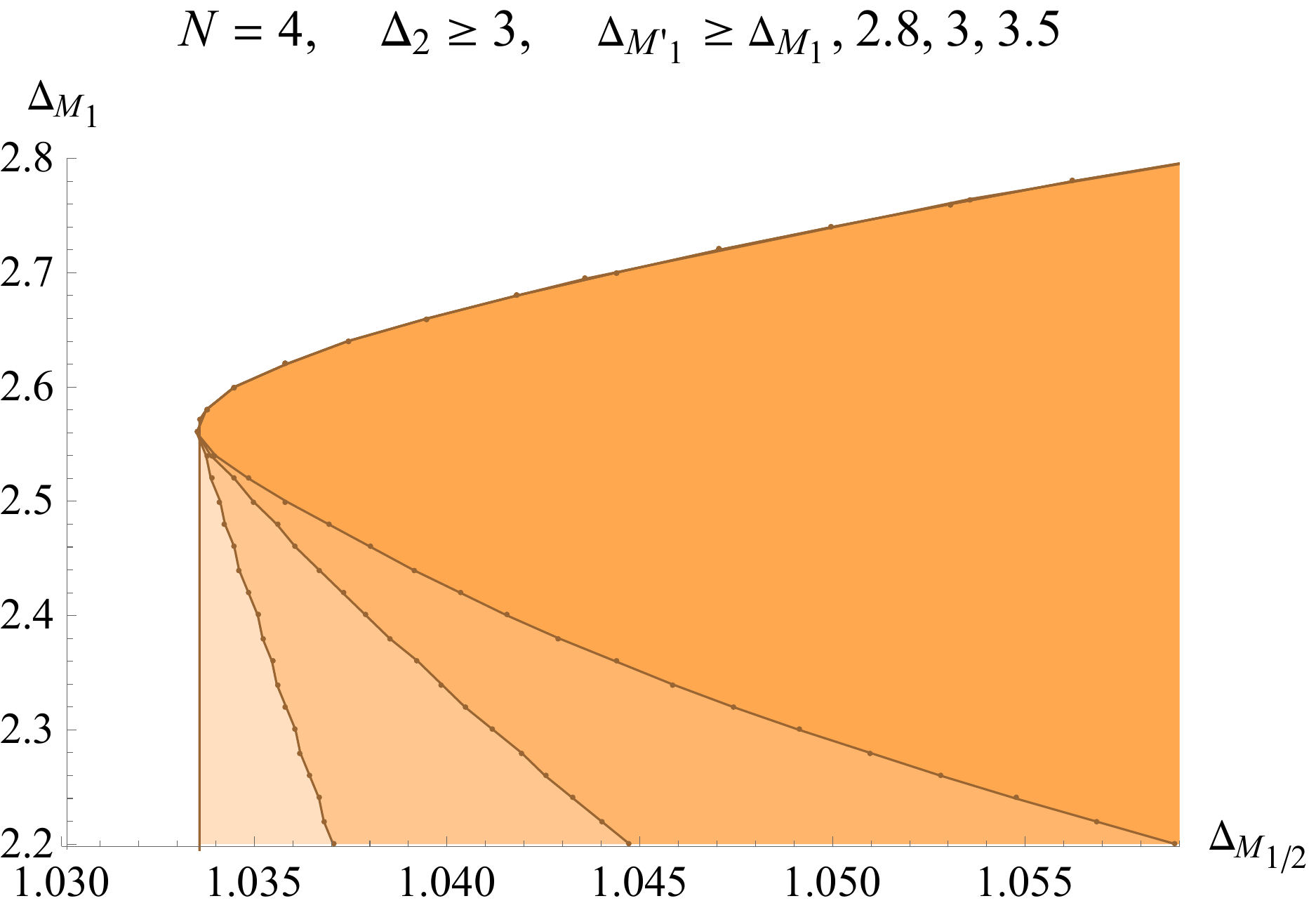}(b)
    \caption{\label{fig:QED3-monopole4} (Color online) (a) is the analogue of Fig.~\ref{fig:QED3-monopole26} for $N_f=4$. (b) starts from the $\Delta_2\geq3$ case of (a), and shows that placing an additional gap $\Delta_{M'_1}$ above $\Delta_{M_1}$ turns the kink into a peninsula \cite{Chester:2016wrc}.}
  \end{figure}

The bootstrap for 4pt functions of $M^I_{\pm1/2}$ was studied in \textcite{Chester:2016wrc} for $N_f = 2, 4, 6$, where they focused on placing bounds on the dimension of the second monopole operator $\Delta_{M_{1}}$, making various assumptions about gaps in the uncharged $(q=0)$ sector. These bounds are shown in Figs.~\ref{fig:QED3-monopole26}, \ref{fig:QED3-monopole4}, where for $N_f=4,6$ they can be compared with the large $N_f$ estimate (black cross). Intriguingly, there is a kink-like discontinuity in the bound which comes close to the large $N_f$ estimate for certain values of the gap in the uncharged sector for operators in the same $SU(N_f)$ representation. By increasing the gap above $M_1$, the allowed region could also be turned into a peninsula around the kink. Similar bounds for the lightest spinning monopoles in the case $N_f = 4$, along with a comparison to the large $N_f$ predictions, were presented in \textcite{Chester:2017vdh}.

While these results are not definitive, they seem promising and show that the bootstrap for QED${}_3$ has a reasonable chance to be successful, perhaps after a few more ingredients are added. Some possible directions would be to consider a multiple correlator bootstrap involving $M_{\pm1/2}$, $M_{\pm1}$, and/or $\bar{\psi}_i \psi^j$. It may also be fruitful to combine these with constraints from 4pt functions containing the $U(1)_T$ current, the $SU(N_f)$ current, or the stress tensor. 

\subsubsection{Bosonic QED${}_3$ and deconfined quantum critical points}
\label{bQED3}

Finally we would like to review the rich physics of bosonic QED${}_3$, where some bootstrap insights have recently been obtained. Bosonic QED${}_3$ is obtained by coupling the $U(1)$ gauge field to $N$ complex scalars $\phi_i$ with an $SU(N)$ invariant potential $m^2 |\phi|^2 +\lambda (|\phi|^2)^2$. This is also known as the $N$-component abelian Higgs model, and is believed to flow to a CFT for large enough $N$. Unlike for fermions, the boson mass term preserves all the symmetries and has to be fine-tuned to reach the fixed point.

This model has been much discussed in the condensed matter literature as the ``non-compact CP$^{N-1}$ model'' (NCCP$^{N-1}$) in connection with the phenomenon of ``deconfined criticality" \cite{deconfined}. To briefly review this connection, the physical systems of interest are certain quantum antiferromagnets in $(2+1)$ dimensions, which have a quantum phase transition between N\'eel and Valence-Bond-Solid (VBS) phases.\footnote{The absence of a disordered phase in such transitions can be understood using `t Hooft anomalies, see \textcite{Komargodski:2017dmc}. This perspective also gives insight into the rich physics of interfaces in these theories \cite{Komargodski:2017smk}.} The transition can be described by the $O(3)$ nonlinear sigma model (NLSM) for the N\'eel order parameter, modified by the inclusion of Berry phase effects which suppress topological defects (hedgehogs), which will play an important role below. 

The $O(3)$ NLSM can be written as the CP$^1$ model, which has two complex vectors $\mathbf{z}=(z_1,z_2)$ subject to the constraint $|z_1|^2+|z_2|^2=1$ and a $U(1)$ gauge invariance $\mathbf{z}\sim e^{i\phi} \mathbf{z}$. This explains the emergence of the gauge field. Replacing the constraint by a quartic potential, and adding a Maxwell kinetic term for the gauge field (expected to be generated by the RG flow), one obtains bosonic QED$_{3}$ with $N=2$. 

In the language of QED${}_{3}$, the above-mentioned topological defects are the monopole operators of quantized charge, similar to the ones in Sec.~\ref{sec:QED3-monopole}. Of course the dimensions of monopole operators differ in bosonic and fermionic QED. Also here we will normalize the topological charge to be integer $q\in \bZ$. 

If a monopole of charge $q$ appears in the action, it breaks the global topological $U(1)_T$ symmetry to the $\bZ_q$ subgroup. Microscopic descriptions of quantum antiferromagnets may realize a discrete subgroup of $U(1)_T$ at the lattice level. On cubic lattices, a $\bZ_{q_0}$ with $q_0=4$ is preserved, while for the hexagonal and rectangular lattices we have $q_0=3$ and $q_0=2$. The $\bZ_{q_0}$ symmetry is also visible in the VBS phase where it permutes the vacua. This microscopic symmetry means that only monopoles with charges multiple of $q_0$ appear. Monopoles with different charges have their fugacity killed by the above-mentioned Berry phases~\cite{Read-Sachdev}.

In light of the above discussion, the analysis of the critical behavior of QED${}_3$ can be split into two parts. First, does bosonic QED${}_3$, with all monopoles suppressed, have a fixed point?
If the answer is yes, then one can ask: can this fixed point be reached provided that one allows monopoles with charges in multiples of $q_0$? For this to happen, the monopole of charge $q_0$ has to be irrelevant.
 
One can study these questions analytically at large $N$: one finds a fixed point and computes the critical exponents in the $1/N$ expansion.\footnote{See \textcite{Murthy:1989ps}, \textcite{Kaul:2008xw}, \textcite{Metlitski:2008dw}, and \textcite{Dyer:2015zha}} At small $N$ one resorts to Monte Carlo simulations. The bootstrap at present cannot by itself resolve the question of the fixed point existence. However, it can provide valuable consistency checks on the other studies. Suppose that a certain Monte Carlo simulation is done on a lattice preserving a $\bZ_{q_0}$ subgroup, finds a second order phase transition, and measures the scaling dimensions $\Delta_q$ of monopole operators $M_q$ for a subset of charges $q$ (we denote by $M_0$ the relevant singlet scalar driving the transition). We have the following OPE algebra in the scalar sector, omitting the OPE coefficients ($M_{-q}=M_q^\dagger$):
\beq
M_{q} \times M_{q'} \sim \delta_{q+q'}\mathds{1}+ M_{q+q'}+\ldots\,.
\eeq
By the above discussion, the operator $M_{q_0}$ has to be irrelevant, as well as the higher charge monopoles. We can use the bootstrap to study the consistency of this algebra given the measured operator dimensions. 

\subsubsection{Aside: constraints on symmetry enhancement}
\label{sec:enhancement}

What we just presented is an instance of a more general question: under which conditions can the global symmetry of the fixed point $G$ be larger than the microscopically realized symmetry $H$? The case of interest for the previous section is $G=U(1)$ and $H=\bZ_{q_0}$. For the symmetry enhancement to happen, operators which break $G$ to $H$ must be irrelevant. The bootstrap is a powerful tool to study whether this irrelevance assumption is consistent with conformal symmetry and with other information which may be available about the fixed point. We will see further applications of this philosophy in Secs.~\ref{sec:4D-bsm} and~\ref{sec:4Dwindow}.

\begin{figure}[t!]
    \centering
\includegraphics[width=\figwidth]{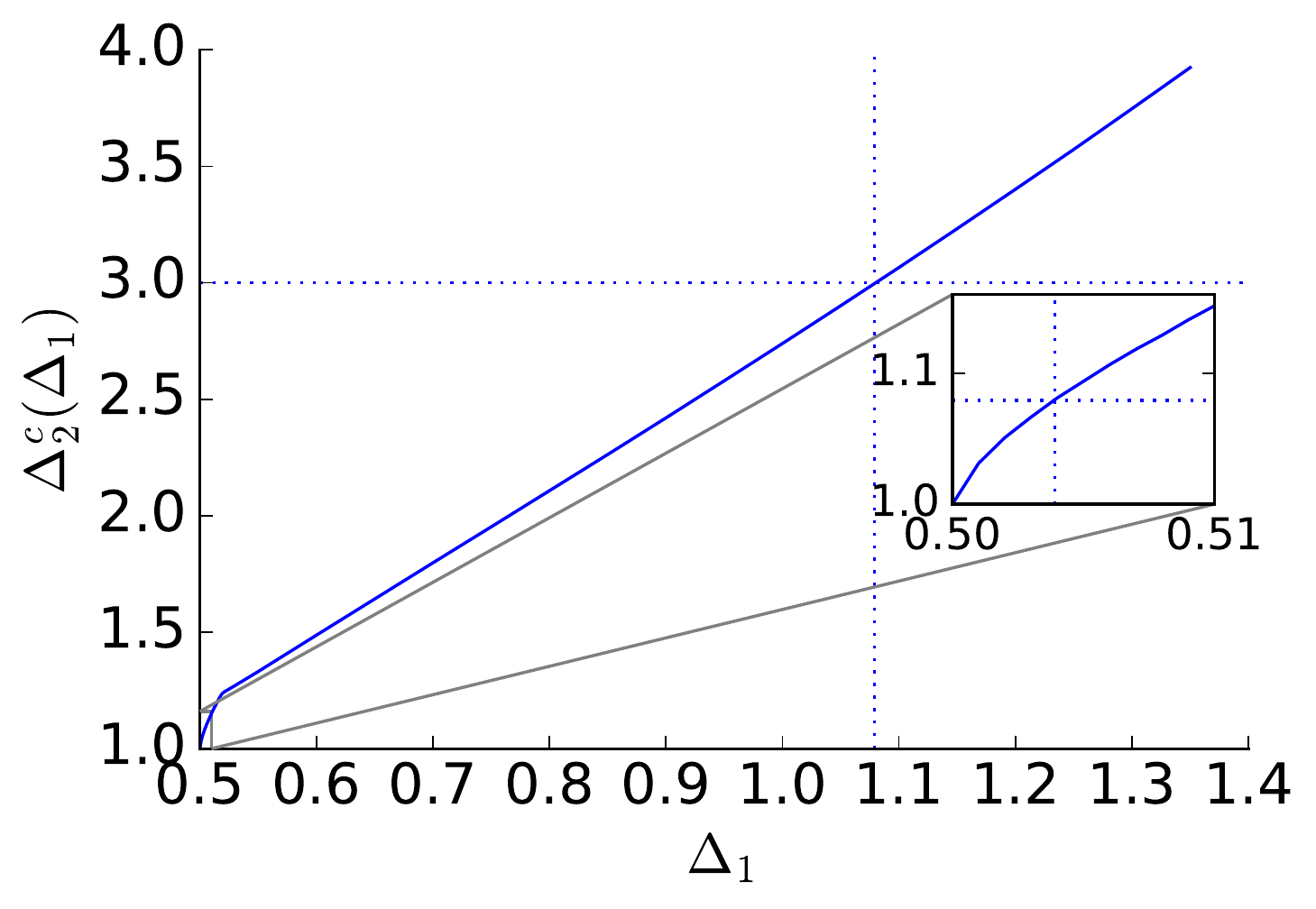}
    \caption{\label{fig:QED3-emergent} (Color online) The 3d upper bound on $\Delta_2$ as a function of $\Delta_1$~\cite{Nakayama:2016jhq}. It may be possible to improve this bound if $\Delta_0$ is known. The same bound applies to $M_2\times M_2\sim M_4$.}
  \end{figure}

We will now describe bootstrap constraints on the symmetry enhancement from $\bZ_{q_0}$ to $U(1)$ derived by~\textcite{Nakayama:2016jhq}. Enhancement from $\bZ_2$ requires that $M_2$ is irrelevant. Since $M_1\times M_1\sim M_2$, one can bound $\Delta_2$ given $\Delta_1$, by studying the 4pt function $\langle M_1 M_1^\dagger M_1 M_1^\dagger\rangle$. The resulting bound is given in Fig.~\ref{fig:QED3-emergent}. Imposing $\Delta_2>3$, one gets a necessary condition \mbox{$\Delta_1>1.08$} for enhancement from $\bZ_2$ to $U(1)$. 

 The same plot in Fig.~\ref{fig:QED3-emergent} can be used to derive rough necessary conditions on the enhancement from $\bZ_4$ to $U(1)$. Indeed, the bound applies also to $M_2\times M_2\sim M_4$. If $M_4$ is irrelevant, then we must have $\Delta_2>1.08$, which in turn implies $\Delta_1>0.504$. 
  
\begin{figure}[t!]
    \centering
\includegraphics[width=\figwidth]{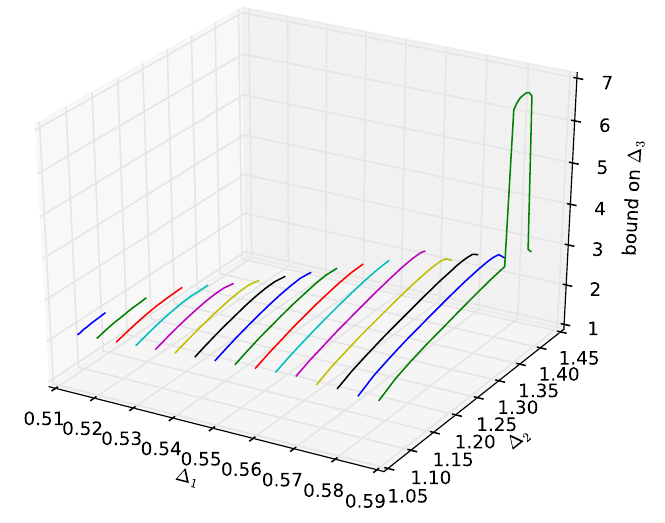}
    \caption{\label{fig:QED3-emergent-ch3} (Color online) An upper bound on $\Delta_3$ as a function of $\{\Delta_1,\Delta_2\}$ under the assumptions that $\Delta_0>1.044$, $\Delta_4>3$ \cite{Nakayama:2016jhq}. It follows from Fig.~\ref{fig:QED3-emergent} that the range of $\Delta_2$ is restricted by the latter assumption from below, and, for fixed $\Delta_1$, from above.}
  \end{figure}

To study enhancement from $\bZ_3$, one analyzes simultaneously three 4pt functions $\langle M_1 M_1^\dagger M_1 M_1^\dagger\rangle$, $\langle M_1 M_1^\dagger M_2 M_2^\dagger\rangle$, $\langle M_2 M_2^\dagger M_2 M_2^\dagger\rangle$. It is reasonable to assume that $M_4$ is irrelevant (as would be the case if $M_3$ is irrelevant and $\Delta_q$ is monotonic in $q$), and to impose $\Delta_0>1.044$ (which follows from an assumption that the fixed point is critical and not multicritical, see Sec.~\ref{sec:multicrit}). Under these assumptions, the upper bound on $\Delta_3$ as a function of $\{\Delta_1,\Delta_2\}$ is shown in Fig.~\ref{fig:QED3-emergent-ch3}. From this bound, irrelevance of $M_3$ requires $\Delta_1>0.585$. 
  
\subsubsection{Back to deconfined criticality: is the transition second order?}
 
 The necessary conditions described in the previous section have been compared with available Monte Carlo and large $N$ data on the N\'eel-VBS transition which claim to see a second-order transition and measure some critical exponents. For square and hexagonal lattices, there is nice consistency, as for rectangular lattices for $N\le 4$ and $N\ge 6$, while some $N=5$ simulations are inconsistent with the bootstrap. The conclusion is that there must either be an error in the $N=5$ Monte-Carlo measurement or in the assumption that the transition is second-order. See \textcite{Nakayama:2016jhq} for this survey and for further details. 

It should be emphasized that while the bootstrap results may point out an inconsistency in Monte Carlo simulations, they cannot, at present, validate them and prove that the phase transition is indeed second order. It is still possible that even in the above cases when there is a nice agreement between Monte Carlo results and the bootstrap necessary conditions, the transition is still very weakly first order and not second order. 

Let us focus on the case $N=2$ which presents a controversy. Large-scale Monte Carlo simulations for $N=2$ were performed in \textcite{Nahum:2015jya}, using a loop model on a cubic lattice which is in the same universality class as NCCP${}^1$ and has monopole suppression up to $q_0=4$, and going up to very large lattices of linear size up to $L=640$.\footnote{See also \textcite{Harada2013, Sreejith:2018ypg} for simulations of other microscopic models in the same universality class.} While they have not seen signs of a finite correlation length or a conventional first order transition, and observed scaling behavior of correlation functions at distances $1\ll r \ll L$, they have seen scaling violation for observables at larger distances $r\sim L$, inconsistent with a conventional second order transition. 

So, is the transition second order or weakly first order? Assuming a second order transition, \textcite{Nahum:2015jya} extracted the scaling dimension of the monopole operator $\Delta_1=0.625(15)$, which is consistent with the bootstrap condition $\Delta_1>0.504$ necessary for the enhancement from $\bZ_4$ to $U(1)$. However there is an extra piece of information which allows one to set up an even more stringent bootstrap test: further symmetry enhancement at the transition from $SO(3)\times U(1)$ to $SO(5)$. Here $SO(3)$ acts on the N\'eel order parameter $N_a = \mathbf{z}^\dagger\sigma_a \mathbf{z}$. Empirically, the scaling dimension of $N$ is very close to $\Delta_1$~\cite{Nahum:2015jya} and, moreover, the joint probability distribution of $(N, M_1)$ is very close to the spherical one after a rescaling~\cite{Nahum:2015vka}, which can be explained if $N$ and $M_1$ belong to a vector multiplet $\Phi$ of $SO(5)$ of dimension $\Delta_\Phi=\Delta_1$. 

In this description, the relevant scalar which drives the transition is a component of the symmetric traceless tensor (roughly $\Phi_A\Phi_B-\text{trace}$).\footnote{\textcite{Nahum:2015vka} measured its scaling dimension to be $\sim 1.5$.} For the $SO(5)$ enhancement to happen, any other scalar which breaks $SO(5)$ back to $SO(3)\times U(1)$ must be irrelevant. In addition, the $SO(5)$ singlet $S$ (roughly $\Phi_A\Phi^A$) must be irrelevant for the transition to be second order, since otherwise the fixed point will not be reached. See \textcite{Wang:2017txt} for further discussion. Given the dimension $\Delta_\Phi=\Delta_1$ as above, it is straightforward to compute an upper bound on the dimension of $S$ which occurs in the OPE $\Phi\times\Phi$. This is the same bound as for $N=5$ in Fig.~\ref{fig:ONsymtrace} except the plot has to be extended to larger $\Delta_\Phi$. \textcite{Nakayama:2016jhq} and~\textcite{DSD2016} performed this analysis and report that $\Delta_S>3$ is excluded for $\Delta_\Phi$ as above. In fact $\Delta_S>3$ requires $\Delta_\Phi>0.76$~\cite{Nakayama2016}.

To summarize, the bootstrap excludes a second-order phase transition described by a unitary 3d CFT with symmetry enhanced to $SO(5)$ and the order parameter scaling dimension taking the above value suggested by the Monte Carlo simulations. 
In our opinion, the most compelling interpretation of available data is a weakly first-order transition due to walking RG behavior which ensues when the RG flow has no fixed points for a real value of the coupling but two complex conjugate fixed points with small imaginary parts. This is the same mechanism as for the weakly first-order transition in the 2d Potts model with $Q\gtrsim 4$. As discussed in \textcite{Nahum:2015jya}, this scenario may resolve the observed scaling violations at distances $r\sim L$. It can also accommodate the enhancement to $SO(5)$~\cite{Wang:2017txt}. In this scenario, there is no unitary 3d CFT (the complex fixed points being nonunitary), and the bootstrap bounds do not apply, resolving the contradiction.

Finally, let us note that a similar analysis can constrain another scenario outlined in \textcite{Wang:2017txt}, in which a variant called the easy-plane NCCP${}^1$ model is conjectured to have a fixed point with enhanced $O(4)$ symmetry and be dual to $N_f=2$ fermionic QED${}_3$. In this scenario, the conjectured fixed point cannot contain any fully $O(4)$-invariant scalar perturbations. However, recent Monte Carlo simulations~\cite{Qin:2017cqw} point to a dimension for the $O(4)$ vector order parameter $\Delta_\Phi = 0.565(15)$ which seems incompatible with the bootstrap bound assuming irrelevance of the $O(4)$ singlet (Fig.~\ref{fig:ONsinglet} extended further to the right), which requires $\Delta_\Phi > 0.868$~\cite{PolandUnpublished}. In the future it will be interesting to further study the fate of these models using both bootstrap and Monte Carlo data.

\subsection{Current and stress-tensor bootstrap}
\label{sec:JandT}

In the previous sections we discussed results following from 4pt functions of scalars and fermions. Here we wish to discuss interesting results which have recently been obtained from 4pt functions of tensor operators, specifically of conserved currents and stress tensors. Namely, a number of numerical bounds on scaling dimensions and OPE coefficients from such correlators in parity-preserving 3d CFTs were recently computed in \textcite{Dymarsky:2017xzb} and \textcite{Dymarsky:2017yzx}, building on important analytical developments for spinning correlators, as reviewed in Sec.~\ref{sec:spinning}.\footnote{See also \textcite{Dymarsky:2013wla} for a discussion about the general properties of these correlation functions.} Such studies are well-motivated because they probe the general space of local CFTs, and may even lead to the discovery of new theories.

A concrete application of these constraints is to study bounds on current and stress-tensor 3pt function coefficients under various assumptions. These coefficients are known to satisfy various analytical bounds following from the averaged null energy condition (see Sec.~\ref{sec:ANEC}), as was originally argued in the context of conformal collider physics in~\textcite{Hofman:2008ar}, with 3d bounds worked out in~\textcite{Buchel:2009sk,Chowdhury:2012km}. One application of the numerical bootstrap is to study how the conformal collider bounds change as a function of gaps. Another application is to determine these coefficients in various CFTs, e.g.~the critical 3d Ising and $O(N)$ models. These studies also allow one to probe parity-odd operators in the spectrum which have previously been inaccessible from the perspective of scalar 4pt functions (although they could be accessed from the fermionic correlators in Sec.~\ref{sec:Fermions}).

\begin{figure}[t!]
    \centering
\includegraphics[width=\figwidth]{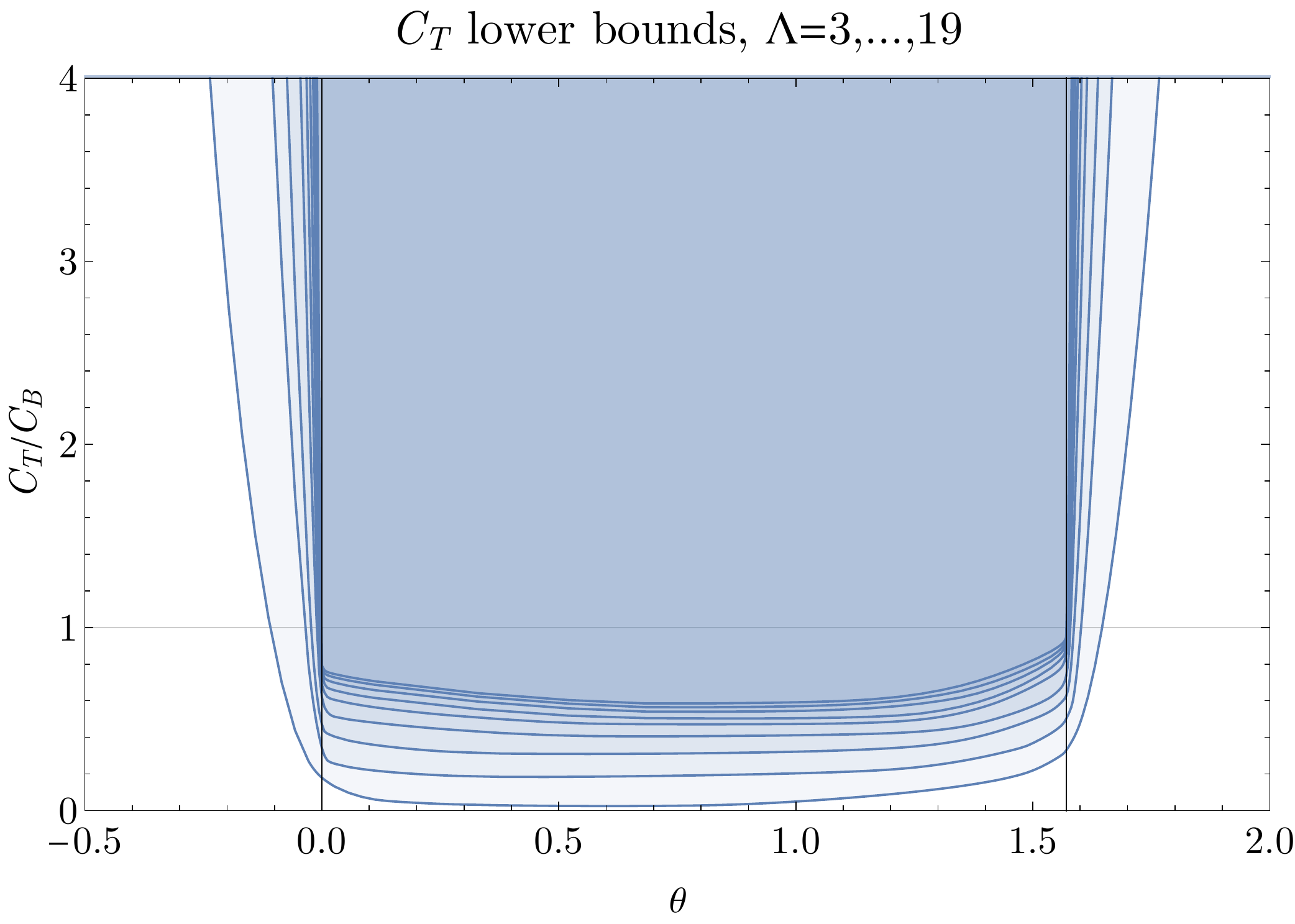}
    \caption{\label{fig:JandT-ctMinLambda3To19Plot}
(Color online) Bounds on $C_T$ as a function of the $\<TTT\>$ 3pt function parameter $\theta$~\cite{Dymarsky:2017yzx}.}
  \end{figure}

\begin{figure}[t!]
    \centering
\includegraphics[width=\figwidth]{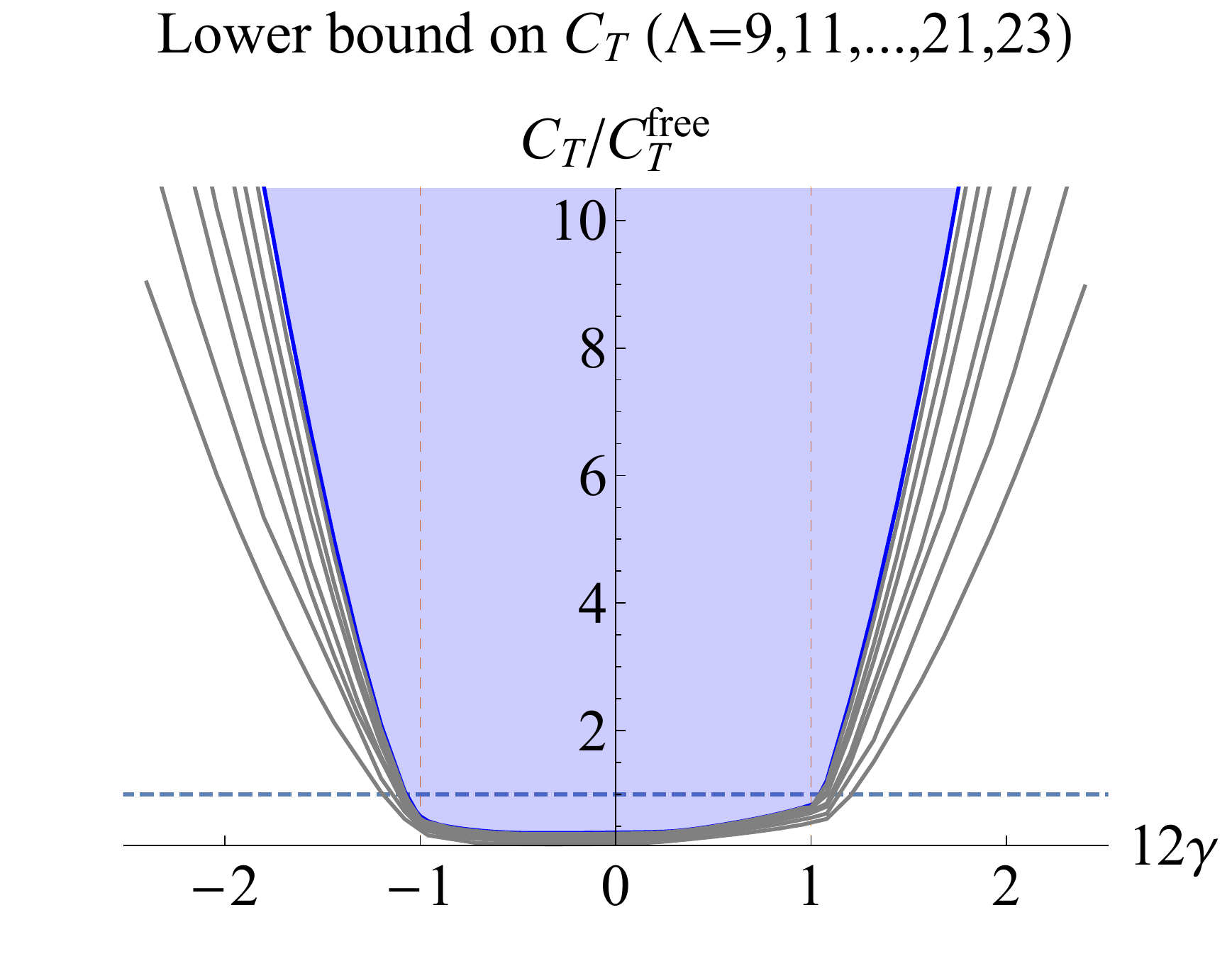}
    \caption{\label{fig:JandT-CTvsGamma_nmax6-12_linear}
(Color online) Bounds on $C_T$ as a function of the $\<JJT\>$ 3pt function parameter $\gamma$~\cite{Dymarsky:2017xzb}.}
  \end{figure}

In Fig.~\ref{fig:JandT-ctMinLambda3To19Plot} we show general lower bounds on the central charge (in units of the free boson central charge $C_B$) from the bootstrap applied to $\<TTTT\>$, as a function of the independent parity-preserving $\<TTT\>$ 3pt function coefficient which is parametrized by the variable $\theta$.\footnote{\label{footnote:pv}In fact, Figs.~\ref{fig:JandT-ctMinLambda3To19Plot} and~\ref{fig:JandT-CTvsGamma_nmax6-12_linear} should also apply to parity-violating theories since all needed conformal blocks have been included (e.g., contributions from parity-violating $\<TTT\>$ or $\<JJT\>$ couplings are accounted for by allowing parity-odd spin-2 contributions at the unitarity bound). This also holds for bounds where identical gaps have been imposed in a given parity-even and parity-odd sector simultaneously (e.g., Figs.~\ref{fig:JandT-scalarGapsExclusionPlot} along the diagonal). Stronger bounds may hold in parity-violating theories after adding crossing relations from parity-violating 4pt structures and assuming nonzero values of parity-violating coefficients.} In this notation the conformal collider bound is \mbox{$0 \leq \theta \leq \pi/2$}, where a free scalar has $\theta = 0$ and free Majorana fermion has $\theta = \pi/2$. It can be readily seen that the numerical bootstrap is able to reproduce this constraint in addition to giving general lower bounds on $C_T$. A similar set of bounds from the $\<JJJJ\>$ bootstrap is shown in Fig.~\ref{fig:JandT-CTvsGamma_nmax6-12_linear}, where in this plot $C_T$ is normalized to the central charge of a free \emph{complex} scalar $C_T^{\text{free}}$ and the independent parity-preserving $\<JJT\>$ coupling is parametrized by $\gamma$, where the conformal collider bound is given by $-\frac{1}{12} \leq \gamma \leq \frac{1}{12}$.
 In this case the free complex scalar has $\gamma = -\frac{1}{12}$ while the free Dirac fermion has $\gamma = \frac{1}{12}$. 

\begin{figure}[t!]
    \centering
\includegraphics[width=\figwidth]{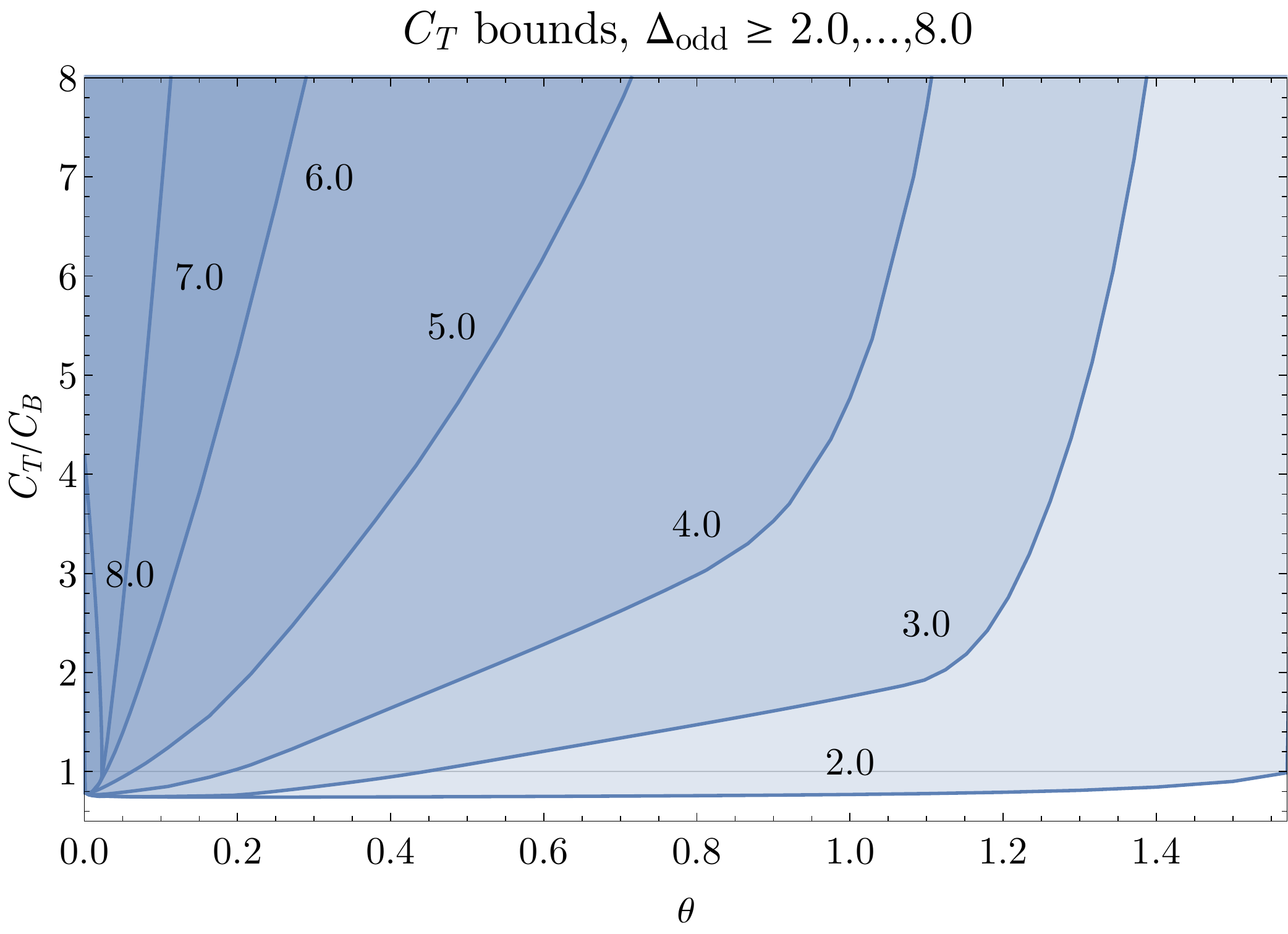}
    \caption{\label{fig:JandT-deltaOddGapsLargePlot}
(Color online) Bounds on $C_T$ as a function of the $\<TTT\>$ 3pt function parameter $\theta$ for different values of the parity-odd scalar gap $\Delta_{\text{odd}}$~\cite{Dymarsky:2017yzx}. }
  \end{figure}

As mentioned above, the advantage of the numerical bootstrap is that it can readily probe how constraints on these couplings depend on gaps in the spectrum. For example, in Fig.~\ref{fig:JandT-deltaOddGapsLargePlot} we show how the lower bound on $C_T$ as a function of $\theta$ in the $\<TTTT\>$ bootstrap varies as one increases the gap in the parity-odd scalar sector $\Delta_{\text{odd}}$ from $2$ to $8$. It can be seen that increasing the parity-odd gap forbids the ``fermion" end of the range for $\theta$ but allows the ``scalar" end. This is consistent with the fact that the free scalar has a very large parity odd gap $\Delta_{\text{odd}} = 11$, while the free Majorana fermion has a small gap $\Delta_{\text{odd}} = 2$. 
 
\begin{figure}[t!]
    \centering
\includegraphics[width=\figwidth]{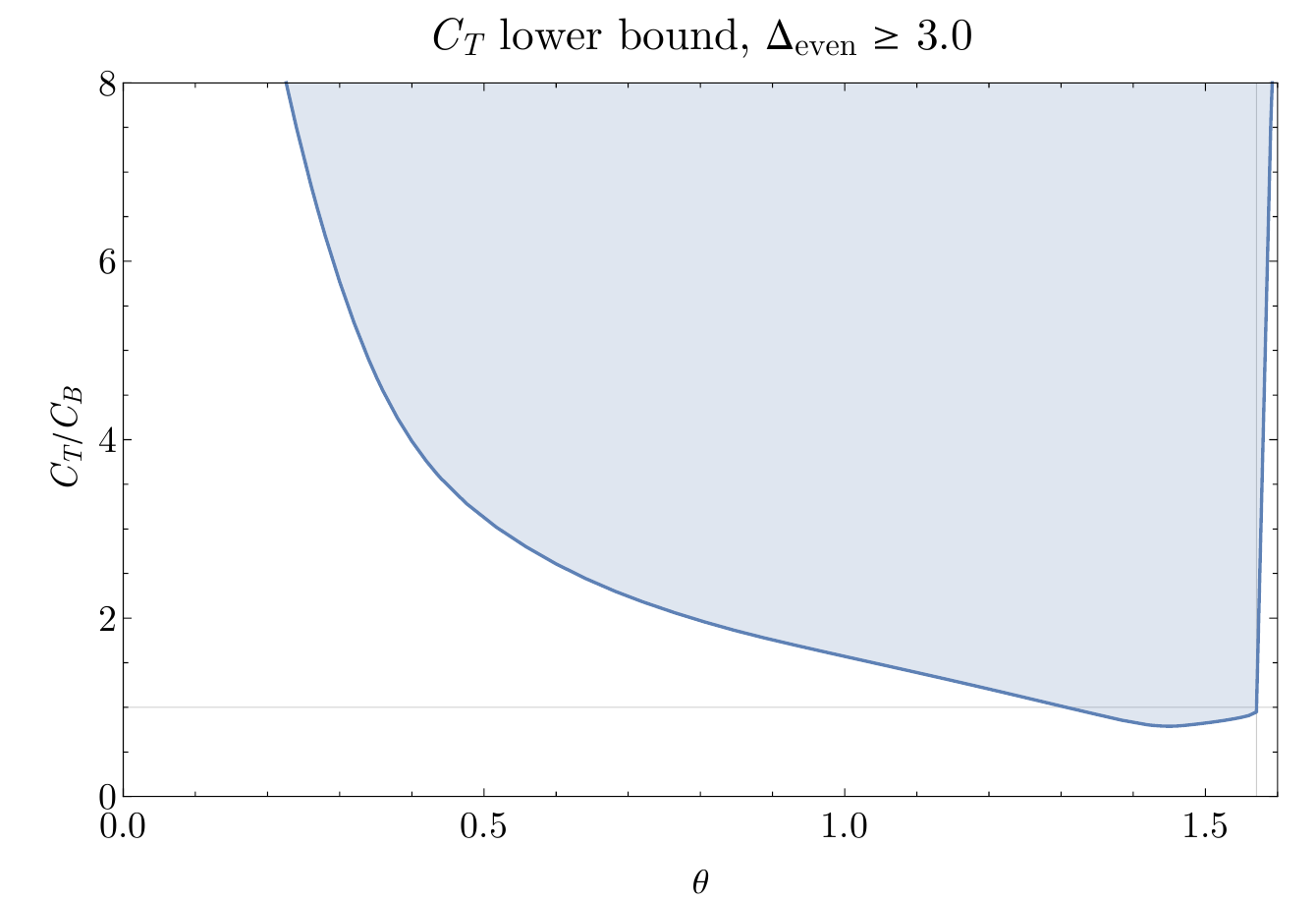}
    \caption{\label{fig:JandT-evenGap3Plot}
(Color online) Bounds on $C_T$ as a function of the $\<TTT\>$ 3pt function parameter $\theta$ assuming that the leading parity-even scalar is irrelevant~\cite{Dymarsky:2017yzx}.}
  \end{figure}

Similarly, imposing a parity-even gap forbids the ``scalar" end. In Fig.~\ref{fig:JandT-evenGap3Plot} we illustrate this by showing what happens when the leading parity-even scalar is irrelevant, corresponding to ``self-organized" CFTs (see Sec.~\ref{sec:multicrit}). This lower central charge bound applies for instance to fermionic QED${}_3$ in the conformal window from Sec.~\ref{sec:QED3}, with $N_f$ even for parity invariance.

Furthermore, imposing gaps on both parity-even and parity-odd scalars forces one to live with intermediate values of $\theta$, at least for moderately small values of the central charge. These and other bounds with different gap assumptions (including upper bounds on $C_T$ for gaps excluding large-$N$ theories) can be found in \textcite{Dymarsky:2017yzx}.
 
\begin{figure}[t!]
    \centering
\includegraphics[width=\figwidth]{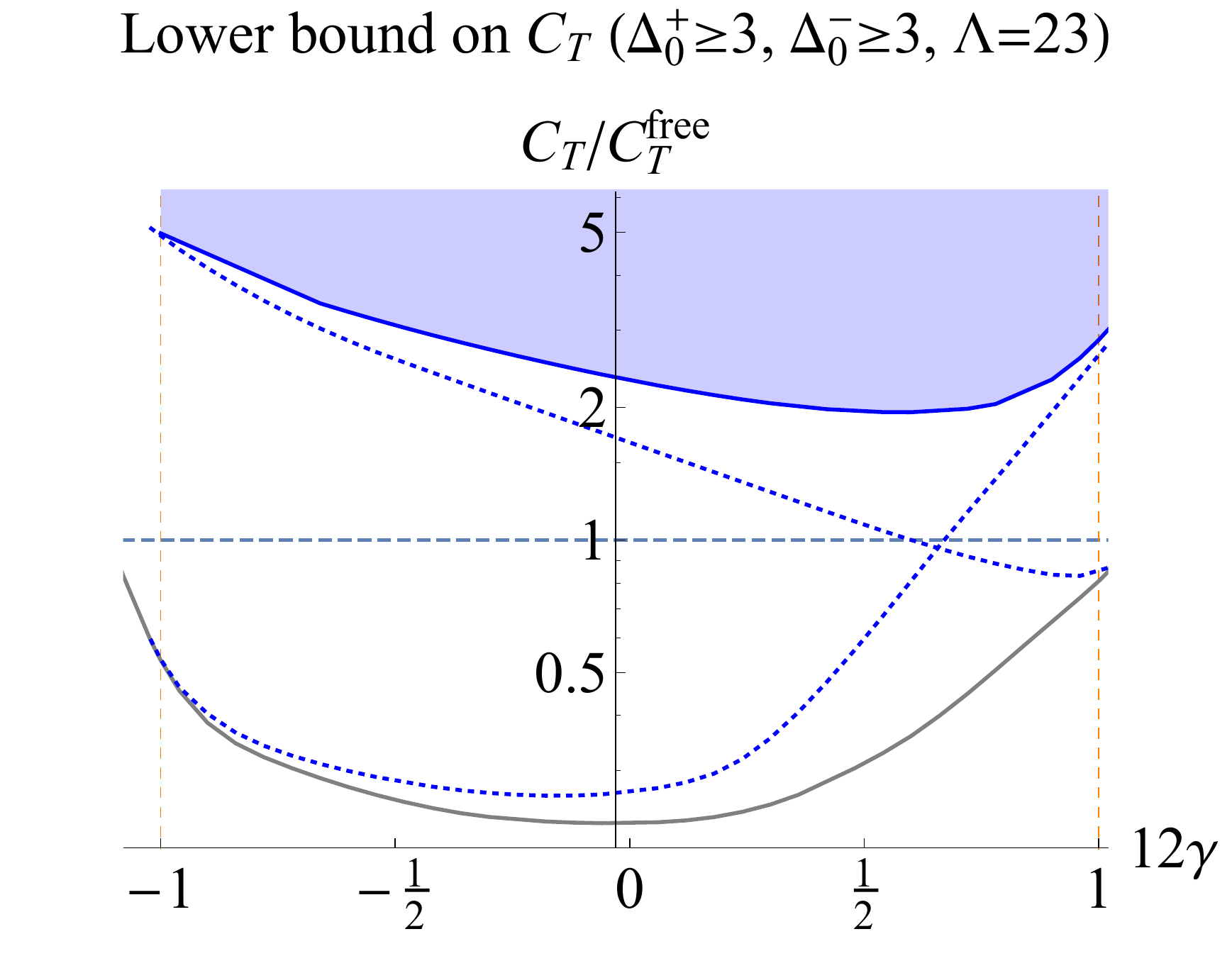}
    \caption{\label{fig:JandT-CTvsGamma_Norelevant_nmax12}
(Color online) Bounds on $C_T$ as a function of the $\<JJT\>$ 3pt function parameter $\gamma$ with no assumptions (lower solid curve), parity-odd scalars irrelevant (lower dashed curve), parity-even scalars irrelevant (upper dashed curve), and all scalars irrelevant (upper solid curve)~\cite{Dymarsky:2017xzb}.}
  \end{figure}

A similar story holds for bounds from the $\<JJJJ\>$ bootstrap, where Fig.~\ref{fig:JandT-CTvsGamma_Norelevant_nmax12} shows how the lower bound on $C_T$ changes when either parity-even or parity-odd scalars are irrelevant (dashed blue lines), or when both are irrelevant (solid blue line). These bounds are consistent with the gaps for a free complex scalar ($\Delta_0^+ = 2$, $\Delta_0^- = 7$) and free Dirac fermion ($\Delta_0^+ = 4$, $\Delta_0^- = 2$). Similar bounds with different gap assumptions can be found in \textcite{Dymarsky:2017xzb}. In the future, by generalizing this analysis to {$SU(N_f)$ (or $SU(N_f) \times U(1)$)} global symmetry one may be able to place interesting bounds on the fermionic QED${}_3$ central charge using the same argument as the one based on Fig.~\ref{fig:JandT-evenGap3Plot}.
  
\begin{figure}[t!]
    \centering
\includegraphics[width=\figwidth]{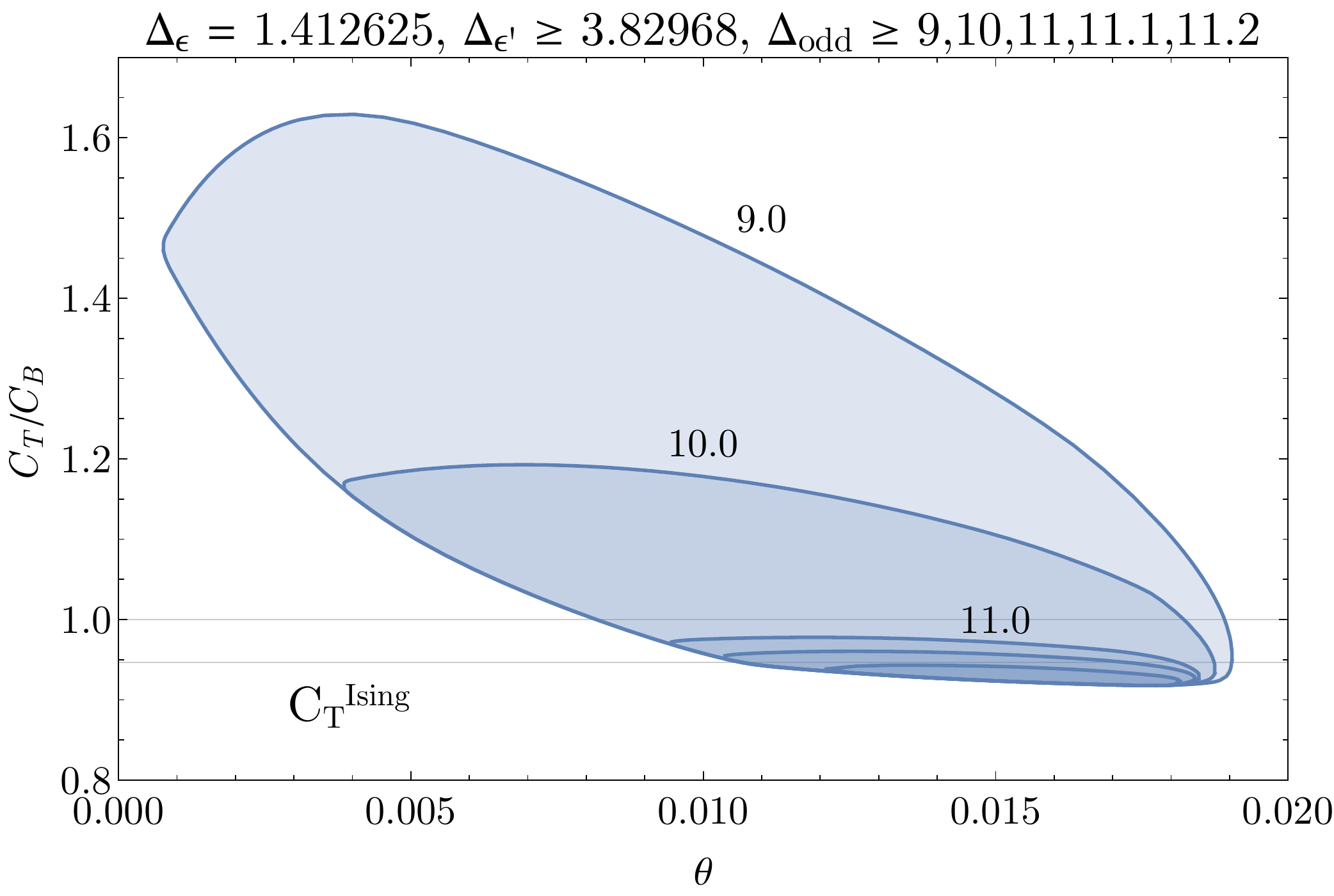}
    \caption{\label{fig:JandT-IsingparityOddGap91011Plot}
(Color online) Bounds on $C_T$ as a function of the $\<TTT\>$ 3pt function parameter $\theta$ with gap assumptions plausible for the Ising model~\cite{Dymarsky:2017yzx}.}
  \end{figure}

It is interesting to ask if one can use these general bootstrap constraints to determine $\theta$ or $\gamma$ in some CFTs of interest, e.g.~the Ising or $O(2)$ models. In the case of the Ising model, it is a plausible assumption (e.g., from the $\epsilon$-expansion) that its leading parity-odd (but $\mathbb{Z}_2$-even) operator has a very large dimension as in the free scalar theory. Using known parity-even data the $\<TTTT\>$ bootstrap yields an upper bound $\Delta^{\text{Ising}}_{\text{odd}} < 11.2$, and one can obtain small closed regions in the $\{\theta, C_T\}$ plane that are consistent with the known Ising central charge, Eq.~\reef{eq:central-charge}. These regions, shown in Fig.~\ref{fig:JandT-IsingparityOddGap91011Plot}, yield the determination $0.01 < \theta < 0.018-0.019$ if $\Delta_{\text{odd}}$ is close to saturating its bound. Note that if one makes the weaker assumption of irrelevance $\Delta_{\text{odd}} > 3$, then there is still a reasonably tight range $0.01 < \theta < 0.05$ consistent with $C_T^{\text{Ising}}$.
  
\begin{figure}[t!]
    \centering
\includegraphics[width=\figwidth]{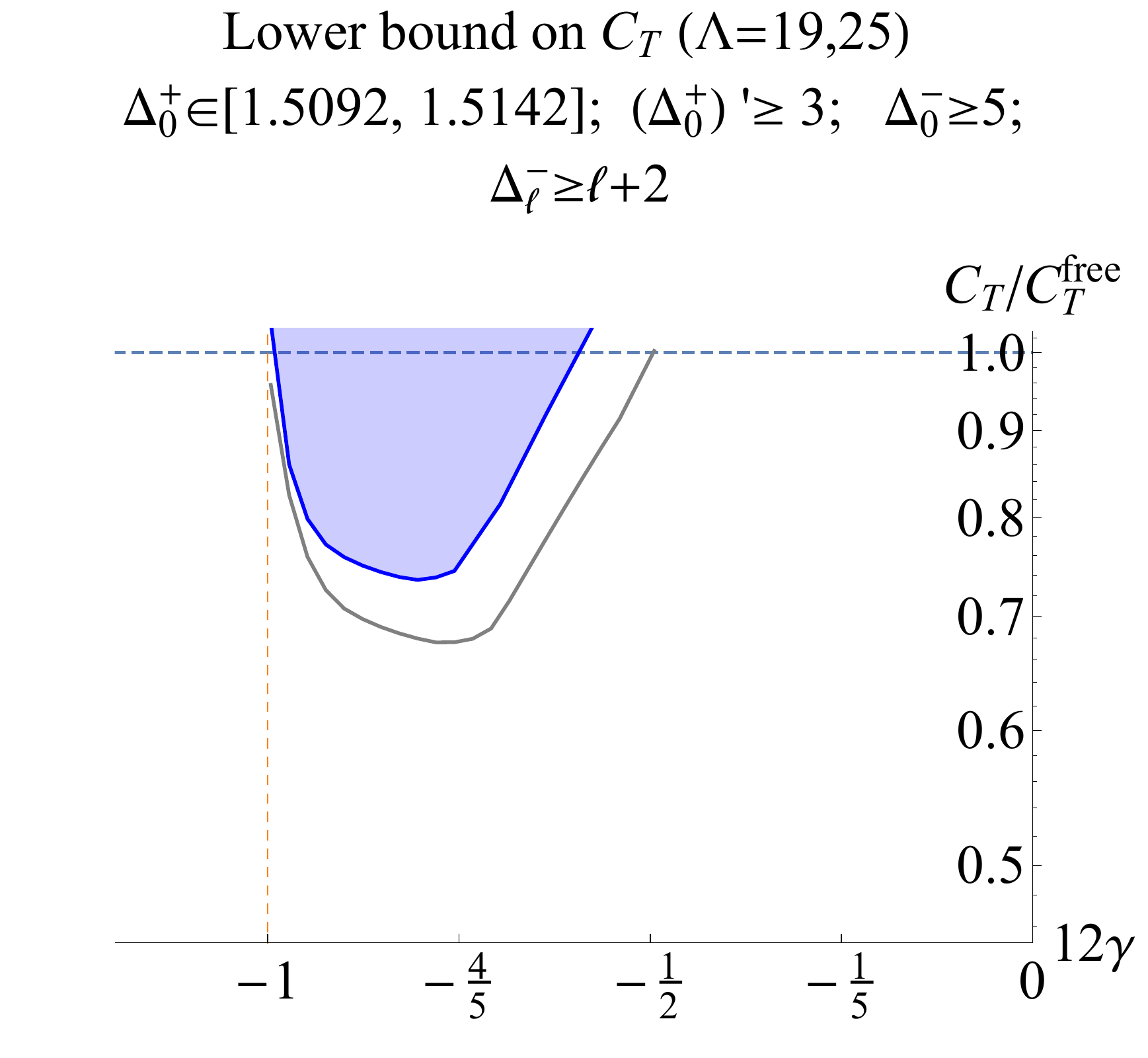}
    \caption{\label{fig:JandT-CTvsGamma_O2_Extrapolation}
(Color online) Bounds on $C_T$ as a function of the $\<JJT\>$ 3pt function parameter $\gamma$ with gap assumptions plausible for the critical $O(2)$ model~\cite{Dymarsky:2017xzb}.}
  \end{figure}

\begin{figure}[t!]
    \centering
\includegraphics[width=\figwidth]{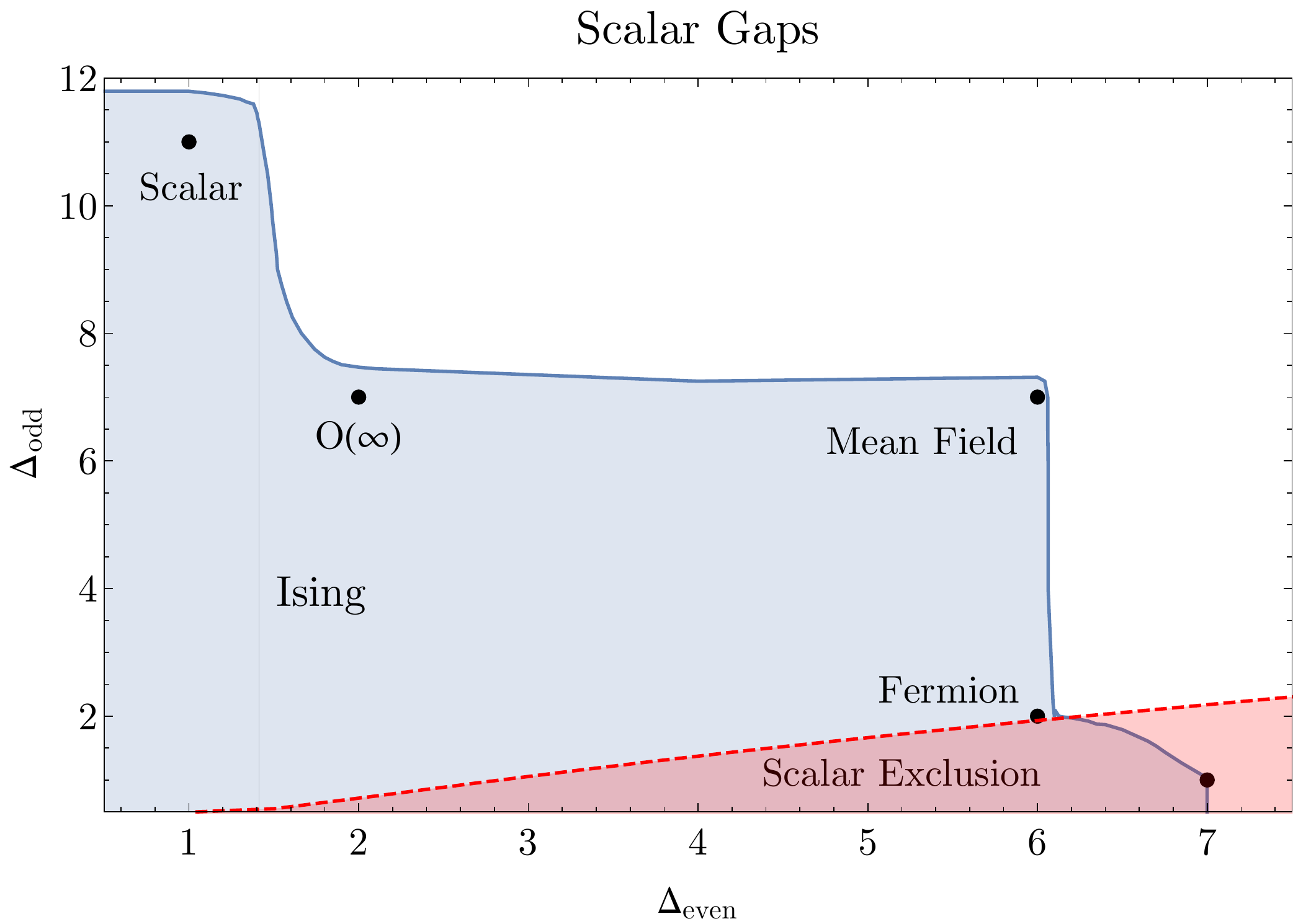}(a)
\includegraphics[width=\figwidth]{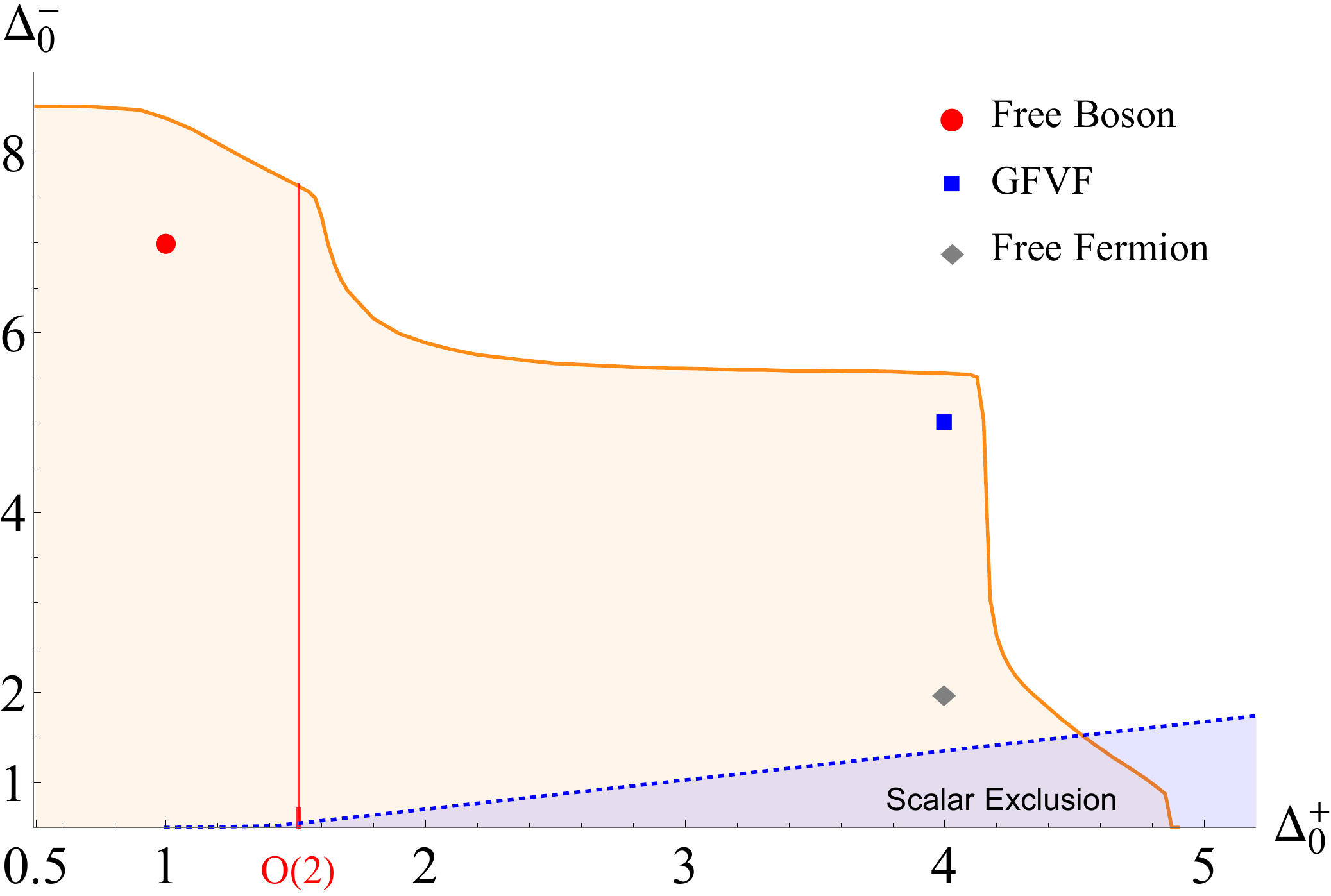}(b)
    \caption{\label{fig:JandT-scalarGapsExclusionPlot}
(Color online) Allowed region for parity-even and parity-odd scalar gaps from (a) $\<TTTT\>$ bootstrap~\cite{Dymarsky:2017yzx} and (b) $\<JJJJ\>$ bootstrap~\cite{Dymarsky:2017xzb}. }
  \end{figure}

In the case of the $\<JJJJ\>$ bootstrap, one can similarly try to determine $\gamma$ for the $O(2)$ model. In this case after inputting the known dimension of the leading $O(2)$ parity-even singlet, irrelevance of the second $O(2)$ parity-even singlet, and the plausible parity-odd gaps $\Delta_0^- \geq 5$ and $\Delta_{\ell}^- \geq \ell+2$, Fig.~\ref{fig:JandT-CTvsGamma_O2_Extrapolation} yields the range $-0.0824 < \gamma < -0.0494$. Based on a linear extrapolation, \textcite{Dymarsky:2017xzb} also estimated the more restrictive range $-0.080 < \gamma < -0.061$. A negative value of $\gamma$ in the $O(2)$ model also appears to be favored by the results of quantum Monte Carlo simulations \cite{Katz:2014rla}. In future work it will be interesting to find ways to improve these determinations, extend them to other $O(N)$ models, and perhaps connect the smallness of the deviations of $\theta$ and $\gamma$ from their free values to the existence of approximate higher-spin currents in these theories.

Finally, in Fig.~\ref{fig:JandT-scalarGapsExclusionPlot} we show a more global picture of the allowed region of parity-odd and parity-even scalar gaps from both the $\<TTTT\>$ and $\<JJJJ\>$ bootstrap, making no additional assumptions. These regions satisfy a number of consistency checks, e.g. being consistent with known gaps in free theories, MFTs, and critical $O(N)$ models. They additionally show fairly sharp features near the scaling dimensions in the Ising and $O(2)$ models. It will be interesting to improve these maps in future studies and identify the locations of other CFTs of interest. The lower ``scalar exclusion" regions of these plots are ruled out from 4pt functions of the leading parity-odd scalar (assuming the parity-even scalar appears in both OPEs as would be generically expected), an example of how the scalar and the stress-tensor/current bootstraps can yield complementary information.
  
There are a number of directions for future work, which include considering mixed systems containing stress tensors/currents together with scalars, studying the implications of parity violation (see footnote~\ref{footnote:pv}), studying in more detail the conditions for large $N$ and holographic theories, and generalizing these studies to other dimensions. Recent progress on how to compute spinning conformal blocks in 4d and in general dimensions\footnote{See Sections~\ref{sec:3pt} and~\ref{sec:spinning} for a summary.} should make these analyses tractable in the future outside of 3d. {Current and stress-tensor multiplets can also be considered in superconformal theories, where the bootstrap analysis can be simplified using the fact that they reside in multiplets with operators of lower spin.}\footnote{{However to do this requires knowledge of the superconformal blocks. Some studies where this has been pursued are mentioned in Sections~\ref{sec:4Dsusy} and~\ref{sec:other}.}}

\subsection{Future targets}
\def\mywidth{0.6\columnwidth}

\label{sec:targets}

In this section we will collect some additional 3d models which in our opinion represent interesting future targets for the bootstrap.

\subsubsection{Multifield Landau-Ginzburg models}

There exists rich phenomenology of fixed points arising from Lagrangians with multiple scalar fields transforming under product group symmetries, e.g.~$SO(n) \times SO(m)$. One can consider Lagrangians involving two coupled scalar multiplets, one transforming in the fundamental of $SO(n)$ and another of $SO(m)$. {Alternatively,} one can consider a field transforming in the bifundamental of $SO(n) \times SO(m)$. Such Lagrangians have been invoked to describe phase transitions in many physical systems{; see \textcite{Vicari:2007ma} for further details.}

{When studying these fixed points using the RG,} a recurrent feature is that many of {them} do not exist in the $4-\eps$ expansion and have to be studied directly in 3d. Since such computations lack a manifestly small expansion parameter, there seems to be no consensus about the existence of these fixed points. So this appears to be a perfect target for a nonperturbative approach like the bootstrap. Some preliminary bootstrap studies of 3d CFTs with \mbox{$SO(n) \times SO(m)$} were carried out in \textcite{Nakayama:2014lva,Nakayama:2014sba}, but in our opinion more work is needed before firm conclusions can be drawn.

\subsubsection{Projective space models}

An interesting 3d lattice model is the CP$^{n}$ model, where microscopic lattice variables belong to CP$^{n}$ and have ferromagnetic interactions preserving the symmetry (see below for the antiferromagnetic case). Recall that CP$^{n}$ can be realized by starting with $(n+1)$-dimensional complex vectors $\mathbf{z}=(z_1,\ldots,z_{n+1})$ and imposing the constraint $\mathbf{z}^\dagger \cdot \mathbf{z} =1$, preserved up to the equivalence $\mathbf{z}\sim e^{i\phi}\mathbf{z}$. A simple lattice Hamiltonian is 
\beq
H=-J\sum_{\langle i j\rangle} |\mathbf{z}^\dagger_i \cdot \mathbf{z}_j|^2\,,
\eeq
with $J>0$ in the considered ferromagnetic case. The physics of this model is influenced by defects (hedgehogs), which are possible because $\pi_2(CP^{n})=\mathbb{Z}$. Here we consider the CP$^{n}$ model with defects allowed. It should be distinguished from the ``non-compact CP$^{n}$ model" which results when defects are suppressed, see Sec.~\ref{bQED3}.

The CP${}^1$ model is equivalent to the $O(3)$ model, with the order parameter $N_a=\mathbf{z}^\dagger \sigma_a \mathbf{z}$, and it has a second-order phase transition described by the same CFT.

The CP${}^2$ model has an internal symmetry $SU(3)$ (modulo global issues), with traceless hermitian matrix $Q_{ab}=z_a \bar{z}_b-\delta_{ab}$ as an order parameter. The Landau-Ginzburg description contains a cubic invariant ${\rm Tr}(Q^3)$ and would suggest a first-order transition, but Monte Carlo simulations \cite{Nahum:2013qha} indicate that the phase transition is continuous. This is similar to what happens for the 3-state Potts model in 2d and can be explained as an effect of fluctuations. Monte Carlo results for the critical exponents are $\eta=0.23(2)$ and $\nu=0.536(13)$, translating into the dimensions of $Q$ and of the relevant singlet scalar. Can this model be isolated using the numerical bootstrap?

One can also consider antiferromagnetic projective space models, taking $J<0$ in the above Hamiltonian.
Antiferromagnetic CP$^{n}$ models \cite{Delfino:2015gba} don't give rise to new universality classes.\footnote{The ACP$^{1}$ model on a cubic lattice is equivalent to the ferromagnetic model and, as the latter, has a phase transition in the $O(3)$ universality class. For higher $n$ there is no equivalence between the antiferromagnetic and ferromagnetic models. The ACP${}^2$ model has a second-order transition which belongs to the $O(8)$ universality class (and so is different from CP${}^2$). For still higher $n$ the transition is first order.} On the other hand, a new class is observed for the antiferromagnetic RP${}^4$ model \cite{Pelissetto:2017pxb}, and it could constitute a target for the bootstrap.\footnote{The RP${}^n$ models are versions of $O(n)$ models with a gauged $\bZ_2$ symmetry. Their second-order phase transitions for $n=2,3$ belong to the $O(2)$ and $O(5)$ classes respectively, but the $n=4$ class is mysterious.}

\subsubsection{Nonabelian gauge and Chern-Simons matter theories}

While we have focused our attention on QED${}_3$, there is a whole landscape of 3d gauge theories coupled to various types of matter. An interesting case is QCD${}_3$ with a simple gauge group $G$ coupled to $N_f$ fundamental fermions. Such theories may for example play a role in the physics of cuprate superconductors~\cite{Chowdhury:2014efa,Chowdhury:2014jya}. A fixed point can be established and the properties studied at large $N_f$~\cite{Appelquist:1989tc}. For example, in \textcite{Dyer:2013fja} a systematic study of monopole operators in such theories was performed, allowing for estimates of the bottom of the conformal window for different choices of $G$ by imposing irrelevance of the monopole operators~\cite[{Table 4}]{Dyer:2013fja}. QCD${}_3$ coupled to both fermions and scalars was also proposed to describe the critical point of the `orthogonal semi-metal' (OSM) confinement transition in \textcite{2018arXiv180401095G}, with critical exponents extracted in quantum Monte Carlo simulations. It would be very interesting to understand how to isolate these theories using bootstrap techniques and test these estimates. 

Another natural set of targets consists of Chern-Simons gauge fields coupled to matter. Such theories are known to have conformal fixed points and sit in an intricate web of dualities.\footnote{See for example \textcite{Aharony:2015mjs}, \textcite{Aharony:2016jvv}, \textcite{Seiberg:2016gmd}, \textcite{Hsin:2016blu}, \textcite{Benini:2017dus}, and \textcite{Gomis:2017ixy}. Duality means that two different microscopic descriptions lead to the same IR CFT (perhaps after tuning some parameters). Why should dualities exist? One reason may be the paucity of CFTs. If so, some dualities may perhaps be explained by the bootstrap, providing evidence that there is a single CFT satisfying certain constraints (symmetry, the number of relevant operators, etc). Then any microscopic theory satisfying these constraints should flow to this CFT at criticality. In this sense, the results of Section \ref{sec:O2} provide an explanation for the particle-vortex duality of the Abelian Higgs model, originally proposed by \textcite{Peskin:1977kp} and \textcite{Dasgupta:1981zz}.} Some possible experimental realizations of these theories as transitions between fractional quantum Hall states were proposed in \textcite{Lee:2018udi}. A hallmark of these theories is the existence of parity-violation; it would be interesting to see if they can be found after introducing parity-violating couplings into the bootstrap. Monopole operators in these theories were also recently studied in~\textcite{Chester:2017vdh} and would constitute natural targets for the bootstrap.

\subsubsection{Other models}

Another theory briefly mentioned in Sec.~\ref{sec:O3} is the Gross-Neveu-Heisenberg (GNH) model, a variant of the GN models with a 3-component scalar order parameter. For a pedagogical review of the model, its applications, and its connection to the lattice Hubbard model, see \textcite{Sachdev:2010uz}. This constitutes another interesting target for the bootstrap.

A 3d CFT with $SU(4)$ global symmetry and an order parameter in the symmetric tensor representation was considered in \textcite{Basile:2004wa,Basile:2005hw}. It was proposed to describe a continuous chiral phase transition in 4d $SU(N)$ gauge theory coupled to $N_f=2$ massless quarks in the adjoint representation at finite temperature. The existence of this CFT and some information about critical exponents was found using RG methods; it would be interesting to explore it using the bootstrap.

\section{Applications in $d=4$}
\label{sec:appl4d}

In this section we now turn to numerical bootstrap applications in unitary 4d CFTs. We first present general constraints in Sec.~\ref{sec:4D-general}, and then discuss more specific physical applications. In Sec.~\ref{sec:4D-bsm} we review applications to high energy physics beyond the Standard Model. Applications to 4d conformal gauge theories will be discussed in Sec.~\ref{sec:4Dwindow}. Supersymmetric 4d CFTs will be presented in the next section. 

\subsection{General results}
\label{sec:4D-general}

This section will follow the same logic as Secs.~\ref{sec:Z2-general} and~\ref{sec:ON-general} devoted to 3d CFTs. Historically however the very first attempt to study crossing relations using numerical techniques focused on 4d CFTs. This analysis, pioneered in \textcite{Rattazzi:2008pe} and then refined in \textcite{Rychkov:2009ij}, was spurred by high energy physics motivations which will be reviewed in Sec.~\ref{sec:4D-bsm}. But first let us discuss general conformal bootstrap results for 4d CFTs with various global symmetries.

\begin{figure}[t!]
    \centering
\includegraphics[width=\figwidth]{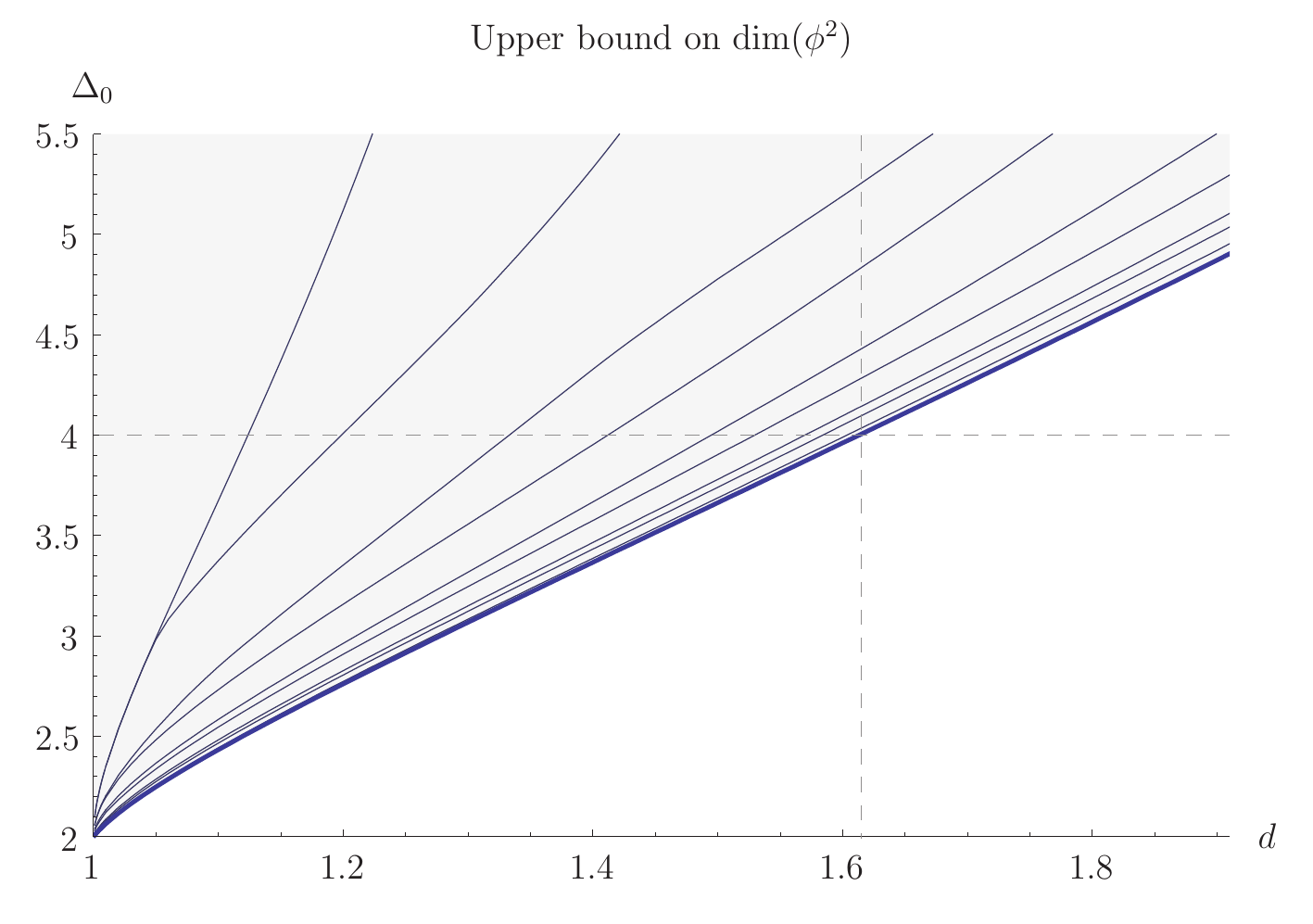}(a)
\includegraphics[width=\figwidth]{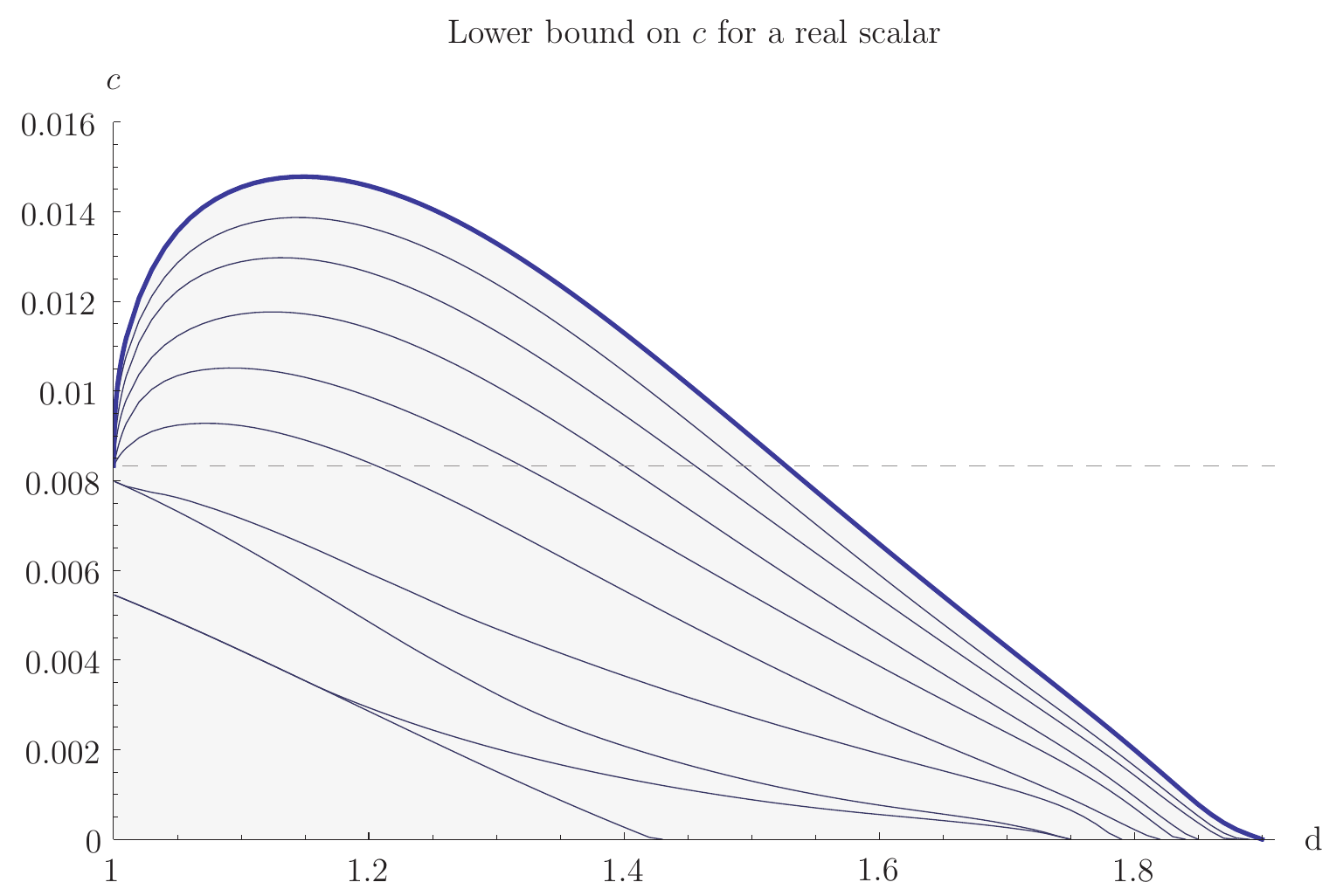}(b)
    \caption{\label{fig:4D-singlecorr} (Color online) (a) Upper bound on the dimension of the first scalar in the $\phi\times \phi$ OPE as a function of $\Delta_\phi$ in 4d unitary CFTs; (b) lower bound on the central charge $C_T$, computed by maximizing the OPE coefficient $\lambda_{\phi\phi T}$~\cite{Poland:2011ey}.
}
  \end{figure}

Consider first the simple case of a 4d CFT containing a scalar operator $\phi$ with dimension $\Delta_\phi$. We further assume that it is charged under a global symmetry (e.g., a $\mathbb{Z}_2$ symmetry) so that the OPE $\phi\times\phi$ does not contain $\phi$. Then it is interesting to ask how high can one push the dimension of the first scalar operator in this OPE. It is also interesting to ask how large the OPE coefficient of the stress tensor $\lambda_{\phi\phi T}\propto \Delta_\phi/\sqrt{C_T}$ is allowed to be~\cite{Poland:2010wg,Rattazzi:2010gj}, which translates into a lower bound on the central charge $C_T$. The best bounds to date were computed in \textcite{Poland:2011ey} and are shown in Fig.~\ref{fig:4D-singlecorr}.\footnote{The bound on $C_T$ can be somewhat strengthened by incorporating the assumption that $\phi$ is the lowest dimension scalar, as in \textcite{Rattazzi:2010gj}.} When $\Delta_\phi$ approaches the unitarity bound, both bounds approach the free theory value for $\Delta_{\phi^2}$ and $C_T$. This is consistent with the fact that a scalar with dimension $(d-2)/2$ satisfies $\del^2\phi=0$ whenever inserted in a correlation function, and must therefore be a free scalar.

\begin{figure}[t!]
    \centering
    \includegraphics[width=\figwidth]{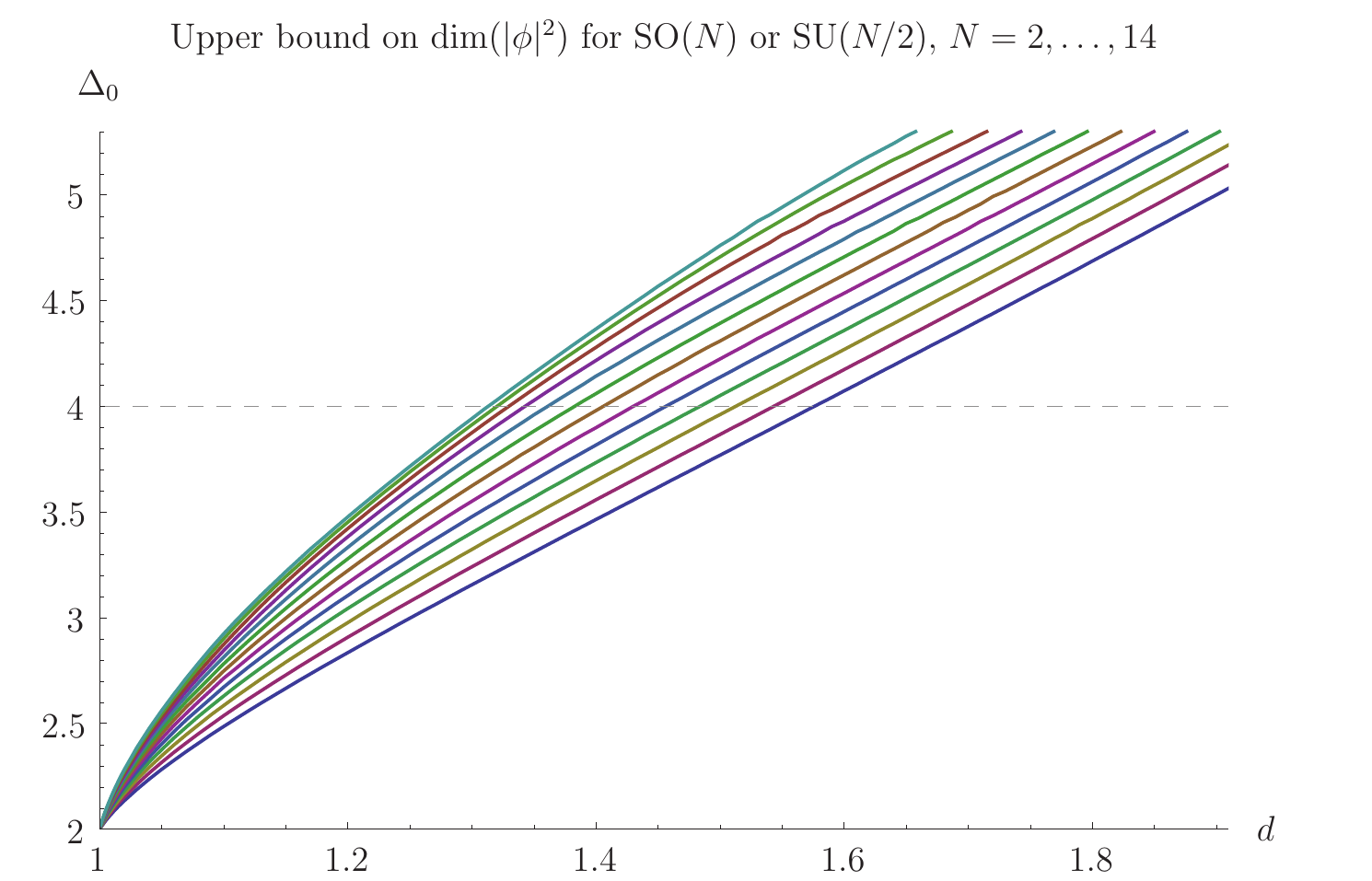}(a)
\includegraphics[width=\figwidth]{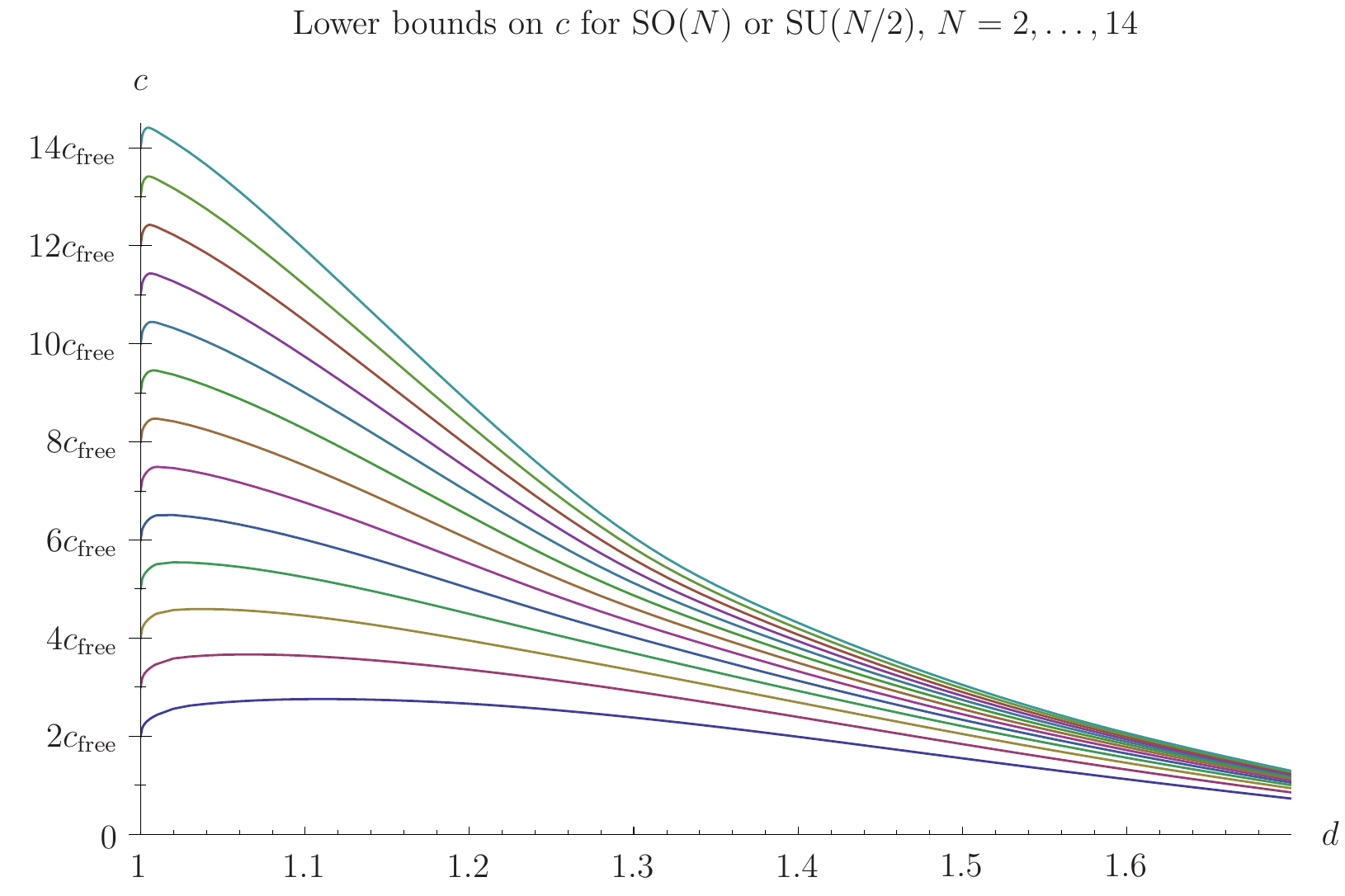}(b)
    \caption{\label{fig:4D-SON} (Color online) (a) Upper bounds on the singlet scalar dimension in $SO(N)$ and $SU(N)$ symmetric 4d CFTs, as a function of $\Delta_\phi$ in the fundamental; (b) lower bounds on $C_T$ in the same theories \cite{Poland:2011ey}.}
  \end{figure}

Analogous bounds have been obtained for CFTs assuming various continuous global symmetries. \textcite{Poland:2011ey} studied the 4pt functions of a scalar $\phi$ transforming in the fundamental representation of $SO(N)$ or $SU(N)$, deriving an upper bound on the dimension of the lowest singlet scalar in the OPEs {$\phi_i \times \phi_j$} (or  {$\phi^{\dagger i}\times \phi_j$} in the case of $SU(N)$), as well as a lower bound on the central charge, shown in Fig.~\ref{fig:4D-SON}.\footnote{Numerics indicate that $SU(N)$ and $SO(2N)$ singlet and central charge bounds coincide~\cite{Poland:2011ey}. A priori, because $SU(N)\subset SO(2N)$, and because only singlets give rise to singlets when representations are reduced, these $SO(2N)$ bounds must be at least as strong as for $SU(N)$, but the exact coincidence is unexpected and remains unexplained.} As expected, the bounds scale with $N$, the size of the fundamental representation, at least when the dimension of the external scalar approaches the free value. It should be possible to extend this analysis to obtain upper bounds on the dimensions of operators transforming in other representations. For scalars in the symmetric traceless representation of $SO(4)$ this was done in \textcite{Poland:2011ey}.

\begin{figure}[t!]
    \centering
\includegraphics[width=\figwidth]{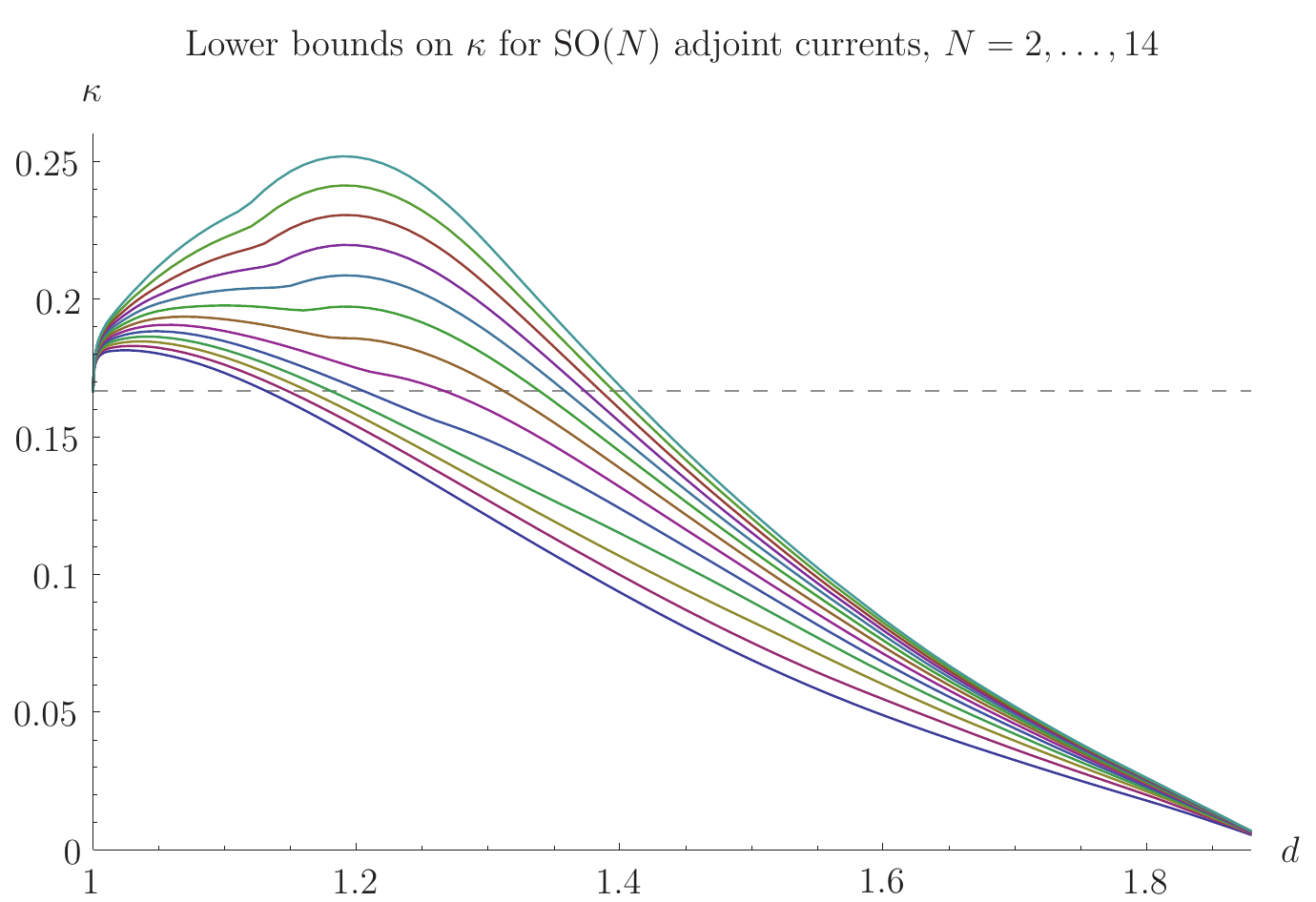}
    \caption{\label{fig:4D-CJ-SON} (Color online) Lower bound on $C_J$ in $SO(N)$-symmetric unitary 4d CFTs as a function of the dimension of a scalar in the fundamental~\cite{Poland:2011ey}. $SU(N/2)$ adjoint currents satisfy the same bound~\cite{Caracciolo:2014cxa}.}
  \end{figure}

It is also possible to place upper bounds on the OPE coefficients of conserved vectors {of dimension 3} in the OPE of $\phi$ with its conjugate. This class of operators includes the conserved currents of the considered global symmetry $G=SO(N)$ or $SU(N)$, transforming in the adjoint representation of $G$. Upper bounds on their OPE coefficients translate into the lower bounds on the central charges $C_J$. These bounds are shown in Fig.~\ref{fig:4D-CJ-SON}. Once again the $SU(N/2)$ bounds coincide with $SO(N)$ ones~\cite{Caracciolo:2014cxa}. For $\Delta_\phi$ close to the free value, these bounds smoothly approach the free $SO(N)$ value.

\begin{figure}[t!]
    \centering
\includegraphics[width=\figwidth]{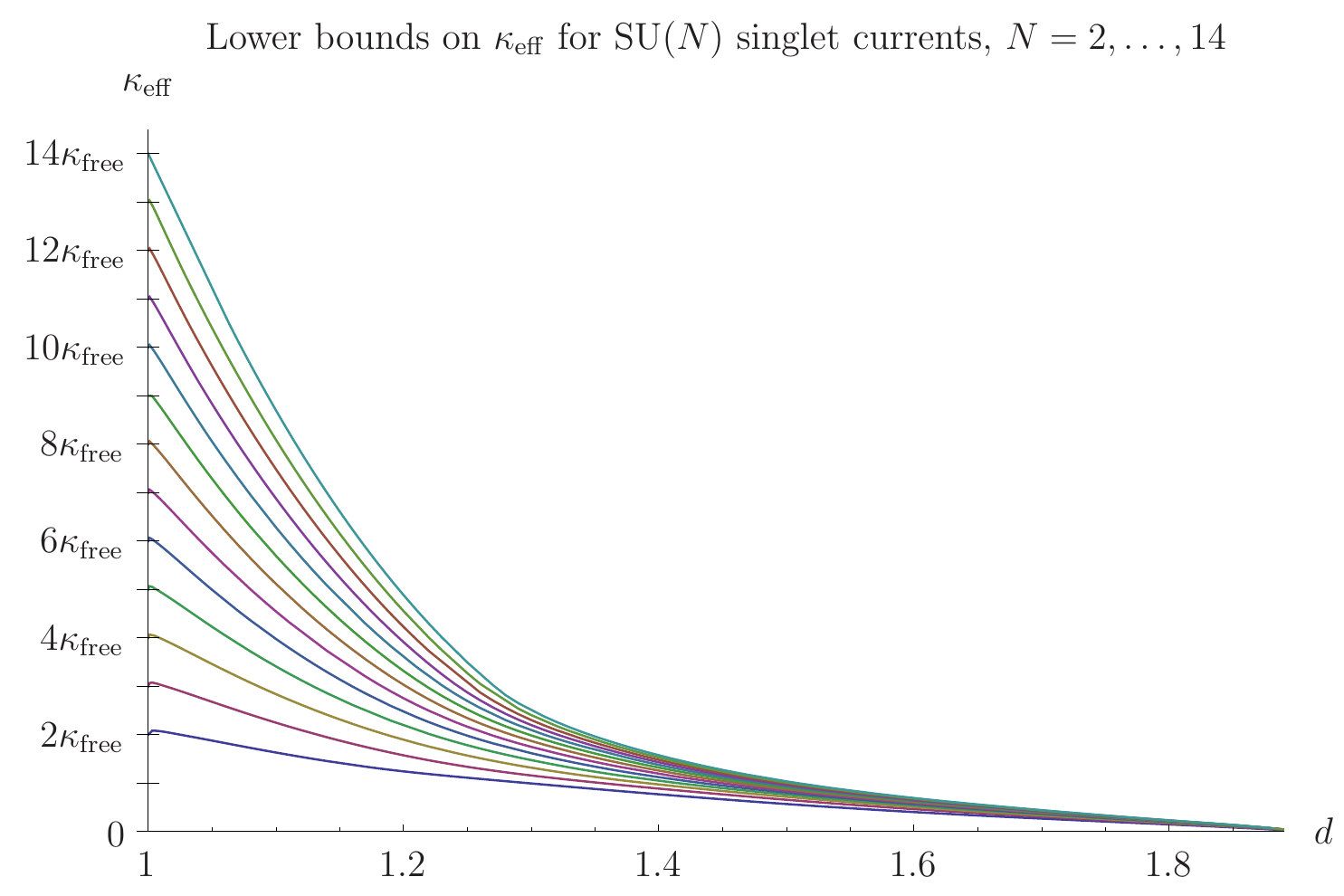}
    \caption{\label{fig:4D-CJ-singlet-SUN}
    (Color online) Lower bound on the inverse square OPE coefficient of a singlet current in $SU(N)$-symmetric unitary 4d CFTs as a function of dimension of a scalar in the fundamental~\cite{Poland:2011ey}.}
  \end{figure}

In addition, for $G=SU(N)$ the OPE {$\phi^{\dagger i}\times\phi_j$} may also contain conserved currents of some other global symmetry which may exist in the theory, which are singlets under $G$. The lower bound on their inverse-square OPE coefficient is given in Fig.~\ref{fig:4D-CJ-singlet-SUN}. Close to the free theory dimension, these bounds approach the value corresponding to the theory of $N$ massless complex scalar fields, whose full symmetry $SO(2N)$ is indeed larger that $SU(N)$. 

Additionally,~\textcite{Caracciolo:2014cxa} derived lower bounds on $C_J$ in the presence of a gap in the scalar singlet sector, as well as for extended global symmetries $SO(N)\times SO(M)$ and $SO(N)\times SU(M)$.

Unlike in 3d, most of the 4d bounds computed so far do not display any prominent kink or other dramatic feature, suggesting that  existing 4d CFTs may lie inside the allowed regions and not on the boundary. Note however that some unexplained features are visible in the $C_J$ lower bounds in Fig.~\ref{fig:4D-CJ-SON}, as well as in the bounds on supersymmetric CFTs discussed in Sec.~\ref{sec:4Dsusy}. 

The bounds discussed in this section have been obtained by studying a 4pt function {$\langle \phi_i\phi_j\phi_k\phi_l\rangle$ or $\langle \phi_i\phi^{\dagger j}\phi_k\phi^{\dagger l}\rangle$}, where {$\phi_i$} is a single primary operator or a global symmetry multiplet of primary operators. As far as we are aware, a systematic study of numerical bootstrap constraints from mixed correlators in 4d CFTs has not yet been performed outside of the supersymmetric context~\cite{Lemos:2015awa,Li:2017ddj}. It will be important to do so in the future, and to study the impact on such bounds of assuming only a limited set of relevant operators. 
  
\subsection{Applications to the hierarchy problem}
\label{sec:4D-bsm}
Next we will review some bootstrap results which shed light on the attempts to alleviate the hierarchy problem of the Standard Model (SM) of particle physics, which historically was one of the motivations for the development of the numerical bootstrap in 4d.

For the purposes of our discussion, the hierarchy problem can be briefly summarized as follows. The SM is certainly not the complete description of fundamental interactions, as it doesn't account for dark matter, baryogenesis, neutrino masses, and gravity. Instead it can be regarded as an effective description, valid at least up to the electroweak scale, where it has been extensively tested, including at the ongoing Large Hadron Collider (LHC) experiments. According to the effective field theory paradigm, the leading effects in this description are captured by the relevant and marginal operators, while all higher-dimensional operators correspond to subleading effects and are suppressed by powers of the electroweak scale ($\Lambda_{\text{IR}}\sim $ {100 GeV}) over the scale of new physics ($\Lambda_{\text{UV}}$). The incredible success of the SM in precisely describing all phenomena observed so far is elegantly explained by simply pushing the scale of new physics to high values. In particular,  electroweak precision tests and more importantly bounds from flavor physics (in particular from $K$-$\bar K$ mixing) generically require $\Lambda_{\text{UV}}\gtrsim 10^{5}\text{ TeV}$. 

This simple assumption creates however a tension (called the hierarchy problem) with the other energy scale in the theory, namely the scale associated with the only relevant operator present in the SM---the Higgs mass term $H^\dagger H$. Indeed, whenever a relevant deformation exists, it is generically expected to be generated at the fundamental scale with order one strength, unless some symmetry prevents this from happening. The contrary is usually considered an unnatural tuning of the model, similarly to how, in condensed matter systems, one typically needs to adjust a control parameter to approach a critical point.

The quest for a solution to the hierarchy problem 
has been and remains an important goal in theoretical high energy physics. Strategies for solving it can be broadly divided in two categories: the first makes use of an additional symmetry that prevents the Higgs mass term from appearing, and then slightly breaks it in order to generate a scale parametrically smaller than $\Lambda_{\text{UV}}$. The second strategy instead removes altogether the dangerous relevant deformation by increasing the scaling dimension of the Higgs mass term. An example of the first strategy is low-energy supersymmetry, while the second one is realized in technicolor, which replaces the Higgs field with a fermion bilinear operator, of scaling dimension close to three. 

While technicolor solves the hierarchy problem by making the Higgs mass term irrelevant, it also raises the SM Yukawa operator dimensions from 4 to 6. To generate heavy quark masses of needed size, these operators need to originate at an energy scale not much above $\Lambda_{\text{IR}}$. This leads to a tension with flavor observables, due to four-fermion operators expected to originate at about the same scale unless yet additional structure is added. To elegantly solve this problem, \textcite{Luty:2004ye} proposed the ``conformal technicolor" scenario, in which the Higgs field has a scaling dimension close to the free value, while the Higgs mass term is close to marginality or irrelevant. 

More precisely, to realize this scenario one would need a unitary CFT which contains a scalar operator $H$ replacing the SM Higgs field. To preserve the SM custodial symmetry, the CFT must have an $SO(4)$ global symmetry, with $H$ transforming in the fundamental. The scaling dimension requirements are as follows: $\Delta_H$ has to be close to 1, while $\Delta_{S}\gtrsim 4$, where $S$ is the first scalar $SO(4)$-singlet operator in the OPE $H^\dagger \times H$, playing the role of the Higgs mass term in this setup. Given the scaling dimension requirements, this hypothetical CFT must necessarily be strongly coupled, while its coupling to the rest of the SM (gauge fields and fermions) can be treated as a small perturbation. 

The 4d numerical bootstrap grew out from the attempts to show that the most optimistic requirements $\Delta_H\to 1$, $\Delta_S>4$ are impossible to realize. A proof of this theorem about unitary 4d CFTs is visible in the upper bound on $\Delta_S$ provided by the $N=4$ curve in Fig.~\ref{fig:4D-SON}, which approaches 2 for $\Delta_H\to 1$. 

\begin{figure}[t!]
    \centering
    \includegraphics[width=\figwidth]{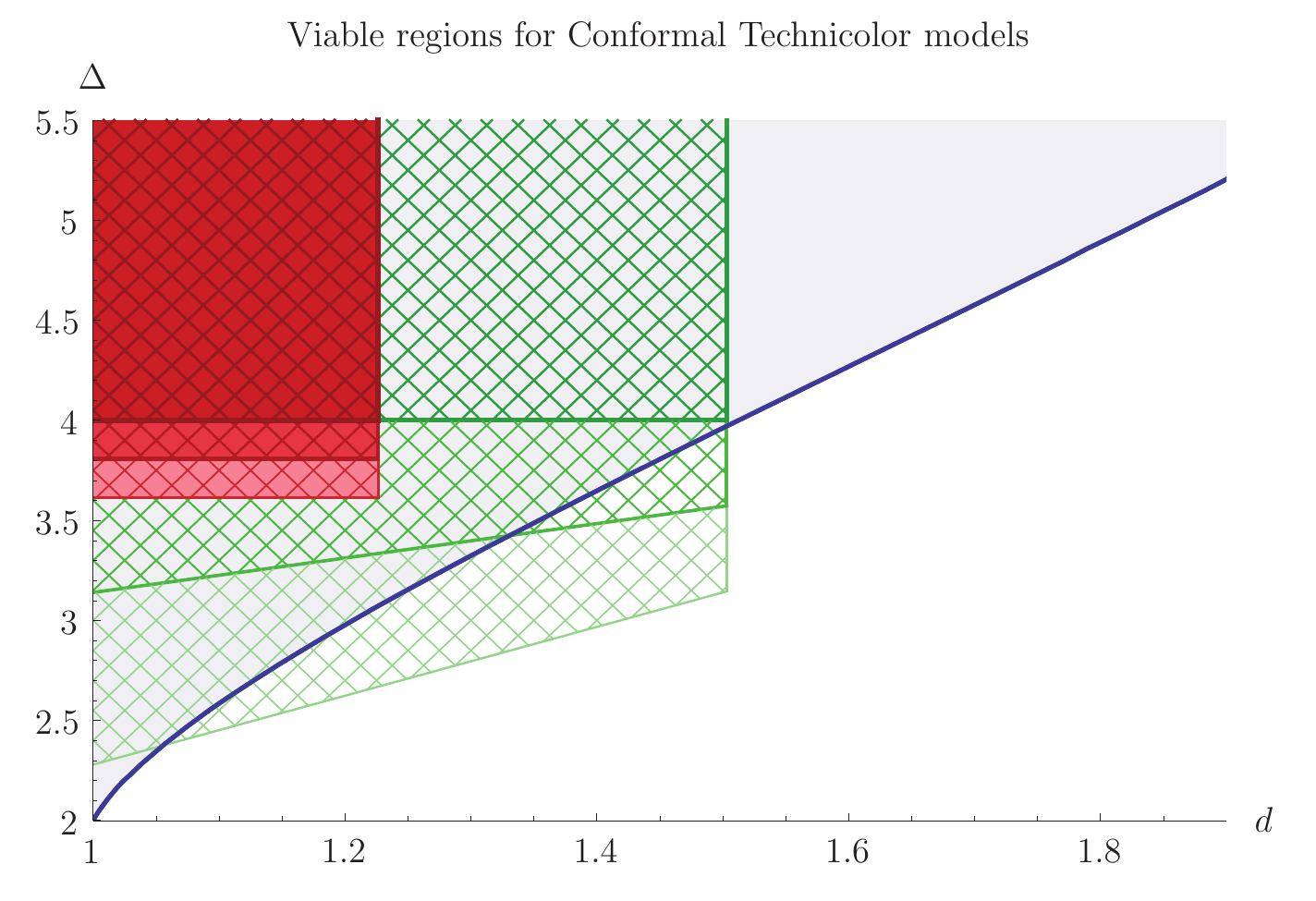}
    \caption{\label{fig:4D-ConformalTechnicolor}
    (Color online) Viable regions in the $\{\Delta_H,\Delta_S\}$ plane for conformal technicolor models in the flavor-generic (red) and flavor-optimistic (cross-hatched green) cases, superimposed with the $SO(4)$ bound. Regions for no tuning, $10\%$, and $1\%$ tuning are shown in successively lighter shades of each color, with the largest region corresponding to  $1\%$ tuning in each case. Flavor-generic models are ruled out~\cite{Poland:2011ey}.}
  \end{figure}

It is phenomenologically acceptable to have $\Delta_H$ slightly deviate from 1 without violating flavor constraints, and to allow $\Delta_S$ somewhat below 4 at the price of some moderate tuning~\cite{Luty:2004ye,Rattazzi:2008pe,Rychkov-talk}. Although this freedom helps to alleviate bootstrap constraints, some tension remains. Fig.~\ref{fig:4D-ConformalTechnicolor} from \textcite{Poland:2011ey} shows the regions of $\{\Delta_H,\Delta_S\}$ allowed under different degrees of tuning and different assumptions about the structure of the flavor sector. The conclusion is that compatibility with the bootstrap bound can be achieved only under optimistic flavor assumptions and with a moderate tuning.

An additional phenomenological constraint on conformal technicolor comes from the existence of the Higgs boson particle. While a SM-like Higgs boson may appear in conformal technicolor as a resonance of the strong dynamics at the electroweak scale associated with breaking of conformal invariance~\cite{Luty:2004ye}, it is expected to be somewhat heavier than the experimentally observed value 125 GeV, and to have some deviations in its coupling to the top quark, which were not seen so far. This further reduces the likelihood that the conformal technicolor scenario is realized in nature. Still, the above analysis, performed prior to the Higgs boson discovery, remains a beautiful example of how theoretical investigations can lead to first-principles constraints on strongly coupled scenarios for particle physics beyond the SM.

\subsection{Constraints on the QCD${}_4$ conformal window}
\label{sec:4Dwindow}

Perhaps the most famous class of unitary 4d CFTs are the IR fixed points of asymptotically free nonabelian gauge theories coupled to massless fermions, often referred to as Banks-Zaks fixed points~\cite{Banks:1981nn} though they were first considered by~\textcite{Caswell:1974gg} and~\textcite{Belavin:1974gu}. Depending on the number of fermion representations $N$, this IR conformal behavior is realized in an interval of $N$ called the ``conformal window". These CFTs are of great interest theoretically, and historically have also been discussed because of their relation to walking technicolor models of electroweak symmetry breaking.\footnote{Walking behavior is expected to be realized for $N$ just below the lower end of the conformal window~\cite{Kaplan:2009kr}, but detailed discussion of this physics is beyond our scope. {See~\textcite{Gorbenko:2018ncu,Gorbenko:2018dtm} for a recent CFT perspective and~\textcite{Appelquist:2017wcg,Appelquist:2017vyy} for a recent lattice perspective using effective field theory.}} Close to the upper end of the conformal window these theories can be studied perturbatively, see e.g. \textcite{Ryttov:2017kmx}. They have also been studied actively using lattice Monte Carlo techniques; see \textcite{Nogradi:2016qek,Svetitsky:2017xqk} for recent reviews. 

Here we will describe what the bootstrap so far has to say about these CFTs. For concreteness we will {give our discussion} for a QCD-like theory with $N$ massless Dirac fermions in the fundamental representation ($\Box$) of an $SU(N_c)$ gauge group, with $N_c\ge 3$, {though the results will also apply to the case $N_c = 2$.} The global symmetry $G$ of this theory is
\beq
G = U(1)_V \times H\,,\qquad H=SU(N)_L \times SU(N)_R\,, 
\eeq
where $H$ rotates left and right Weyl components $\psi_L$ and $\psi_R$ of the fermions separately, while the vectorial $U(1)_V$ rotates them simultaneously; its axial counterpart is instead anomalous. The theory also preserves $P$ and $C$, which interchange left and right fermions.

If an IR fixed point is reached, the above global symmetry remains unbroken. This implies that all operators of the would-be CFT must organize in irreducible representation of $G$.  We will be interested in particular in gauge-invariant scalar operators (``mesons") which are fermion bilinears:
\begin{equation}
\label{eq:fermionbilinears}
\Phi^{\bar k}_i = \bar{\psi}_L^{\bar{k}} \psi_{Ri},
\end{equation}
which transform in $\overline{\Box}\times\Box$ under $H$. The mesons are not charged under the $U(1)_V$, which will play no role below. Parity maps $\Phi$ into its complex conjugate $\bar\Phi$. The scaling dimension of $\Phi$ is an interesting observable, often expressed in terms of the anomalous dimension $\gamma_\Phi=3-\Delta_\Phi$. See \textcite{Giedt:2015alr} for a review of lattice measurements of $\gamma_\Phi$. The bootstrap will give lower bounds on $\Delta_\Phi$, translating into upper bounds on $\gamma_\Phi$.

\begin{figure}[t!]
    \includegraphics[width=\figwidth]{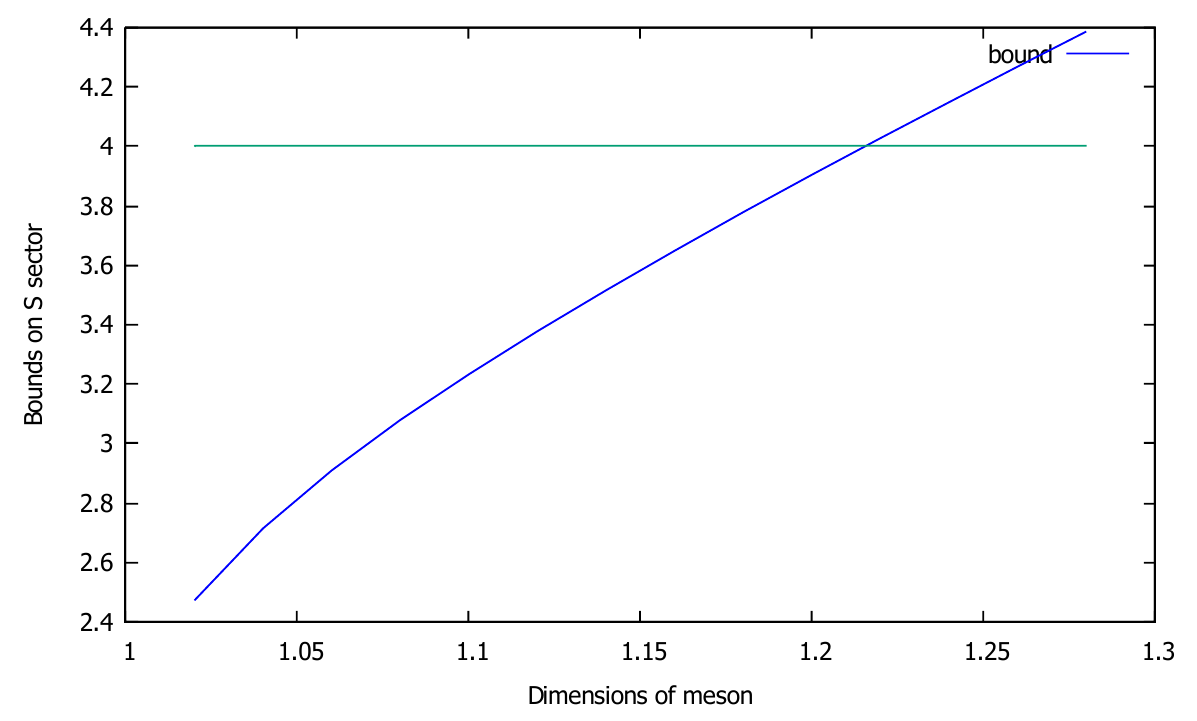}
    \caption{\label{fig:SSbound}
   (Color online) Upper bound on the dimension of the first singlet operator appearing in the OPE $\Phi\times\bar\Phi$ for $N=8$, as a function of $\Delta_\Phi$ \cite{Nakayama:2016knq}.}
  \end{figure}

\textcite{Nakayama:2016knq} carried out a bootstrap analysis of the 4pt function $\langle\Phi \Phi \bar\Phi \bar\Phi\rangle$ using the global symmetry $H$.\footnote{In his notation $\Phi$ was a bifundamental of $H$ under a different (but equivalent) convention for the transformation of left-handed fermions.} Of particular interest is an upper bound on the dimension of the lowest scalar in the OPE $\Phi\times\bar\Phi$ which is a singlet under $H$, shown in Fig.~\ref{fig:SSbound} for $N=8$. Such scalars are parity even, with an example being ${\rm Tr}[F_{\mu\nu}F^{\mu\nu}]$, where $F_{\mu\nu}$ is the Yang-Mills field strength.\footnote{On the other hand, the instanton density operator ${\rm Tr}[F_{\mu\nu}\widetilde{F}^{\mu\nu}]$ is parity odd and does not appear in the OPE $\Phi\times \bar\Phi$. We note in passing that this operator is also expected to be irrelevant, since at the fixed point it becomes a descendant of the anomalous axial current: $\partial^\mu J_\mu^A \sim {\rm Tr}[F_{\mu\nu}\widetilde{F}^{\mu\nu}]$.} A necessary condition for reaching a fixed point is that all such scalars must be irrelevant. Indeed, the Banks-Zaks fixed point is an example of a ``self-organized" CFT in the terminology of Sec.~\ref{sec:multicrit}. Using this crucial observation and the bound in Fig.~\ref{fig:SSbound}, we conclude that if $N=8$ belongs to the conformal window, then necessarily 
\begin{equation}
\label{eq:N=8}
\Delta_\Phi>1.21\quad(N=8)\,.
\end{equation}
For $N=16$ the analogous bound is $\Delta_\Phi>1.71$ \cite{Nakayama:2016knq}, and for other $N$ the bounds can be derived analogously but have not been published.

\begin{figure}[t!]
    \includegraphics[width=\figwidth]{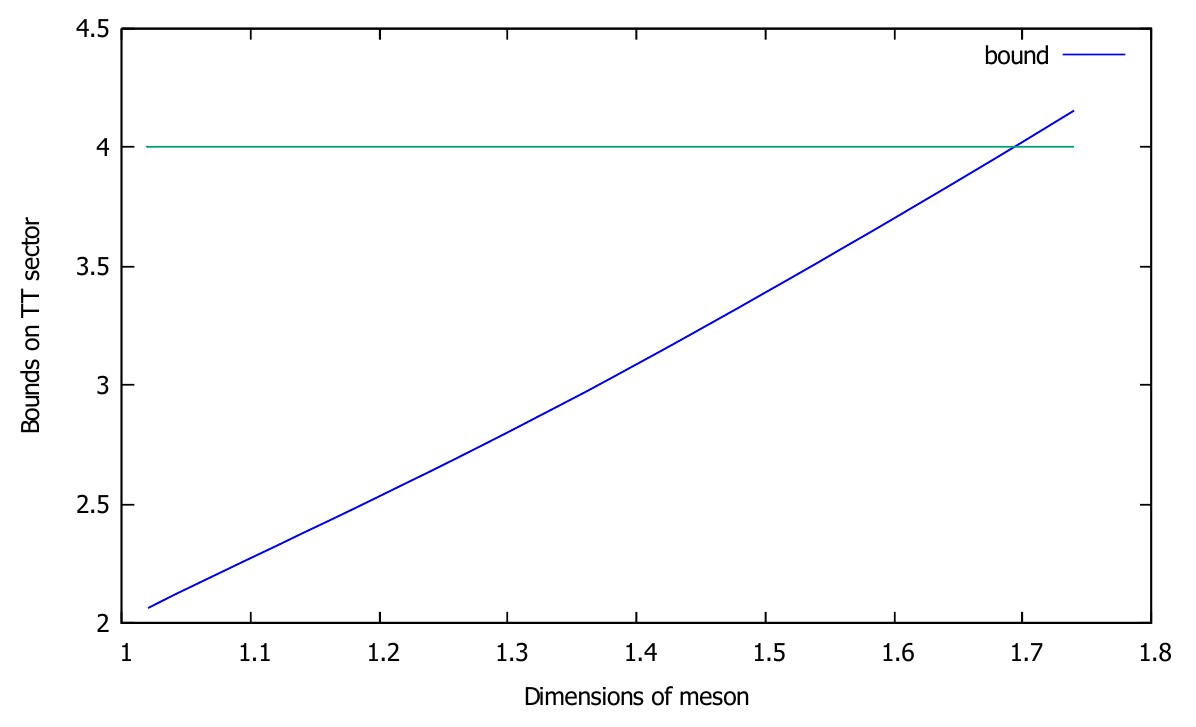}
    \includegraphics[width=\figwidth]{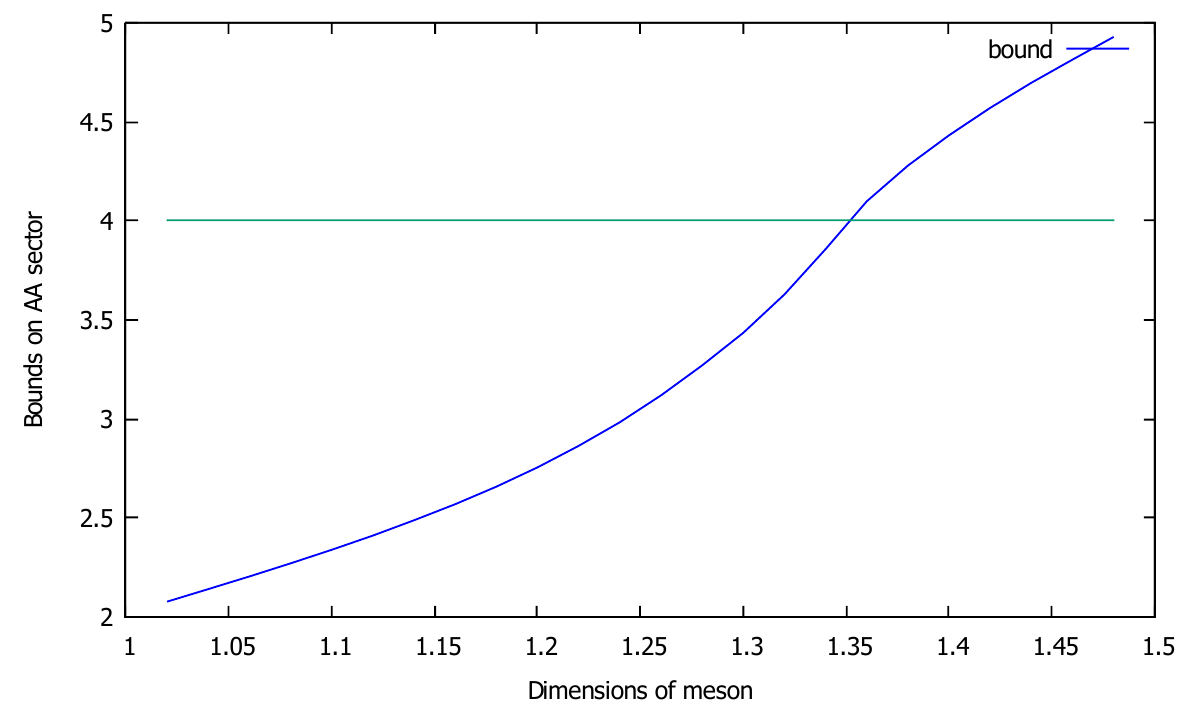}
    \caption{\label{fig:TTAAbound}
   (Color online) Upper bounds on the dimensions of the first TT and AA operators appearing in the OPE $\Phi\times\bar\Phi$ for $N=8$, as a function of $\Delta_\Phi$ \cite{Nakayama:2016knq}.}
  \end{figure}

\textcite{Nakayama:2016knq} also derived, under the same assumptions, upper bounds on the lowest scalars in $\Phi\times\bar\Phi$ which transform as $TT$ or $AA$ under $H$, where $T/A$ stands for the symmetric/antisymmetric traceless tensor representation. See Fig.~\ref{fig:TTAAbound} for the bound at $N=8$. By an idea of~\textcite{Iha:2016ppj},\footnote{See also \textcite{Hasenfratz:2017mdh,Hasenfratz:2017qyr} for a lattice perspective.} such bounds constrain possible global symmetry enhancements, in parallel with Sec.~\ref{sec:enhancement}. Namely, imagine that we are trying to reach the CFT describing the Banks-Zaks fixed point in an RG flow from a microscopic description which at short distances preserves only a subgroup $H'$ of $H$, as well as parity. For example, this would be true for lattice studies with Wilson or domain wall fermions, which only realize the diagonal subgroup 
\beq
H'_{\rm Wilson} = SU(N)_V\,,
\label{eq:Wilson}
\eeq
as in \textcite{Ishikawa:2013tua,Ishikawa:2015iwa}. {A second} example is furnished by staggered fermions, another lattice fermion realization commonly used to study the QCD${}_4$ conformal window. Being defined for $N$ a multiple of 4, these preserve microscopically a subgroup 
\beq
H'_{\rm staggered} = SU(N/4)\times SU(N/4)\,.
\label{eq:staggered}
\eeq

Observing the Banks-Zaks fixed point in a lattice Monte Carlo simulation using a fermion realization with a reduced symmetry implies a symmetry enhancement. For this to be possible, all parity-even scalar operators which are singlets under $H'$ (but not necessarily under $H$) must be irrelevant. Since the $TT$ and $AA$ representations of $H$ contain a singlet when reduced under either Eqs.~\reef{eq:Wilson} or~\reef{eq:staggered}, we obtain a necessary condition that the $TT$ and $AA$ scalars must be irrelevant in both cases. So, from the $TT$ bound in Fig.~\ref{fig:TTAAbound}, enhancement from $H'_{\rm Wilson}$ or $H'_{\rm staggered}$ requires
\begin{gather}
\label{eq:N=8enh}
\text{Wilson or staggered}\ \Rightarrow\ \Delta_\Phi>1.69\ (N=8)\,.
\end{gather}

Compared to \textcite{Nakayama:2016knq}, the earlier analysis of symmetry enhancement bounds for lattice QCD by \textcite{Iha:2016ppj} did not use the full symmetry $H$ but only the vectorial subgroup $SU(N)_V$, grouping the operators (\ref{eq:fermionbilinears}) into the irreducible representations of $SU(N)_V\times P$,
\begin{eqnarray}
\label{eq:qcdoperators}
& S =  \sum_{j=1}^N  \bar{\psi}^{\bar{j}}\psi_j \,,\quad \quad & S^{\bar{k}}_i = \bar{\psi}^{\bar{k}}\psi_i - \frac1N \delta^{\bar{k}}_i  S\,,\\
& \phi =  \sum_{j=1}^N  \bar{\psi}^{\bar{j}}\gamma_5\psi_j\,, \quad \quad  & \phi^{\bar{k}}_i = \bar{\psi}^{\bar{k}}\gamma_5 \psi_i - \frac1N \delta^{\bar{k}}_i  \phi \,,\nonumber
\end{eqnarray}
which transform as singlets or adjoints under $SU(N)_V$, and have $P=\pm$. This simplifies the analysis, as the system of crossing relations for representations of $SU(N)_V$ is easier than for the full $H$. Of course, by not using the full symmetry one loses information, although this can be partially remedied by imposing by hand the constraint that all operators in Eq.~\reef{eq:qcdoperators} have the same dimension.  

In this setup \textcite{Iha:2016ppj} could extract bounds from the symmetry enhancement for staggered fermions (but not for Wilson fermions). To do this they analyzed the 4pt function of the adjoint pseudoscalar $\phi^{\bar{k}}_i$ for $N=8,12,16$. They derived upper bounds on the lowest dimension scalar in the $\phi\times\phi$ OPE transforming in the representation $[N-1,N-1,1,1]$\footnote{The notation gives a list of the number of boxes in each successive column of the Young tableau.} of $SU(N)_V$. They argued that symmetry enhancement requires that this scalar be irrelevant. 

To illustrate this more clearly, using results of~\textcite{Lee:1999zxa} they identified explicitly some four-fermion operators which are singlets under \reef{eq:staggered} and which have nonzero overlap with $[N-1,N-1,1,1]$ when the symmetry is reduced to $SU(N)_V$. The resulting necessary condition for staggered fermion enhancement is:
\begin{eqnarray}
\label{eq:gamma}
\text{Staggered}\ \Rightarrow\ \Delta_\Phi >1.67, 1.71, 1.71\ (N=8,12,16)\,.\ \qquad
\end{eqnarray}
That for $N=8$ this bound is somewhat weaker than Eq.~\reef{eq:N=8enh} is explained partly by not using the full symmetry, and partly by working at a lower derivative order.

It should be mentioned that most studies of the QCD conformal window by lattice Monte Carlo or RG methods point to rather small anomalous dimensions $\gamma_\Phi$. Hence the above bootstrap bounds on $\Delta_\Phi$ are probably not optimal. Finding better bootstrap constraints on the QCD conformal window is an interesting open problem. Some possibilities to make further progress are to pursue mixed correlator studies, to include global symmetry currents and the stress-tensor in the bootstrap (whose 3pt functions contain known anomaly coefficients), and/or to study baryon operators.

\section{Applications to superconformal theories}
\label{sec:4Dsusy}

Conformal symmetry admits a supersymmetric extension into a superalgebra containing the standard anticommuting supercharges $\{\mathcal{Q}, \bar{\mathcal{Q}}\} \sim P$ as well as $R$-symmetry generators and the anticommuting analog of SCTs $\{\mathcal{S}, \bar{\mathcal{S}}\} \sim K$ (called special superconformal transformations). Superconformal field theories (SCFTs) are extraordinarily rich, and have been studied intensively from many viewpoints. In the last decades the zoo of known theories has grown in size: new constructions have been made, including many Lagrangian models, but remarkably also many theories that appear to admit no such description have been found. Also, a great number of strongly coupled SCFTs are known to exist, in different dimensions and with different number of supercharges, some of which can be understood using field theory or holographic dualities. 

SCFTs represent an important playground to test our understanding of CFTs and the effectiveness of bootstrap techniques. The presence of supersymmetry allows one to exactly compute some interesting CFT data, even in a strongly interacting regime, which in turn can be compared with bootstrap predictions.

From the point of view of the conformal bootstrap, supersymmetry has essentially three consequences: 1) relating the OPE coefficients and dimensions of operators belonging to the same supersymmetric multiplet, creating superconformal blocks;  2) acting as a selection rule for the operators entering a given OPE and imposing stronger constraints from unitarity; 3) fixing the dimensions of certain short multiplets. By virtue of these constraints, crossing symmetry is expected to be more effective in SCFTs.

\subsection{Theories with 4d $\mathcal{N}=1$ supersymmetry}

Before discussing the numerical bootstrap results, let us spend a few words on the structure of representations of the superconformal algebra. For concreteness we will give this discussion for SCFTs with 4d $\mathcal{N}=1$ supersymmetry.\footnote{For similar results in other dimensions or with extended supersymmetry see \textcite{Minwalla:1997ka} and the summary of recent progress in Sec.~\ref{sec:other}. Many results described here can also be treated in a uniform way across dimensions for algebras with the same number of supercharges, see \textcite{Bobev:2015jxa}.} Superconformal primary operators (annihilated by the special superconformal generators $\cal S, \bar{\cal S}$) are labelled by four numbers $(q,\bar{q},\ell,\bar{\ell})$, where $\ell,\bar{\ell}$ are the usual Lorentz quantum numbers and $q,\bar{q}$ are related to the scaling dimension $\Delta$ and $R-$charge of the superconformal primary operator:
\begin{equation}
\Delta = q+\bar{q} \qquad R = \frac23(q-\bar{q})\,.
\end{equation}
Unitarity bounds on these operators were worked out by~\textcite{Flato:1983te} and~\textcite{Dobrev:1985qv}, taking the form
\begin{align}
&q \geq \frac12 \ell+1\,,\, \bar{q}\geq \frac12 \bar{\ell}+1 &(\ell\bar{\ell}\neq0)\,,\nonumber\\
&q \geq \frac12 \ell+1  &(\bar{q} = \bar{\ell} =0)  \,,\\
&\bar{q} \geq \frac12 \bar{\ell}+1  &(q = \ell =0)  \,.\nonumber
\end{align}
The second and third lines in the above expression identify chiral ($\Phi_{\alpha_1....\alpha_\ell}$) or antichiral ($\bar{\Phi}_{\dot{\alpha}_1....\dot{\alpha}_{\bar{\ell}}}$) operators, which are annihilated by the supercharge $\bar{\mathcal{Q}}$ or $\mathcal{Q}$, respectively.

Finally, we would like to mention a few theoretical results for superconformal blocks present in the literature, focusing on those relevant for the 4d $\mathcal{N}=1$ bootstrap. Superconformal blocks for correlation function of scalar superconformal primaries can be expressed in terms of finite linear combinations of ordinary scalar conformal blocks with suitable dimensions and spin; however, computing these coefficients can be a challenging task. The work of \textcite{Poland:2010wg,Vichi:2011ux} obtained the superconformal blocks for 4pt functions of a scalar chiral supermultiplet $\Phi$. Shortly after, \textcite{Fortin:2011nq} computed superconformal blocks for 4pt functions of the multiplet associated to global symmetry conserved currents, whose lowest component is again a scalar field.\footnote{Some incorrect coefficients and missing blocks were later pointed out in \textcite{Berkooz:2014yda,Khandker:2014mpa}.} A similar analysis applicable to 4pt functions of $R$-current multiplets (containing the stress tensor) was also recently carried out in \textcite{Manenti:2018xns}. The general approach in these works was to classify the possible 3-point functions in superspace using the formalism of \textcite{Osborn:1998qu} and then expand in the Grassmann variables $\theta_i$ to compute relations between OPE coefficients of conformal primaries. In this approach one must also carefully compute the norm of each conformal primary in the multiplet.\footnote{Such norms were worked out for general multiplets in \textcite{Li:2014gpa}.} 

The work of \textcite{Fitzpatrick:2014oza} developed alternate techniques based on either solving the super-Casimir equation or writing the blocks as superconformal integrals using a super-embedding formalism. The latter approach was employed in \textcite{Khandker:2014mpa} to find the blocks appearing in the more general correlation function $\langle\Phi_1\bar{\Phi}_2 \Psi_1\bar{\Psi}_2\rangle$, where $\Phi_i$ and $\Psi_i$ are scalar superconformal primary operators with arbitrary dimension and $R$-charge,\footnote{The first and second pair have the same conformal weights $q,\bar q$, hence the notation.} with the restriction that the exchanged operator is neutral under $R$-symmetry. This analysis was later extended in \textcite{Li:2016chh} to the more general case of four distinct scalar superconformal primary operators with arbitrary scaling dimensions and $R$-charges, with no restriction on the exchanged operators besides those imposed by superconformal symmetry. However, this analysis was missing a particular class of superconformal blocks, associated to exchanged primaries in representations of the Lorentz group with $\ell\neq\bar\ell$. In this case the corresponding superconformal primary does not enter the OPE of the external operators, but some of its superconformal descendants do. This issue was fixed in \textcite{Li:2017ddj}.

\subsubsection{Bounds without global symmetries}

Now we will summarize numerical results for correlation functions involving scalar chiral superfields. The first numerical studies, starting with \textcite{Poland:2010wg} and improved in \textcite{Vichi:2011ux} and \textcite{Poland:2011ey}, focused on 4pt functions containing a single scalar chiral supermultiplet $\langle\Phi\bar{\Phi}\Phi\bar{\Phi}\rangle$. Crossing symmetry for this correlation function involves two OPE channels. Because of the chirality conditions, the $\Phi\times\bar{\Phi}$ OPE only receives contributions from traceless symmetric tensor superconformal primaries together with their $\mathcal Q\bar{\mathcal{Q}}$ and  $\mathcal Q^2 \bar{\mathcal{Q}}^2$ superdescendants, giving rise to the superconformal blocks described above.  The $\Phi\times\Phi$ OPE on the other hand is more subtle and can receive three different contributions: 1) the chiral superfield $\Phi^2$; 2) $\bar{\mathcal{Q}}$ descendants of semi-short multiplets; 3) $\bar{\mathcal{Q}}^2$ descendants of generic (long) multiplets. As a result, this channel allows conformal blocks of even spin $\ell = \bar{\ell}$ at either the protected dimensions $\Delta=2\Delta_\Phi+\ell$ or at unprotected dimensions satisfying the unitarity bound $\Delta\geq|2\Delta_\Phi-3|+\ell+3$.

\begin{figure}[t!]
    \centering
\includegraphics[width=\figwidth]{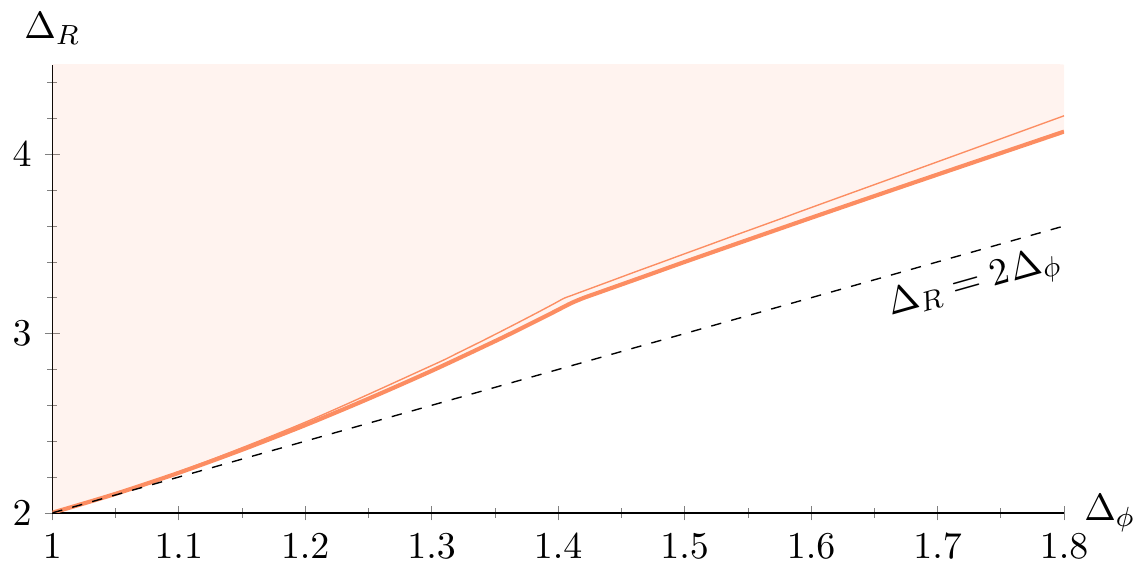} (a)
\includegraphics[width=\figwidth]{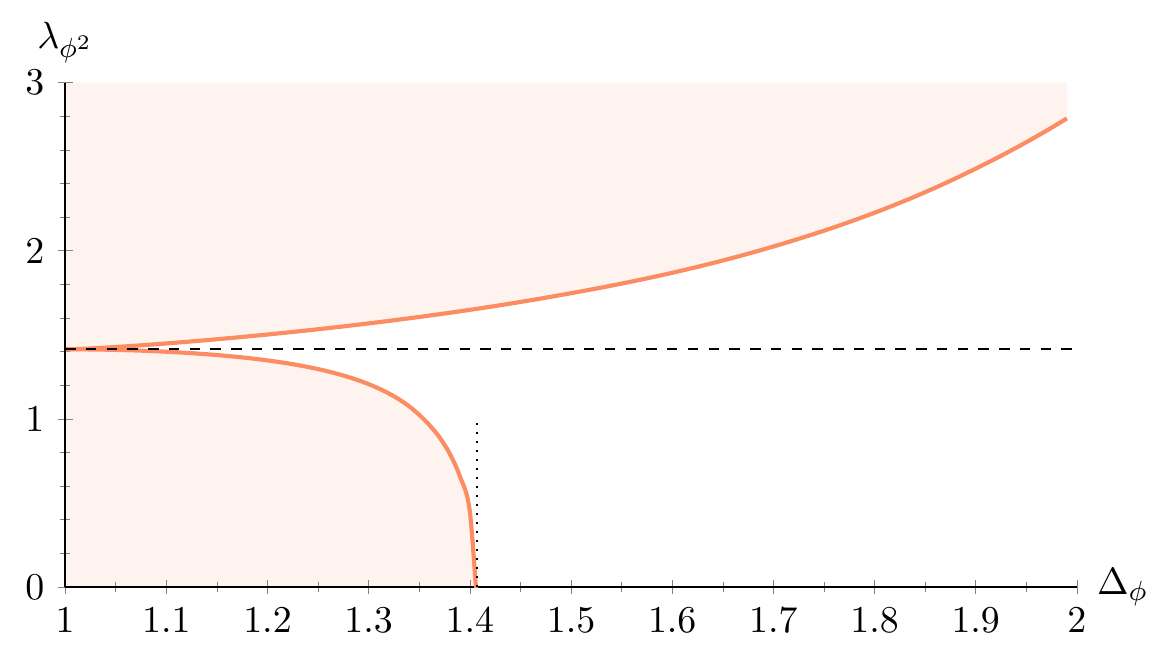} (b)
    \caption{
    \label{fig:4chiral_longscalar} (Color online) (a) Upper bound on the dimension of the operator $\mathcal R$ as a function of $\Delta_\Phi$~\cite{Poland:2011ey,Li:2017ddj}.  The shaded area is excluded. The dashed line at $\Delta_{\mathcal R}=2 \Delta_\Phi$ corresponds to generalized free theories. (b) Lower and upper bounds on the OPE coefficient of the chiral operator $\Phi^2$ entering the $\Phi\times\Phi$ OPE. The vertical dotted line is at $\Delta_\Phi = 1.407$ and the horizontal dashed line is at the free theory value $\lambda_{\Phi\Phi\Phi^2}\equiv\lambda_\phi^2=\sqrt{2}$~\cite{Poland:2015mta}.
}
  \end{figure}

  \begin{figure}[t!]
    \centering
\includegraphics[width=\figwidth]{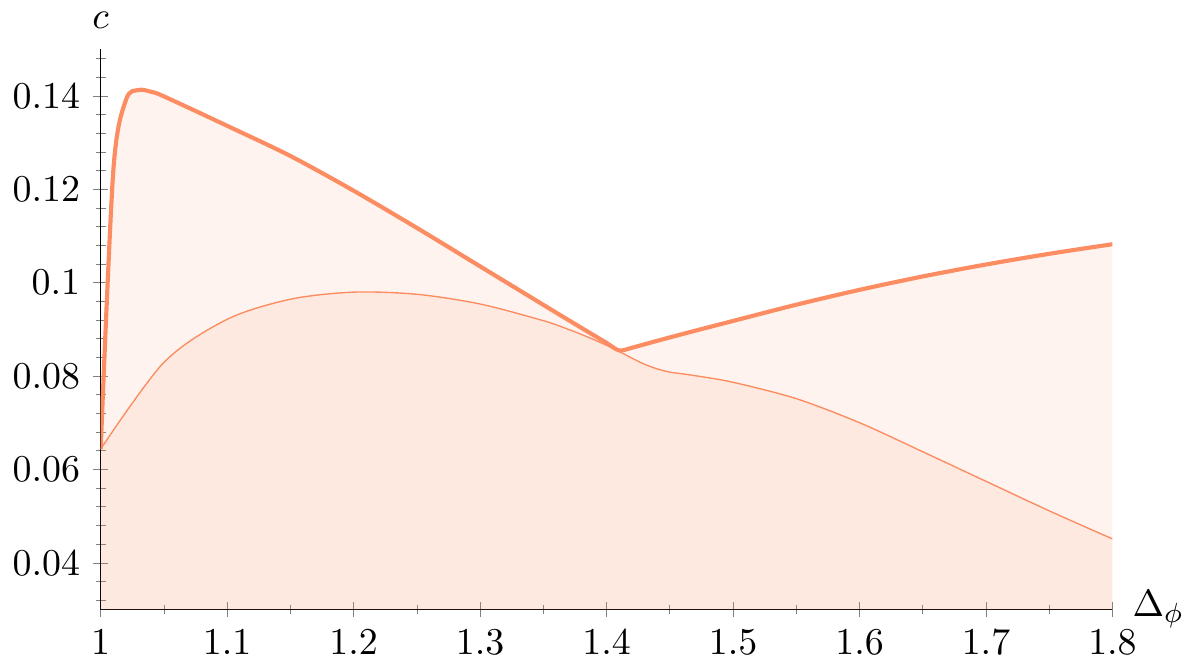} (a)
\includegraphics[width=\figwidth]{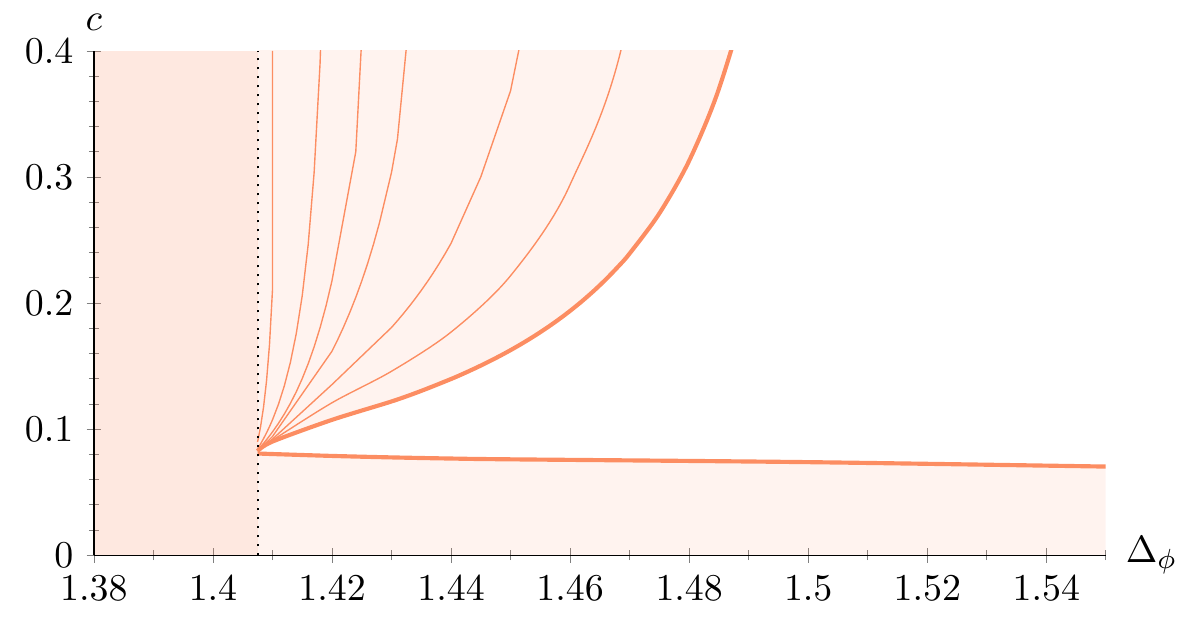} (b)
    \caption{
    \label{fig:4chiral_ct} 
    (Color online) (a) Lower bound on the central charge as a function of $\Delta_\Phi$ assuming that $\Delta_{\mathcal R}$ is consistent with the unitarity bound (thin line) or it saturates the upper bound in Fig.~\ref{fig:4chiral_longscalar} (thick line). The shaded area is excluded~\cite{Li:2017ddj}. (b) Lower and upper bounds on the central charge as a function of $\Delta_\Phi$, with the assumption that there is no $\Phi^2$ operator. The upper bounds correspond to different gaps until the second spin-1 superconformal primary $\Delta_{\ell=1}\geq  3.1, 3.3, 3.5, 3.7, 3.9, 4, 4.1$ (from left to right). The shaded area is excluded~\cite{Poland:2015mta}.
}
  \end{figure}

In Fig.~\ref{fig:4chiral_longscalar} we show an upper bound on the dimension of the first real scalar supermultiplet $\mathcal R$ entering the OPE $\Phi\times\bar{\Phi}$. A first important consequence of this result is that in any perturbative SCFT, the combination $2\Delta_\Phi -\Delta_{\mathcal R}$ must be positive (or very suppressed) to satisfy the bound. Secondly, one can observe  a minor kink-like feature on the boundary of the allowed region. The same feature appears in the lower bound on the central charge, Fig~\ref{fig:4chiral_ct}, and it also coincides with the minimal value of $\Delta_\Phi$ consistent with the absence of the chiral operator $\Phi^2$, as shown in the bottom panel of Fig~\ref{fig:4chiral_longscalar}.  
  
  In light of this, it is tempting to conjecture the existence of a ``minimal SCFT" that realizes the chiral ring relation $\Phi^2=0$ and saturates these bounds. This conjecture has been seriously addressed by~\textcite{Poland:2015mta} and \textcite{Li:2017ddj}, who studied the properties of this hypothetical theory. Notice that the minimal value of $\Delta_\Phi$ consistent with the chiral ring assumption, let us call it $\Delta_\Phi^\text{mSCFT}$, represents an extremal solution, and it is therefore uniquely determined. In addition, to coincide with the kinks it should agree with the solution obtained from maximizing the dimension of the first neutral unprotected operator and the solution obtained from minimizing the central charge, at the same value of $\Delta_\Phi$. 
  
Fig~\ref{fig:4chiral_ct} (a) shows that the two extremization procedures generically lead to two different solutions, except at $\Delta_\Phi^\text{mSCFT}$. This confirms our expectation of a unique solution coinciding with the kinks. Furthermore, by inputting a gap between the stress-tensor multiplet (whose lowest component is the spin-1 $U(1)_R$ current) and the next spin-1 supermultiplet, one is able to extract an upper bound on the central charge. While this bound is gap-dependent at generic $\Delta_\Phi$, it almost coincides with the lower bound at $\Delta_\Phi^\text{mSCFT}$, as shown in Fig~\ref{fig:4chiral_ct} (b). By extrapolating these results at large $\Lambda$,~\textcite{Poland:2015mta} obtained the prediction \mbox{$\Delta_\Phi^\text{mSCFT}  \approx 1.428,\, c \approx 0.111$}, perhaps consistent with \mbox{$\Delta_\Phi^\text{mSCFT} = 10/7,\, c = 1/9$}.

Recently a few theories have been proposed as mSCFT candidates \cite{Xie:2016hny,Buican:2016hnq}, which implement the chiral ring condition $\Phi^2=0$; however, they don't quite match the bootstrap predictions presented here. In particular the central charge is much larger than $1/9$.

It is also worth noticing that, in any solution saturating the dimension bound of Fig.~\ref{fig:4chiral_longscalar}, the chiral operator $\Phi$ is not charged under any global symmetry. If it was, in fact, the solution would contain a spin-$1$ conserved current, which in $\mathcal N=1$ SUSY happens to be the superdescendant of a dimension-2 real scalar which would appear in the $\Phi\times\bar{\Phi}$ OPE. 

To conclude this section, let us mention that the work of \textcite{Li:2017ddj} also performed a bootstrap study of a system of mixed correlators involving a scalar chiral superfield $\Phi$ and a long real scalar superfied $\mathcal R$, identified with the first scalar operator appearing in the $\Phi\times\bar{\Phi}$ OPE. Unlike in 3d, this analysis didn't seem to allow one to easily isolate a closed region. A preliminary inspection of the extremal solution doesn't reveal any obvious low-lying operators decoupling from the spectrum, but rather it involves a rearrangement of higher dimensional operators~\cite{Stergiou2017}. It will be interesting to study this rearrangement further and understand how to robustly isolate the conjectured mSCFT in future work. 
  
\subsubsection{Bounds with global symmetries}

As mentioned in the previous section, conserved currents of global symmetries $j_\mu^a $ sit in real supermultiplets $\mathcal J^a$ whose lowest component is a dimension-2 scalar $J^a$. In addition, the multiplet satisfies the conservation condition $D^2 \mathcal J^a =\bar{D}^2 \mathcal J^a =0$. Bootstrapping correlation functions of the scalars $J^a$ allows one access to the space of local SCFTs with a given global symmetry. Hence, due to supersymmetry, one can apply the same machinery encountered so far, with no need to deal with spinning conformal blocks. 

Bounds on OPE coefficients of $SU(N)$ currents were explored in \textcite{Berkooz:2014yda} and dimension bounds (and coefficient bounds assuming gaps) from single 4pt functions $\<JJJJ\>$ were explored in \textcite{Li:2017ddj}. The latter work also studied the case of mixed correlators involving $J$ and a chiral field $\Phi$ charged under the global symmetry. Note that this charge necessarily differs from the $R$-symmetry, which instead is part of the conformal algebra: the conserved current associated with the latter is the lowest component of the Ferrara-Zumino multiplet which contains the stress-tensor and supercurrents.

\begin{figure}[t!]
    \centering
\includegraphics[width=\figwidth]{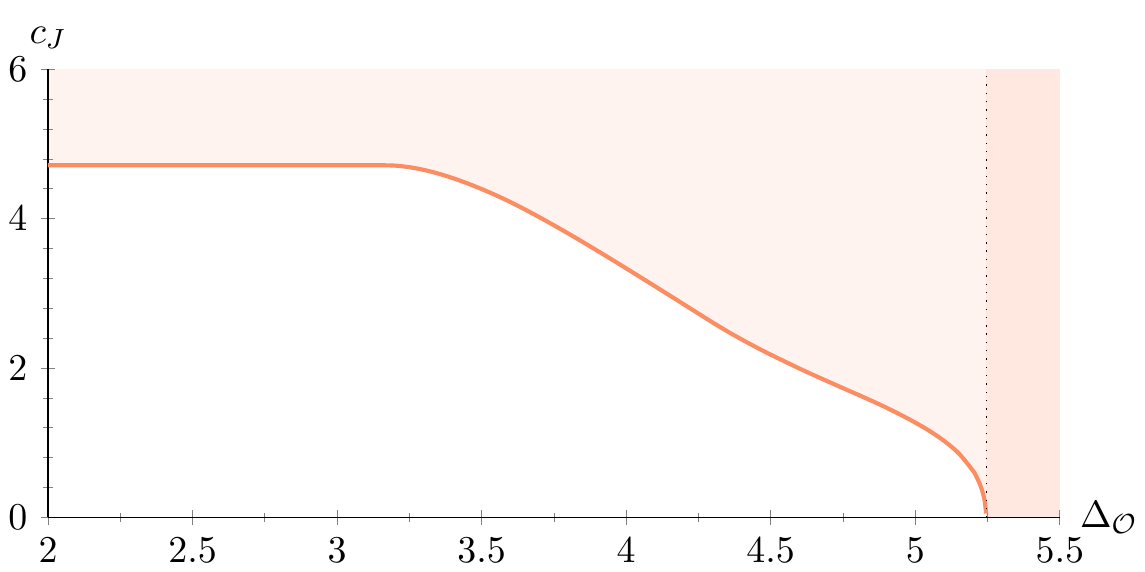} (a)
\includegraphics[width=\figwidth]{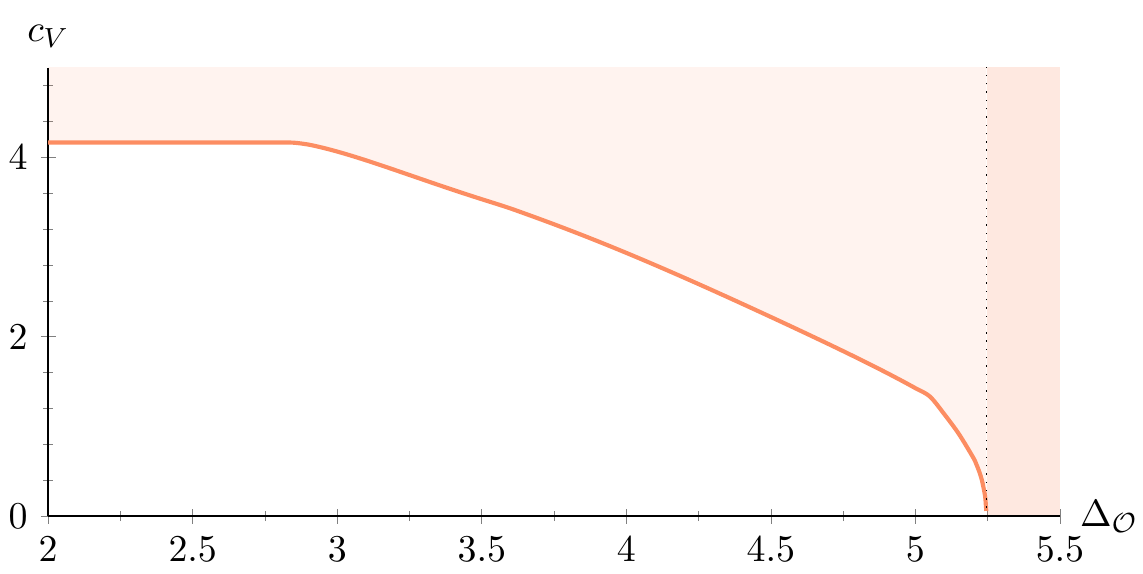} (b)
    \caption{
    \label{fig:JJJJ_ct} 
    (Color online) Upper bounds on the OPE coefficients $c_{\mathcal{O}} \equiv 2^{-\ell/2} \lambda_{JJ\mathcal{O}}$ appearing in the $J\times J$ OPE arising from (a) $J$ itself or (b) the stress-tensor supermultiplet $V$, as a function of the dimension of the first unprotected scalar $\mathcal O$ in the $J\times J$ OPE. The region to the right of the dotted vertical line at $\Delta_\mathcal{O} = 5.246$ is not allowed~\cite{Li:2017ddj}.
}
  \end{figure}

A key result, shown in Fig.~\ref{fig:JJJJ_ct}, shows that any local SCFT with a continuous global symmetry must contain a real scalar multiplet $\mathcal O$ with dimension $\Delta_\mathcal{O} \leq 5.246$.  The same figure also shows upper bounds on the OPE coefficient associated to $J$ itself as well as the one associated to the stress tensor multiplet, denoted as $V$. Interestingly, both bounds on $c_J$ and $c_V$ show plateaus for small values of the gap in the scalar sector. These are perhaps consistent with the existence of SCFT solutions shaping the bounds. On the other hand, the values extracted from Fig.~\ref{fig:JJJJ_ct} are much smaller then the limits one obtains by inspecting the correlation functions of chiral superfields. For instance, the relation between $c_V$ and Fig.~\ref{fig:4chiral_ct} is $c_V^2=1/(90c)$, making the bound on the central charge very weak.\footnote{The OPE coefficient bounds obtained in \textcite{Berkooz:2014yda} for $SU(N)$ current 4pt functions were also relatively weak.}

An alternative method to study SCFTs with global symmetries is to consider external scalar operators in nontrivial representations of the symmetry. An important target is to make contact with supersymmetric QCD theories, e.g.~supersymmetric gauge theories with gauge group $SU(N_c)$ and $N_f$ flavors of quarks \mbox{$Q_i, \overline{Q}^{\bar{j}}$}, with $N_f$ in the conformal window \mbox{$3N_c/2 \leq N_f \leq 3N_c$}~\cite{Seiberg:1994pq}. The simplest gauge-invariant operators are the mesons $M_i^{\bar j}=Q_i \overline{Q}^{\bar{j}}$, which transform as bi-fundamentals of \mbox{$SU(N_f)_L\times SU(N_f)_R$} and have dimension \mbox{$\Delta_M = 3(1-N_c/N_f)$}. Due to supersymmetry, both the central charge and current central charge can be exactly computed due to their relation to anomaly coefficients.

\begin{figure}[t!]
    \centering
\includegraphics[width=\figwidth]{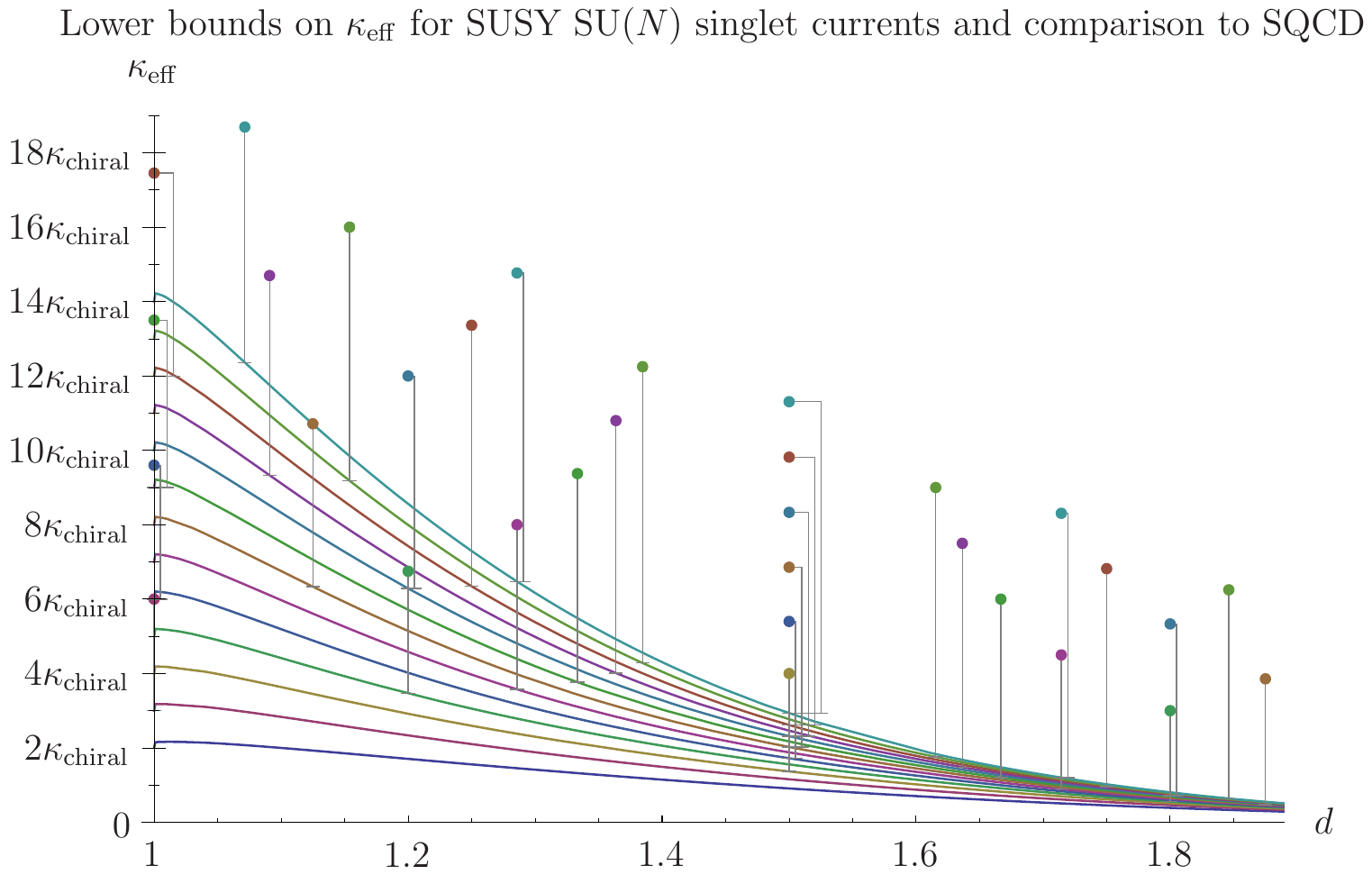}
    \caption{
    \label{fig:SQCD} 
    (Color online) Lower bounds on the effective 2pt function coefficient $\kappa_{\text{eff}} = 1/\lambda_{\Phi\Phi J}^2$ of $SU(N)$ singlet currents appearing in $\Phi \times \bar{\Phi}$, where $\Phi$ is a chiral scalar of dimension $d$ in the fundamental of $SU(N)$, for $N = 2, \ldots, 14$. The bounds are normalized to the value $\kappa_{\text{chiral}}$ corresponding to a free chiral superfield. Each dot connected to a bound corresponds to the exact value in an SQCD theory with the same symmetry~\cite{Poland:2011ey}.
}
  \end{figure}

A partial bootstrap analysis applicable to meson 4pt functions was performed in \textcite{Poland:2011ey}, which considered chiral scalar multiplets transforming in the fundamental representation of $SU(N)$ and obtained bounds on the OPE coefficients associated to conserved currents transforming in both the singlet and the adjoint representations of $SU(N)$. As can be seen in Fig.~\ref{fig:SQCD}, these bounds are still somewhat far from the exact results of supersymmetric QCD (SQCD) theories, most likely because this study didn't utilize the full symmetry. It will be very interesting in future work to extend these analyses of chiral 4pt functions to use the whole SQCD global symmetry, together with mixed correlators containing the $SU(N_f)_{L/R}$ current multiplets and/or the stress-tensor multiplet.

\subsection{Theories with 3d $\mathcal{N}=2$ supersymmetry}
\label{sec:Fermions-WZ}  

Another interesting set of targets for the conformal bootstrap are the zoo of 3d CFTs with $\mathcal{N}=1$ or $\mathcal{N}=2$ supersymmetry. We made initial contact with the former in Sec.~\ref{sec:Fermions-GN}, where there were no constraints from supersymmetry used other than relations between scaling dimensions {(see however footnote~\ref{footnote:3dN1})}. The superconformal representation theory of the latter has a similar structure to that of 4d $\mathcal{N}=1$ SCFTs; for details see \textcite{Minwalla:1997ka} and \textcite{Bobev:2015jxa}. Perhaps the simplest such theory is the $\mathcal{N}=2$ supersymmetric Wess-Zumino model described in Sec.~\ref{sec:FermionModels}. This CFT can be thought of as the IR fixed point of a theory of a single chiral superfield $\Phi = \phi + \psi \theta + F \theta^2$ and superpotential $W = \lambda \Phi^3$. The fixed point has a $U(1)_R$ symmetry under which $\Phi$ has charge 2/3, implying exact dimensions for the complex scalar $\phi$ and the Dirac fermion $\psi$: $\Delta_{\phi} = q_{\phi} = 2/3$, $\Delta_{\psi} = \Delta_{\phi} + 1/2 = 7/6$.

     \begin{figure}[t!]
    \centering
\includegraphics[width=\figwidth]{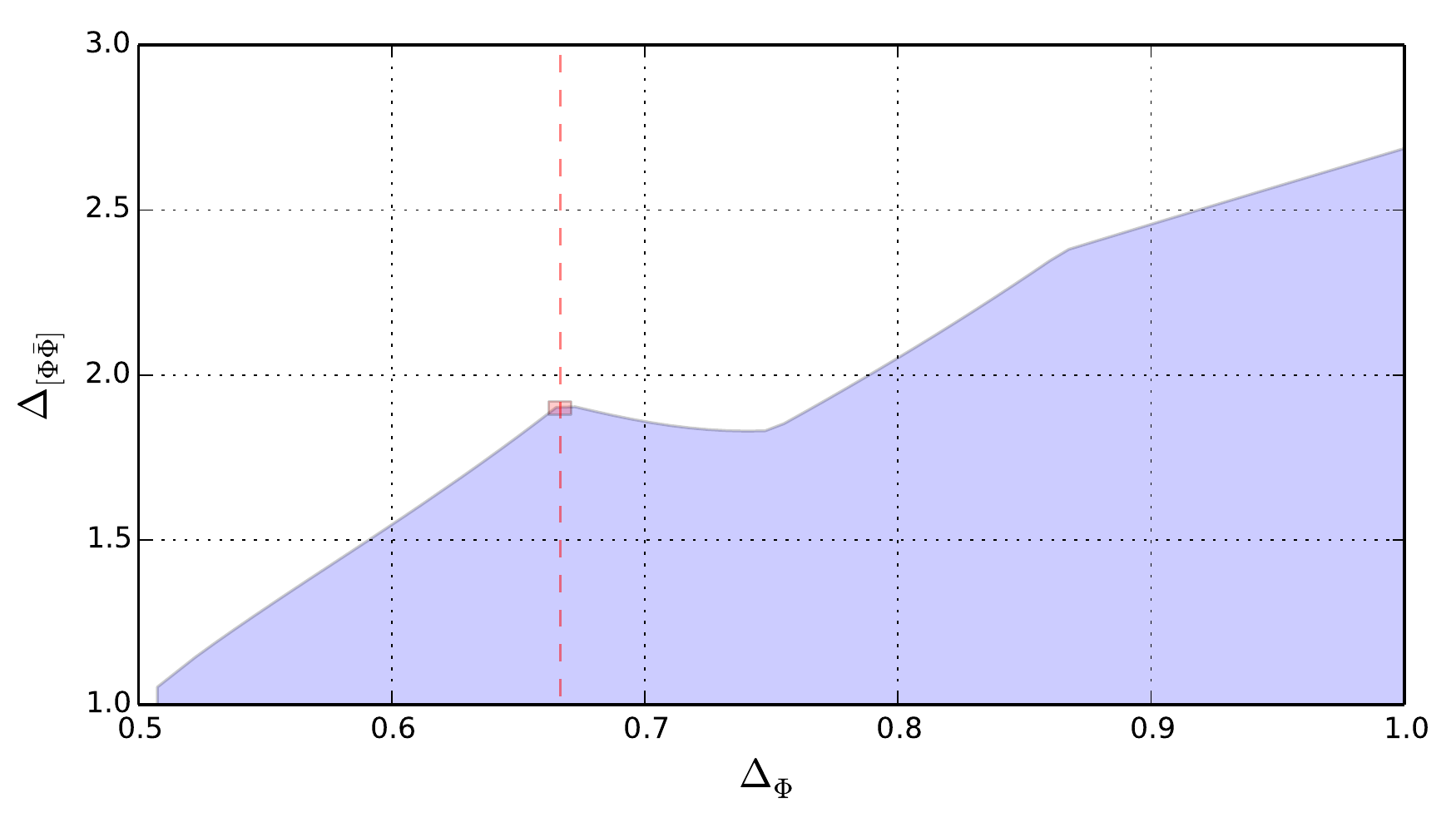}
    \caption{\label{fig:Fermions-SUSY3kinks} (Color online) Bound on the dimension of the first unprotected scalar $\Phi \bar{\Phi}$ in the $\Phi\times\bar{\Phi}$ OPE in 3d SCFTs with $\mathcal{N}=2$ supersymmetry \cite{Bobev:2015vsa}.
  }
  \end{figure}

Applying the numerical bootstrap to the 4pt function $\<\Phi \bar{\Phi} \Phi \bar{\Phi}\>$ and incorporating the unitarity bounds and superconformal blocks of $\mathcal{N}=2$ superconformal symmetry,~\textcite{Bobev:2015vsa,Bobev:2015jxa} and \textcite{Li:2017kck} studied general bounds on the dimension of the leading unprotected scalar operator $\Phi \bar{\Phi}$, with the basic result shown in Fig.~\ref{fig:Fermions-SUSY3kinks}. Curiously, the resulting bound has three distinct features, the first of which occurs at a scaling dimension $\Delta_{\Phi} \simeq 2/3$.  This gives a sharp upper bound $\Delta_{\Phi\bar{\Phi}} < 1.91$ for the $\mathcal{N}=2$ supersymmetric Wess-Zumino model and a plausible conjecture that the model saturates the optimal version of this bound. Further analysis of the extremal spectrum of this kink can be found in \textcite{Bobev:2015vsa,Bobev:2015jxa}, while \textcite{Li:2017kck} found that an isolated island around $\{\Delta_{\Phi}, \Delta_{\Phi \bar{\Phi}}\} = \{0.6678(13), 1.903(10)\}$ could be obtained by assuming a modest gap in the spectrum of spin-1 superconformal primaries $\Delta_{J'} \geq 3.5$.

The middle kink occurs near $\Delta_{\Phi} = 3/4$, and coincides with a kinematic threshold beyond which superconformal descendants of anti-chiral operators $Q^2 \bar{\Psi}$ can no longer appear in the $\Phi \times \Phi$ OPE. It is not yet clear if any CFT sits at this kink. The right-most kink, occurring near $\Delta_{\Phi} \sim .86$, also still lacks a clear interpretation, but seems to interpolate to the kink in the 4d $\mathcal{N}=1$ bounds discussed above. Notably, the extremal spectrum of this kink seems to satisfy the chiral ring relation \mbox{$\Phi^2 = 0$}~\cite{Bobev:2015jxa} and an island around the point can also be isolated using a set of gap assumptions~\cite{Li:2017kck}, making it a plausible candidate for a new CFT.  Finally let us mention that this analysis was also extended to 3d $\mathcal{N}=2$ supersymmetric CFTs with $O(N)$ global symmetry by~\textcite{Chester:2015qca,Chester:2015lej}, who found similar features at each value of $N$. A related 3d $\mathcal{N}=2$ theory with multiple interacting chiral superfields and a conformal manifold was also recently studied using bootstrap methods in~\textcite{Baggio:2017mas}.

\section{Applications to nonunitary models}
\label{sec:nonunitary}

The great majority of numerical conformal bootstrap applications considered to date have concerned unitary CFTs.
This limitation was mainly due to the fact that the main two rigorous numerical bootstrap methods, linear programming and semidefinite programming (Sec.~\ref{sec:numerical}) require positivity of the squares of OPE coefficients or positive-definiteness of the matrices made of their pairwise products, which only hold in unitary theories. Nevertheless there have been some promising attempts to apply conformal bootstrap methods to nonunitary theories, which we wish to briefly describe here.

One naturally occurring class of nonunitary CFTs are theories analytically continued from integer to noninteger space dimensions $d$, the prime example being the Wilson-Fisher family of fixed points in $2\le d<4$. It was only understood very recently that these CFTs are nonunitary for noninteger $d$.\footnote{See~\textcite{Hogervorst:2014rta,Hogervorst:2015akt} for the original observation and~\textcite{DiPietro:2017vsp} for further work.}

Fortunately, the violation of unitarity in these theories seems to be rather mild, as the negative-norm operators have rather high dimension~\cite{Hogervorst:2015akt}. So it is believed that the standard linear and semidefinite programming methods, while non-rigorous in this context, should still give reasonable results. This explains the success of~\textcite{El-Showk:2013nia} who found good agreement between the numerical bootstrap and the $\eps$-expansion in the whole range $2\le d< 4$ using the 4pt function $\<\sigma\sigma\sigma\sigma\>$ and analytically continuing conformal blocks to noninteger $d$. Similarly,~\textcite{Behan:2016dtz} generalized to noninteger $d$ the multiple-correlator analysis leading to the 3d Ising model island in Fig.~\ref{fig:Z2-mixed-differentgaps}. A related successful study by~\textcite{Bobev:2015jxa} analyzed the analytic continuation to $2\le d \le 4$ of theories with 4 supercharges, which for $d=3$ reduces to the $\calN=2$ Wess-Zumino model from Sec.~\ref{sec:Fermions-WZ}.\footnote{See also~\textcite{Chester:2014gqa,Chester:2015qca,Chester:2015lej} and~\textcite{Pang:2016xno} for related studies.} 

Leaving aside the physical interpretation, the $\bZ_2$-invariant Wilson-Fisher fixed points can actually be analytically continued even below $d=2$ and perhaps all the way to $d=1^+$~\cite{Holovatch:1993qt}. Interestingly, the analytic continuation seems to no longer be reproduced by the linear programming bootstrap in $d<2$, likely because violations of unitarity are stronger in this case than for $d\ge 2$~\cite{Golden:2014oqa}. This serves as a warning to keep in mind when applying the linear and semidefinite methods to nonunitary theories.

As described in Sec.~\ref{sec:truncation}, the truncation method should in principle be more suitable for analyzing nonunitary theories. \textcite{Gliozzi:2013ysa} and~\textcite{Gliozzi:2014jsa}\footnote{See also~\textcite{Hikami:2017hwv}. Other nonunitary models of interest to statistical physics tackled by the truncation method include the self-avoiding walk, branched polymers, random field Ising model, and percolation~\cite{Hikami:2017sbg,Hikami:2018mrf,LeClair:2018edq}. The analytic continuation of the $O(N)$ model to noninteger $N$ was also studied using the linear programming method by~\textcite{Shimada:2015gda}, but the unitarity violation effects may not be sufficiently small to allow this (see footnote~\ref{note:ONnonunitary}).} successfully applied the truncation method to the Lee-Yang CFT, which is a nonunitary CFT describing the IR fixed point of the $\phi^3$ scalar theory in $2\le d<6$ dimensions~\cite{Fisher:1978pf}. Truncating the $\phi \times \phi$ OPE to the identity operator, $\phi$ itself, the stress tensor, and two more operators, estimates of $\Delta_\phi$ were obtained in good agreement with RG and lattice predictions, and with the exact solution available for $d=2$.

Finally we mention that bootstrap methods can also be straightforwardly applied to nonunitary CFTs if a reasonable conjecture for the spectrum is available, as happens in the case of 2d percolation~\cite{Picco:2016ilr}. In this special situation, one simply solves crossing relations for the squares of OPE coefficients, with no restriction on sign.

\section{Other applications}

\label{sec:other}

We would like to finish our review by briefly describing some of the many related topics that we have not been able to cover. Many of these topics have also seen significant recent progress and would merit their own reviews. We hope that we can at a minimum give the reader some useful entry points into the literature.

A traditional approach to learning about CFTs has been to use perturbation theory, often in the context of $\epsilon$-expansions or $1/N$-expansions. There is an older literature about using bootstrap-like techniques to reproduce $1/N$ expansions, see e.g.~\textcite{Lang:1991kp,Petkou:1994ad}, involving setting up self-consistency equations using a sum over dressed Feynman diagrams sometimes called a ``skeleton expansion.'' More recently, there have been a number of recent works which use more modern analytical bootstrap techniques to study perturbative expansions, e.g. studying bootstrap equations in a $1/N$ expansion,\footnote{See~\textcite{Heemskerk:2009pn}, \textcite{Heemskerk:2010ty}, \textcite{Alday:2014tsa}, and \textcite{Aharony:2016dwx}.} using conformal invariance and the appearance of null states to reproduce $\epsilon$-expansions,\footnote{See~\textcite{Rychkov:2015naa}, \textcite{Basu:2015gpa}, \textcite{Ghosh:2015opa}, \textcite{Raju:2015fza}, \textcite{Roumpedakis:2016qcg}, \textcite{Gliozzi:2016ysv, Gliozzi:2017hni}, \textcite{Gliozzi:2017gzh}, and \textcite{Liendo:2017wsn}.} and using a formulation of the bootstrap in Mellin space\footnote{The Mellin transformation of CFT correlation functions was introduced by \textcite{Mack:2009mi}, \textcite{Penedones:2010ue}, \textcite{Paulos:2011ie}, and \textcite{Fitzpatrick:2011ia}, and developed in many subsequent works. While it is not yet known if and how it can be used for the numerical bootstrap, this formalism can be used to study many related questions. E.g.,~recently \textcite{Sleight:2017fpc, Sleight:2018epi} constructed spinning CPWs in Mellin space for external traceless symmetric tensors in general $d$ (see Sec.~\ref{sec:spinning} for the discussion in real space).} to reproduce $\epsilon$ and large-$N$ expansions by reviving an old idea by \textcite{Polyakov:1974gs} to make crossing symmetry manifest and impose unitarity.\footnote{See~\textcite{Sen:2015doa}, \textcite{Gopakumar:2016wkt,Gopakumar:2016cpb}, \textcite{Dey:2016mcs}, \textcite{Dey:2017fab}, and \textcite{Dey:2017oim}.} 

A related analytical approach has been to study bootstrap equations in various Lorentzian limits. One such limit is the lightcone limit developed in \textcite{Fitzpatrick:2012yx,Komargodski:2012ek}, which has allowed for a systematic study of CFT data in a large spin expansion\footnote{See~\textcite{Fitzpatrick:2014vua}, \textcite{Kaviraj:2015cxa,Kaviraj:2015xsa}, \textcite{Alday:2015eya}, \textcite{Alday:2015ewa}, \textcite{Li:2015rfa,Li:2015itl}, \textcite{Dey:2016zbg}, \textcite{Hofman:2016awc}, \textcite{Alday:2016mxe}, \textcite{Alday:2016njk}, \textcite{Simmons-Duffin:2016wlq}, \textcite{Dey:2017fab}, \textcite{vanLoon:2017xlq}, \textcite{Elkhidir:2017iov}, and \textcite{Henriksson:2018myn}.} or with slightly-broken higher spin symmetry in the works \textcite{Alday:2015ota} and \textcite{Alday:2016jfr}. Another limit where recent progress has been made is the Regge limit.\footnote{See~\textcite{Li:2017lmh}, \textcite{Costa:2017twz}, \textcite{Meltzer:2017rtf}, \textcite{Alday:2017gde}, and \textcite{Kulaxizi:2017ixa}, which built on  earlier work developing conformal Regge theory in~\textcite{Brower:2006ea}, \textcite{Cornalba:2006xk,Cornalba:2006xm,Cornalba:2007zb}, \textcite{Cornalba:2007fs}, and \textcite{Costa:2012cb}.} Both of these limits can be connected to constraints from Lorentzian causality, where nontrivial bounds and sum rules can be derived,\footnote{See~\textcite{Hartman:2015lfa,Hartman:2016dxc}, \textcite{Hofman:2016awc}, \textcite{Hartman:2016lgu}, and \textcite{Afkhami-Jeddi:2016ntf, Afkhami-Jeddi:2017rmx}.} yielding new arguments for the conformal collider bounds of~\textcite{Hofman:2008ar}, \textcite{Buchel:2009sk}, and \textcite{Chowdhury:2012km}, as well as the more stringent constraints of~\textcite{Camanho:2014apa} in holographic theories.  Finally we should mention the recent development of a powerful Lorentzian inversion formula\footnote{See \textcite{Caron-Huot:2017vep}, \textcite{Simmons-Duffin:2017nub}, \textcite{Cardona:2018nnk}, \textcite{Kravchuk:2018htv}, and \textcite{Cardona:2018dov}.} as well as related work on higher-dimensional crossing kernels,\footnote{See~\textcite{Gadde:2017sjg}, \textcite{Hogervorst:2017sfd}, \textcite{Hogervorst:2017kbj}, \textcite{Karateev:2017jgd}, \textcite{Sleight:2018epi, Sleight:2018ryu}, and \textcite{Liu:2018jhs}.} which are another promising route for further analytical progress.

The conformal bootstrap in two dimensions has a long history, for example the seminal applications to rational CFTs in \textcite{Belavin:1984vu}.\footnote{See also~\textcite{Knizhnik:1984nr}, \textcite{Gepner:1986wi}, and \textcite{Bouwknegt:1992wg}.} The modern numerical bootstrap has been re-applied to 2d CFTs in a number of works.\footnote{See~\textcite{Rattazzi:2008pe}, \textcite{Rychkov:2009ij}, \textcite{Vichi:2011zza}, \textcite{Liendo:2012hy}, \textcite{ElShowk:2012hu}, \textcite{Gliozzi:2013ysa}, \textcite{Gliozzi:2014jsa}, \textcite{El-Showk:2014dwa}, \textcite{Bobev:2015jxa}, \textcite{Lin:2015wcg}, \textcite{Esterlis:2016psv}, \textcite{Lin:2016gcl}, \textcite{Collier:2017shs}, \textcite{Chen:2017yze}, \textcite{Li:2017agi, Li:2017ukc}, \textcite{Behan:2017rca}, and \textcite{Cornagliotto:2017dup}.} A related direction is the modular bootstrap, which sets up consistency conditions arising from modular invariance. A number of recent studies\footnote{See~\textcite{Hellerman:2009bu}, \textcite{Hellerman:2010qd}, \textcite{Friedan:2013cba}, \textcite{Qualls:2013eha}, \textcite{Hartman:2014oaa}, \textcite{Keller:2014xba}, \textcite{Qualls:2014oea, Qualls:2015bta}, \textcite{Chang:2015qfa}, \textcite{Kim:2015oca}, \textcite{Benjamin:2016fhe}, \textcite{Collier:2016cls, Collier:2017shs}, \textcite{Keller:2017iql}, \textcite{Cho:2017fzo}, \textcite{Cardy:2017qhl}, \textcite{Apolo:2017xip}, \textcite{Bae:2017kcl}, \textcite{Dyer:2017rul}, and \textcite{Anous:2018hjh}.} have also looked at these constraints using modern numerical bootstrap techniques. A proper summary of these results and related analytical progress in the context of both the history of the 2d bootstrap and holography would merit its own review; see e.g.~\textcite{Yin-TASI}.

One can also study the conformal bootstrap in more than four dimensions or with extended supersymmetry. One application of the numerical bootstrap has e.g. been to the 6d $(2,0)$ SCFT~\cite{Beem:2015aoa}, interesting in part because of its non-Lagrangian nature and ability to teach us about new dualities. Another application in $d>4$ has been to probe the existence of $O(N)$ vector models in 5d.\footnote{See~\textcite{Nakayama:2014yia}, \textcite{Bae:2014hia}, \textcite{Chester:2014gqa}, and \textcite{Li:2016wdp}.} Progress has also been made placing constraints on a variety of other 5d and 6d SCFTs,\footnote{See~\textcite{Pang:2016xno}, \textcite{Nakayama:2017vdd}, \textcite{Chang:2017cdx}, and \textcite{Chang:2017xmr}.} 4d $\mathcal{N}=2$ and $\mathcal{N}=3$ SCFTs,\footnote{See~\textcite{Beem:2014zpa}, \textcite{Lemos:2015awa}, \textcite{Lemos:2016xke}, and \textcite{Cornagliotto:2017dup,Cornagliotto:2017snu}.} 4d $\mathcal{N}=4$ supersymmetric Yang-Mills theory,\footnote{See~\textcite{Beem:2013qxa}, \textcite{Alday:2013opa, Alday:2014qfa}, \textcite{Beem:2016wfs}, and \textcite{Liendo:2016ymz}.} and 3d $\mathcal{N}=8$ SCFTs~\cite{Chester:2014fya} including ones inspired by M-theory~\cite{Agmon:2017xes}. Related analytical progress at solving sub-sectors of SCFTs with extended supersymmetry was also made in \textcite{Beem:2013sza, Beem:2014kka, Chester:2014mea} and used in a number of follow-up studies.\footnote{See~\textcite{Beem:2014zpa,Beem:2014rza}, \textcite{Lemos:2014lua}, \textcite{Liendo:2015ofa,Liendo:2015cgi}, \textcite{Lemos:2015orc}, \textcite{Nishinaka:2016hbw}, \textcite{Xie:2016evu}, \textcite{Beem:2016cbd}, \textcite{Dedushenko:2016jxl}, \textcite{Song:2016yfd}, \textcite{Creutzig:2017qyf}, \textcite{Fredrickson:2017yka}, \textcite{Cordova:2017mhb}, \textcite{Agmon:2017lga}, \textcite{Beem:2017ooy}, \textcite{Pan:2017zie}, \textcite{Fluder:2017oxm}, and \textcite{Dedushenko:2017avn}.}

Work on the superconformal bootstrap (particularly with extended supersymmetry) is not possible without analytical computations of superconformal blocks, which have been developed in a number of works using both superembedding space methods and by solving the superconformal Casimir equation.\footnote{See~\textcite{Dolan:2001tt,Dolan:2004iy,Dolan:2004mu}, \textcite{Poland:2010wg}, \textcite{Fortin:2011nq}, \textcite{Goldberger:2011yp,Goldberger:2012xb}, \textcite{Khandker:2012pa}, \textcite{Fitzpatrick:2014oza}, \textcite{Khandker:2014mpa}, \textcite{Li:2014gpa}, \textcite{Beem:2014zpa}, \textcite{Liendo:2015ofa}, \textcite{Bobev:2015jxa}, \textcite{Beem:2015aoa}, \textcite{Bissi:2015qoa}, \textcite{Doobary:2015gia}, \textcite{Lemos:2016xke}, \textcite{Ramirez:2016lyk}, \textcite{Li:2016chh}, \textcite{Cornagliotto:2017dup}, \textcite{Bobev:2017jhk}, \textcite{Li:2017ddj}, \textcite{Chang:2017xmr}, \textcite{Li:2018mdl}, and \textcite{Rong:2018okz}.}

Another intriguing line of research is the development of a direct relation between conformal blocks and the wave-functions of integrable Hamiltonians.\footnote{See~\textcite{Isachenkov:2016gim}, \textcite{Schomerus:2016epl}, \textcite{Chen:2016bxc}, \textcite{Schomerus:2017eny}, and \textcite{Isachenkov:2017qgn}.} Finally, while we cannot review here all of the interesting developments in the AdS/CFT correspondence, it is worth mentioning that there has been an abundance of activity in connecting both global blocks\footnote{See~\textcite{Hijano:2015zsa}, \textcite{Nishida:2016vds}, \textcite{Bhatta:2016hpz}, \textcite{Dyer:2017zef}, \textcite{Castro:2017hpx}, \textcite{Sleight:2017fpc}, \textcite{Chen:2017yia}, and \textcite{Tamaoka:2017jce}.} and semi-classical Virasoro blocks\footnote{See~\textcite{Fitzpatrick:2014vua}, \textcite{Hijano:2015qja, Hijano:2015rla}, \textcite{Fitzpatrick:2015zha, Fitzpatrick:2015foa}, \textcite{Alkalaev:2015wia, Alkalaev:2015lca, Alkalaev:2015fbw}, \textcite{Fitzpatrick:2015dlt}, \textcite{Banerjee:2016qca}, \textcite{Besken:2016ooo}, \textcite{Fitzpatrick:2016ive}, \textcite{Alkalaev:2016ptm}, \textcite{Maloney:2016kee}, \textcite{Alkalaev:2016rjl}, \textcite{Hulik:2016ifr}, \textcite{Alkalaev:2016fok}, \textcite{Fitzpatrick:2016mtp}, \textcite{Fitzpatrick:2016mjq}, \textcite{Belavin:2017atm}, \textcite{Kraus:2017ezw}, \textcite{Alkalaev:2017bzx}, \textcite{Chen:2017yze}, and \textcite{Lencses:2017dgf}.} to geodesics in AdS and bulk semi-classical physics.

All of the conformal bootstrap constraints and numerical bounds we have considered in this review have applied to either Euclidean CFT or relativistic Lorentzian CFT. It is also of great interest to learn about nonrelativistic conformal field theories due to their many experimental realizations, e.g. to ultracold atomic gases near the unitary limit. While nonrelativistic conformal symmetries are inherently less constraining, some important theoretical groundwork on correlation functions and the OPE has been laid for systems governed by the nonrelativistic Schr\"{o}dinger algebra,\footnote{See~\textcite{Nishida:2007pj}, \textcite{Golkar:2014mwa}, \textcite{Goldberger:2014hca}, \textcite{Gubler:2015iva}, and \textcite{Pal:2018idc}.} which is a necessary precursor to any bootstrap analysis. 

Another interesting situation that we have not discussed are systems governed by logarithmic CFTs, a class of nonunitary CFTs describing e.g.~models of percolation, self-avoiding random walks, and systems with quenched disorder. While such theories have been considered for a long time in two dimensions (see \textcite{Creutzig:2013hma} for a review), logarithmic CFTs in higher dimensions have received less attention. A general theoretical analysis of correlation functions in such theories was recently developed in \textcite{Hogervorst:2016itc}, building on earlier work~\cite{Ghezelbash:1997cu}. Attempts to apply direct numerical bootstrap to such theories have been mentioned in Sec.~\ref{sec:nonunitary}.\footnote{It should be kept in mind that the conformal blocks in logarithmic theories are in general more complicated than in usual CFTs~\cite[{section 5.2.2}]{Hogervorst:2016itc}.} Let us also mention the study \textcite{Komargodski:2016auf} of the Ising model with quenched disorder which made extensive use of bootstrap data to develop an approach based on conformal perturbation theory.

\section{Outlook}
\label{sec:outlook}

The conformal bootstrap is still in its infancy and there remains much low-hanging fruit to pick along with many important open questions. For instance, can we use the bootstrap to fully classify the space of critical CFTs with a given symmetry, placing universality on a rigorous footing? Can the bootstrap solve the conformal windows of QED${}_3$ and QCD${}_4$? Can it be used as a discovery tool to find new, perhaps non-Lagrangian, CFTs? Is there an analytical understanding of the kinks in numerical bounds or why certain CFTs such as the 3d Ising model live in them? Which CFTs can be found using extremal spectrum or truncation methods? And is there a fruitful way to incorporate developments in the analytical bootstrap with rigorous numerical methods?

For newcomers to the numerical bootstrap who want to quickly get started, after learning CFT basics we recommend becoming familiar with the available software tools,\footnote{Many of the currently available tools are collected at the webpage:~\url{http://bootstrapcollaboration.com/activities/}.} particularly \texttt{SDPB} {which is under active development,} along with one of the efficient methods to compute conformal blocks in the dimension of interest as described in Sec.~\ref{sec:cb}. Then one can start reproducing numerical bounds and thinking about how they can be generalized to say something new about situations of physical interest. For this purpose it is also helpful to get used to restating the physical properties of critical systems using symmetries and the spectrum of scaling dimensions, so questions can be sharply rephrased in the language of the bootstrap. Good luck!

\begin{acknowledgments}

We would like to thank Connor Behan, John Chalker, Shai Chester, Subham Dutta Chowdhury, Luigi Del Debbio, Mykola Dedushenko, Yin-Chen He, Kuo-Wei Huang, Denis Karateev, Zohar Komargodski, Petr Kravchuk, Adam Nahum, Yu Nakayama, Miguel Paulos, Silviu Pufu, Leonardo Rastelli, Subir Sachdev, David Simmons-Duffin, Andreas Stergiou, Tin Sulejmanpasic, Senthil Todadri, Emilio Trevisani, Ettore Vicari, and William Witczak-Krempa for useful discussions and comments. 

DP is supported by NSF grant PHY-1350180 and Simons Foundation grant 488651 (Simons Collaboration on the Nonperturbative Bootstrap). 
SR is supported by the Simons Foundation grant 488655 (Simons Collaboration on the Nonperturbative Bootstrap), and by Mitsubishi Heavy Industries as an ENS-MHI Chair holder.
AV is supported by the Swiss National Science Foundation under grant no. PP00P2-163670 and by the ERC-STG under grant no. 758903.

\end{acknowledgments}

\appendix

\section{Embedding formalism}

\label{sec:embedding}

As we saw in Sec.~\ref{sec:conformaltransformations}, the conformal group is isomorphic to $SO(d+1,1)$.
This suggests that we can avoid complications due to the SCTs acting nonlinearly on $\bR^d$ by embedding this space into $\bR^{d+1,1}$ where the whole conformal group will act linearly. This main idea of the embedding formalism (also called the projective null cone formalism) goes back to \textcite{Dirac:1936fq}. Here we will only review the main logic and a few fundamental results, see \textcite{Costa:2011mg} for details and \textcite{Rychkov:2016iqz} for a pedagogical introduction.

\begin{figure}[h!]
\begin{centering}
\includegraphics[width=0.6\columnwidth]{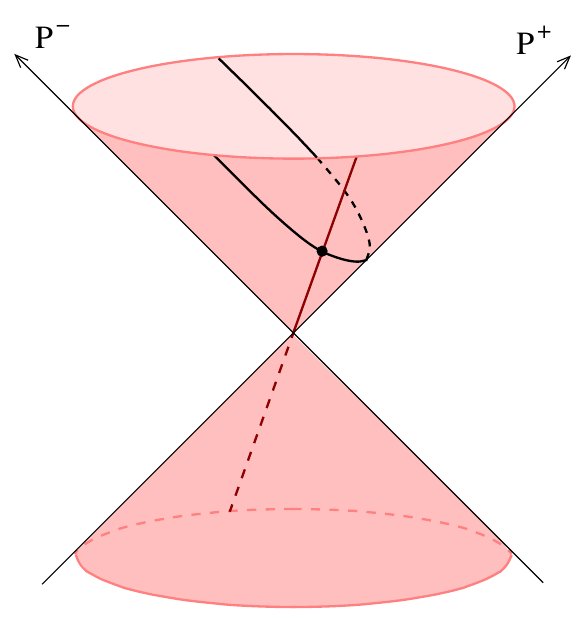}
\caption{\label{fig:embedding}
(Color online) The lightcone in the embedding space; light rays are in one-to-one correspondence with points of $\bR^d$. The Poincar\'e section of the cone is also shown. Figure from \textcite{Costa:2011mg}.
}
\end{centering}
\end{figure}

We denote points in the ``embedding space" $\mathbb{R}^{d+1,1}$ by $P^A$, and use the lightcone coordinates:
\begin{gather}
\label{eq:hyperboloid}
P^A=(P^+,P^-,P^\mu), \quad P^\pm = P^{d+2}\pm P^{d+1},\nn\\
P^2 = P^\mu P_\mu - P^+P^-\,.
\end{gather}
In this space we consider the ``null cone" defined by the equation $P^2=0$. The physical space $\bR^d$ is identified with the ``Poincar\'e section" of the cone given by the equation:
\begin{eqnarray}
P^A=(1,x^2,x^\mu) \,,\quad x^\mu\in\bR^d\,.
\end{eqnarray}
This identification is natural because, as is easy to check, the flat Minkowski metric in $\bR^{d+1,1}$ induces a flat $d$-dimensional metric on the Poincar\'e section (see Fig. \ref{fig:embedding}).

$SO(d+1,1)$ acts naturally on the null rays forming the null cone, and this defines an action on the Poincar\'e section by picking the intersection point. One can verify that this action realizes a conformal transformation of $\bR^d$.

Similarly, operators in $d$ dimensions can be lifted to the embedding space. Focusing on traceless symmetric tensors, $\mathcal O_{\Delta,\ell}^{\mu_1\ldots\mu_\ell}(x)$ is promoted to a traceless symmetric tensor $\widehat{\mathcal{O}}_{\Delta,\ell}^{A_1\ldots A_\ell}(P)$, transforming linearly under $SO(d+1,1)$. The latter operator is required to be homogeneous of degree $-\Delta$:
 \begin{eqnarray}
 \label{eq:homogeneus}
\widehat{\mathcal{O}}_{\Delta,\ell}^{A_1\ldots A_\ell}(\lambda P) = \lambda^{-\Delta} \widehat{\mathcal{O}}_{\Delta,\ell}^{A_1\ldots A_\ell}(P)\,.
\end{eqnarray}
The relation between the two operators is obtained by the projection
\begin{eqnarray}
	\mathcal{O}_{\Delta,\ell}^{\mu_1\ldots\mu_\ell}(x) 
	= \frac{\partial P_{A_1}}{\partial x_{\mu_1}}
	\cdots\frac{\partial P_{A_\ell}}{\partial x_{\mu_\ell}}  
	\widehat{\mathcal{O}}_{\Delta,\ell}^{A_1\ldots A_\ell}(P)\,,
\end{eqnarray}
where $P$ is restricted to the Poincar\'e section so that $\frac{\partial P_{A}}{\partial x_{\mu}} = (0,2x_\mu,\delta_\mu^\alpha)$\,.
This is consistent with the symmetric traceless condition. Notice as well that two embedding space tensors which differ by anything proportional to $P^A$ project to the same physical space tensor, because $P^A \partial P_{A}/\partial x_{\mu} =0$. This is sometimes referred to as ``gauge freedom", and it ensures that both representations have the same number of physical components.

The main advantage of this formalism is that the embedding space operators transform linearly under the conformal group. Thus, the problem of classifying correlation functions in embedding space is reduced to finding covariant tensors of $SO(d+1,1)$. In the index free notation, this is equivalent to constructing invariant polynomials depending on the position vectors $P_A$ and the polarization vectors $Z_A$, with the correct homogeneity properties \reef{eq:homogeneus}. For traceless tensors it is enough to work with null polarization vectors $Z^2=0$. Due to the above mentioned ``gauge freedom", it is also enough to restrict to transverse polarizations: $Z\cdot P = 0$. In these conventions, all correlation functions can be built of the basic building blocks \cite{Costa:2011mg}
\begin{align}
H_{ij} &\equiv \frac{(Z_i \cdot Z_j )( P_i\cdot P_j) - (Z_i \cdot P_j )( Z_j\cdot P_i)}{(P_i \cdot P_j )}\,, \nonumber
\\
V_{i,jk} &\equiv \frac{(Z_i \cdot P_j )(P_i\cdot P_k )- (Z_i \cdot P_k )( P_i\cdot P_j)}{\sqrt{-2 (P_i \cdot P_j)(P_i \cdot P_k)(P_j \cdot P_k)}}\,.
\label{eq:HV} 
\end{align}
In particular we have $(P_{ij} \equiv -2P_i\cdot P_j)$
\begin{align}
\langle  \widehat{O}_{\Delta,\ell}(P_1,Z_1)  \widehat{O}_{\Delta,\ell}&(P_2,Z_2) \rangle = \frac{(H_{12})^\ell }{P_{12}^\Delta}\,,\\
\langle  \widehat{O}_{\Delta_1}(P_1)  \widehat{O}_{\Delta_2}(P_2)  \widehat{O}_{\Delta_3}&(P_3,Z_3)  \rangle = \\
  &\lambda_{123} \frac{(V_{3,12})^\ell }{P_{12}^{h_{123}+\ell} P_{13}^{h_{132}-\ell}\,
P_{23}^{h_{231}-\ell}}\,.\nn
\end{align}
Projecting to $\bR^d$ gives Eqs.~\reef{eq:2points} and \reef{eq:3points}. 

Operators transforming in other $SO(d)$ representations can also be lifted to the embedding space, see e.g.~\textcite{Costa:2014rya} for mixed-symmetry tensors. One can also construct other types of embedding spaces which may be more convenient for dealing with fermions, for supersymmetric CFTs, or in specific $d$.\footnote{{For some examples see} \textcite{Weinberg:2010fx}, \textcite{Goldberger:2011yp}, \textcite{SimmonsDuffin:2012uy}, \textcite{Siegel:2012di}, \textcite{Goldberger:2012xb}, \textcite{Khandker:2012pa}, \textcite{Fitzpatrick:2014oza}, \textcite{Elkhidir:2014woa}, and \textcite{Iliesiu:2015qra}.}

\bibliography{Bootstrap-rmp}

\end{document}